\pdfoutput=1
\documentclass{cernyrep}
\usepackage{color}
\usepackage{amsmath}

\newcommand{\vr}{\bm{r}}
\newcommand{\ton}{T_{\rm on}}

\newcommand{\Enu}{E_{\overline{\nu}}}

\newcommand{\Enubar}{\epsilon_{\overline{\nu}}}

\newcommand{\tbar}{\overline{\tau}_n}

\newcommand{\edeex}{\epsilon_\gamma^{\rm dxn}}
\newcommand{\elab} {\epsilon_\gamma^{\rm LAB}}
\newcommand{\egdr} {\omega_r^{\rm GDR}}

\usepackage{epsfig}
\newcommand{\postscript}[2]{\setlength{\epsfxsize}{#2\hsize}
   \centerline{\epsfbox{#1}}}

\def\es{{\rm erg\ s}^{-1}}

\long\def\symbolfootnote[#1]#2{\begingroup%
\def\thefootnote{\fnsymbol{footnote}}\footnote[#1]{#2}\endgroup}

\def\Offline{\mbox{$\overline{\textrm%
{Off}}$\hspace{.05em}\protect\raisebox{.4ex}%
{$\protect\underline{\textrm{line}}$}}\xspace}  

\def\lsim{\mathrel{\rlap{\lower4pt\hbox{\hskip1pt$\sim$}}
    \raise1pt\hbox{$<$}}}         
\def\gsim{\mathrel{\rlap{\lower4pt\hbox{\hskip1pt$\sim$}}
    \raise1pt\hbox{$>$}}}         

\begin{document}

\title{\LARGE Ultrahigh Energy Cosmic Rays: Facts, Myths, and Legends}

\author{Luis Alfredo Anchordoqui}
\institute{
University of Wisconsin-Milwaukee,  USA}
\maketitle

\begin{abstract}
  This is a written version of a series of lectures aimed at graduate students in astrophysics/particle theory/particle experiment. In the first part, we explain the important progress made in recent years towards understanding the experimental data on cosmic rays with energies $\gtrsim 10^{8}~{\rm GeV}$. We begin with a brief survey of the available data, including a description of the energy spectrum, mass composition, and arrival directions. At this point we also give a short overview of experimental techniques. After that, we introduce the fundamentals of acceleration and propagation in order to discuss the conjectured nearby cosmic ray sources, and emphasize some of the prospects for a new (multi-particle) astronomy. Next, we survey the state of the art regarding the ultrahigh energy cosmic neutrinos which should be produced in association with the observed cosmic rays.
In the second part, we summarize the phenomenology of cosmic ray air showers. We explain the hadronic interaction models used to extrapolate results from collider data to ultrahigh energies, and describe the prospects for insights into forward physics at the Large Hadron Collider (LHC). We also explain the main electromagnetic processes that govern the longitudinal shower evolution. Armed with these two principal shower ingredients and motivation from the underlying physics, we describe  the different methods proposed to distinguish primary species. In the last part, we outline how ultrahigh energy cosmic ray interactions can be used to probe new physics beyond the electroweak scale.\symbolfootnote[4]{Lecture notes from the  6th CERN-Latin-American School
of High-Energy Physics, Natal, Brazil, March - April, 2011. {\tt http://physicschool.web.cern.ch/PhysicSchool/CLASHEP/CLASHEP2011/}.}

\end{abstract}
{\footnotesize
\tableofcontents
} 
\section{Setting the stage}
\label{sec:intro}

For  biological reasons our perception of the Universe is based on the observation of photons, most trivially by staring at the night-sky with our bare eyes. Conventional astronomy covers many orders of magnitude in photon wavelengths, from $10^4$ cm radio-waves to $10^{-14}$ cm gamma rays of GeV energy.  This 60 octave span in photon frequency allows for a dramatic expansion of our observational capacity beyond the approximately one octave perceivable by the human eye.
Photons are not, however, the only messengers of astrophysical processes; we can also observe cosmic rays and neutrinos (and maybe also gravitons in the not-so-distant future). Particle astronomy may be feasible for neutral particles or possibly charged particles with energies high
enough to render  their trajectories  magnetically rigid.  This new astronomy can probe the extreme high energy behavior of distant sources and perhaps provide a window into optically opaque regions of the Universe. In these lectures we will focus attention on the highest energy
particles ever observed. These ultrahigh energy cosmic rays (UHECRs) carry about seven orders of magnitude  more energy than the LHC  beam. We will first discuss the requirements for cosmic ray acceleration and propagation in the intergalactic space. After that we will discuss the properties of the particle cascades initiated when UHECRs interact in the atmosphere.  Finally we outline strategies to search for physics beyond the highly successful but conceptually incomplete Standard Model (SM) of weak, electromagnetic, and strong interactions.

Before proceeding, we pause to present our notation. Unless otherwise stated, we work with natural (particle physicist's) Heaviside-Lorentz (HL) units with 
\begin{equation*}
\hbar=c=k =\varepsilon_0=\mu_0=1\,.
\end{equation*}
The fine structure constant is $\alpha=e^2/(4\pi\varepsilon_0\hbar c)\simeq1/137$.
All SI units can then be expressed in electron Volt (eV), namely
\begin{align*}
1~{\rm m} &\simeq 5.1\times10^6~{\rm eV}^{-1}\,,&
1~{\rm s} & \simeq 1.5\times10^{15}~{\rm eV}^{-1}\,,&
1~{\rm kg} & \simeq 5.6\times10^{35}~{\rm eV}\,, \\
1~{\rm A} & \simeq 1244~{\rm eV}\,, &
1~{\rm G} &\simeq1.95\times10^{-2}{\rm eV}^2\,,&
1~{\rm K} &\simeq8.62\times10^{-5}~{\rm eV}\,.
\end{align*}
Electromagnetic processes in astrophysical environments are often described in terms of Gauss (G) units. For a comparison of formulas used in the literature we note the conversion, $(e^2)_{\rm HL} = 4\pi(e^2)_{\rm G}$, $(B^2)_{\rm HL} = (B^2)_{\rm G}/4\pi$ and $(E^2)_{\rm HL} = (E^2)_{\rm G}/4\pi$. To avoid confusion we will present most of the formulas in terms of `invariant' quantities, {\it i.e.}~$eB$, $eE$ and the fine-structure constant $\alpha$. Distances are generally measured in Mpc $\simeq 3.08 \times 10^{22}~{\rm m}$. Extreme energies are sometimes expressed in EeV $\equiv 10^{18}$~eV. The symbols and units of most common quantities are summarized in Table~\ref{tienda_de_cafe}.

\begin{table}
\caption{Symbols and units of most common quantities. \label{tienda_de_cafe}}
\begin{tabular}{ccc}
\hline
\hline
symbol ~~~~~~~~&~~~~~~~~ quantity ~~~~~~~~&~~~~~~~~ unit \\
\hline
$J$ ~~~~~~~~&~~~~~~~~ diffuse flux ~~~~~~~~&~~~~~~~~ ${\rm GeV}^{-1}~{\rm cm}^{-2}~{\rm s}^{-1}~{\rm sr}^{-1}$ \\
$\Phi$ ~~~~~~~~&~~~~~~~~ diffuse neutrino flux (upper bound) ~~~~~~~~&~~~~~~~~ ${\rm GeV}^{-1}~{\rm cm}^{-2}~{\rm s}^{-1}~{\rm sr}^{-1}$ \\
$\phi$ ~~~~~~~~&~~~~~~~~ point-source neutrino flux (upper bound) ~~~~~~~~&~~~~~~~~ ${\rm GeV}^{-1}~{\rm cm}^{-2}~{\rm s}^{-1} $ \\
$L$ ~~~~~~~~&~~~~~~~~ source luminosity ~~~~~~~~&~~~~~~~~ ${\rm erg}\,{\rm s}^{-1}$ \\
$\mathcal{L}$ ~~~~~~~~&~~~~~~~~ integrated luminosity ~~~~~~~~&~~~~~~~~ ${\rm fb}^{-1} \equiv 10^{-39}~{\rm cm}^2 $ \\
$Q$ ~~~~~~~~&~~~~~~~~  source emissivity ~~~~~~~~&~~~~~~~~ ${\rm GeV}^{-1}\,{\rm s}^{-1}$ \\
$\mathcal{Q} $ ~~~~~~~~&~~~~~~~~  source emissivity per volume ~~~~~~~~&~~~~~~~~ ${\rm GeV}^{-1}\,{\rm cm}^{-3}\,{\rm s}^{-1}$\\
\hline
\hline
\end{tabular}
\end{table}


\section{Frontiers of multi-messenger astronomy}
\subsection{A century of cosmic ray observations}

In 1912 Victor Hess carried out a series of pioneering balloon flights during which he measured the levels of ionizing radiation as high as 5 km above the Earth's surface~\cite{Hess}.  His discovery of increased radiation at high altitude revealed that we are bombarded by ionizing particles from above. These CR particles are now known to consist primarily of protons, helium, carbon, nitrogen and other heavy ions up to iron.  Measurements of energy and isotropy showed conclusively that one obvious source, the Sun, is not the main source. Only below 100~MeV kinetic energy or so, where the solar wind shields protons coming from outside the solar system, does the Sun dominate the observed proton flux. Spacecraft missions far out into the solar system, well away from the confusing effects of the Earth's atmosphere and magnetosphere, confirm that the abundances around 1~GeV are strikingly similar to those found in the ordinary material of the solar system. Exceptions are the overabundance of elements like lithium, beryllium, and boron, originating from the spallation of heavier nuclei in the interstellar medium.

Above $\sim 10^{5}~{\rm GeV},$ the rate of CR primaries is less than one particle per square meter per year and direct observation in the upper layers of the atmosphere (balloon or aircraft), or even above (spacecraft) is inefficient. Only ground-based experiments with large apertures and long exposure times can hope to acquire a significant number of events. Such experiments exploit the atmosphere as a giant calorimeter.  The incident cosmic radiation interacts with the atomic 
nuclei of air molecules and produces air showers which spread out over large areas. 
Already in 1938, Pierre Auger concluded from the size of the air showers that the spectrum extends up to and perhaps beyond $10^{6}$~GeV~\cite{Auger:1938,Auger:1939}. In recent years, substantial progress has been made in measuring the extraordinarily low flux at the high end of the spectrum.

There are two primary detection methods that have been successfully used in high exposure experiments. In the following paragraph we provide a terse overview of these approaches.  For an authoritative review on experimental techniques and historical perspective see~\cite{Nagano:ve}. For a more recent comprehensive update including future prospects see~\cite{LetessierSelvon:2011dy}.

\begin{figure}
\postscript{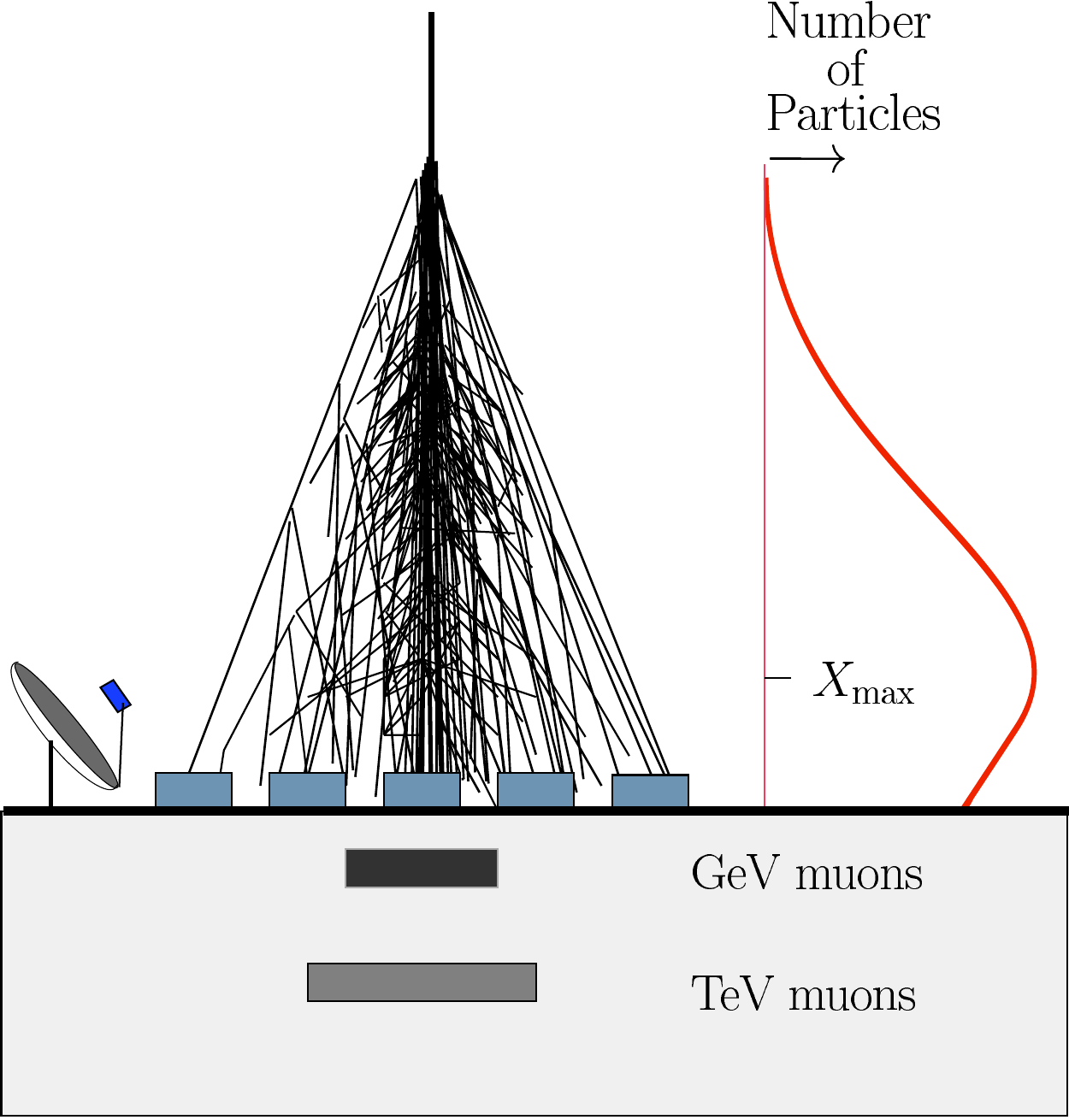}{0.70}
\caption{Particles interacting near the top of the atmosphere initiate 
an electromagnetic and hadronic cascade. Its profile is shown on the right. 
The different detection methods are illustrated. Mirrors collect the 
\v{C}erenkov and nitrogen fluorescent light, arrays of detectors sample the 
shower reaching the ground, and underground detectors identify the muon 
component of the shower. From Ref.~\cite{Anchordoqui:2002hs}.}
\label{detM}
\end{figure}

The size of an extensive air shower (EAS) at sea-level depends on the primary energy and arrival direction. For showers of UHECRs, the cascade is typically several hundreds of meters in diameter and contains millions of secondary particles. Secondary electrons and muons produced in the decay of pions may be observed in scintillation counters or alternatively by the Cherenkov light emitted in water tanks. The separation of these detectors may range from 10~m to 1~km depending on the CR energy and the optimal cost-efficiency of the detection array. The shower core and hence arrival direction can be estimated by the relative arrival time and density of particles in the grid of detectors. The lateral particle density of the showers can be used to calibrate the primary energy.  Another well-established method of detection involves measurement of the shower longitudinal development (number of particles versus atmospheric depth, shown schematically in Fig.~\ref{detM}) by sensing the fluorescence light produced via interactions of the charged particles in the atmosphere.  The emitted light is typically in the 300 - 400~nm ultraviolet range to which the atmosphere is quite transparent. Under favorable atmospheric conditions, EASs can be detected at distances as large as 20~km, about 2 attenuation lengths in a standard desert atmosphere at ground level. However, observations can only be done on clear Moon{\it less} nights, resulting in an average 10\% duty cycle.

In these lectures we concentrate on  results from the High Resolution Fly's Eye experiment and the Pierre Auger Observatory to which we will often refer as HiRes and Auger.  HiRes~\cite{AbuZayyad:2000uu} was an up-scaled version of the pioneer Fly's Eye experiment~\cite{Baltrusaitis:1985mx}. The facility was comprised of two air fluorescence detector sites separated by 12.6~km. It was located at the US Army Dugway Proving Ground in the state of Utah at 40.00$^\circ$~N, 113$^\circ$~W, and atmospheric depth of 870~g/cm$^2$. Even though the two detectors (HiRes-I and HiRes-II) could trigger and reconstruct events independently, the observatory was designed to measure the fluorescence light stereoscopically. The stereo mode allows accurate reconstruction of the shower geometry (with precision of 0.4$^\circ$) and provides valuable information and cross checks about the atmospheric conditions at the time the showers  were detected. HiRes-I and HiRes-II collected data until April 2006 for an accumulated exposure in stereoscopic mode of $3,460$~hours . The monocular mode had better statistical power and covered a much wider energy range.

The Pierre Auger Observatory~\cite{Abraham:2004dt} is 
designed to measure the properties of  EASs produced 
by CRs at the highest energies, above about $10^{9}$~GeV.
It features a large aperture to gather a significant sample
of these rare events, as well as complementary detection 
techniques to mitigate some of the systematic
uncertainties associated with deducing properties of CRs from
air shower observables.

The observatory is located on the vast plain of {\em Pampa Amarilla}, near the town of Malarg\"ue in Mendoza Province, Argentina ($35.1^\circ - 35.5^\circ$~S, $69.6^\circ$~W, and atmospheric depth of 875~g/cm$^2$). The experiment began collecting data in 2004, with construction of the baseline design completed by 2008.  As of October 2010, Auger had collected in excess of $20,000~{\rm km}^2~\rm{sr}~\rm{yr}$ in exposure, significantly more exposure than other cosmic ray observatories combined.  Two types of instruments are employed.  Particle detectors on the ground sample air shower fronts as they arrive at the Earth's surface, while fluorescence telescopes measure the light produced as air shower particles excite atmospheric nitrogen.

The surface array~\cite{Abraham:2010zz} comprises $1,600$ 
surface detector (SD) stations, each consisting of a tank 
filled with 12 tons of water and instrumented with 3 
photomultiplier tubes which detect the Cherenkov light 
produced as particles traverse the water.
The signals from the photomultipliers are read out with flash analog
to digital converters at 40~MHz and timestamped by a GPS unit, allowing
for detailed study of the arrival time profile of shower particles.
The tanks are arranged on a triangular grid with a 1.5~km spacing, 
covering about $3,000~{\rm km}^2$. 
The surface array operates with close to a 100\% duty cycle,
and the acceptance for events above $10^{9.5}$~GeV
is nearly 100\%.

The fluorescence detector (FD) system~\cite{Abraham:2009pm} 
consists of 4 buildings, each housing 6
telescopes which overlook the surface array.
Each telescope employs an $11~{\rm m}^2$ segmented
mirror to focus the fluorescence light entering through a 2.2~m diaphragm
onto a camera which pixelizes the image using 440 photomultiplier tubes.
The photomultiplier signals are digitized at 10~MHz, providing
a time profile of the shower as it develops in the atmosphere.
The FD can be operated only when the sky is dark and clear,
and has a duty cycle of  10-15\%.  In contrast to the SD acceptance,
the acceptance of FD events depends strongly on 
energy~\cite{:2010zzl}, extending down to about $10^{9}$~GeV.

The two detector systems provide complementary information, 
as the SD measures the lateral distribution
and time structure of shower particles arriving at the ground, and 
the FD measures the longitudinal development of the shower in
the atmosphere.
A subset of showers is observed simultaneously by the SD and FD.
These ``hybrid'' events are very precisely measured and
provide an invaluable calibration tool.  In particular,
the FD allows for a roughly colorimetric measurement of the
shower energy since the amount of fluorescence light generated 
is proportional to the energy deposited along the shower path; 
in contrast, extracting the shower energy 
via analysis of particle densities at the ground
relies on predictions from hadronic interaction models 
describing physics at energies beyond those accessible
to current experiments.
Hybrid events can therefore be exploited to set
a model-independent energy scale for the SD array, which in turn
has access to a greater data sample than the FD due to
the greater live time.

The Extreme Universe Space Observatory  (EUSO) is currently being considered by the Japan Aerospace Exploration Agency for a possible payload on the Japanese Experimental Module (JEM) of the International Space Station~\cite{Takahashi:2009ng}. The mission is currently  scheduled for 2017~\cite{Tom}. The launch will be provided by the H-II Transfer Vehicle {\em Kounotori}. Looking straight down, the UV telescope of  JEM-EUSO will have 1~km$^2$ resolution on the Earth's surface, which provides an angular resolution of $2.5^\circ$. The surface area expected to be covered on Earth is about $160,000~{\rm km}^2$, with a duty cicle of order of 10\%.  The detector will also operate
in a tilted mode that will increase the viewing area by a factor of up to 5, but decreasing its resolution. The telescope will be equiped with devices that measure the transparency of the atmosphere and the existence of clouds. 

In the remainder of the section, we describe recent results from
HiRes and Auger, including the measurement of
the cosmic ray energy spectrum, composition, and searches
for anisotropy in the cosmic ray arrival directions.

\subsubsection{Energy spectrum}

\begin{figure}[t]
\begin{center}
\includegraphics[width=0.8\textwidth]{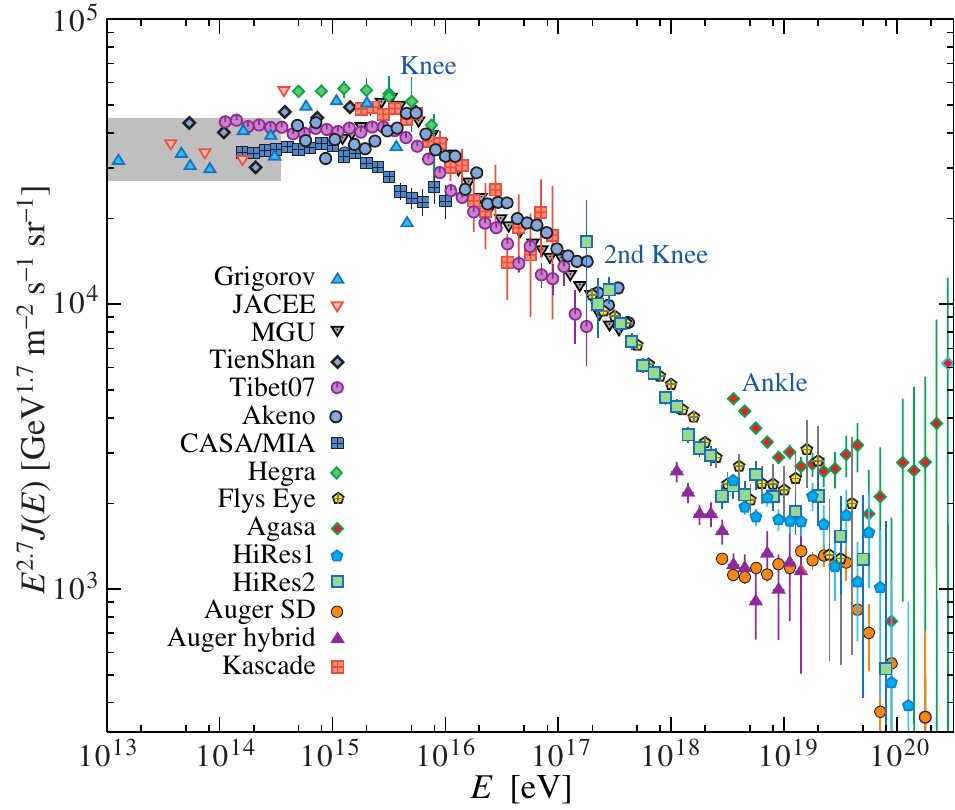}
\end{center}
\caption{All-particle spectrum of cosmic rays. From Ref.~\cite{Amsler:2008zzb}.}
\label{CRspectrum}
\end{figure}

The CR spectrum spans over roughly 11 decades of energy. Continuously running monitoring using sophisticated equipment on high altitude balloons and ingenious installations on the Earth's surface encompass a plummeting flux that goes down from $10^4$ m$^{-2}$ s$^{-1}$ at $\sim 1$~GeV to $10^{-2}$ km$^{-2}$ yr$^{-1}$ at $\sim 10^{11}$~GeV.  Its shape is remarkably featureless, with little deviation from a constant power law ($J \propto E^{-\gamma}$, with $\gamma \approx 3$) across this large energy range.  The small changes in the power index, $\gamma$, are conveniently visualized taking the product of the flux with some power of the energy. In this case the spectrum reveals a leg-like structure as it is sketched in Fig.~\ref{CRspectrum}. The anatomy of this {\it cosmic leg} -- its changes in slope, mass composition or arrival direction -- reflects the various aspects of CR propagation, production and source distribution.

A steepening of the spectrum ($\gamma\simeq2.7\to3.1$) at an energy of about $10^{6.5}$~GeV is known as the cosmic ray {\it knee}. Composition measurements in cosmic ray observatories indicate  that this feature of the spectrum is composed of the subsequent fall-off of Galactic nuclear components with maximal energy $E/Z$~\cite{Aglietta:2004np,Antoni:2005wq,Apel:2008cd}.  This scaling with atomic number $Z$ is expected for particle acceleration in a magnetically confining environment, which is only effective as long as the particle's Larmor radius is smaller than the size of the accelerator. If this interpretation holds, the Galactic contribution in CRs can not extend much further than $10^8$~GeV, assuming iron ($Z=26$) as the heaviest component. However, the observational data at these energies is inconclusive and the end-point of Galactic CRs has not been pinned down so far. For a survey of spectral features at lower energies see~\cite{Bluemer:2009zf}.

The onset of an extragalactic contribution could be signaled by the so-called {\it second knee} -- a further steepening ($\gamma\simeq3.1\to3.2$) of the spectrum at about $10^{8.7}$~GeV. Note that extragalactic CRs are subject to collisions with the interstellar medium during their propagation over cosmological distances. Depending on the initial chemical composition, these particle-specific interactions will be imprinted in the spectrum observed on Earth. It has been argued~\cite{Hillas:1967,Berezinsky:2002nc}  that an extragalactic proton population with a simple power-law injection spectrum may reproduce the spectrum above the second knee. In these models the flattening ($\gamma\simeq3.2\to2.7$) in the spectrum at around $10^{9.5}$~GeV -- the so-called {\it ankle} -- can be identified as a ``dip'' from $e^+e^-$ pair production together with a ``pile-up'' of protons from pion photoproduction. However, this feature relies on a proton dominance in extragalactic cosmic rays since heavier nuclei like oxygen or iron have different energy loss properties in the cosmic microwave background (CMB) and mixed compositions,  in general will not reproduce the spectral features~\cite{Aloisio:2006wv}.  A cross-over at higher energies  is also feasible: above the ankle the Larmor radius of a proton in the galactic magnetic field exceeds the size of the Galaxy and it is generally assumed that an extragalactic component dominates the spectrum at these energies~\cite{Cocconi:et}.  Moreover, the galactic-extragalactic transition ought to be accompanied by the appearence of spectral features, {\em e.g.} two power-law contributions would naturally produce a {\em flattening}  in the spectrum if the harder component dominates at lower energies. 
Hence, the ankle seems to be the natural candidate for this transition.

Shortly after the CMB was discovered~\cite{Penzias:1965wn}, Greisen~\cite{Greisen:1966jv}, Zatsepin, and Kuzmin~\cite{Zatsepin:1966jv} (GZK)  pointed out that the relic photons make the universe opaque to CRs of sufficiently high energy. This occurs, for example, for protons with energies beyond the photopion  production threshold, 
\begin{equation}
E_{p\gamma_{\rm CMB}}^{\rm th} = \frac{m_\pi \, (m_p + m_\pi/2)}{\omega_{\rm CMB}} 
\approx 6.8 \times 10^{10}\,
\left(\frac{\omega_{\rm CMB}}{10^{-3}~{\rm eV}}\right)^{-1} \,\,{\rm GeV}\,,
\label{1}
\end{equation}
where $m_p$ ($m_\pi$) denotes the proton (pion) mass and $\omega_{\rm CMB} \sim 10^{-3}$~eV is a typical CMB photon energy. After pion production, the proton (or perhaps, instead, a neutron) emerges with at least 50\% of the incoming energy. This implies that the nucleon energy changes by an $e$-folding after a propagation distance~$\lesssim (\sigma_{p\gamma}\,n_\gamma\,y_\pi)^{-1} \sim 15$~Mpc. Here, $n_\gamma \approx 410$~cm$^{-3}$ is the number density of the CMB photons, $\sigma_{p \gamma} > 0.1$~mb is the photopion production cross section, and $y_\pi$ is the average energy fraction (in the laboratory system) lost by a nucleon per interaction. For heavy nuclei, the giant dipole resonance (GDR) can be excited at similar total energies and hence, for example, iron nuclei do not survive fragmentation over comparable distances. Additionally, the survival probability for extremely high energy ($\approx 10^{11}$~GeV) $\gamma$-rays (propagating on magnetic fields~$\gg 10^{-11}$~G) to a distance $d$, \mbox{$p(>d) \approx \exp[-d/6.6~{\rm Mpc}]$}, becomes less than $10^{-4}$ after traversing a distance of 50~Mpc. All in all, as our horizon shrinks dramatically for energies $\gtrsim 10^{11}$~GeV, one would expect a sudden suppression in the energy spectrum if the CR sources follow a cosmological distribution.

\begin{figure}[t]
\begin{center}
\includegraphics[width=0.8\textwidth]{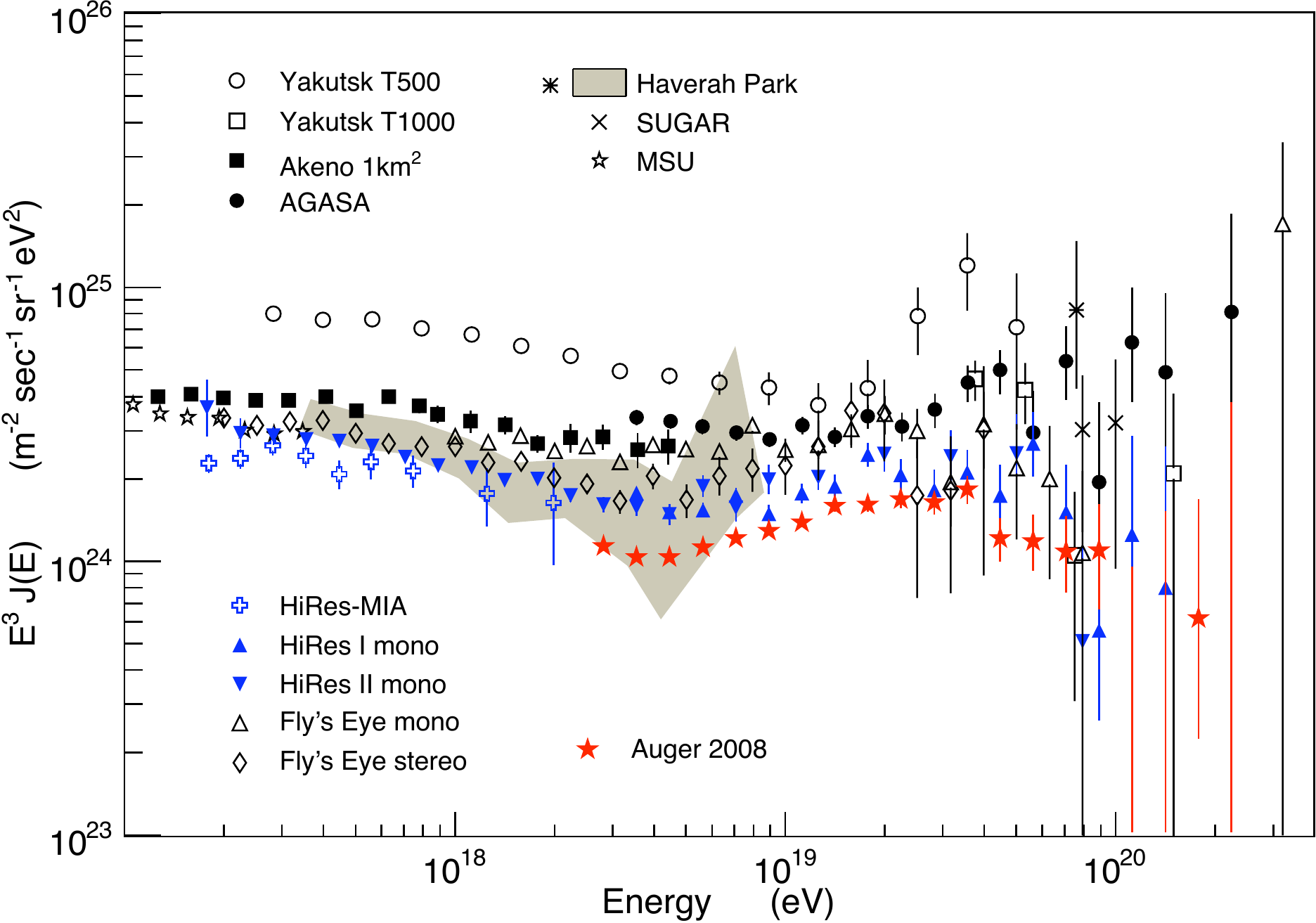}
\end{center}
\caption{Comparison of flux measurements scaled by $E^3$. Only statistical errors are shown.  Shown are the data of AGASA, Akeno, Auger, Fly's Eye, Haverah Park, HiRes-MIA, HiRes Fly's Eye, MSU, SUGAR, and Yakutsk. Yakutsk T500 (trigger 500) refers to the smaller sub-array of the experiment with 500m detector spacing and T1000 (trigger 1000) to the array with 1000m detector distance. In case several analyses of the same data set are available, only the most recent results are included in the plot.  The shaded area, depicting the results of the analysis of the Haverah Park data, accounts for some systematic uncertainties by assuming extreme elemental compositions, either fully iron or proton dominated.  The highest energy point (Fly's Eye monocular observation) corresponds to the highest energy event. For sake of clarity, upper limits are not shown. The data of the MSU array are included to show the
connection of the high-energy measurements to lower energy data covering
the knee of the cosmic-ray spectrum.  From Ref.~\cite{Bluemer:2009zf}.}
\label{fig:flux3-all}
\end{figure}

At the beginning of summer 2002, in a pioneering paper Bahcall and Waxman~\cite{Bahcall:2002wi} noted that the energy spectra of CRs reported by the AGASA, the Fly's Eye, the Haverah Park, the HiRes, and the Yakutsk collaborations (see Fig.~\ref{fig:flux3-all}) are consistent with the expected GZK suppression at $\sim 3.5\sigma$ level according to the Fly's Eye normalization, increasing up to $\sim 8\sigma$ if the selected normalization is that of Yakutsk. Five years later, the HiRes Collaboration reported a suppression of the CR flux above $E = [5.6 \pm 0.5 ({\rm stat})  \pm 0.9  ({\rm syst})] \times 10^{10}~{\rm GeV}$, with 5.3$\sigma$ significance~\cite{Abbasi:2007sv}. The spectral index of the flux  steepens from $2.81\pm 0.03$ to $5.1 \pm 0.7$. The discovery of the GZK  suppression has been confirmed by the Pierre Auger Collaboration, measuring  $\gamma = 2.69 \pm 0.2 ({\rm stat}) \pm 0.06 ({\rm syst})$  and  $\gamma = 4.2 \pm 0.4 ({\rm stat}) \pm 0.06 ({\rm syst})$ below and above $E = 4.0 \times 10^{10}~{\rm GeV}$, respectively (the systematic uncertainty in the energy determination is estimated as 22\%)~\cite{Abraham:2008ru}.

\begin{figure}[htbp]
\centering
\includegraphics[width=0.99\textwidth]{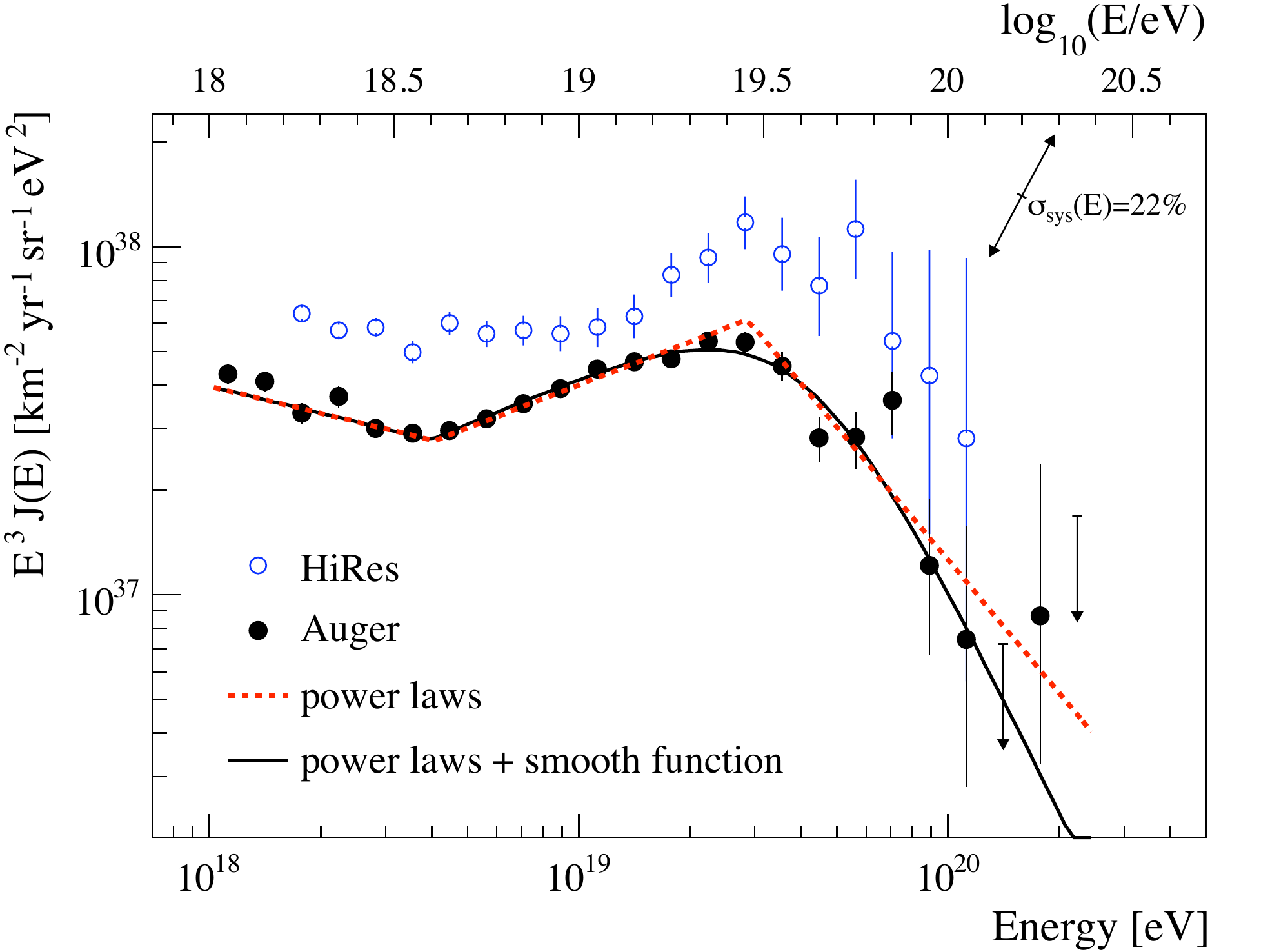}
\caption{Combined spectrum from Auger (hybrid and SD events) and the stereo spectrum HiReS.  The Auger systematic uncertainty of  the flux scaled by $E^3$, due to the uncertainty of the energy scale of 22\%, is  indicated by arrows. The results of the two experiments are consistent within systematic uncertainties. From Ref.~\cite{Abraham:2010mj}.}
\label{fig:cal}
\end{figure}

Last year, an updated Auger measurement of the energy spectrum
was published~\cite{Abraham:2010mj}, 
corresponding to a surface array exposure
of $12,790~{\rm km}^2~{\rm sr~yr}$.  This measurement, combining
both hybrid and SD-only events, is shown in Fig.~\ref{fig:cal}.
The ankle feature and flux suppression are clearly visible.
A broken power law fit to the spectrum shows that the
break corresponding to the ankle
is located at $\log_{10}(E/{\rm eV}) = 18.61 \pm 0.01$
with $\gamma = 3.26 \pm 0.04$ before the break and
$\gamma = 2.59 \pm 0.02$ after it.  The break corresponding
to the suppression is located at 
$\log_{10}(E/{\rm eV}) = 19.46 \pm 0.03$.  Compared to a power
law extrapolation, the significance of the suppression is
greater than $20\sigma$.

\subsubsection{Primary species}

Unfortunately, because of the highly indirect method of measurement, extracting precise information from EASs has proved to be exceedingly difficult. The most fundamental problem is that the first generations of particles in the cascade are subject to large inherent fluctuations and consequently this limits the event-by-event energy resolution of the experiments. In addition, the center-of-mass (c.m.) energy of the first few cascade steps is well beyond any reached in collider experiments. Therefore, one needs to rely on hadronic interaction models that attempt to extrapolate, using different mixtures of theory and phenomenology, our understanding of particle physics.

The longitudinal development has a well defined maximum, usually referred to as $X_{\rm max}$, which increases with primary energy as more cascade generations are required to degrade the secondary particle energies. Evaluating $X_{\rm max}$ is a fundamental part of many of the composition studies done by detecting air showers. For showers of a given total energy $E$, heavier nuclei have smaller $X_{\rm max}$ because the shower is already subdivided into $A$ nucleons when it enters the atmosphere. The average depth of maximum $\langle X_{\rm max} \rangle$ scales approximately as $\ln(E/A)$~\cite{Linsley:gh}. Therefore, since $\langle X_{\rm max} \rangle$ can be determined directly from the longitudinal shower profiles measured with a fluorescence detector, the composition can be extracted after estimating $E$ from the total fluorescence yield. Indeed, the parameter often measured is $D_{10}$, the rate of change of $\langle X_{\rm max} \rangle$ per {\it decade} of energy.

Photons penetrate quite deeply into the atmosphere due to decreased secondary multiplicities and suppression of cross sections by the Landau-Pomeranchuk-Migdal (LPM)
effect~\cite{Landau:um,Migdal:1956tc}. Indeed, it is rather easier to distinguish photons from protons and iron than protons and iron are to distinguish from one another. For example, at $10^{10}~{\rm GeV}$, the $\langle X_{\rm max} \rangle$ for a photon is about 1000~g/cm$^2$, while for protons and iron the numbers are 800~g/cm$^2$ and 700~g/cm$^2$, respectively. 

\begin{figure}[t]
\begin{center}
\includegraphics[width=0.85\textwidth]{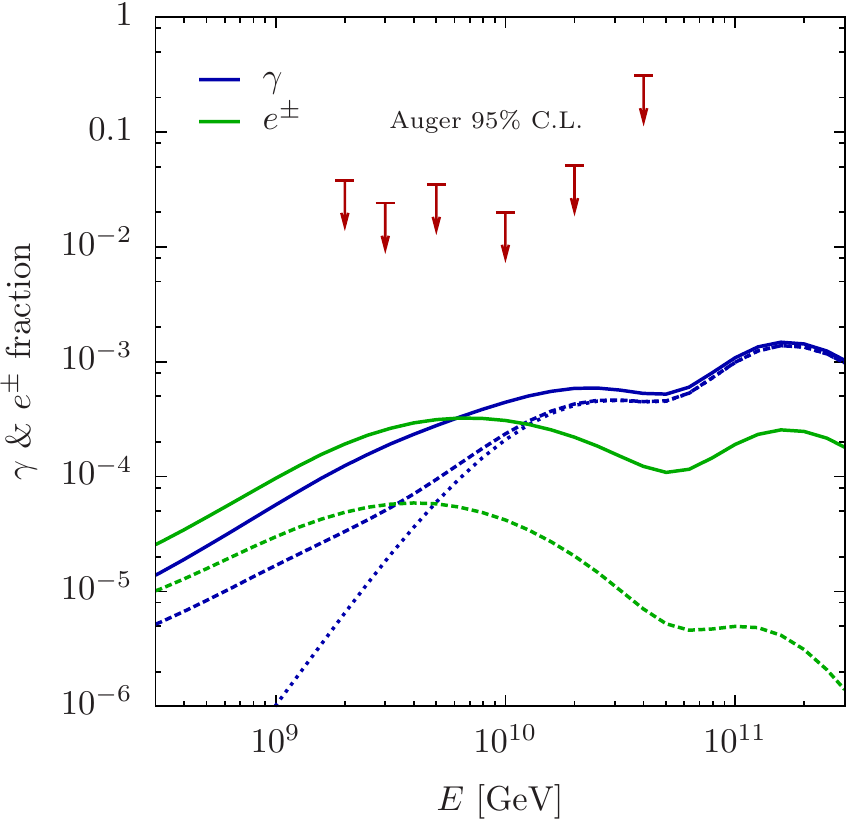}
\end{center}
\caption[]{Upper limits on the photon fraction reported by the Pierre Auger Collaboration. A prediction for the cosmogenic photon flux is also shown for comparison (details on the 
calculation are given Sec.~\ref{Boltzmann}) The color and line coding is the same as the one in Fig.~\ref{fig:Bsamples}. This figure is courtesy of Markus Ahlers.}
\label{fig:pfraction}
\end{figure}

Searches for photon primaries have been conducted using both the surface and fluorescence istruments of Auger.  While analysis of the fluorescence data exploits the direct view of shower development, analysis of data from the surface detector relies on measurement of quantities which are indirectly related to the $X_{\rm max}$, such as the signal risetime at 1000~m from the shower core and the curvature of the shower front. Presently, the 95\% CL upper limits on the fraction of CR photons above $2,\ 3,\ 5,\ 10,\ 20,$ and $40 \times 10^{9}~{\rm GeV}$ are $3.8\%, 2.4\%, 3.5\%, 2.0\%, \ 5.1\%,$ and $31\%,$ respectively. Further details on the analysis procedures can be found in~\cite{Abraham:2006ar,Aglietta:2007yx,Abraham:2009qb}. 
In Fig.~\ref{fig:pfraction} these upper limits are compared with predictions of the cosmogenic photon flux.

\begin{figure}[t]
\begin{center}
\includegraphics[width=0.85\textwidth]{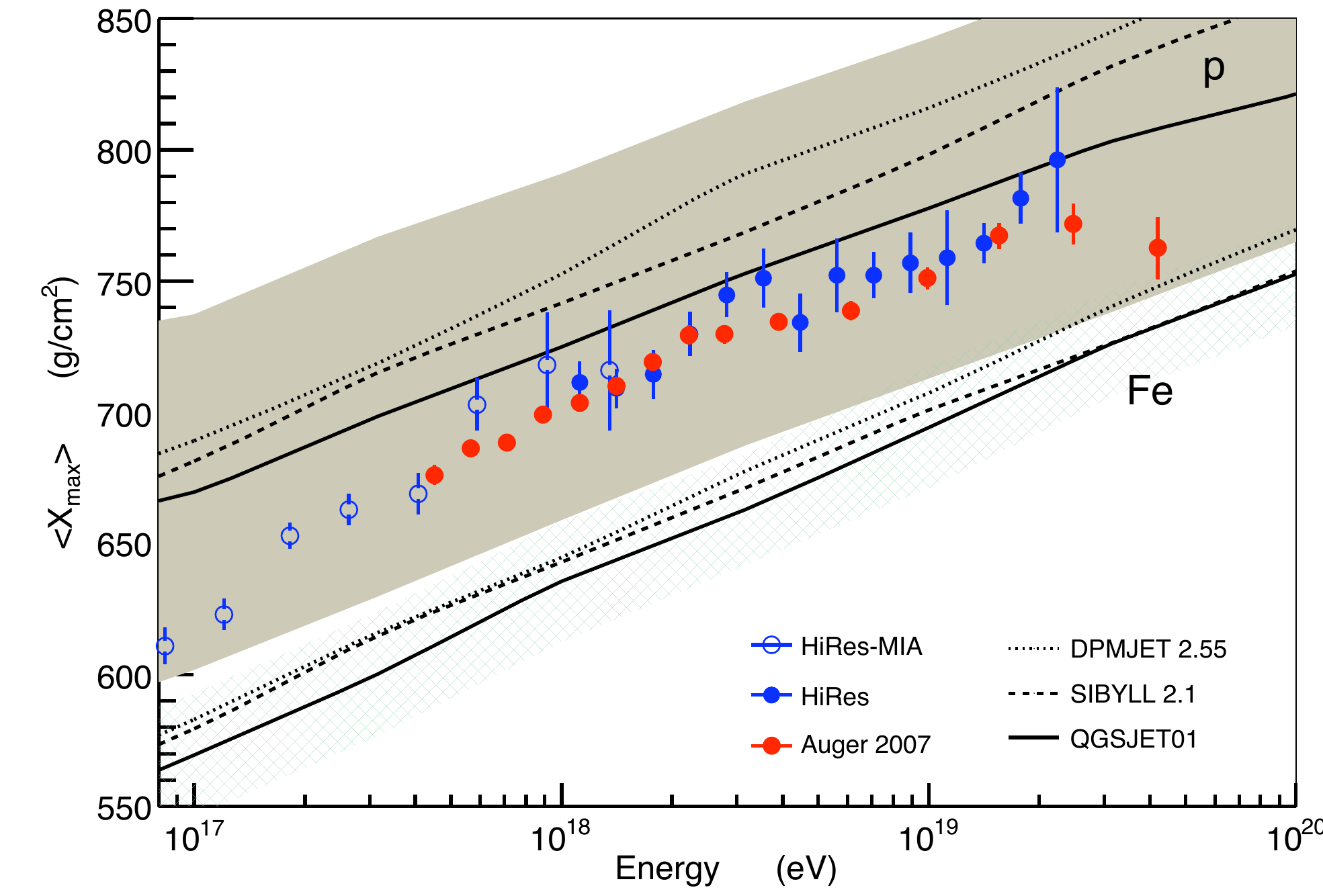}
\end{center}
\caption{The compilation of
  fluorescence-based measurements of the mean $X_{\rm max}$ of various experiments is compared with predictions from EAS simulations using three event generators of hadronic interactions ({\sc sibyll}, {\sc qgsjet}, and {\sc dpmjet}). The {\sc qgsjet} predictions on the
  shower-to-shower fluctuations of the depth of maximum are indicated
  by the shaded (cross-hatched) area for proton (iron) primaries. From Ref.~\cite{Bluemer:2009zf}.}
\label{fig:X} 
\end{figure}

In Fig.~\ref{fig:X} we show the variation of $\langle X_{\rm max} \rangle$ with primary energy 
as measured by several experiments. 
Interpreting the results of these measurements relies on comparison to the predictions of high energy hadronic interaction models. As one can see in Fig.~\ref{fig:X}, there is considerable variation in predictions among the different interaction models.  For $1.6 \times 10^{9}~{\rm GeV} < E < 6.3 \times 10^{10}~{\rm GeV}$, the HiRes data are consistent with a constant elongation rate $ d\langle X_{\rm max}\rangle/d(\log(E)) = 47.9 \pm 6.0({\rm stat}) \pm 3.2 ({\rm sys})~{\rm g}/{\rm cm}^2/{\rm decade}$~\cite{Abbasi:2009nf}.  The inference from the HiRes data is therefore a change in cosmic ray composition, from
heavy nuclei to protons, at $E \sim 
10^{8.6}$~GeV~\cite{Bergman:2004bk}. This is an order of magnitude lower
in energy than the previous crossover deduced from the Fly's Eye
data~\cite{Bird:1993yi}.  On the other hand, Auger measurements, interpreted with current hadronic interaction models, seem to favor a mixed (protons + nuclei) composition at energies above $10^{8.6}~{\rm GeV}$~\cite{Abraham:2010yv}. However, uncertainties in the extrapolation of the proton-air interaction -- cross section~\cite{Ulrich:2009yq}  and elasticity and multiplicity of secondaries~\cite{Anchordoqui:1998nq} -- from accelerator measurements to the high energies characteristic for air showers are large enough to undermine any definite conclusion on the chemical composition. 

\subsubsection{Distribution of arrival directions}
\label{DoAD}

There exists ``lore'' that convinces us that the highest energy CRs observed should exhibit trajectories which are {\em relatively} unperturbed by galactic and intergalactic magnetic fields. Hence, it is natural to wonder whether anisotropy begins to emerge at these high energies.  Furthermore, if the observed flux suppression is the GZK effect, there is necessarily some distance, {\cal O}(100~Mpc), beyond which cosmic rays with energies near $10^{11}$~GeV will not be seen.  Since the matter density within about 100~Mpc is not isotropic, this compounds the potential for anisotropy to emerge in the UHECR sample. On the one hand, if the distribution of arrival directions exhibits a large-scale anisotropy, this could indicate whether or not certain classes of sources are associated with large-scale structures (such as the Galactic plane  or the Galactic halo).  On the other hand, if cosmic rays cluster within a small angular region or show directional alignment with powerful compact objects, one might be able to associate them with isolated sources in the sky.

CR air shower detectors which experience stable operation over a
period of a year or more can have a uniform exposure in right ascension,
$\alpha$. A traditional technique to
search for large-scale anisotropies is then to fit the right ascension
distribution of events to a sine
wave with period $2\pi/m$ ($m^{\rm th}$ harmonic) to determine the
components ($x, y$) of the Rayleigh
vector~\cite{Linsley:1975kp}
\begin{equation}
x = \frac{2}{{\cal N}} \sum_{i=1}^{N}  \, w_i \, \cos(m\, \alpha_i) \,, \,\,\,\,\,y =
\frac{2}{{\cal N}} \sum_{i=1}^{N}  \, w_i\,\,
\, \sin( m\, \alpha_i)\,,
\label{eqn:fh}
\end{equation}
where the sum runs over the number of $N$ events in the considered energy range,  
 ${\cal N} = \sum_{i=1}^{N}  w_i$ is  
the normalization factor, and the weights, $w_i = \omega^{-1}(\delta_i)$, are the reciprocal of 
the relative exposure, $\omega$, given as a function of the declination, $\delta_i$~\cite{Sommers:2000us}. The $m^{\rm th}$ harmonic amplitude of $N$ measurements of $\alpha_i$ is given by the Rayleigh vector length ${\cal R}~=~(x^2~+~y^2)^{1/2}$, and the phase is $\varphi = {\rm arctan} (y/x)$. The expected length of such a vector for values randomly sampled from a uniform phase distribution is ${\cal R}_0~=~2/\sqrt{{\cal N}}$.  The chance probability of obtaining an amplitude with length larger than that measured is $p(\geq~{\cal R})~=~e^{-k_0},$ where $k_0~=~{\cal R}^2/{\cal R}_0^2.$ To give a specific example, a vector of length $k_0~\geq~6.6$ would be required to claim an observation whose probability of arising from random fluctuation was 0.0013 (a ``$3\sigma$'' result)~\cite{Sokolsky:rz}. For a given CL, upper limits on the amplitude can 
be derived using a  distribution drawn from a population characterized by an anisotropy of 
unknown amplitude and phase 
\begin{equation}
\label{eqn:ul}
{\rm CL} = \sqrt{\frac{2}{\pi}}\frac{1}{I_0({\cal R}^2/4\sigma^2)}\int_0^{{\cal R}_{UL}} \frac{ds}{\sigma}\,I_0\bigg(\frac{{\cal R}s}{\sigma^2}\bigg)\,\exp{\bigg(-\frac{s^2+{\cal R}^2/2}{2\sigma^2}\bigg)},
\end{equation}
where $I_0$ is the modified Bessel function of the first kind with order zero and $\sigma=\sqrt{2/\mathcal{N}}$~\cite{Linsley:1975kp}.

The first harmonic amplitude of the CR right ascension distribution can be  directly related to the amplitude $\alpha_d$ of a dipolar distribution of the form 
$J(\alpha,\delta) = J_0 (1+ \alpha_d \ \hat d \cdot \hat u)$, where $\hat u$ and $\hat d$   respectively  denote the 
unit vector in the direction of an arrival direction and in the direction of the dipole. Setting $m =1$, we can rewrite $x$, $y$ and $\mathcal{N}$ as:
\begin{eqnarray}
x&=& \frac{2}{\mathcal{N}} \int_{\delta_{min}}^{\delta_{max}} d\delta \int_0^{2\pi}
d\alpha \cos \delta \ J(\alpha,\delta) \ \omega(\delta) \cos \alpha, \nonumber \\  
y&=& \frac{2}{\mathcal{N}} \int_{\delta_{min}}^{\delta_{max}} d\delta \int_0^{2\pi}
d\alpha \cos \delta \ J(\alpha,\delta) \ \omega(\delta) \sin \alpha, \label{corolo} \\ 
\mathcal{N}&=& \int_{\delta_{min}}^{\delta_{max}} d\delta \int_0^{2\pi}
d\alpha \cos \delta \ J(\alpha,\delta) \ \omega(\delta) \, .\nonumber
\end{eqnarray}
In (\ref{corolo}) we have neglected  the small dependence on right 
ascension in the exposure. Next, we write the angular dependence in 
$J(\alpha,\delta)$ as $ \hat d \cdot \hat u = \cos \delta \cos \delta_0 
\cos (\alpha-\alpha_0) + \sin \delta \sin \delta_0$, where $\alpha_0$ and $\delta_0$
are the right ascension and declination of the apparent origin of the dipole, respectively. Performing the $\alpha$
integration  in (\ref{corolo}) it follows that
\begin{equation}
\label{eqn:amplitudes}
{\cal R}=\left| \frac{Ad_\perp}{1+Bd_\parallel} \right|
\end{equation}
where
$$A =\frac{\int d\delta\,\omega(\delta) \cos^2 \delta}
{\int d\delta\,\omega(\delta) \cos \delta} \,, \hspace{1cm}  
B =\frac{\int d\delta\,\omega(\delta) \cos \delta \sin \delta}
{\int d\delta\,\omega(\delta) \cos \delta} \,,$$ 
$d_\parallel=\alpha_d\sin{\delta_0}$ is the component of the dipole along the 
Earth rotation axis, and  $d_\perp=\alpha_d\cos{\delta_0}$ is the component in the 
equatorial plane~\cite{Aublin:2005am}. The coefficients $A$ and $B$ can be
estimated from the data as the mean values of the cosine 
and the sine of the event declinations. For example, for the Auger data sample we have 
$A=\left<\cos{\delta}\right>\simeq 0.78$ and $B=\left<\sin{\delta}\right>\simeq -0.45$. 
For a dipole amplitude $\alpha_d$, the measured amplitude of the first harmonic
 in right ascension ${\cal R}$ thus depends on the 
region of the sky observed, which is essentially a function of the latitude 
of the observatory $\ell_{\rm site}$, and the range of zenith angles considered. In the case of a 
small $B d_\parallel$ factor, the dipole  component in the equatorial plane $d_\perp$ 
is obtained as $d_\perp\simeq {\cal R}/\langle\cos{\delta}\rangle$.
The phase $\varphi$ corresponds to the right ascension of the dipole direction
$\alpha_0$.  For a fixed number of arrival directions, the RMS error in the amplitude, $\Delta \alpha_d \approx 1.5 N^{-1/2},$ has little dependence on the amplitude~\cite{Sommers:2000us}.\footnote{A point worth noting at this juncture: A pure dipole distribution is not possible because the cosmic ray intensity cannot be negative in half of the sky.  A ``pure dipole deviation from isotropy'' means a superposition of monopole and dipole, with the intensity everywhere $\geq 0$. An approximate dipole deviation from isotropy could be caused by a single strong source if magnetic diffusion or dispersion distribute the arrival directions over much of the sky. However, a single source would produce higher-order moments as well.  An example is given in Sec.~\ref{diffusion}.}

\begin{figure}[t]
  \centering	
  \includegraphics[width=13cm]{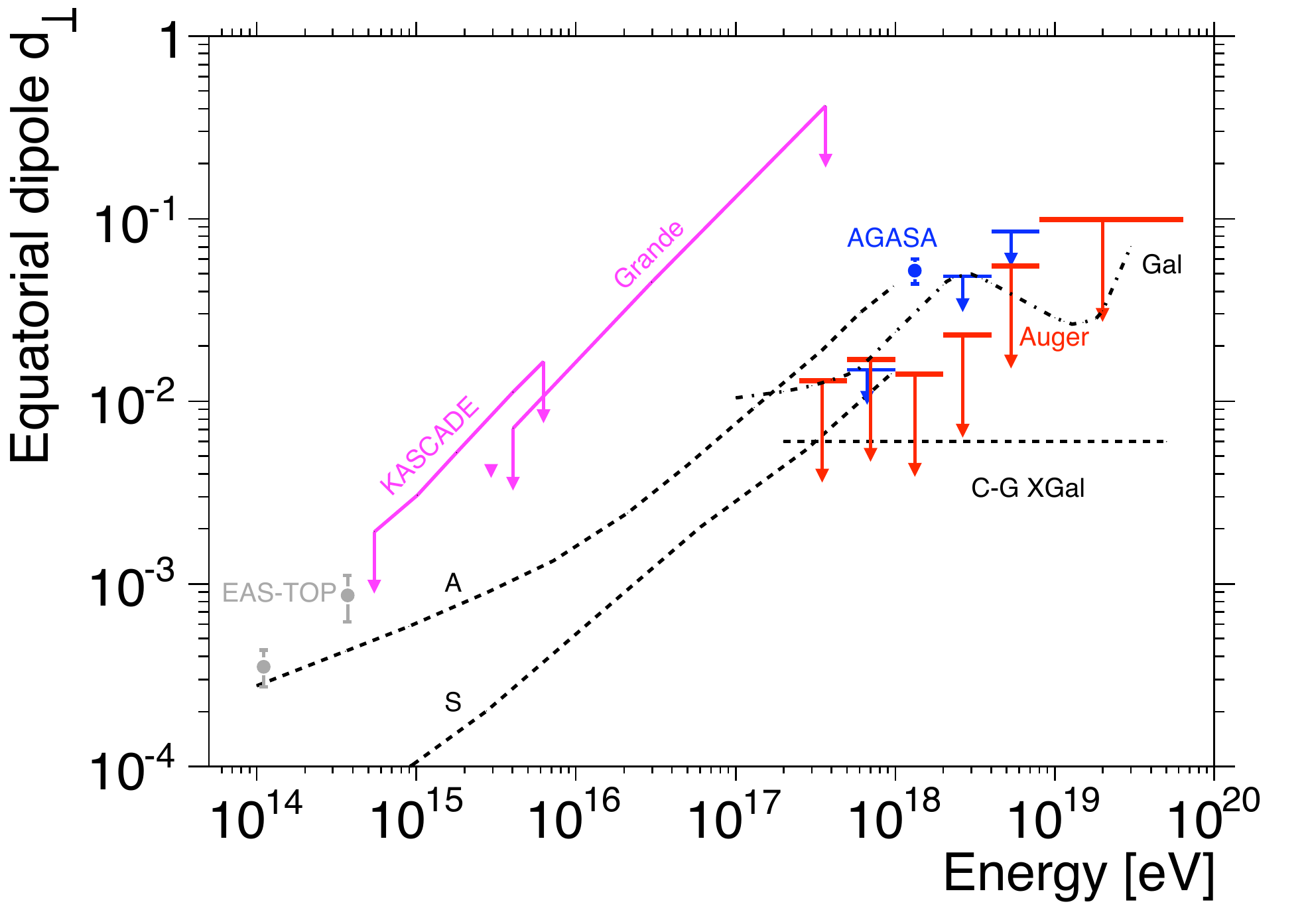}
  \caption{Upper limits on the anisotropy amplitude of first harmonic as a function of 
energy from Auger, EAS-TOP, AGASA, KASCADE and KASCADE-Grande experiments. 
Also shown are the predictions up to 1~EeV from two different galactic magnetic field models with 
different symmetries ($A$ and $S$), the predictions for a purely galactic origin of UHECRs up to a few times $10^{10}\,$~GeV ($Gal$), and the expectations from the Compton-Getting effect for an extragalactic component isotropic in the CMB rest frame ($C$-$G\,XGal$). 
Below 1~EeV, due to variations of the event counting rate arising from atmospheric effects, the Pierre Auger Collaboration adopted the  {\em East/West method}~\cite{Bonino}, which is two times less efficient but doesn't require correction for trigger efficiency. From Ref.~\cite{Abreu:2011zz}}.
  \label{UL}
\end{figure}

In Fig.~\ref{UL} we show upper limits and measurements of $d_\perp$ from various experiments together with some predictions from UHECR models of both galactic and extragalactic origin. The AGASA Collaboration reported a correlation of the CR arrival directions to the Galactic plane at the $4\sigma$ level~\cite{Hayashida:1998qb}. The energy bin width which gives the maximum $k_0$-value corresponds to the region $10^{8.9}$~GeV -- $10^{9.3}$~GeV where $k_0 = 11.1,$ yielding a chance probability of $p(\geq~{\cal R}_{_{E\sim {\rm EeV}}}^{^{\rm AGASA}}) \approx 1.5 \times 10^{-5}.$ The recent results reported by the Pierre Auger Collaboration are inconsistent with those reproted by the AGASA Collaboration~\cite{Abreu:2011zz}.  If the galactic/extragalactic transition occurs at the ankle, UHECRs at $10^{9}$~GeV are predominantly of galactic origin and their escape from the Galaxy by diffusion and drift motions are expected to induce a modulation in this energy range.  These predictions depend on the assumed galactic magnetic field model as well as on the source distribution and the composition of the UHECRs. Two alternative models are displayed in Fig.~\ref{UL}, corresponding to different geometries of the halo magnetic fields~\cite{Stanev:1996qj}.  The bounds reported by the Pierre Auger Collaboration already exclude the particular model with an antisymmetric halo magnetic field ($A$) and are starting to become sensitive to the predictions of the model with a symmetric field ($S$). The predictions shown in Fig.~\ref{UL}  are based on the assumption of predominantly heavy composition in the galactic component~\cite{Candia:2003dk}. Scenarios in which galactic protons dominate at $10^{9}$~GeV  would typically predict a larger anisotropy.  Alternatively, if the structure of the magnetic fields in the halo is such that the turbulent component dominates over the regular one, purely diffusion motions may confine {\em light} elements of galactic origin up to $\simeq 10^{9}$~GeV, and may induce an ankle-like feature at higher energy due to the longer confinement of heavier elements~\cite{Calvez:2010uh}. Typical signatures of such a scenario in terms of large scale anisotropies are also shown in Fig.~\ref{UL} (dotted line). The corresponding amplitudes are challenged by the  current sensitivity of Auger. On the other hand, if the transition is taking place at lower energies, say around the second knee, UHECRs above $10^9$~GeV are dominantly of extragalactic origin and their large scale distribution could be influenced by the relative motion of the observer with respect to the frame of the sources. If the frame in which the UHECR distribution is isotropic coincides with the CMB rest frame, a small anisotropy is expected due to the Compton-Getting effect~\cite{Compton:g}. Neglecting the effects of the galactic magnetic field, this anisotropy would be a dipolar pattern pointing in the direction $\alpha_0\simeq 168^\circ$ with an amplitude of about 0.6\%~\cite{Kachelriess:2006aq}, close to the upper limits set by the Pierre Auger Collaboration. The statistics required to detect an amplitude of 0.6\% at 99\% CL is $\simeq 3$ times the published Auger sample~\cite{Abreu:2011zz}.

The right harmonic analyses are completely blind to intensity variations which depend only on declination.  Combining anisotropy searches in $\alpha$ over a range of declinations could dilute the results, since significant but out of phase Rayleigh vectors from different declination bands can cancel each other out. An unambiguous interpretation of anisotropy data requires two ingredients: {\it exposure to the full celestial sphere and analysis in terms of both celestial coordinates}~\cite{Sommers:2000us}.

\begin{figure}[t]
\begin{center}
\includegraphics[width=0.8\textwidth]{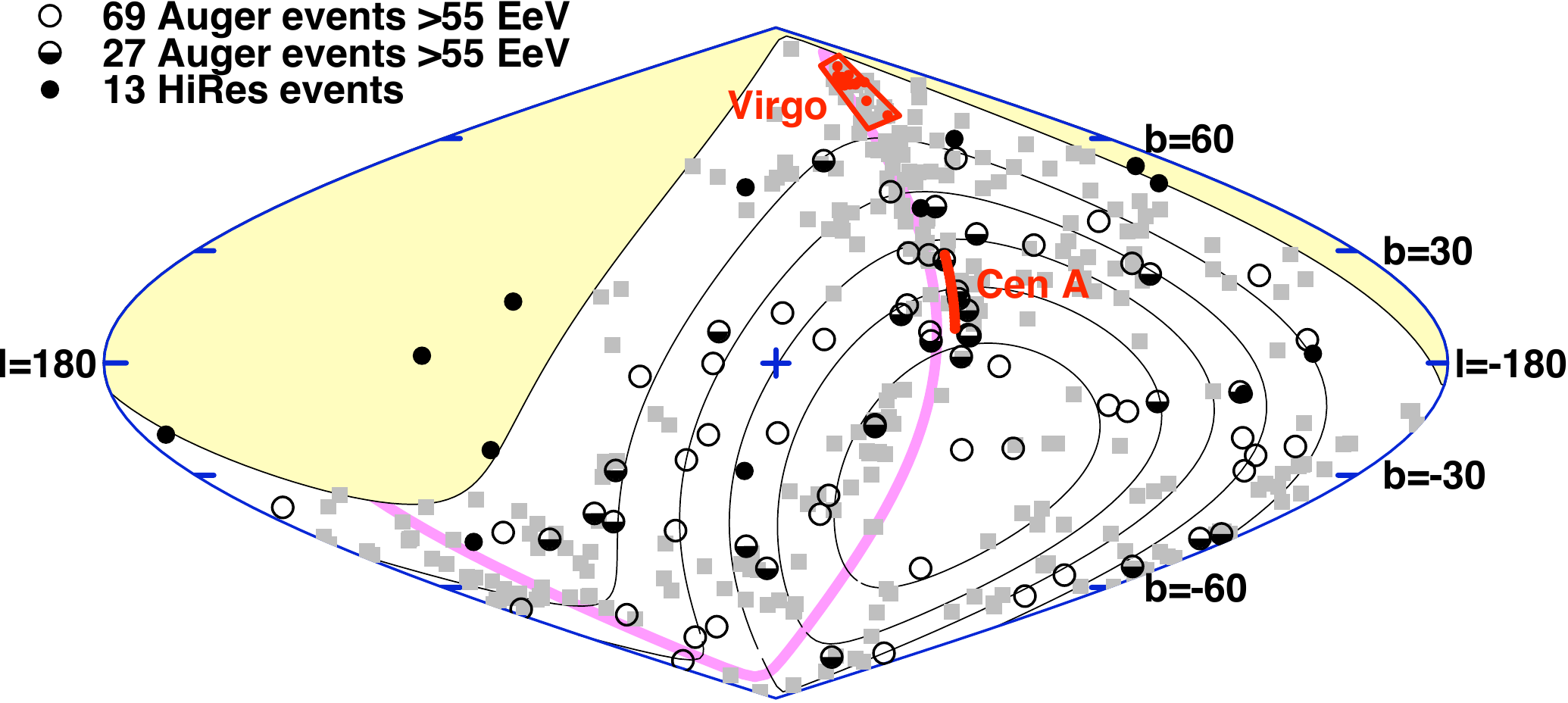}
\end{center}
\caption{Correlation of the arrival directions of UHECR with AGNs from
 the VCV catalog. The shaded part of the sky is not visible by Auger.
 The gray squares are the AGNs within $z$ less than 0.018. The Auger 
 events are shown with circles. The first 27 events are half 
 filled. The 13 HiRes events are shown with black dots. The thin lines show
 the six regions of the sky to which Auger has equal exposure.
 The wide gray line is the supergalactic plane.  From Ref.~\cite{LetessierSelvon:2011dy}.
}
\label{69}
\end{figure}

One way to increase the chance of success in finding out the sources of UHECRs is to check for correlations between CR arrival directions and known candidate astrophysical objects.  To calculate a meaningful statistical significance in such an analysis, it is important to define the search procedure {\it a priori} in order to ensure it is not inadvertently devised especially to suit the particular data set after having studied it. With the aim of avoiding accidental bias on the number of trials performed in selecting the cuts, the Auger anisotropy analysis scheme followed a pre-defined process. First an exploratory data sample was employed for comparison with various source catalogs and for tests of various cut choices.  The results of this exploratory period were then used to design prescriptions to be applied to subsequently gathered data.

Based on the results of scanning an exposure of $4,390~{\rm km}^2$ sr yr, a prescription was designed to test the correlation of events having energies $E > 5.5 \times 10^{10}$~GeV with objects in the Veron-Cetty \& Veron (VCV) catalog of Active Galactic Nuclei (AGNs).  The prescription called for a search of $3.1^\circ$ windows around catalog objects with redshifts $z < 0.0018$.  The significance threshold set in the prescription was met in 2007, when the exposure more than doubled and the total number of events reached 27, with 9 of the 13 events in the post-prescription sample correlating~\cite{Abraham:2007si, Cronin:2007zz}. For a sample of 13 events from an isotropic distribution, the probability that 9 or more correlate by chance with an object in the AGN catalogue (subject to cuts on the exposure weighted fraction of the sky within the opening angles and the redshift) is less than 1\%. This corresponds to roughly a $2.5\sigma$ effect. In the summer of 2008, the HiRes Collaboration applied the Auger prescription to their data set and found no significant correlation~\cite{Abbasi:2008md}.  One has to exercise caution when comparing results of different experiments with potentially different energy scales, since the analysis involves placing an energy cut on a steeply falling spectrum. More recently, the Pierre Auger Collaboration published an update on the correlation results from an exposure of $20,370~{\rm km}^2$ sr yr (collected over 6~yr but equivalent to 2.9~yr of the nominal exposure/yr of the full Auger), which contains 69 events with $E > 5.5 \times 10^{10}$~GeV~\cite{:2010zzj}.  A skymap showing the locations of all these events is displayed in Fig.~\ref{69}.  For a {\it physical} signal one expects the significance to increase as more data are gathered. In this study, however, the significance has not increased.  A $3\sigma$ effect is not neccesarily cause for excitement; of every 100 experiments, you expect about one $3\sigma$ effect. Traditionally, in particle physics there is an unwritten $5\sigma$ rule for ``discovery.'' One should keep in mind though that in the case of CR physics we do not have the luxury of controlling the luminosity.

\begin{figure}[h]
\begin{center} 
\includegraphics[width=0.75\linewidth]{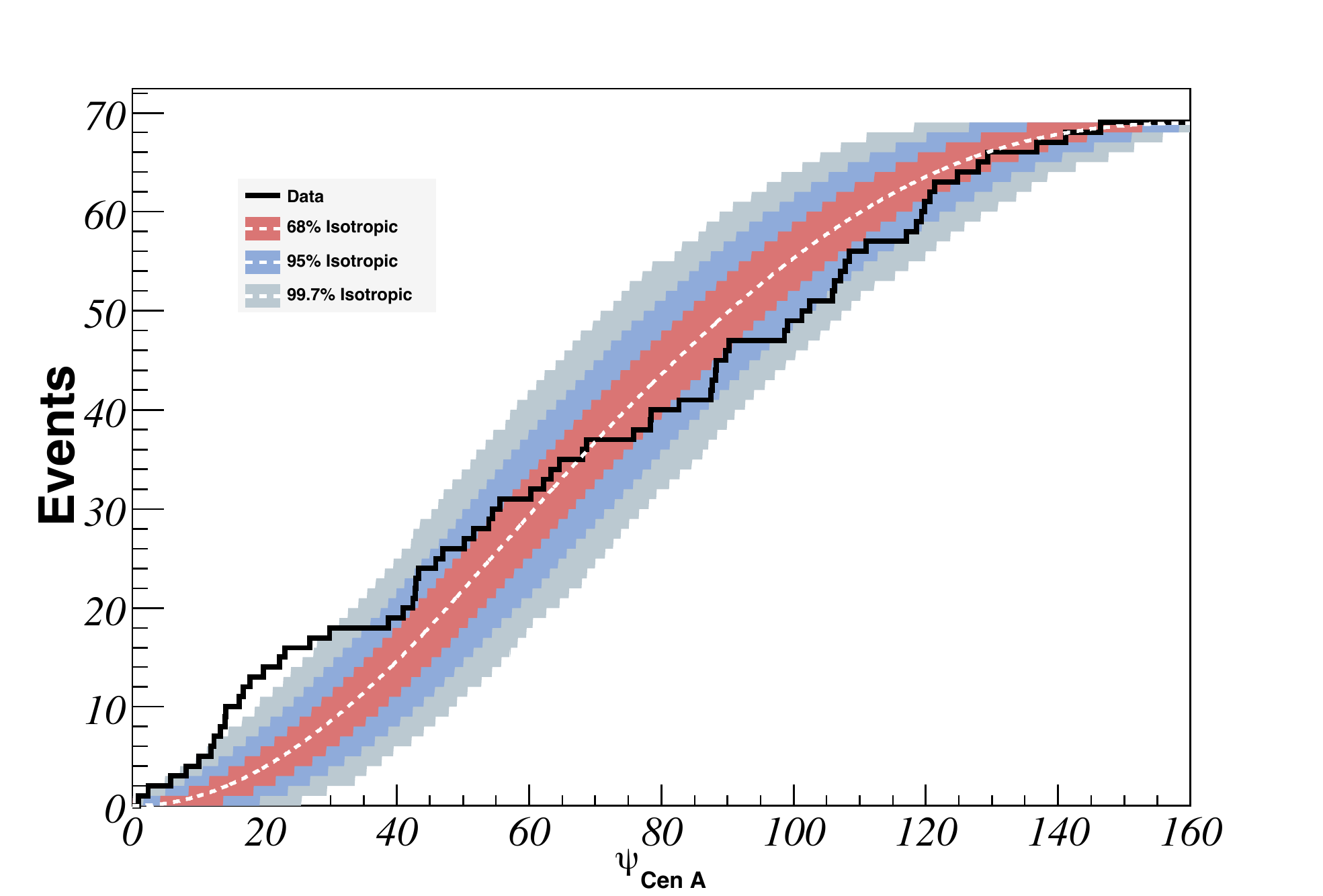} 
\end{center} 
\caption{Cumulative number of events with $E\geq55$~EeV as a function
of angular distance from the direction of Cen A. The bands correspond to the 68\%, 95\%, and 99.7\% dispersion 
expected for an isotropic flux. From Ref.~\cite{:2010zzj}.}\label{cena_69}
\end{figure}

A number of other interesting observations are described in~\cite{:2010zzj}, including comparisons with other catalogs as well as a specific search around the region of the nearest active galaxy, Centaurus A (Cen A).  It is important to keep in mind that these are all {\em a posteriori} studies, so {\em one cannot use them to determine a confidence level for anisotropy as the number of trials is unknown}. A compelling concentration of events in the region around the direction of Cen A has been observed.  As one can see in Fig.~\ref{cena_69}, the maximum departure from isotropy occurs for a ring of $18^\circ$ around the object, in which 13 events are observed compared to an expectation of 3.2 from isotropy. There are no events coming from less than $18^\circ$ around M87, which is almost 5 times more distant than Cen A and lies at the core of the Virgo cluster. As shown in Fig.~\ref{69} the Auger exposure is 3 times smaller for M87 than for Cen A. Using these two rough numbers and assuming equal luminosity, one expects 75 times fewer events from M87 than from Cen A. Hence, the lack of events in this region is not completely unexpected.

The Centaurus cluster lies 45~Mpc behind Cen A. An interesting question then is whether   some of the events in the $18^\circ$ circle could come from the Centaurus cluster rather than from Cen A. This does not appear likely because the Centaurus cluster is farther away than the Virgo cluster and for comparablel CR luminosities one would expect a small fraction of events coming from Virgo. Furthermore, the events emitted by Cen A and deflected by magnetic fields could still register
as a correlation due to the overdense AGN population lying behind Cen A, resulting in a 
spurious signal~\cite{Gorbunov:2008ef}.\\

In summary, the inaugural years of data taking at the Pierre Auger Observatory have yielded a large, high-quality data sample.  The enormous area covered by the surface array together with an excellent fluorescence system and hybrid detection techniques have provided us with large statistics, good  energy resolution, and solid control of systematic uncertainties. Presently, Auger is collecting some $7,000~{\rm km}^2~{\rm sr}~{\rm yr}$ of exposure each year, and is expected to run for 2 more decades.  New detector systems are being deployed, which will lower the energy detection threshold down to $10^{8}$~GeV.  An experimental radio detection program is also co-located with the observatory  and shows promising results.  As always, the development of new analysis techniques is ongoing, and interesting new results can be expected.

\subsection{Origin of ultrahigh energy cosmic rays}

It is most likely that the bulk of the cosmic radiation is a result of some very general magneto-hydrodynamic (MHD) phenomenon in space which transfers kinetic or magnetic energy into cosmic ray energy. The details of the acceleration process and the maximum attainable energy depend on the particular physical situation under consideration. There are basically two types of mechanism that one might invoke. The first type assumes the particles are accelerated directly to very high energy (VHE) by an extended electric field~\cite{Hillas:1985is}. This idea can be traced back to the early 1930's when Swann~\cite{Swann} pointed out that betatron acceleration may take place in the increasing magnetic field of a sunspot. These so-called ``one-shot'' mechanisms have been worked out in greatest detail, and the electric field in question is now generally associated with the rapid rotation of small, highly magnetized objects such as neutron stars (pulsars) or AGNs. Electric field acceleration has the advantage of being fast, but suffers from the circumstance that the acceleration occurs in astrophysical sites of very high energy density, where new opportunities for energy loss exist. Moreover, it is usually not obvious how to obtain the observed power law spectrum in a natural way, and so this kind of mechanism is not widely favored these days. The second type of process is often referred to as statistical acceleration, because particles gain energy gradually by numerous encounters with moving magnetized plasmas. These kinds of models were mostly pioneered by Fermi~\cite{Fermi:1949ee}. In this case the $E^{-2}$ spectrum emerges very convincingly. However, the process of acceleration is slow, and it is hard to keep the particles confined within the Fermi engine.  In this section we first provide a summary of statistical acceleration based on the simplified version given in Ref.~\cite{Gaisser:vg}. For a more detailed and rigorous discussion, the reader is referred to~\cite{Blandford:1987pw}.  After reviewing statistical acceleration, we turn to the issue of the maximum achievable energy within diffuse shock acceleration and explore the viability of some proposed UHECR sources.

\subsubsection{Fermi acceleration at shock waves}
\label{fermi_acceleration}

\begin{figure}[t]
\begin{center}
\includegraphics[width=0.45\textwidth]{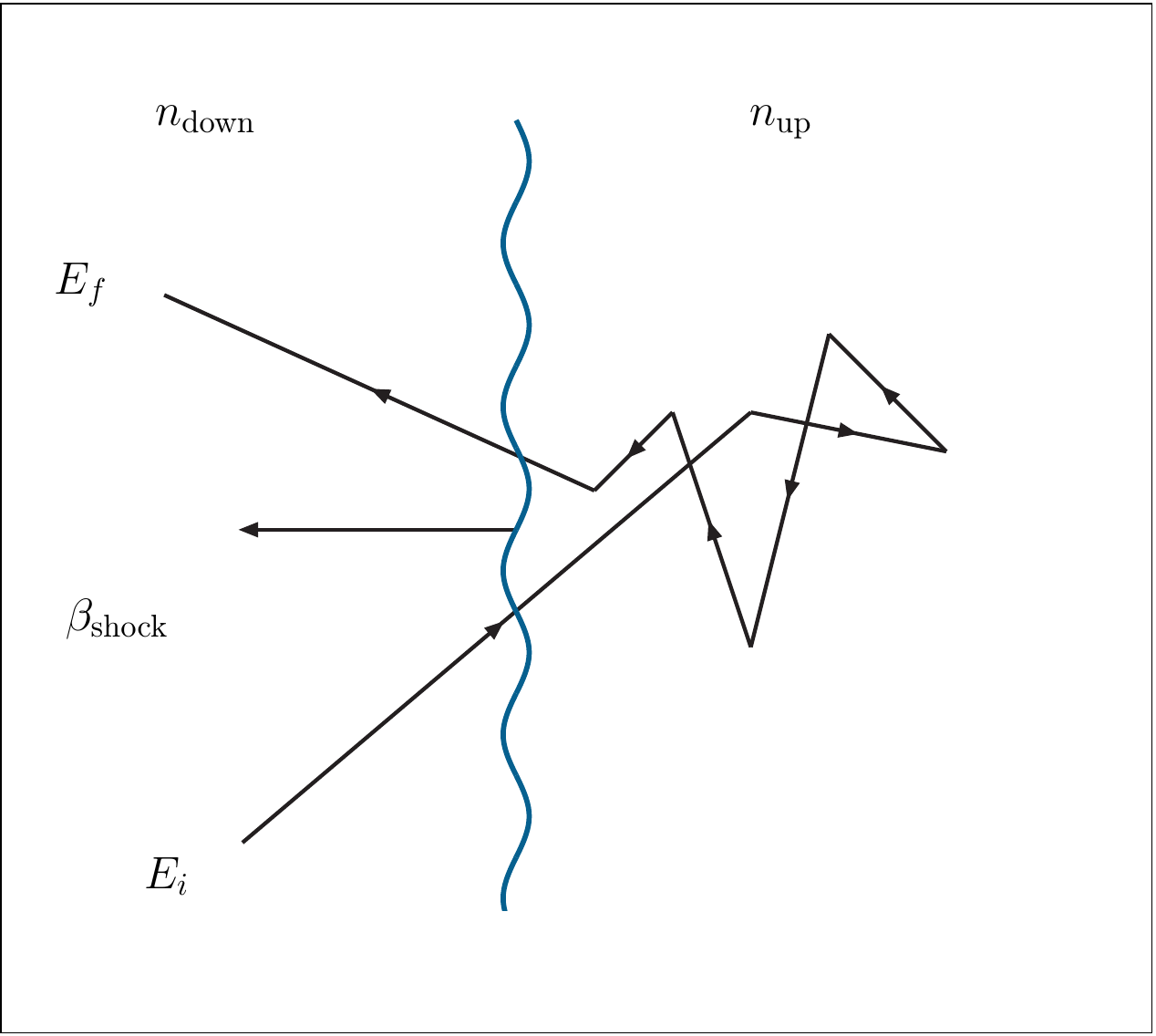}
\includegraphics[width=0.455\textwidth]{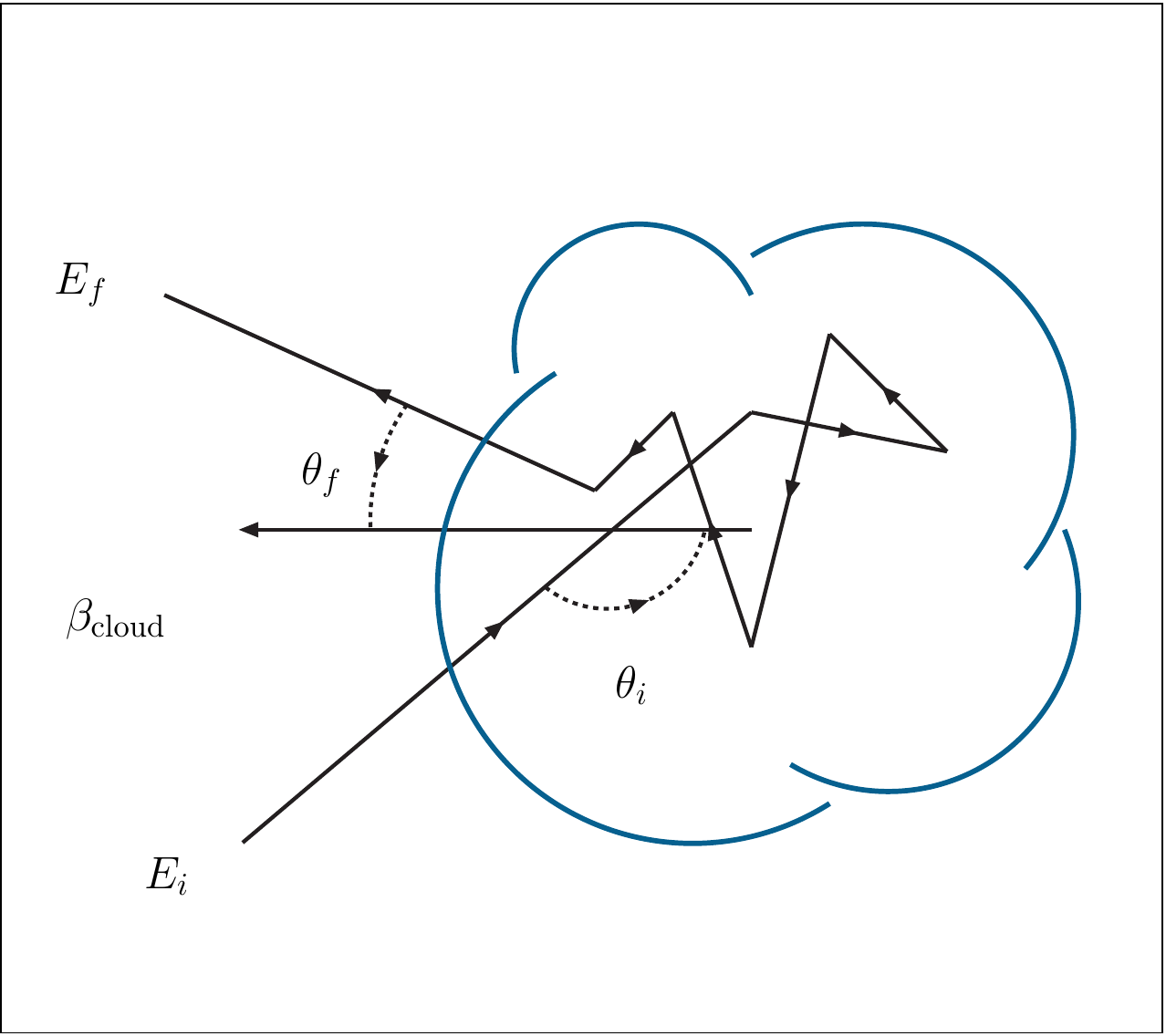}
\end{center}
\caption{A sketch of 1st and 2nd order Fermi acceleration by scattering off plasma shocks (right) and magnetic clouds (left), respectively.  From Ref.~\cite{yellowbook}. \label{FERMI}}
\end{figure}

In his original analysis, Fermi~\cite{Fermi:1949ee} considered the scattering
of CRs on moving magnetized clouds. The right panel of
Fig.~\ref{FERMI} shows a sketch of these encounters. Consider a CR
entering into a single cloud with energy $E_i$ and incident angle
$\theta_i$ with the cloud's direction undergoing diffuse scattering on
the irregularities in the magnetic field.  After diffusing inside the
cloud, the particle's average motion coincides with that of the gas
cloud. The energy gain by the particle, which emerges at an angle
$\theta_f$ with energy $E_f$, can be obtained by applying Lorentz
transformations between the laboratory frame (unprimed) and the cloud
frame (primed).  In the rest frame of the moving cloud, the CR
particle has a total initial energy
\begin{equation}
  E_i' = \Gamma_{\rm cloud} \,E_i\, (1 - \beta_{\rm cloud}\, \cos \theta_i)\,, 
\end{equation}
where $\Gamma_{\rm cloud}$ and $\beta_{\rm cloud} = V_{\rm cloud}/c$ are the
Lorentz factor and velocity of the cloud in units of the speed of
light, respectively.  In the frame of the cloud we expect no change in
energy ($E_i' = E_f'$), because all the scatterings inside the cloud
are due only to motion in the magnetic field (so-called collisionless
scattering). There is elastic scattering between the CR and the cloud
as a whole, which is much more massive than the CR. Transforming to
the laboratory frame we find that the energy of the particle after its
encounter with the cloud is
\begin{equation}
E_f = \Gamma_{\rm cloud} \,E_f'\, (1 + \beta_{\rm cloud} \cos \theta_f)\,.
\end{equation}
The fractional energy change in the 
laboratory frame is then
\begin{equation}
\frac{\Delta E}{E} = \frac{E_f-E_i}{E_i} = \frac{1 - \beta_{\rm cloud} \cos \theta_i + \beta_{\rm cloud} \cos \theta_f - \beta_{\rm cloud}^2 \cos \theta_i \cos \theta_f}{1 - \beta_{\rm cloud}^2} - 1 \,.
\label{flash0}
\end{equation}
Inside the cloud the CR direction becomes randomized and so 
$\langle \cos \theta_f\rangle = 0.$  The collisionless scattered particle will gain energy in a
head-on collision ($\theta_i>\pi/2$) and lose energy by tail-end
($\theta_i<\pi/2$) scattering. The net increase of its energy is a
statistical effect. The average value of 
$\cos \theta_i$ depends on the relative 
velocity between the cloud and the particle. The 
probability $P$ per unit solid angle $\Omega$ of having a collision at angle 
$\theta_i$ is proportional to $(v - V_{\rm cloud} \cos \theta_i)$, where $v$ is the CR 
speed. In the 
ultrarelativistic limit, i.e., $v \sim c$ (as seen in the laboratory frame),
\begin{equation}
\frac{dP}{d \Omega_i} \propto (1 - \beta_{\rm cloud} \cos\theta_i)\,,
\end{equation}
so
\begin{equation}
\langle \cos \theta_i \rangle  = -\frac{\beta_{\rm cloud}}{3}\,.
\label{flash}
\end{equation}
Now, inserting Eq.~(\ref{flash}) into Eq.(\ref{flash0}), one obtains for 
$\beta_{\rm cloud} \ll 1$, 
\begin{equation}
\frac{\langle\Delta E\rangle}{E} = \frac{1 +\beta_{\rm cloud}^2/3}{1-\beta_{\rm cloud}^2} - 1 \approx \frac{4}{3} \,\beta_{\rm cloud}^2\,.
\label{ja}
\end{equation}
Note that $\langle \Delta E \rangle/E \propto \beta_{\rm cloud}^2$, so even though
the average magnetic field may vanish, there can still be a net
transfer of the macroscopic kinetic energy from the moving cloud to
the particle. However, the average energy gain is very small, because
$\beta_{\rm cloud}^2 \ll 1$.  This acceleration process is very similar to a
thermodynamical system of two gases, which tries to come into thermal
equilibrium~\cite{Drury:1994fg}. Correspondingly, the spectrum of CRs
should follow a thermal spectrum which might be in conflict with the
observed power-law.

A more efficient acceleration may occur in the vicinity of plasma shocks occurring in astrophysical environments~\cite{Bell:1978zc,Blandford:ky}.  Suppose that a strong (nonrelativistic) shock wave propagates through the plasma as sketched in the left panel of Fig.~\ref{FERMI}. Then, in the rest frame of the shock the conservation relations imply that the upstream velocity $u_{\rm up}$ (ahead of the shock) is much higher than the downstream velocity $u_{\rm down}$ (behind the shock). The compression ratio $r= u_\text{up}/u_\text{down}=n_\text{down}/n_\text{up}$ can be determined by requiring continuity of particle number, momentum, and energy across the shock; here  $n_\text{up}$  ($n_\text{down}$) is the particle density of the upstream (downstream) plasma.  For an ideal gas the compression ratio can be related to the specific heat ratio and the Mach number of the shock. For highly supersonic shocks,  $r=4$~\cite{Blandford:1987pw}.  Therefore, in the primed frame stationary with respect to the shock, the upstream flow approaches with speed $u_{\rm up} = \beta_{\rm up}  c = 4 \beta c/3$ and the downstream flow recedes with speed $u_{\rm down} = \beta_{\rm down} c = \beta c/3$. When measured in the stationary upstream frame,  the quantity $u = u_{\rm up} - u_{\rm down} = \beta  c$ is the speed of the shocked fluid and $u_{\rm up} = \beta_{\rm shock}$ is the speed of the shock. 
Hence,  because of the converging flow -- whichever side of the shock you are on, if you are moving with the plasma, the plasma on the other side of the shock is approaching you with velocity $u$ -- to first order there are only head-on collisions for particles crossing the shock front. The acceleration process, although stochastic, always leads to a gain in energy. In order to work out the energy gain per shock crossing, we can visualize magnetic irregularities on either side of the shock as clouds of magnetized plasma of Fermi's original theory.  By considering the rate at which CRs cross the shock from downstream to upstream,
and upstream to downstream, one finds $\langle \cos \theta_{i}
\rangle = -2/3$ and $\langle \cos \theta_{f} \rangle =
2/3$. Hence, Eq.~(\ref{flash0}) can be generalized to
\begin{equation}
{\langle \Delta E \rangle \over E}  \simeq {4 \over 3} \beta = \frac{4}{3} 
\frac{u_{\rm up} -u_{\rm down}}{c}.
\end{equation}
Note this is first order in $\beta=u/c$, and is therefore more efficient than Fermi's original mechanism.

An attractive feature of Fermi acceleration is its prediction of a power-law flux of CRs. Consider a test-particle with momentum $p$ in the rest frame of the upstream fluid (see Fig.~\ref{FERMI}). The particle's momentum distribution is isotropic in the fluid rest frame. For pitch angles $\pi/2<\theta_i<\pi$ relative to the shock velocity vector (see Fig.~\ref{FERMI}) the particle enters the downstream region and has  {\it on average} the relative momentum $p[1+2(\beta_{\rm up}-\beta_{\rm down})/3]$. Subsequent diffusion in the downstream region `re-isotropizes' the particle's momentum distribution in the fluid rest frame. As the particle diffuses back into the upstream region (for pitch angles $0<\theta_f<\pi/2$) it has {\it gained} an average momentum of $\langle\Delta p\rangle/p \simeq 4(\beta_\text{up}-\beta_\text{down})/3$. This means that 
the momentum gain of a particle per time is proportional to its momentum, 
\begin{equation}
\dot{p} = p/t_\text{gain} \, .
\label{ganancia}
\end{equation}
On the other hand, the loss of particles from the acceleration region is proportional to their number, 
\begin{equation}
\dot{N} = - N/t_\text{loss} \, . 
\label{perdida}
\end{equation}
Therefore, taking the ratio (\ref{ganancia})/(\ref{perdida}) we first obtain 
\begin{equation}
dN/dp= -\alpha N/p \,,   
\end{equation}
and after integration $N(p) \propto p^{-\alpha},$ with $\alpha=t_\text{gain}/t_\text{loss}$. 
If the acceleration cycle across the shock takes the time $\Delta t$ we have already identified $\Delta t/t_\text{gain} = \langle\Delta p\rangle/p \simeq 4(\beta_\text{up}-\beta_\text{down})/3$. For the loss of relativistic particles one finds $\Delta t/t_\text{loss} \simeq 4\beta_\text{up}$.  Therefore,  $\alpha \simeq 3\beta_\text{up}/(\beta_\text{up}-\beta_\text{down}) = 3r/(r-1)$, yielding $\alpha\simeq4$ for highly supersonic shocks and $\alpha>4$ otherwise. The energy spectrum $N(E) \propto E^{-\gamma}$ is related to the momentum spectrum by ${\rm d}E N(E) = 4\pi p^2{\rm d}p N(p)$ and hence $\gamma = \alpha-2 \gtrsim 2$. The steeply falling spectrum of CRs with $\gamma\simeq3$ seems to disfavor supersonic plasma shocks. However, for the comparison of these injection spectra with the flux of CRs observed on Earth, one has to consider particle interactions in the source and in the interstellar medium. This can have a great impact on the shape, as we will discuss in  Secs.~\ref{diffusion} and \ref{Boltzmann}.

\begin{figure}[t]
\centering
\includegraphics[width=0.9\columnwidth]{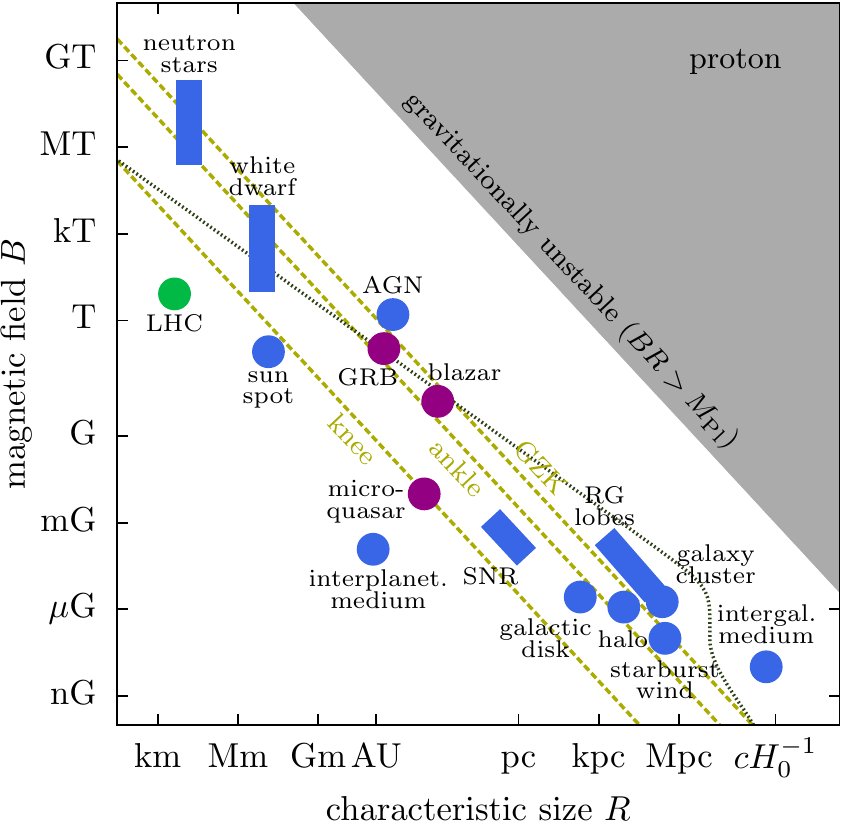}
\caption[]{The ``Hillas plot'' for various CR source candidates (blue areas). Also shown are jet-frame parameters for blazers, gamma-ray bursts, and microquasars (purple areas). The corresponding point for the LHC beam is also shown. The red dashed lines show the {\it lower limit} for accelerators of protons at the CR knee ($\sim 10^{6.5}$~GeV), CR ankle ($\sim 10^{9.5}$~GeV) and the GZK suppression ($\sim 10^{10.6}$~GeV). The dotted gray line is the {\it upper limit} from synchrotron losses and proton interactions in the cosmic photon background ($R\gg1$~Mpc). The grey area corresponds to astrophysical environments with extremely large magnetic field energy that would be gravitationally unstable. From Ref.~\cite{yellowbook}. \label{HILLASPLOT}}
\end{figure}

In general, the maximum attainable energy of Fermi's mechanism is 
determined by the time scale over which particles are able to interact 
with the plasma.  For the efficiency of a ``cosmic cyclotron'' particles have to be confined in the accelerator by its magnetic field $B$ over a sufficiently long time scale compared to the characteristic cycle time. The Larmor radius of a particle with charge $Ze$ increases with its energy $E$ according to
\begin{eqnarray}\label{LARMOR}
r_L & = &\sqrt{\frac{1}{4\pi\alpha}}\frac{E}{ ZeB}  \nonumber \\                                                   & = &\frac{1.1}{Z}\left(\frac{E}{\text{EeV}}\right)\left(\frac{B}{\mu\text{G}}\right)^{-1}\,{\rm kpc}\,.
\end{eqnarray} 
The particle's energy is limited as its Larmor radius approaches the characteristic radial size $R_\text{source}$ of the source 
\begin{equation}
\label{EMAX}
E_\text{max} \simeq  Z \left(\frac{B}{\mu\text{G}}\right)\left(\frac{R_\text{source}}{\text{kpc}}\right)\, \times 10^{9}~{\rm GeV}\,.
\end{equation}
This limitation in energy is conveniently visualized by the `Hillas plot'~\cite{Hillas:1985is} shown in Fig.~\ref{HILLASPLOT} where the characteristic magnetic field $B$ of candidate cosmic accelerators is plotted against their characteristic size $R$. It is important to stress that in some cases the acceleration region itself only exists for a limited period of time; for example, supernovae shock waves dissipate after about $10^{4}$~yr.  In such a case, Eq.~(\ref{EMAX}) would have to be  modifed accordingly.  Otherwise, if the plasma disturbances persist for much longer periods, the maximum energy may be limited by an increased likelihood of escape from the region. A look at Fig.~\ref{HILLASPLOT} reveals that the number of sources for the extremely high energy CRs around $10^{12}$~GeV is very sparse. For protons, only radio galaxy lobes and clusters of galaxies seem to be plausible candidates. For nuclei, terminal shocks of  galactic superwinds  originating in the metally-rich starburst galaxies are  potential sources~\cite{Anchordoqui:1999cu}. Exceptions may occur for sources which move relativistically in the host-galaxy frame, in particular jets from AGNs and gamma-ray bursts (GRBs). In this case the maximal energy might be increased due to a Doppler boost by a factor $\sim30$ or $\sim1000$, respectively. 

For an extensive discussion on  the potential CR-emitting-sources shown in Fig.~\ref{HILLASPLOT}, see e.g.~\cite{Torres:2004hk}. Two of the most attractive examples are discussed next.

\subsubsection{AGNs}

AGNs are composed of an accretion disk around a central super-massive black hole and are sometimes associated with jets terminating in lobes which can be detected in radio. One can classify these objects into two categories: radio-quiet AGN with no prominent radio emission or jets and radio-loud objects presenting jets.

Fanaroff-Riley II (FRII) galaxies~\cite{Fanaroff} are the largest known dissipative objects (non-thermal sources) in the cosmos.  Localized regions of intense synchrotron emission, known as ``hot spots,'' are observed within their lobes. These regions are presumably produced when the bulk kinetic energy of the jets ejected by a central active nucleus is reconverted into relativistic particles and turbulent fields at a ``working surface'' in the head of the jets~\cite{Blandford}. Specifically, the speed $u_{\rm head}\approx \; u_{\rm jet}\,[ 1 + ( n_e /n_{\rm jet})^{1/2}]^{-1}$, with which the head of a jet advances into the intergalactic medium of particle density $n_e$ can be obtained by balancing the momentum flux in the jet against the momentum flux of the surrounding medium; where $n_{\rm jet}$ and $u_{\rm jet}$ are the particle density and the velocity of the jet flow, respectively (for relativistic corrections, see~\cite{Rosen:1999jm}). For $n_e \geq n_{\rm jet}$,
$u_{\rm jet}> u_{\rm head}$ so that that the jet decelerates. The result is the formation of a strong collisionless shock, which is responsible for particle reacceleration and magnetic field amplification~\cite{Begelman}.  The acceleration of particles up to ultrarelativistic energies in the hot spots is the result of repeated scattering back and forth across the shock front, similar to that discussed in Sec.~\ref{fermi_acceleration}. The particle deflection in this mechanism is dominated by the turbulent magnetic field with wavelength $k$ equal to the Larmor radius of the particle concerned~\cite{Drury:1983zz}. A self-consistent (although possibly not unique) specification of the turbulence is to assume that 
 the energy density per unit of wave number of MHD turbulence is of the Kolmogorov type, $I(k) \propto k^{-5/3}$, just as for hydrodynamical turbulence~\cite{Kolmogorov}. With this in mind,  to order of magnitude accuracy using effective quantities averaged over upstream (jet) and downstream (hot spot) conditions (considering that downstream counts a fraction of 4/5~\cite{Drury:1983zz}) the acceleration timescale  at a shock front is found to be~\cite{Rachen:1992pg} 
\begin{equation} \tau_{\rm acc}^{\rm AGN} \approx \frac{20 D_\parallel(E)}{u_{\rm jet}^2} \,, 
\label{acc}
\end{equation} 
where 
\begin{equation}
D_\parallel (E) = \frac{2c}{\pi\, U}\,\left(\frac{E}{eB}\right)^{1/3}\, R^{2/3}
\end{equation}
is the Kolmogorov diffusion coefficient, $U$ is the ratio of turbulent to ambient magnetic energy density in the region of the shock (of radius $R$), and $B$ is the total magnetic field strength.

The subtleties surrounding the conversion of particles kinetic
energy into radiation provide ample material for discussion. The most popular mechanism to date relates
$\gamma$-ray emission to the development of electromagnetic
cascades triggered by secondary photomeson products that cool
instanstaneously via synchrotron radiation. The
characteristic single photon energy in synchrotron radiation
emitted by an electron is
\begin{equation}
E_\gamma^ {\rm syn} = \left(\frac{3}{2}\right)^{1/2} \frac{h\,e\,E_e^2\,B}{2 \,\pi\, m_e^3 \,
c^5} \sim 5.4 \times 10^{-2}\, B_{\mu{\rm G}}\, (E_e/{\rm EeV})^{2} \, {\rm TeV}\ \ .\label{synch}
\end{equation}
For a proton this number is $(m_p/m_e)^3 \sim 6 \times 10^9$ times smaller.

The acceleration process will then be efficient as long as the
energy losses by synchrotron radiation and/or photon--proton
interactions do not become dominant.  The synchrotron loss time for protons is given by~\cite{Rybicki}
\begin{equation}
\tau_{\rm syn} \sim \frac{6\, \pi\, m_p^3\,c}{\sigma_{\rm T}\,m_e^2\,\Gamma\,B^2}\,,
\label{tausyn}
\end{equation}
where  $\sigma_{\rm T}$ and $\Gamma = E/(m_pc^2)$ are the  Thomson cross section  and Lorentz factor, respectively. Considering an average cross
section $\bar{\sigma}_{\gamma p}$~\cite{Armstrong:1971ns} for the three dominant
pion--producing interactions, 
$\gamma p \rightarrow p  \pi^0, \ 
\gamma  p \rightarrow n  \pi^+, \
\gamma  p\rightarrow p  \pi^+  \pi^-,$
the time scale of the energy losses, including synchrotron and photon 
interaction losses, reads~\cite{Biermann:ep}
\begin{equation}
\tau_{\rm loss} \simeq \frac{6\pi\ m_p^4\ c^3}{\sigma_{\rm T}\ m_e^2\ B^2\ (1+Aa)}\ E^{-1}\ = \frac{\tau_{\rm syn}}{1 + Aa} \,,
\end{equation} 
where $a$ stands for the ratio of photon to magnetic energy densities and $A$ gives a measure of the relative strength of $\gamma p $ interactions versus the synchrotron emission.
 Note that the second channel involves the creation of ultrarelativistic neutrons that can readily escape the system.  For typical hot spot conditions, the number density of photons per unit energy interval  follows a power-law spectrum
\begin{equation}
n_\gamma^{\rm AGN} (\omega) = \left\{\begin{array}{cl} (N_0/\omega_0) \ \ (\omega/\omega_0)^{-2} ~~~~&~~~~ \omega_0 \leq \omega \leq \omega^\star \\
0 ~~~&~~~~ {\rm otherwise} \end{array} \right.
\end{equation} 
where $N_0$ is the normalization constant and $\omega_0$ and $\omega^\star$ correspond to radio and gamma rays energies, respectively. The ratio of photon to magnetic energy density is then
\begin{equation}
a = \frac{N_0 \ \omega_0 \ \ln(\omega^\star/\omega_0)}{B^2/8\pi}
\end{equation}
and $A$ is only weakly dependent on the properties of the source
\begin{equation}
A = \frac{\sigma_{\gamma p}}{\sigma_{\rm T}} \, \frac{(m_p/m_e)^2}{\ln (\omega^\star/\omega_0)} \approx \frac{\sigma_{\gamma p}}{\sigma_{\rm T} } \ 1.6 \times 10^5 \approx 200 \, .
\end{equation}
The maximum attainable energy can be obtained by balancing the energy gains and losses~\cite{Anchordoqui:2001bs}
\begin{equation}
E_{20}=1.4\times 10^5\,\,B_{\mu{\rm G}}^{-5/4}\,\,\beta_{\rm jet}^{3/2}
\,\,u^{3/4}\,\,R_{\rm kpc}^{-1/2}\,\,(1+Aa)^{-3/4}\ ,
\label{ab}
\end{equation}
where  $E \equiv 10^{20} \, E_{20}~{\rm eV}$ and $ R \equiv R_{\rm kpc} \, 1~{\rm kpc}$. It is of interest to apply the acceleration conditions to the nearest AGN.

\begin{figure}
\centering
\includegraphics[width=0.7\textwidth]{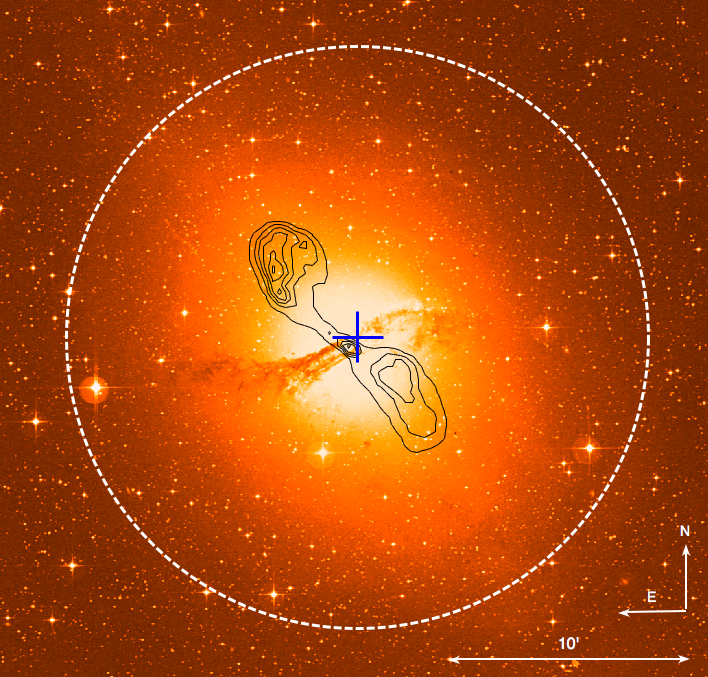}
\caption{Optical image of Cen~A (UK 48-inch Schmidt) overlaid with radio contours (black, VLA~\cite{Condon:1996}). H.E.S.S. VHE best fit position with $1\sigma$ statistical errors (blue cross) and VHE extension 95\% CL upper limit  (white dashed circle) are also shown. From Ref.~\cite{Aharonian:2009xn}.}
\label{cenapic}
\end{figure}

At only 3.4~Mpc distance, Cen A is a complex FRI radio-loud source identified at optical frequencies with the galaxy NGC 5128~\cite{Israel}.  Radio observations at different wavelengths have revealed a rather complex morphology shown in Fig.~\ref{cenapic}. It comprises a compact core, a jet (with subluminal proper motions $\beta_{\rm jet} \sim 0.5$~\cite{Hardcastle:2003ye}) also visible at $X$-ray frequencies, a weak counterjet, two inner lobes, a kpc-scale middle lobe, and two giant outer lobes. The jet would be responsible for the formation of the northern inner and middle lobes when interacting with the interstellar and intergalactic media, respectively. There appears to be a compact structure in the northern lobe, at the extrapolated end of the jet. This structure resembles the hot spots such as those existing at the extremities of FRII galaxies. However, at Cen A it lies at the side of the lobe rather than at the most distant northern edge, and the brightness contrast (hot spot to lobe) is not as extreme~\cite{Burns}.  Estimates of the radio spectral index of synchrotron emission in the hot spot and the observed degree of linear polarization in the same region suggests that the ratio of turbulent to ambient magnetic energy density in the region of the shock is $U \sim 0.4$~\cite{Romero:1995tn}. The broadband radio-to-X-ray jet emission yields an equipartition magnetic field $B_{\mu{\rm G}} \sim 100$~\cite{Honda:2009xd}.\footnote{The usual way to estimate the magnetic field strength in a radio source is to minimize its total energy. The condition of minimum energy is obtained when the contributions of the magnetic field and the relativistic particles are approximately equal (equipartition condition). The corresponding $B$-field is commonly referred to as the equipartion magnetic field.} The radio-visible size of the hot spot can be directly measured from the large scale map $R_{\rm kpc}\simeq 2$~\cite{Junkes}. The actual size can be larger because of uncertainties in the angular projection of this region along the line of sight.\footnote{For example, an explanation of the apparent absence of a counterjet in Cen A via relativistic beaming suggests that the angle of the visible jet axis with respect to the line of sight is at most 36$^{\circ}$~\cite{Burns}, which could lead to a doubling of the hot spot radius. It should be remarked that for a distance of 3.4 Mpc, the extent of the entire source has a reasonable size even with this small angle.} Replacing these fiducial values in (\ref{EMAX}) and (\ref{ab})  we conclude that if the ratio of photon to magnetic energy density $a \lesssim 0.4$, it is plausible that Cen A can accelerate protons up  $E \approx 2 \times 10^{11}$~GeV.

EGRET observations of the gamma ray flux for energies $>100~{\rm MeV}$  allow an estimate $L_{\gamma} \sim 10^{41}\ \es$ for Cen A~\cite{Sreekumar:1999xw}. This value of $L_{\gamma}$ is consistent with an earlier observation of photons in the TeV-range during a period of elevated $X$-ray activity~\cite{Grindlay}, and is considerably smaller than the estimated bolometric luminosity $L_{\rm bol}\sim 10^{43}\es$\cite{Israel}.  Recent data from H.E.S.S. have confirmed Cen A as a TeV $\gamma$-ray emitting source~\cite{Aharonian:2009xn}. Extrapolating the spectrum measured with EGRET in the GeV regime to VHEs roughly matches the H.E.S.S. spectrum, though the softer end of the error range on the EGRET spectral index is preferred. More recent data from Fermi-LAT  established  that a large fraction ($>1/2$) of the total $> 100~{\rm MeV}$ emission from Cen A emanates from the lobes~\cite{Collaboration:2010xd}.  For values of $B$ in the 100~$\mu$G range, substantial proton synchrotron cooling is suppressed, allowing production of high energy electrons through photomeson processes. The average energy of synchrotron photons scales as~\cite{Ginzburg:1965su}
\begin{equation}
\langle E_\gamma^ {\rm syn} \rangle \simeq 0.29 \ E_\gamma^ {\rm syn} \, ,
\label{Ginzburg}
\end{equation}
and therefore, to account for the observed TeV photons Cen A should harbor a population of ultra-relativistic electrons, with $E_e \sim 10^{9}$~GeV. We further note that this would require the presence of protons with energies between one and two orders of magnitude larger, since the electrons are produced as secondaries.

\subsubsection{GRBs}

GRBs are flashes of high energy radiation that can be brighter, during
their brief existence, than any other source in the sky. The bursts
present an amazing variety of temporal profiles, spectra, and
timescales~\cite{Fishman:95}. Our insights into this
phenomenon have been increased dramatically by BATSE observations
of over 2000 GRBs, and more recently, by data from SWIFT.

There are several classes of bursts, from single-peaked events,
including the fast rise and exponential decaying (FREDs) and their
inverse (anti-FREDs), to chaotic structures~\cite{Link:1996ss}. There
are well separated episodes of emission, as well as bursts with
extremely complex profiles. Most of the bursts are time asymmetric, but
some are symmetric. Burst timescales range from about 30~ms to
several minutes.

The GRB angular distribution appears to be isotropic, suggesting a
cosmological origin~\cite{Meegan:xg}. Furthermore, the detection of
``afterglows'' --- delayed low energy (radio to $X$-ray) emission ---
from GRBs has confirmed this via the redshift determination of several
GRB host-galaxies~\cite{Metzger:1997wp}.

The $\gamma$-ray luminosity implied by cosmological distances is astonishing: $L_\gamma \sim 10^{52}$~erg/s. The most popular interpretation of the GRB-phenomenology is that the observable effects are due to the dissipation of the kinetic energy of a relativistic expanding plasma wind, a ``fireball''~\cite{Piran:kx}.  Although the primal cause of these events is not fully understood, it is generally believed to be associated with the core collapse of massive stars (in the case of long duration GRBs) and stellar collapse induced through accretion or a merger (short duration GRBs)~\cite{Meszaros:2006rc}.

The very short timescale observed in the light curves indicates an extreme compactness ({\em i.e.} distance scale comparable to a light-ms: $r_0\sim10^7$~cm) that implies a source which is initially opaque (because of $\gamma \gamma$ pair creation) to $\gamma$-rays
\begin{equation}
\tau_{\gamma\gamma}\sim r_0 n_\gamma^{\rm GRB}\, \sigma_T\sim{\sigma_TL_\gamma
\over4\pi \, r_0 \, c \, \epsilon_\gamma}\sim10^{15}\, , \label{eq:tau-pair}
\end{equation}
where $n_\gamma^{\rm GRB} $ is number density of photons at the source and $\epsilon_\gamma \simeq1$~MeV is the characteristic photon energy.

The high optical depth creates the fireball: a thermal plasma of photons, electrons, and positrons. The radiation pressure on the optically thick source drives relativistic expansion (over a time scale $r_0/c$), converting internal energy into the kinetic energy of the inflating shell.  As the source expands, the optical depth is reduced. If the source expands with a Lorentz factor $\Gamma$, the energy of photons in the source frame is smaller by a factor $\Gamma$ compared to that in the observer frame, and most photons may therefore be below the pair production threshold.  Baryonic pollution in this expanding flow can trap the radiation until most of the initial energy has gone into bulk motion with Lorentz factors of $\Gamma \sim 10^2 - 10^3$~\cite{Waxman:2003vh}.  The kinetic energy can be partially converted into heat when the shell collides with the interstellar medium or when shocks within the expanding source collide with one another. The randomized energy can then be radiated by synchrotron radiation and inverse Compton scattering yielding non-thermal bursts with timescales of seconds at large radii, $r >10^{12}~{\rm cm}$, beyond the Thompson sphere.  Charged particles may be efficiently accelerated to ultrahigh energies in the fireball's internal shocks, hence GRBs are often considered as potential sources of UHECRs~\cite{Waxman:1995vg}.

Coburn and Boggs~\cite{NATURE}  reported the detection of
polarization,  a particular orientation of the electric-field
vector, in the $\gamma$-rays observed from a burst.
 The radiation released through synchrotron
emission is highly polarized, unlike in other previously suggested
mechanisms such as thermal emission or energy loss by
relativistic electrons in intense radiation fields. Thus,
polarization in the $\gamma$-rays from a burst provides direct
evidence in support of synchrotron emission as the mechanism of
$\gamma$-ray production (see also~\cite{Nakar:2003qc}).

Following Hillas criterion,  the Larmor radius $r_L$
should be smaller than the largest scale $l_{\rm GRB}$ over which
the magnetic field fluctuates, since otherwise Fermi acceleration
will not be efficient. One may estimate $l_{\rm GRB}$ as follows.
The comoving time, {\em i.e.} the time measured in the fireball rest
frame, is $t = r/\Gamma c$. Hence, the plasma wind properties
fluctuate over comoving scale length up to $l_{\rm GRB} \sim
r/\Gamma$, because regions separated by a comoving distance larger
than $r/\Gamma$ are causally disconnected. Moreover, the internal
energy is decreasing because of the expansion and thus it is
available for proton acceleration (as well as for $\gamma$-ray
production) only over a comoving time $t$. The typical
acceleration time scale is then~\cite{Waxman:1995vg}
\begin{equation}
\tau_{\rm acc}^{\rm GRB} \sim \frac{r_L}{c}\,  .
\label{accgrb}
\end{equation}
Equation~(\ref{accgrb}) sets a constraint on the
required comoving magnetic field strength, and the Larmor radius
$r_L = E'/eB = E/\Gamma eB$, where $E' = E / \Gamma$ is
the proton energy measured in the fireball frame. This constraint sets a lower limit to the magnetic field carried by the wind, which may be expressed as 
\begin{equation}
\frac{\zeta_B }{\zeta_e } >0.02 \frac{\Gamma_{2.5}^2
E_{20}^2} {L_{ 52}}, \label{larmor}
\end{equation}
where $\Gamma=10^{2.5} \Gamma_{2.5}$,
$L_\gamma=10^{52} L_{52}~{\rm erg} \, {\rm\ s}^{-1}.$ Here, $\zeta_B$ is the fraction of the wind energy density which is carried by the magnetic field, $4 \pi r^2 \Gamma^2 (B^2/8\pi) = \zeta_B L,$ and $\zeta_e$ is the fraction of wind energy carried by shock accelerated electrons. Note that because the electron energy is lost radiatively,  $L_\gamma \approx \zeta_e L$.

 The dominant energy loss process in this case is synchrotron
cooling. Therefore, the condition that the synchrotron loss time
of Eq.~(\ref{tausyn}) be smaller than the acceleration time sets
the upper limit on the magnetic field strength~\cite{Waxman:1995vg}
\begin{equation}
B<3\times10^5 \ \Gamma_{2.5}^{2} \ E_{20}^{-2}~{\rm G}.\label{sync}
\end{equation}
Since the equipartition field is inversely proportional to the
radius $r$, this condition may be satisfied simultaneously with
(\ref{larmor}) provided that the dissipation radius is large
enough, {\em i.e.}
\begin{equation}
r >10^{12} \ \Gamma_{2.5}^{-2 } \ E_{20}^3~{\rm cm}.\label{dis}
\end{equation}
The high energy protons also lose energy   in interaction with the
wind photons (mainly through pion production). It can be shown,
however, that this energy loss is less important than the
synchrotron energy loss~\cite{Waxman:1995vg}.

In summary, a dissipative ultra-relativistic wind, with luminosity and
variability time implied by GRB observations, satisfies the
constraints necessary to accelerate protons to energy $\gtrsim 
10^{11}$~GeV, provided that $\Gamma > 100$, and the magnetic field
is close to equipartition with electrons.

\subsection{Energy losses of baryonic cosmic rays on the pervasive radiation fields}

\subsubsection{Opacity of the CMB to UHECR protons}
\label{CMBopacity} 

Ultrahigh energy protons degrade their energy through Bethe-Heitler (BH) pair 
production $(p \gamma \rightarrow p  e^+  e^-$)  and photopion production ($p \gamma \rightarrow \pi  N$), each successively  dominating as the proton energy increases.  The fractional  energy loss due to interactions with the cosmic background radiation at a redshift $z=0$ is determined by the integral of the nucleon  energy loss per collision multiplied by the probability per unit time for a nucleon collision in an isotropic gas of photons~\cite{Stecker:68}. This integral can be explicitly written as follows,
\begin{equation}
-\frac{1}{E} \frac{dE}{dt} =\frac{c}{2 \Gamma^2}\,\sum_j\, \int_0^{\omega_m} d\omega_r \,\, 
y_j  \, \,  
\sigma_j (\omega_r)\, \, \omega_r \, \int_{\omega_r/2 \Gamma}^{\omega_m} d\omega \, 
\frac{n_\gamma(\omega)}{\omega^2}   \,,
\label{conventions}
\end{equation} 
where $\omega_r$ is the photon energy in the rest frame of the nucleon, and $y_j$ is the inelasticity, {\em i.e.}  the average fraction of the energy lost by the photon to the nucleon in the laboratory frame for the $j$th reaction channel. (Here the laboratory frame is the one in which the CMB  is at a temperature $\approx 2.7$~K.) The sum is carried out over all channels, $n_\gamma(\omega)d\omega$ stands for the number density of photons with energy between $\omega$ and $d\omega$, $\sigma_j(\omega_r)$ is the total cross section of the $j$th interaction channel, $\Gamma$ is the usual Lorentz factor of the nucleon, and $\omega_m$ is the maximum energy of the photon in the photon gas.

Pair production and photopion production processes are only of importance 
for interactions with the 2.7~K  blackbody background radiation~\cite{Berezinsky:wi}. Collisions 
with optical and infrared photons give a negligible contribution. Therefore, 
for interactions with a blackbody field of temperature $T$, the photon 
density is that of a Planck spectrum, so the 
fractional energy loss is given by
\begin{equation}
-\frac{1}{E} \frac{dE}{dt} = - \frac{ckT}{2 \pi^2 \Gamma^2 (c \hbar)^3}
\sum_j \int_{\omega_{0_j}}^{\infty}  d\omega_r \,
\sigma_j (\omega_r) \,y_j \, \omega_r \, \ln ( 1 -  e^{-\omega_r / 2 \Gamma kT}) \,,
\label{phds!}
\end{equation}
where $\omega_{0_j}$ is the threshold energy for the $j$th reaction 
in the rest frame of the nucleon.

At energies $E\ll m_e\,m_p/kT = 2.1 \times 10^{9}$~GeV 
(i.e., $\omega_r/m_e -2 \ll 1$), when the reaction takes place on the photons 
from the high energy tail of the 
Planck distribution, the fraction of energy lost in one collision and the 
cross section can be approximated by the threshold values
\begin{equation}
y_{_{\rm BH}} = 2\,\frac{m_e}{m_p}\,,
\end{equation}
and
\begin{equation}
\sigma_{_{\rm BH}}(\omega_r) = \frac{\pi}{12}\, \alpha\,\, r_0^2 \,\,
\left(\frac{\omega_r}{m_e} - 2\right)^3\,,
\end{equation}
where $\alpha$ is the fine structure constant  
and $r_0$ is the classical radius of the electron~\cite{Berezinsky:wi}.
The fractional energy loss due to pair production for $E \lesssim 10^{9}$~GeV is 
then, 
\begin{equation}
-\frac{1}{E}\, \left(\frac{dE}{dt}\right)_{\rm BH} = 
\frac{16 c}{\pi}\, \frac{m_e}{m_p}\, 
\alpha\, r_0^2\,
\left(\frac{kT}{hc}\right)^3 \, \left(\frac{\Gamma k T}{m_e}\right)^2\, 
\exp \left(-\frac{m_e}{\Gamma kT}\right).
\label{uniden}
\end{equation}
At higher energies ($E > 10^{10}$~GeV) the characteristic time for the energy 
loss due to pair production is $t \approx 5\times 10^9$~yr~\cite{Blumenthal:1970nn}.  In this energy regime, the photopion 
reactions $p \gamma \rightarrow p \pi^0$ and
$p \gamma \rightarrow \pi^+  n$ on the tail of the Planck distribution give 
the main contribution to proton energy loss. The cross sections of these 
reactions are well known and the kinematics is simple.

Photopion production turns on at a photon energy in the proton rest frame of 
145~MeV with a strongly increasing cross section at the $\Delta (1232)$ 
resonance, which decays into the one pion channels $\pi^+ n$ and $\pi^0 p$.  
With increasing energy, heavier baryon resonances occur and the proton 
might reappear only after successive decays of resonances. The most important 
channel of this kind is $p\gamma \rightarrow \Delta^{++} \pi^-$ with 
intermediate $\Delta^{++}$ states leading finally to 
$\Delta^{++} \rightarrow p\pi^+$. $\Delta^{++}$ examples in this category are
the $\Delta (1620)$ and $\Delta(1700)$ resonances. The cross section in this 
region can be described by either a sum or a product of Breit-Wigner 
distributions over the main 
resonances produced in $N \gamma$ collisions considering $\pi N$, 
$\pi \pi N$ and $K\Lambda$ ($\Lambda \rightarrow N \pi$) final 
states~\cite{Barnett:1996hr}. At high energies, $3.0\, {\rm GeV} 
< \omega_r < 183 \, {\rm GeV}$, the CERN-HERA and COMPAS Groups have made 
a fit to the $p\gamma$ cross section~\cite{Montanet:1994xu}. 
The parameterization is
\begin{equation}
\sigma_\pi(\omega_r) = A + B \,\,\ln^2\left(\frac{\omega_r}{{\rm GeV}}\right) + C \,\,
\ln \left(\frac{\omega_r}{{\rm GeV}}\right) \,\,{\rm mb}\,,
\end{equation} 
where $A = 0.147 \pm 0.001$, $B = 0.0022\pm 0.0001$, 
and $C = -0.0170 \pm 0.0007$. In this energy range, the 
$\sigma_{{\rm total}} (n\gamma)$ is 
to a good approximation identical to $\sigma_{{\rm total}} (p\gamma)$. 

We turn now to the kinematics of photon-nucleon interactions. The inelasticity $y_\pi$ depends not only on the outgoing particles but also  on the kinematics of the final state. Nevertheless, averaging over final state kinematics leads to a good approximation of $y_\pi$.  The 
c.m. system quantities (denoted by $*$) are determined from the relativistic 
invariance of the square of the total 4-momentum $p_\mu p^\mu$ of 
the photon-proton system. This invariance leads to the relation
\begin{equation}
s=(\omega^* + E^*)^2 = m_p^2 + 2m_p \omega_r.
\end{equation}
The c.m. system energies of the particles are uniquely determined 
by conservation of energy and momentum. 
For reactions mediated by resonances one can assume a decay, which in the 
c.m. frame is symmetric in the forward and backward directions with 
respect to the collision axis (given by the incoming particles). 
For instance, we consider single pion production via the reaction
$p  \gamma \rightarrow \Delta \rightarrow p  \pi.$ Here,
\begin{equation}
E_\Delta^{*} = \frac{(s + m_\Delta^2 - m_\pi^2)}{2\, \sqrt{s}}\,.
\end{equation}
Thus, the mean energy of the outgoing proton is
\begin{equation}
\langle E_p^{*{\rm final}} \rangle = \frac{(s + m_\Delta^2 - m_\pi^2)}{2
\,\sqrt{s}\,m_\Delta} \frac{(m^2_\Delta + m_p^2 - m_\pi^2)}{2\,m_\Delta},
\end{equation}
or in the lab frame
\begin{equation}
\langle E_p^{\rm final} \rangle =\frac{E}{s} \frac{(s-m_\pi^2+m_\Delta^2)}{ 2\,m_\Delta}
\frac{(m_\Delta^2 - m_\pi^2 + m_p^2)}{2\, m_\Delta}.
\label{6}
\end{equation}
The mean inelasticity $y_\pi = 1 - (\langle E^{\rm final}\rangle/E)$ of a reaction that provides a proton after $n$ resonance decays can be obtained by straightforward generalization of Eq.~(\ref{6}), and is given by 
\begin{equation}
y_\pi(m_{R_0}) = 1 - \frac{1}{2^n} \prod_{i=1}^{n} \left( 1 + \frac{m_{R_{_i}}^2 -
m_M^2}{m_{R_{_{i-1}}}^2} \right)\,,
\label{kj}
\end{equation}
where $m_{R_{_i}}$ denotes the mass of the $i^{\rm th}$ resonant system of the decay chain, $m_M$ the mass of the associated meson, $m_{R_{_0}} = \sqrt{s}$ is the total energy of the reaction in the c.m., and $m_{R_{_n}}$ the mass of the nucleon.  For multi-pion production the case is much more complicated because of the non-trivial final state kinematics. However, it is well established experimentally~\cite{Golyak:cz} that, at very high energies ($\sqrt{s} \gtrsim 3$~GeV), the incoming particles lose only one-half their energy via pion photoproduction independently of the number of pions produced, $y_\pi \sim 1/2$. This is the ``leading particle effect''.

\begin{figure}
\postscript{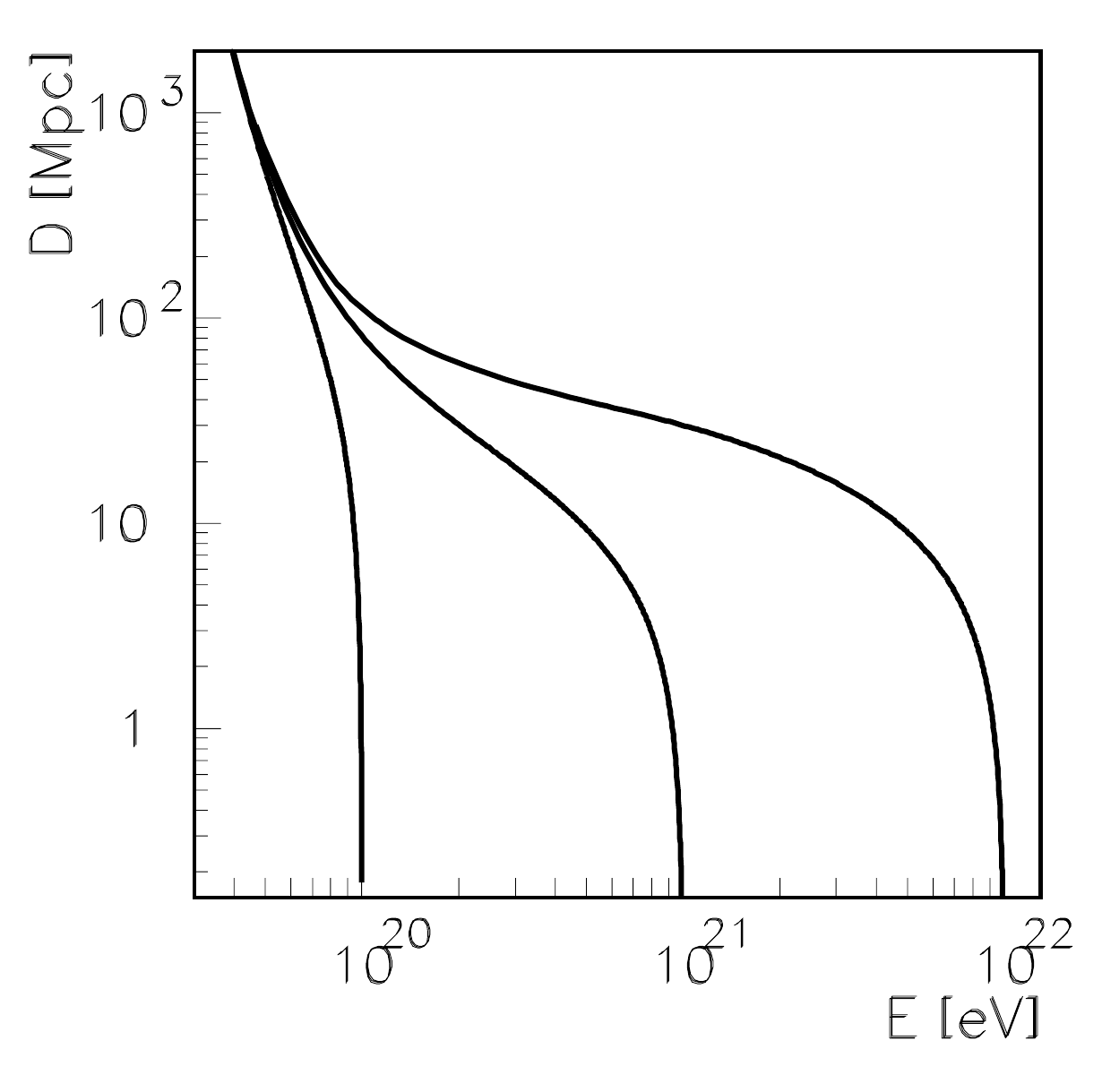}{0.7}
\caption{Energy attenuation length of protons in the intergalactic
medium. Note that after a distance of $\sim 100$~Mpc, or propagation time 
$\sim 3 \times 10^{8}$~yr, the mean energy is essentially independent of the 
initial energy of the protons, with a critical energy around $10^{11}$~GeV. 
From Ref.~\cite{Anchordoqui:1996ru}.} 
\label{gzk}
\end{figure}

For $\sqrt{s} < 2$ GeV, the best maximum likelihood fit to Eq.~(\ref{phds!}) 
with the exponential behavior
\begin{equation} 
- \frac{1}{E}\,\left(\frac{dE}{dt}\right)_\pi = A \, {\rm exp} [ - B / E ]\,,
\label{okop}
\end{equation}
derived from the values of cross section 
and fractional energy loss at threshold, gives~\cite{Anchordoqui:1996ru} 
\begin{equation}
A = ( 3.66  \pm 0.08 )
\times 10^{-8} \, {\rm yr}^{-1}, \,\,\,\,\, B = (2.87 \pm 0.03 )\times
10^{11}\, {\rm GeV} \, .
\label{parame}
\end{equation} 
The fractional energy loss due to production of multipion final
states at higher c.m. energies ($\sqrt{s} \gtrsim 3$~GeV) 
is roughly a constant, 
\begin{equation}
-\frac{1}{E}\,\left(\frac{dE}{dt}\right)_\pi = C = ( 2.42 \pm 0.03 ) 
\times 10^{-8}  \,\, {\rm yr}^{-1} \,.
\label{ce}
\end{equation}
From the values determined for the fractional energy loss, it is 
straightforward to compute the energy degradation of UHECRs in terms of their flight time. This is given by,
\begin{equation}
A \, t \, - \,  {\rm Ei}\,(B/E) 
+ \, {\rm Ei}\, (B/E_0) 
= 0\,,\,\,\,\,\,\,  {\rm for} \,\,10^{10}\,{\rm GeV} \lesssim E \lesssim 10^{12} \,{\rm GeV} \,,
\label{degradacion}
\end{equation}
and
\begin{equation}
E (t) = E_0 \exp[- \,C \ t \,]\,,\,\,\,\,\,\, {\rm for}\,\, E \gtrsim 10^{12} \, 
{\rm GeV}\,,
\end{equation}
where Ei is the exponential integral~\cite{Abramowitz}. Figure~\ref{gzk} shows the proton energy degradation as a function of the mean flight distance. Notice that, independent of the initial energy of the nucleon, the mean energy values approach $10^{11}$~GeV after a distance of $\approx 100~{\rm Mpc}$.

\subsubsection{Photonuclear interactions}
\label{nuclei}

 The relevant mechanisms for the energy loss that extremely high energy nuclei suffer during their trip to Earth are: Compton interactions, pair production in the field of the nucleus, photodisintegration, and hadron photoproduction. The Compton interactions have no threshold energy.  In the nucleus rest-frame, pair production has a threshold at $\sim 1$~MeV, photodisintegration is particularly important at the peak of the GDR  (15 to 25~MeV), and photomeson production has a threshold energy of $\sim 145$~MeV.

Compton interactions result in only a negligibly small energy loss for 
the nucleus given by~\cite{Puget:nz}
\begin{equation}
-\frac{dE}{dt} = \frac{Z^4}{A^2} \rho_\gamma \left( \frac{E}{A m_p c^2}
\right)^2 \, \,\, {\rm eV\, s^{-1}} \,
\end{equation}
where $\rho_\gamma$ is the energy density of the ambient photon field in 
eV cm$^{-3}$, $E$ is the total energy of the nucleus in eV,
and $Z$ and $A$ are the atomic number and weight of the nucleus. 
The energy loss rate due to photopair production is $Z^2/A$ 
times higher than for a proton of the same Lorentz factor~\cite{Chodorowski}, 
whereas the 
energy loss rate due to photomeson production remains roughly the same. 
The latter is true because the cross section for photomeson production by 
nuclei is proportional to the mass number $A$~\cite{Michalowski:eg}, 
while the inelasticity is 
proportional to $1/A$. However, it is photodisintegration rather than 
photopair and photomeson production that determines the energetics of 
UHECR nuclei. During this process some fragments of 
the nuclei are released, mostly single neutrons and protons.  
Experimental data of photonuclear interactions are consistent with a 
two-step process: photoabsorption by the nucleus to form a compound state, 
followed by a statistical decay process involving the emission of one or more 
nucleons.

Following the conventions of Eq.~(\ref{conventions}), the 
disintegration rate with production of $i$ nucleons is given 
by~\cite{Stecker:fw}
\begin{equation}
R_{Ai} = \frac{1}{2 \Gamma^2} \int_0^{\infty} d\omega \,
\frac{n_\gamma(\omega)}{\omega^2} \, \int_0^{2\Gamma \omega} d\omega_r
 \, \omega_r \sigma_{Ai}(\omega_r)
\label{phdsrate}
\end{equation}
with $\sigma_{Ai}$ the cross section for the interaction. 

The photoabsorption cross section roughly obeys the Thomas-Reiche-Kuhn (TRK) dipole sum rule \begin{equation} \Sigma_d \equiv \int_0^\infty \sigma(\omega_r) d\omega_r = 59.8\,\frac{N\,Z}{A}\,\, {\rm MeV\, mb}\,, \end{equation} where $N = A - Z$ is the number of neutrons. (Indeed, this integral is experimentally $\sim 20-30\%$ larger, e.g.  for $^{56}$Fe, $1,020$~mb-MeV for the left hand side, 22\% larger than the right hand side~\cite{Hayward:62}.)  These cross sections contain essentially two regimes. At $\omega_r < 30$~MeV there is the domain of the GDR where disintegration proceeds mainly by the emission of one or two nucleons. A Gaussian distribution in this energy range is found to adequately fit the cross section data~\cite{Puget:nz}. At higher energies, the cross section is dominated by multinucleon emission and is approximately flat up to $\omega_r \sim 150$~MeV. Specifically, \begin{equation}
  \sigma_{Ai}= \frac{\xi_{Ai} \Sigma_d \, \Theta(w_r-2) \,
    \Theta(30 - \omega_r
    )
    \, e^{ -2 ( \omega_r -\epsilon_{0i})^2/\Delta_i^2}}{W \Delta_i}
 + \frac{f_i \Sigma_d \, \Theta(\omega_r- 30)}{120}\,,
\label{cs1}
\end{equation}
for $i$=1, 2, and 
\begin{equation}
\sigma_{Ai} = \frac{f_i \Sigma_d \, \Theta(\omega_r - 30)}{120}\,,
\label{cs2}
\end{equation}
for $i > 2$~\cite{Puget:nz}. 
Here, $W$ is a normalization factor given by
\begin{displaymath}
W = \left(\frac{\pi}{8}\right)^2 \left[ \Phi(\sqrt{2}(30 -
\epsilon_{0i} )/\Delta_i) + \Phi (\sqrt{2}(\epsilon_{0i}-2 )/\Delta_i) \right],
\end{displaymath}
$\Phi(x)$ is the error function, and  $\Theta (x)$
the Heaviside step function. The dependence of the 
width $\Delta_i$, the peak energy $\epsilon_{0i}$, the branching 
ratio $f_i$, and 
the dimensionless integrated cross section $\xi_i$ are given in 
Ref.~\cite{Puget:nz} for isotopes up to $^{56}$Fe. 

The photon background relevant for nucleus disintegration consists 
essentially of photons of the 2.7~K CMB. The background of optical 
radiation turns out to be of (almost) no relevance for UHECR 
propagation. The cosmic infrared background (CIB) radiation~\cite{Salamon:1997ac}
\begin{equation}
\frac{dn_\gamma (\omega)}{d\omega} = 1.1 \times 10^{-4} \,
\left(\frac{\omega}{{\rm eV}}\right)^{-2.5}\,\, {\rm cm}^{-3} \,{\rm eV}^{-1}\,, 
\end{equation}
only leads to sizeable effects far below $10^{11}$~GeV and for time-scales 
${\cal O}$ ($10^{17}$ s)~\cite{Epele:1998ia}. 

By substituting Eqs. (\ref{cs1}) and (\ref{cs2}) into Eq.
(\ref{phdsrate}) the photodisintegration rates on the CMB can be expressed as 
integrals of two basic forms. The first one is
\begin{equation}
I_1=\frac{{\cal A}}{2 \Gamma^2 \pi^2 \hbar^3 c^2}
\left[\int_{1/\Gamma}^{15/\Gamma} d\omega
(e^{\omega/kT}-1)^{-1} {\cal J} + \int_{15/\Gamma}^{\infty} d\omega
(e^{\omega/kT}-1)^{-1} {\cal J'}\right]\,,
\end{equation}
where the functions ${\cal J}$ and ${\cal J'}$ are given by the expressions,
\begin{eqnarray}
{\cal J} & = & \sqrt{\frac{\pi}{8}} \epsilon_{0i} \Delta_i \left[
 \Phi
(\sqrt{2}(2 \Gamma \omega - \epsilon_{0i} )/\Delta_i) +\Phi
(\sqrt{2}(\epsilon_{0i}-2) / \Delta_i) \right] \\  \nonumber
 & + & \left(\frac{\Delta_i}{2} \right)^2 \left\{ e^{ -2 (
(\epsilon_{0i}-2)/\Delta_i)^2} - e^{ -2 ((2\Gamma \omega -\epsilon_{0i})/
\Delta_i)^2} \right\}\,,
\end{eqnarray}
and
\begin{eqnarray}
{\cal J'} & = & \sqrt{\frac{\pi}{8}} \epsilon_{0i} \Delta_i \left[ \Phi
(\sqrt{2}(30 - \epsilon_{0i} )/\Delta_i) +\Phi
(\sqrt{2}(\epsilon_{0i}-2) / \Delta_i) \right] \\ \nonumber
 & + & \left(\frac{\Delta_i}{2} \right)^2 \left\{ e^{ -2 (
(\epsilon_{0i}-2)/\Delta_i)^2} - e^{ -2 ((30 -\epsilon_{0i})/
\Delta_i)^2} \right\}.
\end{eqnarray}
The second basic integral is of the form
\begin{equation}
I_2 = (\pi^2 \hbar^3 c^2)^{-1} \sigma_{Ai}\left[
\int_{15/\Gamma}^{\infty} \frac{\omega^2
d\omega}{e^{\omega/kT} - 1} - \left( \frac{15}{\Gamma}\right)^2
\int_{15/\Gamma}^{\infty} \frac{
d\omega}{e^{\omega/kT} - 1}\right].
\end{equation}
With this in mind, Eq.~(\ref{phdsrate}) can be re-written as~\cite{Anchordoqui:ed}
\begin{eqnarray}
R_{Ai} & = & \frac{1}{\pi^2 \hbar^3 c^2 \Gamma^2}
\left\{ \frac{{\cal A}}{2}  \left(  \frac{\pi}{8} \right)^{1/2}
\epsilon_{0i} \Delta_i \left[
e^{-2 \epsilon_{0i}^2/\Delta_i^2}  {\cal S}_1  + {\cal S}_2  \right] -
 \frac{{\cal A }}{2} \, {\cal J'} k T  \ln (1 - e^{-15/\Gamma k T})
\right.   \nonumber \\
&  &  -  \frac{{\cal A}}{8}\,
e^{-2 \epsilon_{0i}^2/\Delta_i^2} \left(\frac{\pi}{32}\right)^{1/2}
\frac{ \Delta_i^3}{ \Gamma}
   {\cal S}_3 +  \frac{{\cal A}}{2} \,
{\cal K} k T \left[
\ln (1 - e^{-15/\Gamma k T}) - \ln (1 - e^{-1/\Gamma
k T}) \right]  \nonumber \\
 & & + \left. \frac{f_i \Sigma_d }{120}  \left[  \Gamma^2 {\cal S}_4  +
15^2 k T \ln (1 - e^{-15/\Gamma k T})
\right] \right\}\,,
\label{R}
\end{eqnarray}
with ${\cal A}$, ${\cal S}_i$, and ${\cal K}$ as given in Table~\ref{TII}. 
Summing over all the possible channels for a given
number of nucleons, one obtains the effective nucleon loss rate 
$R_A = \sum_i i R_{Ai}$. The effective nucleon loss rate for 
light elements, as well as for those in  the carbon, silicon 
and iron groups
can be scaled as~\cite{Puget:nz}
\begin{equation}
\left. \frac{dA}{dt}\right|_A \sim \left. \frac{dA}{dt}\right|_{^{56}{\rm Fe}} \left(\frac{A}{56}\right) =  R_{56} \left(\frac{A}{56}\right) \, .
\end{equation}
with the photodisintegration rate (\ref{R})  parametrized by~\cite{Anchordoqui:1997rn}
\begin{equation}
R _{56} (\Gamma) =3.25 \times 10^{-6}\, 
\Gamma^{-0.643}                                          
\exp (-2.15 \times 10^{10}/\Gamma)\,\, {\rm s}^{-1} 
\label{oop}
\end{equation}
for $\Gamma \,\in \, [1.0 \times 10^{9}, 36.8 \times 10^{9}]$, and 
\begin{equation}
R_{56}(\Gamma) =1.59 \times 10^{-12} \, 
\Gamma^{-0.0698}\,\, {\rm s}^{-1}   
\end{equation}
for $ \Gamma\, 
\in\, 
[3.68 \times 10^{10}, 10.0 \times 10^{10}]$. \\

{\bf EXERCISE 2.1} Approximating the cross section in Eq.~(\ref{cs1}) by the single pole of the Narrow-Width Approximation~\cite{Anchordoqui:2006pe}
\begin{equation}
\sigma_A(\omega') = \pi \sigma_0  \, \frac{\Gamma_{\rm \! GDR}}{2} \, \delta(\omega' - \omega_0) \,,
\end{equation}
show that for interactions with the CMB photons 
\begin{equation}
R_A \approx \frac{\sigma_0 \, \omega_0' \, \Gamma_{\rm \! GDR} \ T}{4 \Gamma^2 \pi} \, \left|\ln \left(1 - e^{-\omega'_0/2 \Gamma T} \right) \right| \, ,
\end {equation}
where $\sigma_0/A = 1.45\times 10^{-27} {\rm cm}^2$, $\Gamma_{\rm \! GDR} = 8~{\rm
  MeV}$, and  $\epsilon'_0 = 42.65 A^{-0.21} \, (0.925 A^{2.433})~{\rm
  MeV},$ for $A > 4$ ($A\leq 4$)~\cite{Karakula:1993he}. Verify that for $^{56}$Fe this solution agrees to within 20\% with the parametrization given in Eq.~(\ref{oop}). 

\begin{table}
\caption{Series and functions of Eq.~(\ref{R}). \label{TII}}

\hfill

\begin{tabular}{c|c}
\hline
\hline
${\cal A}$ & $W^{-1} \xi_{Ai} \Sigma_d \Delta_i^{-1} $\\
\hline
${\cal S}_1$ &  $  \sum_{_{j=1}}^{^{\infty}} \,\, k T j^{^{-1}}\,
\exp [{\cal B}^{^2}] \,
\{ \Phi( {\cal B} +
15 \sqrt{8}/ \Delta_i) - \Phi ({\cal B} + \sqrt{8} /  \Delta_i)
  \}$ \\
\hline
${\cal S}_2$ & $\sum_{_{j=1}}^{^{\infty}} \,\,k Tj^{^{-1}}\,
\exp\{-j/ \Gamma kT \} [
\Phi (\sqrt{2}(2-\epsilon_{0i})/\Delta_i) -
\Phi (\sqrt{2} ( 30-\epsilon_{0i})/\Delta_i ) ]$ \\
\hline
${\cal S}_3$ &  $  \sum_{_{j=1}}^{^{\infty}} \,\,
\exp [{\cal B}^{^2}] \,
\{ \Phi ({\cal B} + 15 \sqrt{8} /  \Delta_i)
 - \Phi( {\cal B} + \sqrt{8}/ \Delta_i)\}$
\\
\hline
${\cal S}_4$ & $\sum_{_{j=1}}^{^{\infty}}
\exp\{-15j/\Gamma k T\} [ (kT/j)
(15 / \Gamma)^2 +(kT/j)^2
(15/\Gamma) + (kT/j)^3 \,]$ \\
\hline
${\cal B}$ & $j \Delta_i / \Gamma k T \sqrt{32} - 2 \epsilon_{0i} /
\sqrt{2} \Delta_i$\\
\hline
${\cal K}$ & $\sqrt{\frac{\pi}{8}}\,\epsilon_{0i}\,\Delta_i
\,\;\Phi (\sqrt{2}\,(\epsilon_{0i}-2)/\Delta_i) + (\Delta_i/2
)^2 \exp\{-2 (\epsilon_{0i}-2)^2/\Delta_i^2 \}$\\
\hline 
\hline
\end{tabular}
\end{table}

For photodisintegration, the averaged fractional
energy loss results equal the fractional loss in mass number of the 
nucleus, because the nucleon emission is isotropic in the
rest frame of the nucleus. During the photodisintegration process the 
Lorentz factor of the nucleus is conserved, unlike the cases of pair 
production and photomeson production processes which involve the creation 
of new 
particles that carry off energy. The total fractional energy loss 
 is then 
\begin{equation}
-\frac{1}{E} \frac{dE}{dt} = \frac{1}{\Gamma} \frac{d\Gamma}{dt} + \frac{R}{A} \,.
\end{equation}
For $\omega_r \lesssim 145$~MeV the reduction in $\Gamma$ comes from the nuclear 
energy loss due to pair production. The $\gamma$-ray momentum absorbed by the 
nucleus during the formation of the excited compound nuclear state that 
precedes nucleon emission is ${\cal O} (10^{-2})$ times the energy loss by 
nucleon emission~\cite{Stecker:1998ib}. For $\Gamma > 10^{10}$ the  energy loss due to photopair production is 
negligible, and thus
\begin{eqnarray}
  E (t)  & \sim & 938\,\, A(t)\,\, \Gamma\,\,\, {\rm MeV}  \nonumber \\
  & \sim & E_0\, \exp\left[\frac{-  R_{56}(\Gamma)\,t}
  {56}\right]\,.
\label{pats}
\end{eqnarray}
\begin{figure}
  \postscript{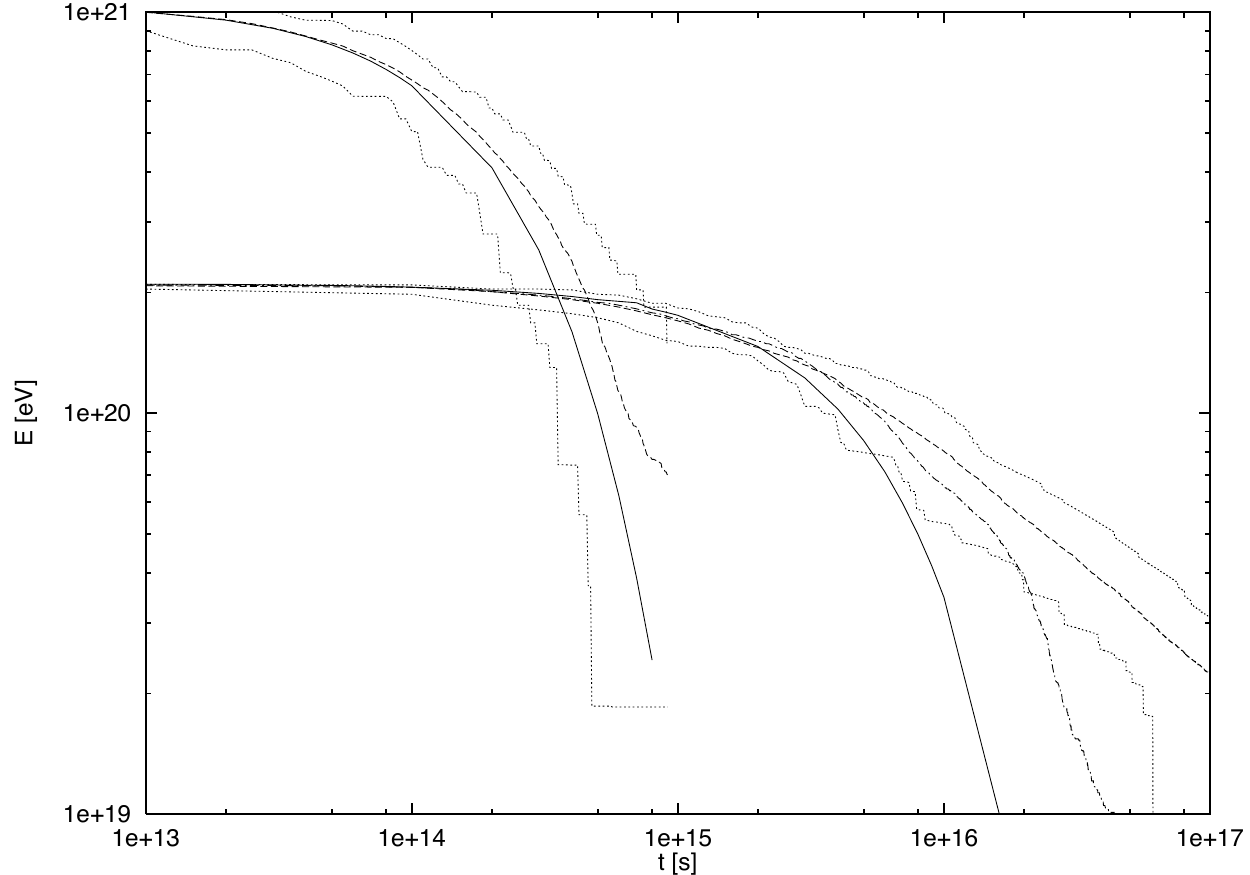}{0.70}
  \caption{The energy of the surviving fragment ($\Gamma_0 = 4 \times
    10^{9}$, $\Gamma_0 = 2 \times 10^{10}$) vs. propagation time    obtained using
    Eq.~(\ref{pats}) is indicated with a solid line. Also included is the energy attenuation
    length obtained from Monte Carlo simulations with (dashed) 
    and without (dotted-dashed) pair creation production, for comparison.
    The region between the two dotted lines includes 95\% of the  simulations. This gives a clear idea of the range of values which can result from fluctuations from the average behaviour. It is important to keep in mind that a light propagation distance of $1.03 \times 10^{14}$~s corresponds to 1~Mpc.
    From Ref.~\cite{Anchordoqui:1998ig}.}
\label{gzk2}
\end{figure}

Figure~\ref{gzk2} shows the energy of the heaviest surviving nuclear
fragment as a function of the propagation time, for initial iron nuclei. 
The solid curves are obtained using Eq.~(\ref{pats}), whereas
the dashed and dotted-dashed curves are obtained by means of
Monte Carlo simulations~\cite{Epele:1998ia}. One can see that nuclei with 
Lorentz factors above $10^{10}$ cannot survive 
for more than 10 Mpc. For these distances, the approximation given in 
Eq.~(\ref{pats}) always lies in the region which includes 95\% of the 
Monte Carlo simulations. When the nucleus is emitted with a Lorentz factor 
$\Gamma_0 < 5 \times 10^9$, pair production
losses start to be relevant, significantly reducing the value of $\Gamma$ 
as the nucleus propagates distances of ${\cal O}(100~{\rm Mpc})$. 
The effect has a maximum for $\Gamma_0 \approx 4 \times 10^{9}$
but becomes small again for $\Gamma_0 \leq 10^{9}$, for which
appreciable effects only appear for cosmological distances ($> 1000$
Mpc), see for instance~\cite{Epele:1998ia}. 

Note that Eq.~(\ref{pats}) imposes a  strong 
constraint on the location of nucleus-sources:  less than 1\%
of iron nuclei (or any surviving fragment of their spallations)
can survive more than $3 \times 10^{14}$~s with an energy $>10^{11.5}$~GeV.
For straight line propagation, this represents a  maximum distance of 
$\sim 3$ Mpc.

\subsection{Diffuse propagation of protons in a magnetized Local Supercluster}
\label{diffusion}

In addition to the interactions with the radiation fields permeating the universe, baryonic CRs suffer deflection and delay in magnetic fields, effects which can camouflage their origins.  For example, the regular component of the Galactic magnetic field can distort the angular images of CR sources: the flux may appear dispersed around the source or globally translated in the sky with rather {\it small} dispersion, {\em viz.} the deflection for CRs of charge $Ze$ and energy $E$ should not exceed $\sim 10^\circ \, Z\, (4\times 10^{10}~{\rm GeV}/E)$~\cite{Harari:1999it}.

One interesting possibility to explain the observed near-isotropy of arrival directions is to envisage a large scale  extragalactic magnetic field that can  provide sufficient bending to the CR trajectories. Surprisingly little is actually known about the extragalactic magnetic field strength. There are some measurements of diffuse radio emission from the bridge area between the Coma and Abell superclusters~\cite{Kim}, which under
assumptions of equipartition allows an estimate of ${\cal
O}(0.2-0.6)\,\mu$G for the magnetic field in this 
region. Fields of ${\cal O}(\mu{\rm G})$ are also indicated in
a more extensive study of 16 low redshift clusters~\cite{Clarke}. It is assumed that the observed $B$-fields result from the amplification of much weaker seed fields. However, the nature of the initial week seed fields is largely unknown. There are two broad classes of models for seed fields: cosmological models, in which the seed fields are produced in the early universe, and astrophysical models, in which the seed fields are generated by motions of the plasma in (proto)galaxies. Of particular interest here is the second class of models.  If most galaxies lived through an active phase in their history, magnetized outflows from their jets and winds would efficiently pollute the extragalactic medium. The resulting $B$-field is expected to be randomly oriented within cells of sizes below the mean separation between galaxies, $\lambda_B \lesssim 1~{\rm Mpc}.$

Extremely weak unamplified extragalactic magnetic fields have escaped detection up to now. Measurements of the Faraday rotation in the linearly polarized radio emission from distant quasars~\cite{Kronberg:1993vk} and/or distortions of the spectrum and polarization properties in the CMB~\cite{Barrow:1997mj,Jedamzik:1999bm} imply upper limits on the extragalactic magnetic field strength  as a function of the reversal scale.  It is important to stress that Faraday rotation measurements (RM)  sample extragalactic magnetic fields of any origin (out to quasar distances), while the CMB analyses set limits {\em only} on primordial magnetic fields. The RM bounds depend significantly on assumptions about the electron density profile as a function of the redshift.  When electron densities follow that of the Lyman-$\alpha$ forest, the average magnitude
of the magnetic field receives an upper limit of $B \sim 10^{-9}$~G for
reversals on the scale of the horizon, and $B \sim 10^{-8}$~G for
reversal scales on the order of 1~Mpc~\cite{Blasi:1999hu}.  As a statistical average over the sky,  an all pervading extragalactic magnetic field  is constrained to be~\cite{Farrar:1999bw}
\begin{equation}
  B \lesssim 3 \times 10^{-7} \,(\Omega_bh^2/0.02)^{-1} \, (h/0.72) \, (\lambda_B/{\rm Mpc})^{1/2}~{\rm G} \, ,
\end{equation}
where $\Omega_b h^2 \simeq 0.02$ is the baryon density and $h \simeq 0.72$ is the present day normalized Hubble expansion rate. This is a conservative bound because $\Omega_b$ has contributions from neutrons and only electrons in ionized gas are relevant to Faraday rotation.

 In the spirit of~\cite{Farrar:2000nw,Anchordoqui:2001nt},  very recently we proposed that neutron emission from Cen A could dominate the observed CR flux above the GZK suppression~\cite{Anchordoqui:2011ks}. Neutrons that are able to decay generate proton diffusion fronts in the intergalactic turbulent magnetic plasma. In our calculations we assume  a strongly turbulent magnetic field: $B = 50~{\rm nG}$, $\lambda_B \sim  1~{\rm Mpc}$, and  largest turbulent eddy $\ell \sim 2 \pi \lambda_B$~\cite{Sigl:1998dd}.   For energies above the GZK supression, $\lambda_B \lesssim r_L \lesssim \ell$ and so  the diffusion coefficient is given by the Bohm formula~\cite{Drury:1983zz}
\begin{equation}
  D(E) = \frac{c r_L}{3} = 0.1  \, \left(\frac{E}{{\rm EeV}}\right) \, \left( \frac{B}{{\rm nG}} \right)^{-1}~{\rm Mpc}^2\ {\rm Myr}^{-1} \, .
\end{equation}
The evolution of the proton spectrum is driven by the so-called ``energy loss-diffusion equation''
\begin{equation}
  \frac{\partial n (E,\vr,t)}{\partial t}  = \frac{\partial[b(E) n (E,\vr,t)]}{\partial E} +
  \bm{\nabla} [D(E,\vr,t) \,\bm{\nabla} n(E,\vr,t)] + Q (E,t) \, \delta^3(x) \,, 
  \label{spectrum}
\end{equation}
where $n(E,\bm{r}, 0) = N_0 \ \delta^3 (x)$.  Here, $b(E) \equiv dE/dt$ is the mean rate at which particles lose energy and $Q(E,t)$ is the number of protons per unit energy and per unit time generated by the source.  For the situation at hand,  $D(E,\vr,t) = D(E)$ and hence the second term becomes $D(E) \nabla^2 n(E,\vr,t)$.
Idealizing the emission to be uniform with a rate $dN_0/dt = N_{\rm tot}/\tau$, we have 
\begin{equation}
  Q (E,t)= \frac{N_{\rm tot}}{\tau}\, \, [\Theta(t- t_{{\rm on}}) - 
  \Theta(t - t_{{\rm off}})],
\end{equation}
where  $\int{\cal  Q}(E,\vr',t') \,d^3x' \,dt' = N_{\rm tot}$, $\Theta$ is the 
Heaviside step function, and $t_{{\rm on}}$ ($t_{{\rm off}}$) is the time 
since the engine turned on (off) its CR production, $t_{{\rm off}} -t_{{\rm on}} = \tau$.  For the energy region of interest, the expected time delay of the diffuse protons, $\tau_{\rm delay} \sim d^2/D(E),$ is significantly smaller than the characteristic time scale for photopion production derived in Eq.~(\ref{okop}).

If the energy loss term  is neglected, 
the solution to Eq.~(\ref{spectrum}) reads,
\begin{equation}
  n (E, \vr, t) = \int dt' \int d^3x' \,\,G(\vr-\vr', t-t') \,\, {\cal Q}(E,\vr', t'),
  \label{ndif}
\end{equation}
where 
\begin{equation}
  G(\vr - \vr', t - t')  =  [4 \,\pi \,D \,(t-t')]^{-3/2}\,\, \Theta (t - t') 
  \exp\{-(\vr - \vr')^2/ 4D(t-t')\}\,,
\end{equation}
is the Green function~\cite{Duff-Naylor}. The density of protons at the present time $t$
of energy $E$
at a distance $\vr$ from Cen A, which is assumed to be continuously emitting at
a constant spectral rate $dN_0/dE\,dt$ from time $t_{\rm on}$ until 
the present, is found to be~\cite{Anchordoqui:2001nt}
\begin{eqnarray}
\frac{dn(E,\vr,t)}{dE} & = &  \frac{dN_0}{dE\,dt}
\frac{1}{[4\pi D(E)]^{3/2}} \int_{t_{\rm on}}^t dt'\,
\frac{e^{-r^2/4D(t-t')}}{(t-t')^{3/2}} \nonumber \\
  & = &  \frac{dN_0}{dE\, dt}  \, \frac{1}{4\pi^{3/2} D(E) r} \int_{v_1}^{v_2} \frac{dv}{v^{3/2}} \, e^{-1/v} \nonumber \\
& = &  \frac{dN_0}{dE\,dt}
\frac{1}{4\pi D(E)r} \,\,I(x)\,,
\label{22}
\end{eqnarray}
where we have used the change of variables
\begin{equation}
u=\frac{r^2}{4D(t-t')} = \frac{1}{v} \,, 
\label{changeofvariables}
\end{equation}
with $x = 4D\ton/r^2$, $\ton=t-t_{\rm on},$ and
\begin{equation}
I(x) = \frac{1}{\sqrt{\pi}} \int_{1/x}^\infty 
\frac{du}{\sqrt{u}} \,\, e^{-u}\ \ . \label{I}
\end{equation}
For $\ton\rightarrow \infty$, the density approaches its time-independent
equilibrium value $n_{\rm eq}$. 

As a result of this diffusion the $J\propto E^{-4}$ behavior of the observed CR spectrum reflects a $dN_0/dEdt\propto E^{-3}$ injection in the region of the source cutoff. For $S = 3000~{\rm km}^2$ detector like Auger, the neutron rate  is
\begin{equation}
\frac{dN_n}{dt} = \frac{S}{4 \pi d^2} \int_{E_1}^{E_2} e^{-d/\lambda(E)} \  \frac{dN_0}{dEdt} \, dE, 
\end{equation}
where $\lambda (E) \simeq 9.2 \times 10^{-3} \, E_{\rm EeV}~{\rm Mpc}$ is the neutron decay length. 
For the energy interval between $E_1 = 55~{\rm EeV}$ and $E_2 = 150~{\rm EeV}$, we  calculate the normalization factor using the observation of 2 neutrons in 3~yr of the nominal  exposure/yr of  Auger.  We then use this normalization factor to calculate the luminosity of the source in the above energy interval. We find
    $L_{\rm CR}^{(E_1, E_2)}    
  =  0.86 \times 10^{40}~ {\rm erg/s} $~\cite{Anchordoqui:2011ks}.
Next, we assume continuity of the spectrum at $E_1$ as it flattens to $E^{-2}.$    
Taking  the lower bound  on the energy to be $E_0 = 10~{\rm EeV}$, we can then fix     
the luminosity for this interval and find                                    
$L_{\rm CR}^{(E_0 , E_1)}   =    
    2.3 \times 10^{40}~{\rm erg/s} . $
Adding these, we find the (quasi) bolometric luminosity to be               
$L_{\rm CR}^{(E_0, E_2)}   
=  3.2 \times 10^{40}~{\rm erg/s},$                                                                                                                          
which is about a factor of 3 smaller than the observed luminosity in  $E > 100~{\rm MeV}$ $\gamma$-rays $L_\gamma \approx 10^{41}~{\rm erg/s}$~\cite{Sreekumar:1999xw}.       
To further constrain the parameters of the model, we evaluate the energy-weighted approximately isotropic proton flux at 70 EeV. If the source actively emitted UHECRs for at least  70~Myr,  from Eq.~(\ref{22}) we obtain 
\begin{eqnarray}
\langle E^4 \, J (E) \rangle & = & \frac{E^4 \, c}{(4 \pi)^2 d D(E)} \, \frac{dN_0}{dE\, dt} \, I(x)  \nonumber \\
 & \approx & 1.6 \times 10^{57}~{\rm eV}^3 \, {\rm km}^{-2}\, \rm {yr}^{-1} \, {\rm sr}^{-1} \, ,
\end{eqnarray}
in agreement with observations~\cite{Abraham:2008ru}.  If we assume circular pixel sizes with $3^\circ$ radii, the neutrons will be collected in a pixel representing a solid angle $\Delta \Omega \simeq 8.6 \times 10^{-3}~{\rm sr}$. The event rate of (diffuse) protons coming from the direction of Cen A  is found to be
\begin{equation}
\frac{dN_p}{dt} = S \ \Delta \Omega \, \int_{E_1}^{E_2} \langle E^4 \, J \rangle \ \frac{dE}{E^4} = 0.08~{\rm events/yr} \, .
\end{equation}
All in all, in the next 9~yr of operation we expect about 6 direct neutron events against an almost negligible background.  Note that our model also predicts  no excess in the direction of M87,
 in agreement with observations (see Fig.~\ref{69}).

We turn now to the discussion of anisotropy. The number of particles with velocity $c$ hitting a unit area in a unit time in a uniform gas of density $n(E, \vr, t)$ is $n(E, \vr, t) \, c$. Due to the gradient in the number density with radial distance from the source, the downward flux at Earth per steradian as a function of the angle $\theta$ to the source is~\cite{Farrar:2000nw}
\begin{equation}
J(\theta, \vr, t) =  \frac{n(E,\vr, t) \, c }{4\pi} (1 + \alpha_d \cos \theta) \, ,
\label{glennys}
\end{equation}
where 
\begin{equation}
\alpha_d \, \cos \theta= \frac{|\bm{j}(E,\vr,t)|}{n(E,\vr,t) \, c} \  .
\end{equation}
The asymmetry parameter $\alpha_d$  can be found by computing the incoming current flux density $\bm{j} = D \bm{\nabla_{r'}} n$ as viewed by an observer on Earth, where $\bm{r} = \bm{R} + \bm{r}'$. We obtain
\begin{eqnarray}
j'_i (E,x'_i,t)& = & D \, \frac{\partial n(E,x'_i,t)}{\partial x'_i} \nonumber \\
     & = & D \frac{dN_0}{dEdt} \frac{1}{(4 \pi D)^{3/2}} \int_{t_{\rm on}}^t \frac{dt'}{(t-t')^{3/2}} e^{-(R^2 + 2 \bm{R} \, .\, \bm{r'} + r'^2)/4 D (t-t')} \, \frac{-(2R_i + 2 x'_i)}{4D(t-t')} \nonumber \\
 & = & - \frac{(R_i + x'_i) }{2 (4 \pi D)^{3/2}} \int_{t_{\rm on}}^t  \frac{dt'}{(t - t')^{5/2}} \, e^{-r^2/[4D (t-t')]} \ .
\end{eqnarray}
Near $x'_i =0$, using the change of variables in Eq.~(\ref{changeofvariables}), we obtain
\begin{eqnarray}
j'_i(E,x'_i,t) & = &  \frac{R_i}{2} \frac{dN_0}{dEdt} \frac{1}{(4 \pi D)^{3/2}} \, \frac{r^2}{4D}  \left(\frac{4D}{r^2}\right)^{5/2} \int_{1/x}^\infty du \, u^{1/2} \, e^{-u} \nonumber \\
 & = & \frac{R_i}{2 \pi}  \, \frac{dN_0}{dEdt} \, \frac{1}{r^3} \,  I'(x) \,,
\end{eqnarray}
where
\begin{equation}
I'(x) = \frac{1}{\sqrt{\pi}} \int_{1/x}^\infty du\ \sqrt{u}
\,\, e^{-u} \, .
\end{equation}
Finally, taking $R_x = R_x =0$ and $R_z = r \, \cos \theta$ we obtain~\cite{Anchordoqui:2001nt}
\begin{equation}
\alpha_d = \frac{2D(E)}{cr} \ \,  \frac{I'(x)}{I(x)} \ .
\label{anisotropy}
\end{equation}
For $\ton\rightarrow \infty$, $0 \leq \alpha \leq 1$. Taking our fiducial values $E=70~ {\rm EeV}$, $B_{\rm nG} = 50$,  and $T_{\rm on} = 70$,  we find $\alpha_d = 0.29$. This is within $\sim 1\sigma$ of the anisotropy amplitude  $\alpha_d = d_\perp/\cos \delta_0 = 0.25 \pm 0.18$ obtained from the 69 arrival directions, assuming a dipole function for a source model with a maximum value at Cen A: $(\alpha_0, \delta_0) = (201.4^\circ, -43.0^\circ)$. 

\begin{figure}
\begin{center}
\postscript{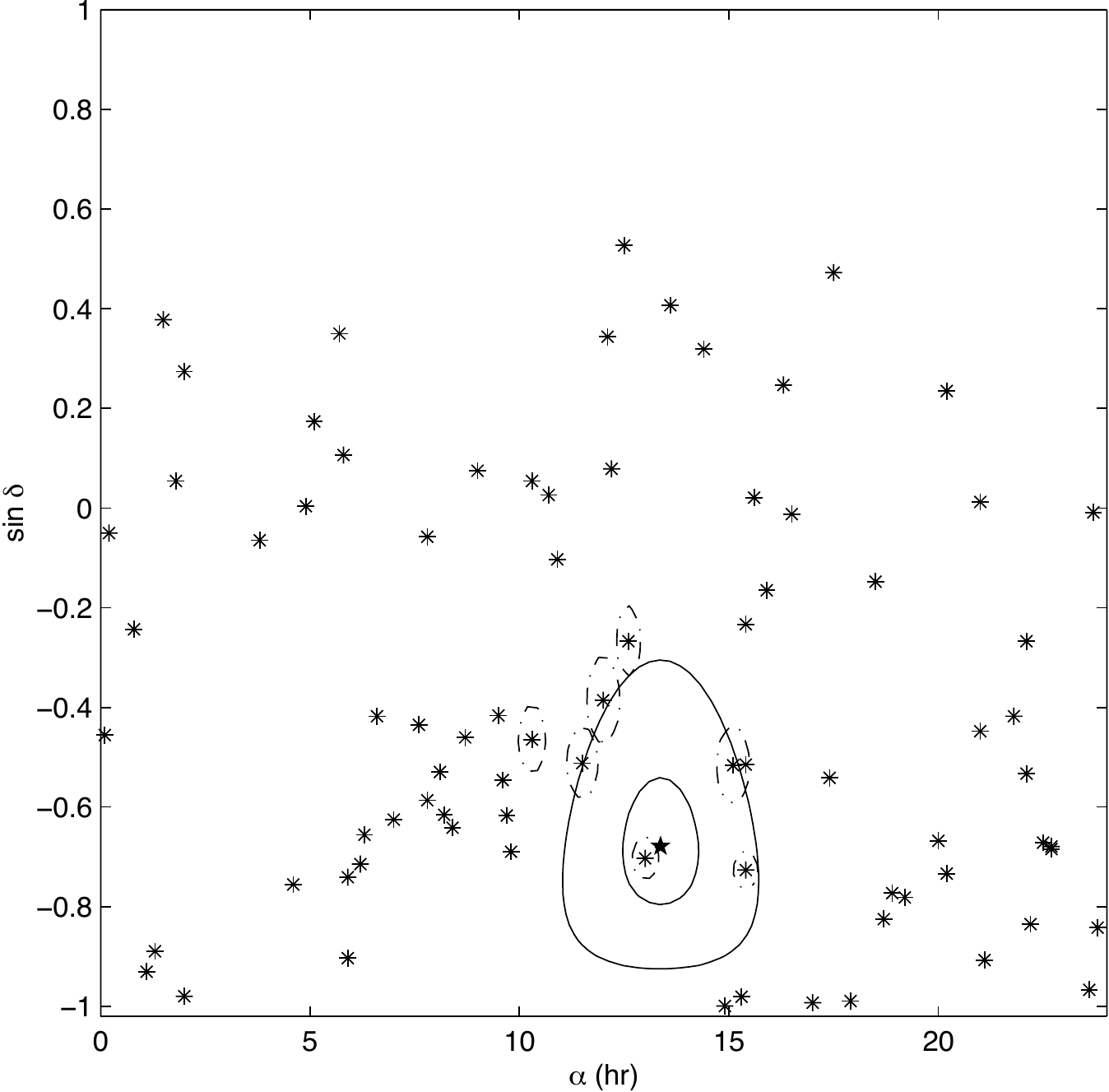}{0.5}
\caption{The
nominal arrival directions ($\alpha =$ right ascension, $\delta =$
declination) of SUGAR events with energies above $4 \times
10^{10}$~GeV~\cite{Winn:1986un}. Also shown in solid lines are contour maps
indicating the circular areas of the celestial sphere centered at
Cen A (indicated by $\star$) with $10^\circ$ and $25^\circ$
radii. The dashed lines surrounding several of the events
indicate the uncertainty of each arrival direction; this was found to be about $3^\circ \ {\rm sec}~\theta,$ where $\theta$ is the shower zenith angle.  From Ref.~\cite{Anchordoqui:2001bs}.  \label{sugar7}}
\end{center}
\end{figure}

One caveat is that we assumed that neutrons completely dominate the ultrahigh energy Cen A emission spectrum; that is 
\begin{equation}
\frac{dN_0}{dE\, dt} = (N_0^n + N_0^p) E^{-3} \,, {\quad} {\rm with} {\quad}  N_0^p/N_0^n \ll 1 \, .
\label{npratio}
\end{equation}
This reduces the number of free parameters in the model. The actual proton-to-neuton fraction depends on the properties of the source, dominantly by the ratio of photon-to-magnetic energy density. The relation (\ref{npratio}) results from $a\sim 0.4,$ which implies that photopion production -- and not proton leakege from the accelerator region -- determines the shape of the cutoff  in the spectrum at the source.\\

{\bf EXERCISE 2.2}  The assymetry parameter given in Eq.~(\ref{anisotropy}) accurately describes the diffuse propagation of CRs if $N_0^n/N_0^p \ll 1$.  Show that asymmetry parameter from the neutron decay shell is given by
\begin{equation}
\alpha_d = \frac{2D(E)}{cr}\, \frac{I''(x)}{I(x)}\  \ \ ,
\label{anisotropya}
\end{equation}
where
\begin{equation}
I''(x) = \frac{1}{4\sqrt{\pi}\kappa} \int_{1/x}^\infty \frac{du}{u^{3/2}}
\left[  \left((1-\kappa)u + \tfrac{1}{2} \right)\ e^{-(1-\kappa)^2u}
 - \left((1+\kappa)u + \tfrac{1}{2}\right)\ e^{-(1+\kappa)^2u}\right]
\label{anisa}
\end{equation}
and $\kappa=\lambda(E)/r$. Show that in spite of the complicated nature of 
Eq.(\ref{anisa}), the results for
$\alpha_d$ are very
similar to the ones for the primary diffusion front given above.\\

An observation demanding of some attention is the distribution of arrival directions of the events collected by the SUGAR experiment, shown in Fig.~\ref{sugar7}.  We take the data at face value keeping in mind that the techniques available at the time the experiment was conducted were not as refined as the techniques currently at our disposal. As one can see in Fig.~\ref{sugar7}, while the Auger and SUGAR distrbutions are not dissimilar, the statistics are rather limited and do not support a firm conclusion.

As an attentive reader should know by now, the observed excess in the direction of Cen A can 
also be explained using proton directional signals and diffusion of heavy nuclei in a $B$-field of about 1 nG~\cite{Piran:2010yg}. However, if this were the case, a larger anisotropy (as yet unobserved)  should be  present at EeV energies~\cite{Lemoine:2009pw}. Regarding the preceeding discussion, one should note that the $3^\circ$ window does not have an underlying theoretical motivation. Recall that this angular range resulted from a scan of parameters to maximize signal significance. Cen A covers an eliptical region spanning about $10^\circ$ along the major axis; see Fig.~\ref{cenapic}. Therefore, some care is required to select the region of the sky which is most likely to maximize the signal-to-noise.

In summary, existing data is consistent with the hypothesis that Cen A dominates the CR sky at the high end of the spectrum~\cite{Farrar:2000nw}.  Auger is in a gifted position to explore Cen A and would provide  in the next 9~yr of operation sufficient statistics to test this hypothesis. The potential detection of neutrons at Auger can subsequently be validated by the larger aperture of the  JEM-EUSO mission.

\subsection{Ultrahigh energy cosmic neutrinos}

\subsubsection{Waxman-Bahcall bound}

It is helpful to envision the CR engines as machines where
protons are accelerated and (possibly) permanently confined by the
magnetic fields of the acceleration region. The production of neutrons
and pions and subsequent decay produces neutrinos, $\gamma$-rays, and
CRs. If the neutrino-emitting source also produces high energy
CRs, then pion production must be the principal agent for the
high energy cutoff on the proton spectrum.  Conversely, since the
protons must undergo sufficient acceleration, inelastic pion
production needs to be small below the cutoff energy; consequently,
the plasma must be optically thin. Since the interaction time for
protons is greatly increased over that of neutrons due to magnetic
confinement, the neutrons escape before interacting, and on decay give
rise to the observed CR flux. The foregoing can be summarized
as three conditions on the characteristic nucleon interaction time
scale $\tau_{\rm int}$; the neutron decay lifetime $\tau_n$; the
characteristic cycle time of confinement $\tau_{\rm cycle}$; and the
total proton confinement time $\tau_{\rm conf}$: $(i)\; \tau_{\rm
  int}\gg \tau_{\rm cycle}$; $(ii)\; \tau_n > \tau_{\rm cycle}$;
$(iii)\; \tau_{\rm int}\ll \tau_{\rm conf}$. The first condition
ensures that the protons attain sufficient energy.  Conditions $(i)$
and $(ii)$ allow the neutrons to escape the source before
decaying. Condition $(iii)$ permits sufficient interaction to produce
neutrons and neutrinos. These three conditions together define an
optically thin source~\cite{Ahlers:2005sn}. A desirable property to
reproduce the almost structureless energy spectrum is that a single
type of source will produce cosmic rays with a smooth spectrum across
a wide range of energy.

The cosmic ray flux above the ankle is often
summarized as ``one $3 \times 10^{10}$~GeV particle per kilometer square per
year per steradian.'' This can be translated into an energy flux~\cite{Gaisser:1997aw}
\begin{eqnarray}
E \left\{ E{J_{\rm CR}} \right\} & = & {3 \times 10^{10}\,{\rm GeV} 
\over \rm (10^{10}\,cm^2)(3\times 10^7\,s) \, sr} \nonumber \\
 & = &  10^{-7}\rm\, GeV\ cm^{-2} \, s^{-1} \, sr^{-1} \,.
\end{eqnarray}
From this we can derive the energy density $\epsilon_{\rm CR}$ in UHECRs using flux${}={}$velocity${}\times{}$density, or
\begin{equation}
4\pi \int  dE \left\{ E{J_{\rm CR}} \right\} =  c \, \epsilon_{\rm CR}\,.
\end{equation}
This leads to
\begin{equation}
  \epsilon_{\rm CR} = {4\pi\over c} \int_{E_{\rm min}}^{E_{\rm max}} { 10^{-7}\over E} 
  dE \, {\rm {GeV\over cm^2 \, s}} \simeq   10^{-19} \, 
{\rm {TeV\over cm^3}} \,,
\end{equation}
taking the extreme energies of the accelerator(s) to be $E_{\rm min} \simeq 10^{10}~{\rm GeV}$ and
$E_{\rm max}  = 10^{12}~{\rm GeV}$. The power required for a population of  sources
to generate this energy density over the Hubble time (${\cal T}_H \approx 10^{10}$~yr) is: 
$\dot \epsilon_{\rm CR}^{[10^{10}, 10^{12}]} \sim 5 \times 10^{44}~{\rm TeV} \, {\rm Mpc}^{-3} \, {\rm yr}^{-1} \simeq 3 \times 10^{37}~{\rm erg} \, {\rm Mpc}^{-3} \, {\rm s}^{-1}$.  This works out to roughly
({\it i}\,) $L \approx 3 \times 10^{39}$ erg s${}^{-1}$ per galaxy, 
({\it ii}\,) $L \approx 3 \times 10^{42}$ erg s${}^{-1}$ per cluster of galaxies, 
({\it iii}\,) $L \approx 2 \times 10^{44}$ erg s${}^{-1}$ per active galaxy, or 
({\it iv}\,) $\int L \, dt\approx  10^{52}$ erg per cosmological GRB~\cite{Gaisser:1997aw}. 
The coincidence between these numbers and
the observed output in electromagnetic energy of these sources
explains why they have emerged as the leading candidates for the
CR accelerators. 

The energy production rate of protons derived professionally, assuming a cosmological distribution of  sources (with injection spectrum  typical of shock acceleration $dN_0/dE \propto E^{-2}$) is~\cite{Waxman:1995dg} 
\begin{equation}
\dot \epsilon_{\rm CR}^{[10^{10}, 10^{12}]} \sim 5 \times 10^{44}~{\rm erg} \, {\rm Mpc}^{-3} \, {\rm yr}^{-1} \, .
\label{professionally}
\end{equation}
This is within a factor of our back-of-the-envelope estimate (1 TeV = 1.6 erg). The energy-dependent generation rate of CRs is therefore given by 
\begin{eqnarray}
E^2  \frac{d \dot n}{dE } 
 & = & \frac{\dot \epsilon_{\rm CR}^{[10^{10}, 10^{12}]}}{\ln(10^{12}/10^{10})} \nonumber \\
& \approx & 10^{44}\,\rm{erg}\,\rm{Mpc}^{-3} \rm{yr}^{-1} \,\, .
\end{eqnarray} 

The energy density of neutrinos produced through $p\gamma$
interactions of these protons can be directly tied to the injection
rate of CRs
\begin{equation}
E^2_{\nu} \frac{dn_{\nu}}{dE_{\nu}}
\approx \frac{3}{8} \epsilon_\pi \, {\cal T}_H \,E^2 \, \frac{d \dot n}{dE} \,,
\end{equation}
where  $\epsilon_\pi$ is the
fraction of the energy which is injected in protons lost into photopion
interactions.  The factor of 3/8 comes from the fact that, close to
threshold, roughly half the pions produced are neutral, thus not
generating neutrinos, and one quarter of the energy of charged pion
decays goes to electrons rather than neutrinos. Namely, resonant $p \gamma$ interactions produce twice as many neutral pions as charged pions. Direct pion production via virtual  meson exchange contributes only about 20\% to the total cross section, but is almost exclusively into $\pi^+$. Hence, $p \gamma$ interactions produce roughly equal numbers of $\pi^+$ and 
$\pi^0$. The average  neutrino energy from the direct pion decay is $\langle E_{\nu_\mu}
  \rangle^\pi = (1-r)\,E_\pi/2 \simeq 0.22\,E_\pi$ and that of the
  muon is $\langle E_{\mu} \rangle^\pi = (1+r)\,E_\pi/2 \simeq
  0.78\,E_\pi$, where $r$ is the ratio of muon to the pion mass
  squared. In muon decay, since the $\nu_\mu$ has about 1/3 of its parent energy,
  the average muon neutrino energy is $\langle E_{\nu_\mu} \rangle^\mu =(1+r)E_\pi/6=0.26 \,
  E_\pi$.

The ``Waxman-Bahcall bound'' is defined by the condition $\epsilon_\pi
=1$
\begin{eqnarray} 
E^2_{\nu} \ \Phi_{\rm WB}^{\nu_{\rm all}}  & \approx & (3/8)
  \,\xi_z\, \epsilon_\pi\, {\cal  T}_H \, \frac{c}{4\pi}\,E^2 \, 
   \frac{d \dot n}{dE}  \nonumber \\ & \approx & 2.3
  \times 10^{-8}\,\epsilon_\pi\,\xi_z\, \rm{GeV}\,
  \rm{cm}^{-2}\,\rm{s}^{-1}\,\rm{sr}^{-1},
\label{wbproton}
\end{eqnarray}
where the parameter $\xi_z$ accounts for the effects of source
evolution with redshift, and is expected to be $\sim
3$~\cite{Waxman:1998yy}. For interactions with the ambient gas ({\em i.e.} 
$pp$ rather than $p \gamma$ collisions), the average fraction of the
total pion energy carried by charged pions is about $2/3$, compared to
$1/2$ in the photopion channel. In this case, the upper bound given
in Eq.~(\ref{wbproton}) is enhanced by 33\%.  Electron antineutrinos can also be 
produced through neutron $\beta$-decay. This contribution, however, turns out to 
be negligible (see Appendix~\ref{AA} for details).

The actual value of the neutrino flux depends on what fraction of the
proton energy is converted to charged pions (which then decay to
neutrinos), {\em i.e.}  $\epsilon_\pi$ is the ratio of charged pion energy to the {\em
  emerging} nucleon energy at the source.  For resonant
photoproduction, the inelasticity is kinematically determined by
requiring equal boosts for the decay products of the
$\Delta^+$, giving $\epsilon_\pi = E_{\pi^+}/E_n
\approx 0.28$, where $E_{\pi^+}$ and $E_n$ are the emerging charged pion
and neutron energies, respectively.  For $pp\rightarrow NN + {\rm
  pions},$ where $N$ indicates a final state nucleon, the inelasticity
is $\approx 0.6$~\cite{Alvarez-Muniz:2002ne}. This then implies that
the energy carried away by charged pions is about equal to the
emerging nucleon energy, yielding (with our definition)
$\epsilon_\pi\approx 1.$

At production, if all muons decay, the neutrino flux consists of equal fractions of
$\nu_e$, $\nu_{\mu}$ and $\bar{\nu}_{\mu}$. Originally, the
Waxman-Bahcall bound was presented for the sum of $\nu_{\mu}$ and
$\bar{\nu}_{\mu}$ (neglecting $\nu_e$), motivated by the fact that
only muon neutrinos are detectable as track events in neutrino
telescopes. Since oscillations in the neutrino sector mix the
different species, we chose instead to discuss the sum of all neutrino
flavors. When the effects of oscillations are accounted for, {\it
  nearly} equal numbers of the three neutrino flavors are expected at
Earth~\cite{Learned:1994wg}. \\

{\bf EXERCISE 2.3}~~~The assumption that GRBs are the sources of the observed UHECRs generates a calculable flux of neutrinos produced when the protons interact with the fireball photons~\cite{Waxman:1997ti}.  In the observer's frame, the spectral photon density (${\rm GeV}^{-1} \, {\rm cm}^{-3}$) can be adequately parametrized by a broken power-law spectrum
$n_\gamma^{\rm GRB} (\epsilon_\gamma) \propto \epsilon_\gamma^{-\beta}$, where $\beta \simeq 1, \ 2$ respectively at energies below and above $\epsilon_\gamma^{\rm break} \simeq   1~{\rm MeV}$~\cite{Band:1993eg}. Show that
\begin{equation}                      
\Phi_{\rm GRB}^{\nu_{\rm all}} (E_\nu > E_\nu^{\rm break}) \sim 10^{-13} \ \left(\frac{E_\nu^{\rm break}}{10^5~{\rm GeV} }\right)^{-1} ~{\rm cm}^{-2} \,          
{\rm s}^{-1} \, {\rm sr}^{-1} \, ,                                                             
\end{equation}      
where $E_\nu^{\rm break} \sim 5 \times 10^{5} \, \Gamma_{2.5}^2 (\epsilon_\gamma^{\rm break}/{\rm MeV})^{-1}~{\rm GeV}.$ Recall that in our convention $\epsilon_\gamma = \Gamma \epsilon'_\gamma$, where $\epsilon'_\gamma$  is the photon energy measured in the fireball frame.  Convince yourself that the non-observation of extraterrestrial neutrinos from sources other than the Sun and SN1987a puts the GRB model of UHECR acceleration on probation~\cite{Ahlers:2011jj}.\\

If the injected cosmic rays include nuclei heavier than protons, then
the neutrino flux expected from the cosmic ray sources may be
modified.  Nuclei undergoing acceleration can produce pions, just as
protons do, through interactions with the ambient gas, so the
Waxman-Bahcall argument would be unchanged in this case. However, if
interactions with radiation fields dominate over interactions with
matter, the neutrino flux would be suppressed if the cosmic rays are
heavy nuclei. This is because the photodisintegration of nuclei
dominates over pion production at all but the very highest energies.
Defining $\kappa$ as the fraction of nuclei heavier than protons in
the observed cosmic ray spectrum, the resulting neutrino flux is then
given by
$  E^2_{\nu} \ \Phi^{\nu_{\rm all}} \approx  (1 - \kappa) \,\,
E^2_{\nu} \ \Phi^{\nu_{\rm all}}_{\rm{WB}} \,\,$~\cite{Anchordoqui:2007tn}. The most up-to-date calculation of $\kappa$ combines a double-fit analysis of the energy and elongation rate measurements to constrain the spectrum and chemical composition of UHECRs at their sources~\cite{Anchordoqui:2007fi}. Injection models with a wide range of chemical composition are found to be consistent with observations. In particular, the data are consistent with a proton dominated spectrum with only a small (1 - 10\%) admixture of heavy nuclei.\footnote{It is important to stress that the essential results of the analysis in Ref.~\cite{Anchordoqui:2007fi} are not altered by the new Auger data~\cite{Abraham:2010yv}.}

By duplicating the Waxman-Bahcall calculation for Cen A,  we obtain an upper limit on the intensity of neutrinos from the direction of the nearest active galaxy,
\begin{eqnarray}
E_\nu^2 \ \phi_{_{\rm Cen\,A}}^{\nu_{\rm all}} & = & \frac{1}{4 \pi d^2} \ L_{\rm CR} \ \frac{3}{8} \ \epsilon_\pi \nonumber \\
 & \approx & 5.0 \times 10^{-9}~{\rm GeV} \ {\rm cm}^{-2} \ {\rm s}^{-1} \, ,
\end{eqnarray}
with $E_\nu  \lesssim 10^{9.5}~{\rm GeV}$. 
For the model introduced in Sec.~\ref{diffusion}, we have $\epsilon_\pi \sim 0.28$ and therefore a  prediction for the all-flavor neutrino flux
\begin{equation}
E_\nu^2 \ \phi_{_{\rm Cen\,A}}^{\nu_{\rm all}}  \approx 1.5 \times 10^{-9}~{\rm GeV} \ {\rm cm}^{-2} \ {\rm s}^{-1} \, ,
\end{equation}
in agreement with the results of~\cite{Halzen:2008vz}. In addition to neutrinos, one also expects a similar flux of photons at the source, which also carries unperturbed directional information. However, photons at these energies are difficult to dig out from the hughe proton background.

Although there are no other nearby FRI of this magnitude which can potentially be detected as point sources, one can integrate over the estimated FRI population out to the horizon to obtain an upper limit for the diffuse FRI neutrino flux. This quantity is given by~\cite{Anchordoqui:2004eu}
\begin{eqnarray}
E_\nu^2  \ \Phi_{\rm FRI}^{\nu_{\rm all}}  & = & \frac{1}{4\pi} \, {\cal R}_H \ n_{_{\rm FRI}} \, L_{\rm CR} \ 
\frac{3}{8} \ \epsilon_\pi \nonumber \\
  & \approx & 1.5 \times 10^{-8}~{\rm GeV} \ {\rm cm}^{-2} \ {\rm s}^{-1} \, {\rm sr}^{-1} \,,
\end{eqnarray}
where ${\cal R}_H \simeq 1$~horizon $\simeq 3~{\rm Gpc}$ and $n_{_{\rm FRI}} \sim 8 \times 10^4~{\rm Gpc}^{-3}$ is the number density~\cite{Padovani}. Note that this flux is about a factor of 3 smaller than the Waxman-Bahcall upper limit. Hence, the reduction in luminosity of the ensemble of
neutrino sources roughly compensates for the presence of distant optically thin sources whose CR components are hidden by  extragalactic magnetic fields. 

The diffuse neutrino flux has an additional component originating in
the energy losses of UHECRs {\em en route} to
Earth~\cite{Beresinsky:1969qj}. The
accumulation of these neutrinos over cosmological time is known as the
cosmogenic neutrino flux.  The GZK reaction chain generating cosmogenic neutrinos is well
known~\cite{Stecker:1978ah}. The intermediate state of the reaction $p
\gamma_{\rm CMB} \to n \pi^+/p\pi^0$ is dominated by the $\Delta^+$
resonance, because the neutron decay length is smaller than the
nucleon mean free path on the CMB. Gamma-rays, produced via $\pi^0$ decay,
subsequently cascade electromagnetically on intergalactic radiation
fields through $e^+ e^-$ pair production followed by inverse Compton
scattering. The net result is a pile up of $\gamma$-rays at GeV-TeV
energies, just below the threshold for further pair production on the
diffuse optical background. Meanwhile each $\pi^+$ decays to 3
neutrinos and a positron; the $e^+$ readily loses its energy through
inverse Compton scattering on the diffuse radio background or through
synchrotron radiation in intergalactic magnetic fields. As we have seen, the neutrinos
carry away about 3/4 of the $\pi^+$ energy, therefore the energy in
cosmogenic neutrinos is about 3/4 of that produced in $\gamma$-rays. The intensity of $\gamma$-ray pile up currently provides the most stringent bound on the flux of cosmogenic neutrinos. It is this that we now turn to study.

\subsubsection{Boltzmann equation,  universal cosmic ray  spectrum, and cosmogenic neutrinos}
\label{Boltzmann}

For a spatially homogeneous distribution of cosmic sources, emitting
UHE particles of type $i$, the comoving number
density, $Y_i (z,E) \equiv n_i(z,E)/ (1+ z)^3,$ is governed by a set of one-dimensional (Boltzman)
continuity equations,
\begin{equation}\label{diff0}
\dot Y_i = \partial_E(HEY_i) + \partial_E(b_iY_i)-\Gamma_{i}\,Y_i+\sum_j\int d E_j\,\gamma_{ji}Y_j+\mathcal{Q}_i\,,
\end{equation}
together with the Friedman-Lema\^{\i}tre equations describing the
cosmic expansion rate $H(z)$ as a function of the redshift
$z$.\footnote{This is given by \mbox{$H^2 (z) =
    H^2_0\,[\Omega_{\mathrm{m} } (1 + z)^3 + \Omega_{\Lambda}]$},
  normalized to its value today of $H_0 \sim 100~h$
  km\,s$^{-1}$\,Mpc$^{-1}$, in the usual ``concordance model'' dominated
  by a cosmological constant with $\Omega_{\Lambda} \sim 0.7$ and a
  (cold) matter component, $\Omega_\mathrm{m} \sim
  0.3$~\cite{Amsler:2008zzb}. The time-dependence of the red-shift can
  be expressed via $\mathrm{d}z = -\mathrm{d} t\,(1+z)H$.}  The first term
on the r.h.s. describes adiabatic energy losses~\cite{Anchordoqui:2002hs}.  The second term describes interactions on the cosmic photon backgrounds which can be approximated by  continuous energy losses. The third and fourth
terms describe more general interactions involving particle losses ($i
\to$ anything) with interaction rate $\Gamma_i$, and particle
generation of the form $j\to i$.  The last term on the r.h.s.,
$\mathcal{Q}_i$, corresponds to the luminosity density per comoving
volume of sources emitting CRs of type $i$. We now
discuss the calculation of these terms and their scaling with
redshift.

The angular-averaged (differential) interaction rate, $\Gamma_i$
($\gamma_{ij}$), appearing on the r.h.s.~of Eq.~(\ref{diff0}) is
defined as
\begin{eqnarray}
\label{Gamma}
\Gamma_{i}(z,E_i) & = & \frac{1}{2}\int\limits_{-1}^1 d \cos\theta\int d \omega\,(1-\beta \cos\theta) n_\gamma(z,\omega)\sigma^\mathrm{tot}_{i\gamma}\,, \\
\gamma_{ij}(z,E_i,E_j) & = & \Gamma_i(z,E_i)\,\frac{d N_{ij}}{d E_j}(E_i,E_j)\,,
\label{gamma}
\end{eqnarray}
where $n_\gamma(z,\omega)$ is the energy distribution of background
photons at redshift $z$ and $d N_{ij}/d E_j$ is the angular-averaged
distribution of particles $j$ after interaction. The factor $(1-\beta\cos\theta)$ takes into account the relativistic Doppler shift of the photon density.

\begin{figure}[t]
\begin{center}
\includegraphics[width=0.5\linewidth]{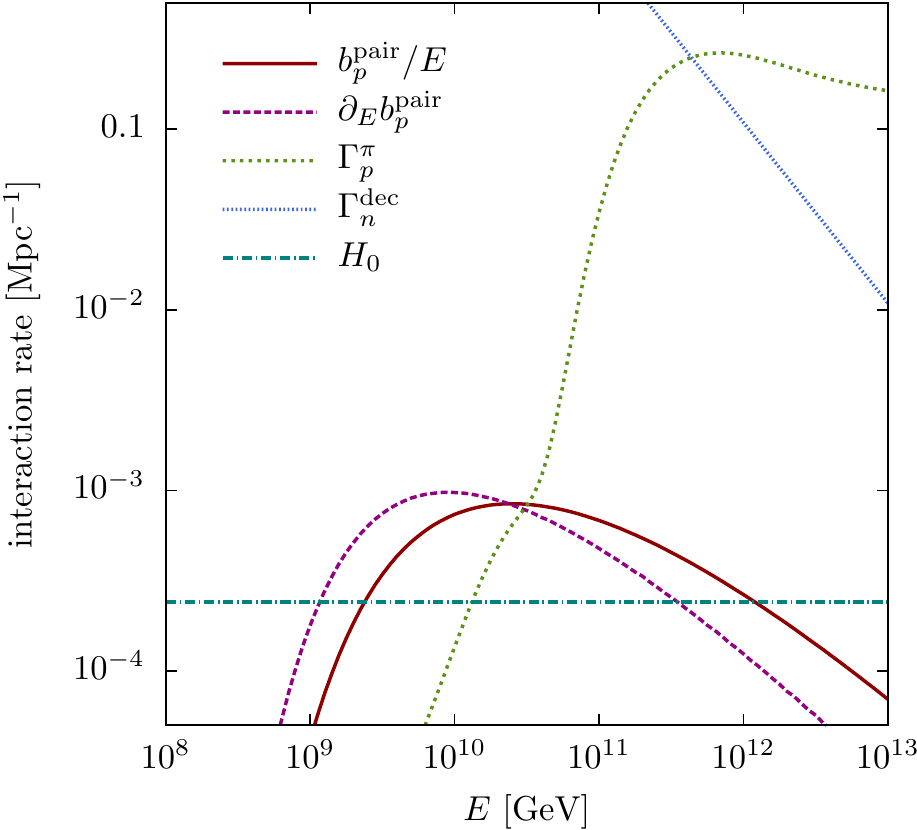}
\hfill
\includegraphics[width=0.465\linewidth]{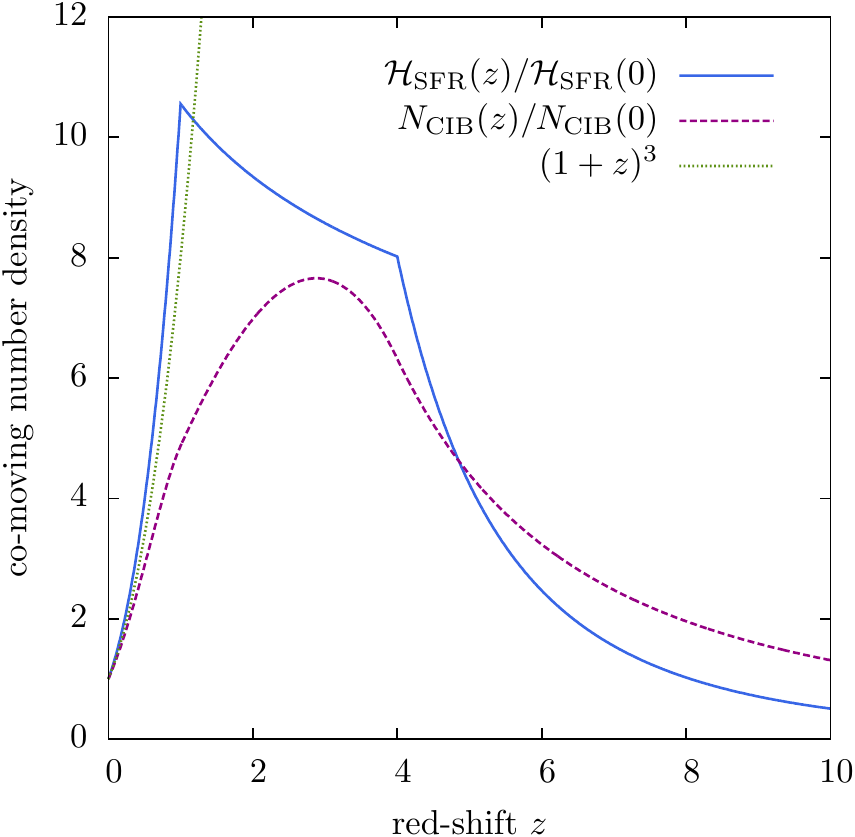}
\end{center}
\vspace{-0.3cm}
\caption[]{{\bf Left:} The interaction and decay rates appearing in
  the Boltzmann equations for the CMB and CIB at
  $z=0$. {\bf Right:} Star formation rate
  (Eq.~(\ref{HSFR}) from Ref.~\cite{Yuksel:2008cu}) compared with our
  approximation (\ref{HIR}) for the CIB photon density scaling with
  redshift.  For comparison, the $\propto (1+z)^3$ scaling of the CMB
  photon density is also shown. From Ref. ~\cite{Ahlers:2009rf}.}
\label{scales}
\end{figure}

In general, any transition $i\to
i$ which can be approximated as
$\gamma_{ii}(E,E')\simeq\delta(E-E'-\Delta E)\Gamma_i(E)$ with
$\Delta E/E \ll 1$ can be replaced in the Boltzmann
equations~(\ref{diff0}) as
\begin{equation}
-\Gamma(E)Y_i(E)+\int d E'\, \gamma_{ii}(E',E)\,Y_i(E')\\\to\partial_E(b_i Y_i)\,, 
\end{equation}
with $\displaystyle b_i \equiv \Delta E\,\Gamma_i \simeq -\dot E$.
The production of electron-positron pairs in the photon background
with a small energy loss is usually approximated as a continuos energy loss (CEL)
process. As we have seen in Sec.~\ref{CMBopacity}, it is also possible to approximate the energy loss in the hadronic cascade
due to photopion production as a CEL 
\begin{equation}\label{eq:CELpgamma}
\frac{d E}{ d t}(z,E) \equiv b_\pi(z,E) 
\simeq E\,\Gamma_{p}(z,E) - \int
dE'E'\gamma_{pp}(z,E,E')\,,
\end{equation}
with $b_\pi(0,E)/E$ given by (\ref{okop}) and (\ref{ce}). Diffractive $p\gamma$ processes at high energies with large final
state multiplicities of neutrons and protons ultimately invalidate the
CEL approximation. However, the relative error below $10^{12}$~GeV is
less than $15\%$, so we will use this approximation for a detailed
numerical scan in the model space of proton spectra. For neutrons with energy less than $10^{11}$~GeV the decay length is
always smaller than the interaction length on the photon backgrounds. It is convenient to approximate the production of neutrons as $\Gamma_{\rm pp}^{\rm eff} \simeq \Gamma_{pp}^\pi + \Gamma_{pn}^\pi$~\cite{Ahlers:2009rf}; we have adopted this in all our calculations. In the left panel of Fig.~(\ref{scales}) we show the
quantities $b^\mathrm{pair}_p/E$, $\partial_Eb^\mathrm{pair}$,
$\Gamma_p$ and $H_0$ for comparison.

The redshift scaling of Eqs.~(\ref{Gamma}) and (\ref{gamma}) depend on how the photon backgrounds  vary with redshift. The CMB spectral density (GeV${}^{-1}$ cm${}^{-3}$) scales
adiabatically as:
\begin{equation}\label{CMB}
n_\gamma(z,\omega) = (1+z)^2\,n_\gamma(0,\omega/(1+z))\,,
\end{equation}
following from \mbox{$\dot Y_\gamma=\partial_E(HEY_\gamma)$} and
$Y_\gamma \propto a^3n_\gamma,$ where $a$ is the cosmic scale factor. The scaling behavior of Eq.~(\ref{CMB})
translates into the following scaling of the quantities $\Gamma_i$ and
$\gamma_{ij}$~\cite{Berezinsky:2002nc}:
\begin{eqnarray}\label{scaling1}
\Gamma_i(z,E_i) &=& (1+z)^3\,\Gamma_i(0,(1+z)E_i)\,,\\\label{scaling2}
\gamma_{ij}(z,E_i,E_j) &=& (1+z)^4\,\gamma_{ij}(0,(1+z)E_i,(1+z)E_j)\,.
\end{eqnarray}
The
scaling behaviour of $b$ and its derivative $\partial_Eb$ do again
depends on the photon background. For the CMB contribution, we have
\begin{eqnarray}\label{scaling3}
b_i(z,E_i) &= &(1+z)^2\,b_i(0,(1+z)E_i)\,,\\\label{scaling4}
\partial_Eb_i(z,E_i) &= &(1+z)^3\,\partial_Eb_i(0,(1+z)E_i)\,.
\end{eqnarray}

\begin{figure}[t]
\begin{center}
\includegraphics[width=0.6\linewidth]{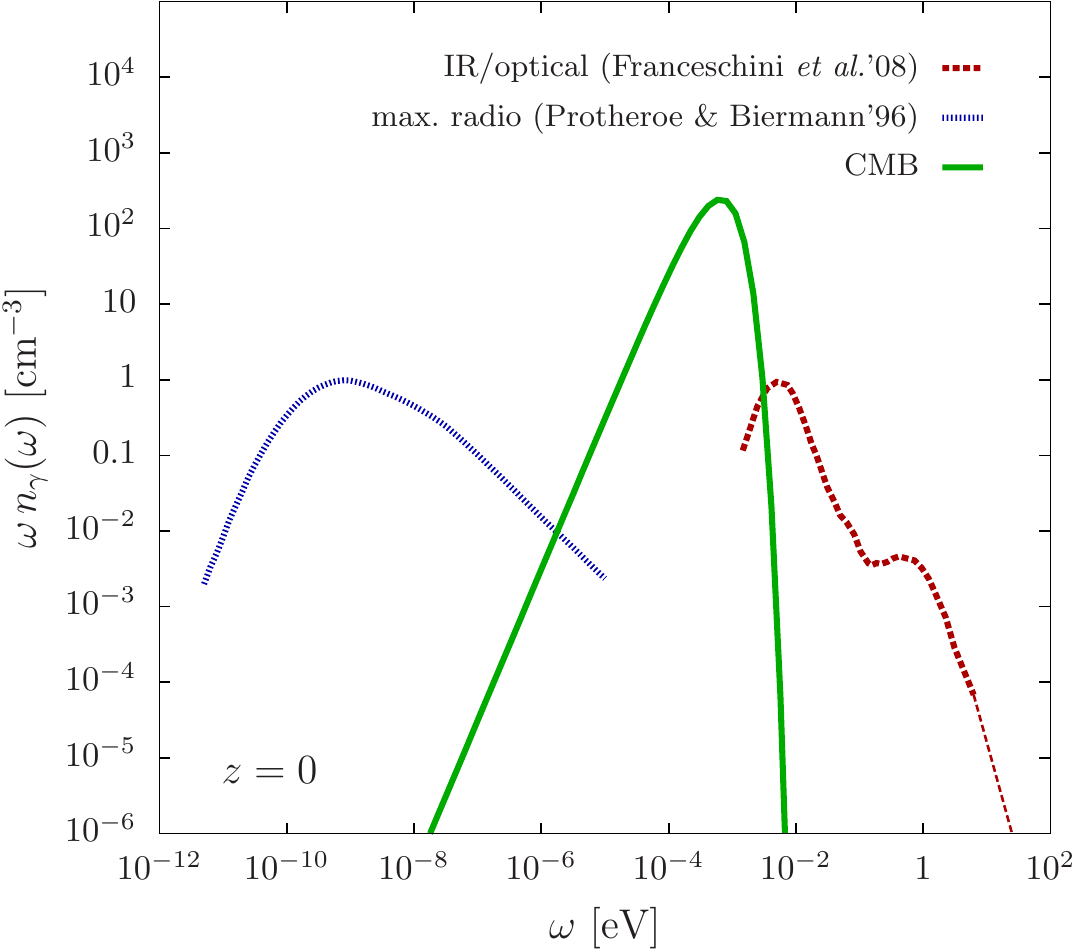}
\end{center}
\vspace{-0.5cm}
\caption[]{The energy spectrum of the CMB and
  the CIB in the IR/optial  and
  radio range at $z=0$. The thin dashed line 
  shows our extrapolation to UV energies. From Ref.~\cite{Ahlers:2010fw}.}
\label{fig:CRB}
\end{figure}

We now discuss the red-shift scaling of the quantities $b_i$,
$\partial_Eb_i$, $\gamma_{ij}$ and $\Gamma_i$ for the case of the
cosmic infrared background. The CIB spectrum has been studied and
tabulated for redshifts up to $z=2$~\cite{Franceschini:2008tp} (which we 
extrapolate slightly to UV energies as seen in Fig.~\ref{fig:CRB}).
This is consistent with the constraints on
the $\gamma$-ray opacity of the universe set by HESS
\cite{Aharonian:2005gh}, MAGIC \cite{Aliu:2008ay} and Fermi-LAT
\cite{LAT:2010kz}  from
non-observation of the expected cutoffs in the $\gamma$-ray spectra of
extragalactic sources.  The redshift
dependence is given by:
\begin{equation}
  n_{\rm CIB}(z,(1+z)E) = (1+z)^2\int_z^\infty d z'
  \frac{1}{H(z')}\mathcal{Q}_{\rm CIB}(z',(1+z')E)\,,
\end{equation}
where $\mathcal{Q}_{\rm CIB}$ is the comoving luminosity density of
the sources and we neglect absorption effects other than expansion. We
assume that this follows the star formation rate: $\mathcal{Q}_{\rm
  CIB}(z,E) \propto \mathcal{H}_{\rm SFR}(z)\,\mathcal{Q}_{\rm
  CIB}(0,E)$. We can then infer the bolometric evolution to be:
\begin{equation}
  \frac{N_{\rm CIB}(z)}{N_{\rm CIB}(0)} = (1+z)^3\frac{\int_z^\infty d z'\,
    \mathcal{H}_{\rm SFR}/(H(z')(1+z'))}{\int_0^\infty d z'\,
    \mathcal{H}_{\rm SFR}/(H(z')(1+z'))}\,,\label{HIR}
\end{equation}
where $N_{\rm CIB}(z)$ is the number of infrared--optical photons per
proper volume at redshift $z$. Following a recent
compilation~\cite{Hopkins:2006bw,Yuksel:2008cu}, we adopt
\begin{equation}\label{HSFR}
\mathcal{H}_{\rm SFR}(z) = \begin{cases}(1+z)^{3.4}&z<1\,,\\N_1\,(1+z)^{-0.3}&1<z<4\,,\\N_1\,N_4\,(1+z)^{-3.5}&z>4\,,\end{cases}
\end{equation}
with appropriate normalization factors, $N_1 = 2^{3.7}$ and $N_4 =
5^{3.2}$ (see the right panel of Fig.~\ref{scales}).  For comparison, the CMB evolves as
$N_{\rm CMB}(z)/N_{\rm CMB}(0) = (1+z)^3$. To simplify the numerical
evaluation we approximate the evolution with redshift as
\begin{equation}\label{NIR}
  n_{\rm CIB}(z,\omega) \simeq \frac{1}{1+z}\,\frac{N_{\rm CIB}(z)}{N_{\rm CIB}(0)}\, n_{\rm CIB}(0,\omega/(1+z))\,,
\end{equation}
and this is shown in the right panel of Fig.~\ref{scales}.
The redshift scaling of the quantities $\gamma_{ij}$, $\Gamma_i$,
$b_i$ and $\partial_Eb_i$ for the CIB is then obtained from the
corresponding scaling given for the CMB in
Eqs.~(\ref{scaling1}/\ref{scaling2}) and
(\ref{scaling3}/\ref{scaling4}), by multiplying the r.h.s.~by a factor
\mbox{$N_{\rm CIB}(z)/N_{\rm CIB}(0)/(1+z)^3$}.

We have little direct knowledge of the cosmic radio background. An
estimate made using the RAE satellite~\cite{Clark} is often used to
calculate the cascading of UHE photons. A
theoretical estimate has been made~\cite{Protheroe:1996si} of the
intensity down to kHz frequencies, based on the observed luminosity
function and radio spectra of normal galaxies and radio galaxies, 
although there are large uncertainties in the assumed evolution. The
calculated values are about a factor of $\sim 2$ above the
measurements and to ensure maximal energy transfer in the cascade we
will adopt this estimate and assume the same redshift scaling as the
cosmic infrared/optical background. We summarize the adopted cosmic
radiation backgrounds in Fig.~\ref{fig:CRB}.

The emission rate of CR protons per comoving volume is assumed, as per usual
practice, to follow a power-law:
\begin{equation}
\mathcal{Q}_p(0,E) \propto (E/E_0)^{-\gamma}\times\begin{cases} f_
-(E/E_{\rm min})&E<E_{\rm min}\,,\\1& E_{\rm min}<E<E_{\rm max}\,,\\
f_+(E/E_{\rm max})& E_{\rm max}<E\,.
\end{cases}
\label{eq:injection}
\end{equation}
We will consider spectral indices $\gamma$ in the range $2\div3$. The
functions $f_\pm(x) \equiv x^{\pm2}\exp(1-x^{\pm2})$ in
Eq.~(\ref{eq:injection}) smoothly turn off the contribution below
$E_{\rm min}$ and above $E_{\rm max}$. We set $E_{\rm
  max}=10^{12}$~GeV in the following and vary $E_{\rm min}$ in the
range $10^{8.5}\div10^{10}$~GeV, corresponding to a
galactic-extragalactic crossover between the second knee and the
ankle in the CR spectrum.

The cosmic evolution of the spectral emission rate per comoving volume
is parameterized as:
\begin{equation}
\mathcal{Q}_p(z,E) = \mathcal{H}(z)\mathcal{Q}_p(0,E)\,.
\label{eq:sourden1}
\end{equation}
For simplicity, we use the standard approximation
\begin{equation}
\mathcal{H}(z) \equiv (1+z)^n\Theta(z_{\rm max}-z)\,,
\label{eq:sourden2}
\end{equation} 
with $z_{\rm max} = 2$. Note that the dilution of the source density
due to the Hubble expansion is taken care of since $\mathcal{Q}$ is
the {\em comoving} density, {\it i.e.} for no evolution we would
simply have $\mathcal{H} = 1$. We consider cosmic evolution of UHECR
sources with $n$ in the range $2\div6$.

We can express the system of partial integro-differential equations~(\ref{diff0}) in terms of
a system of ordinary integro-differential equations~\cite{Ahlers:2009rf},
\begin{align}\label{diff1}
  \dot{\mathcal{E}}_i &= -H\mathcal{E}_i - b_i(z,\mathcal{E}_i)\,,\\
  \dot Z_i &=
  \big[\beta_i(z,\mathcal{E}_i)-\Gamma_{i}(z,\mathcal{E}_i)\big]\,Z_i+(1+z)\mathcal{Q}^\mathrm{eff}_i(z,\mathcal{E}_i)\,,\label{diff2}
\end{align}
where we have defined $\beta_i(z,E) \equiv \partial_Eb_i(z,E)$ and
$Z_i(z,E) \equiv (1+z)Y_i(z,\mathcal{E}_i(z,E))$. The quantity
$\mathcal{E}_i(z,E)$ gives the energy that a particle of type $i$ had
at redshift $z$ if we observe it today with energy $E$ and take into
account CEL. The effective source term in Eq.~(\ref{diff2}) is
\begin{equation}
 \mathcal{Q}^\mathrm{eff}_i(z,\mathcal{E}_i(z,E))
=\mathcal{Q}_i +\sum_j\int d
  E\,\,\partial_E\mathcal{E}_j\,\gamma_{ji}(z,\mathcal{E}_j,E_i)\,\frac{Z_j}{1+z}\,,
\end{equation}
where $\mathcal{E}_j(z\,,E)$ and $Z_j(z,E)$ are subject to the
boundary conditions $\mathcal{E}_j(0,E) = E$ and
$Z_j(z_\mathrm{max},E) = 0$. The flux at $z=0$ can be expressed as
\begin{eqnarray}\label{celflux}
 J_i(E) &= &\frac{1}{4\pi}Z_i(0,E) \nonumber \\ &= &\frac{1}{4\pi}\int_0^\infty \!\!\!\! \!\! d z\exp\left[\int_0^{z} \!\!d z' \frac{\partial_Eb_i(z',\mathcal{E}_i(z',E)) - \Gamma_i(z',\mathcal{E}_i(z',E))}{(1+z')H(z')}\right]  
    \frac{\mathcal{Q}^\mathrm{eff}_i(z,\mathcal{E}_i(z,E))}{H(z)}. 
\end{eqnarray}
For the numerical evaluation of the Boltzmann equations~(\ref{diff0})
it is convenient to solve~\cite{Ahlers:2010fw}
\begin{equation}
-H(z)(1+z)\partial_z {\mathscr Z}_i(z,E)  =  -\Gamma(z, {\mathscr E}) {\mathscr Z_i}(z,E)  +\frac{1}{(1+z)}\partial_E[b(z, {\mathscr E}) {\mathscr Z}_i(z,E)]
 +  (1+z) {\cal Q}_i^{\rm eff}(z, {\mathscr E})\, . 
\label{newdiff}
\end{equation}
where we have defined ${\mathscr E} = (1+z)E$, and ${\mathscr Z}_i(z,E) \equiv
(1+z)Y_i(z, {\mathscr E})$, subject to the boundary condition
${\mathscr Z}_j(z_{\rm max},E) = 0$. The effective source term becomes
${\cal Q}_i^{\rm eff}(z, {\mathscr E}) =\mathcal{Q}_i +\sum_j\int  d
  E_j\,\gamma_{ji}(z, {\mathscr{E}}_j,E_i)\, {\mathscr Z_j}$. 
In our calculation we use
logarithmic energy bins with size $\Delta\log_{10}E = 0.05$ between
$10^5$~GeV and $10^{15}$~GeV. For the numerical evolution of the
differential Eq.~(\ref{diff1})  we use a simple Euler method with
a step-size $\Delta z=10^{-4}$. The corresponding step-size in the
propagation distance $\Delta r = c\Delta t$ is then always smaller
than the proton interaction length. The flux at $z=0$ is given by $J_i(E) = \mathscr{Z}_i(0,E)/(4\pi).$

Electromagnetic (EM) interactions of photons and
leptons with the extra-galactic background light and magnetic field
can happen on time-scales much shorter than their production rates.
It is convenient to account for these contributions during the proton
propagation as fast developing electro-magnetic cascades at a fixed
redshift. We will use the efficient method of ``matrix
doubling''~\cite{Protheroe:1992dx} for the calculation of the
cascades. Since the cascade $\gamma$-ray flux is mainly in the GeV-TeV
region and has an almost universal shape here, it is numerically much
more efficient to calculate the total energy density $\omega_{\rm
  cas}$ injected into the cascade and compare this value to the limit
imposed by Fermi-LAT. The total energy density (eV cm${}^{-3}$) of
EM radiation from proton propagation in the past is given
as
\begin{equation}
\label{eq:omegacas}
\omega_{\rm cas} \equiv \int dE E n_{\rm cas}(0,E) = \int dt\int dE \,\frac{b_{\rm cas}(z,E)}{(1+z)^4}\,n_ p(z,E)\,,
\end{equation}
where $n_p(z,E)$ is the physical energy density at redshift $z$, defined
via $n_p(z,E)\equiv(1+z)^3Y_p(z,E)$; details on the calculation are given in Appendix~\ref{AB}.  
The continuous energy loss of protons into
the cascade, denoted by $b_{\rm cas}$, is in the form of electron,
positron and $\gamma$-ray production in BH ($b_{\rm BH}$) and
photopion ($b_\pi$) interactions. We derive the BH and photopion contribution to
$\omega_{\rm cas}$ separately. For the photopion contribution we
estimate
\begin{equation}\label{eq:bpion}
b_{\pi}(z,E) \simeq \int d E' E'\left[\gamma_{pe^-}(z,E,E') +
\gamma_{pe^+}(z,E,E') + \gamma_{p\gamma}(z,E,E')\right] \, ,
\end{equation}
where the angular-averaged distribution of particle $j$ after the interaction are determined using the Monte Carlo
package {\sc sophia}~\cite{Mucke:1999yb}. For the energy loss via BH pair production we use (\ref{uniden}).  Note, that since the photopion
contribution in the cascade is dominated at the GZK cutoff, its
contribution should {\it increase} relative to BH pair production with
increasing crossover energy and, hence, also the associated neutrino
fluxes after normalization to $\gamma$-ray and CR data.

\begin{figure}[t]
\begin{center}
\includegraphics[width=0.49\linewidth]{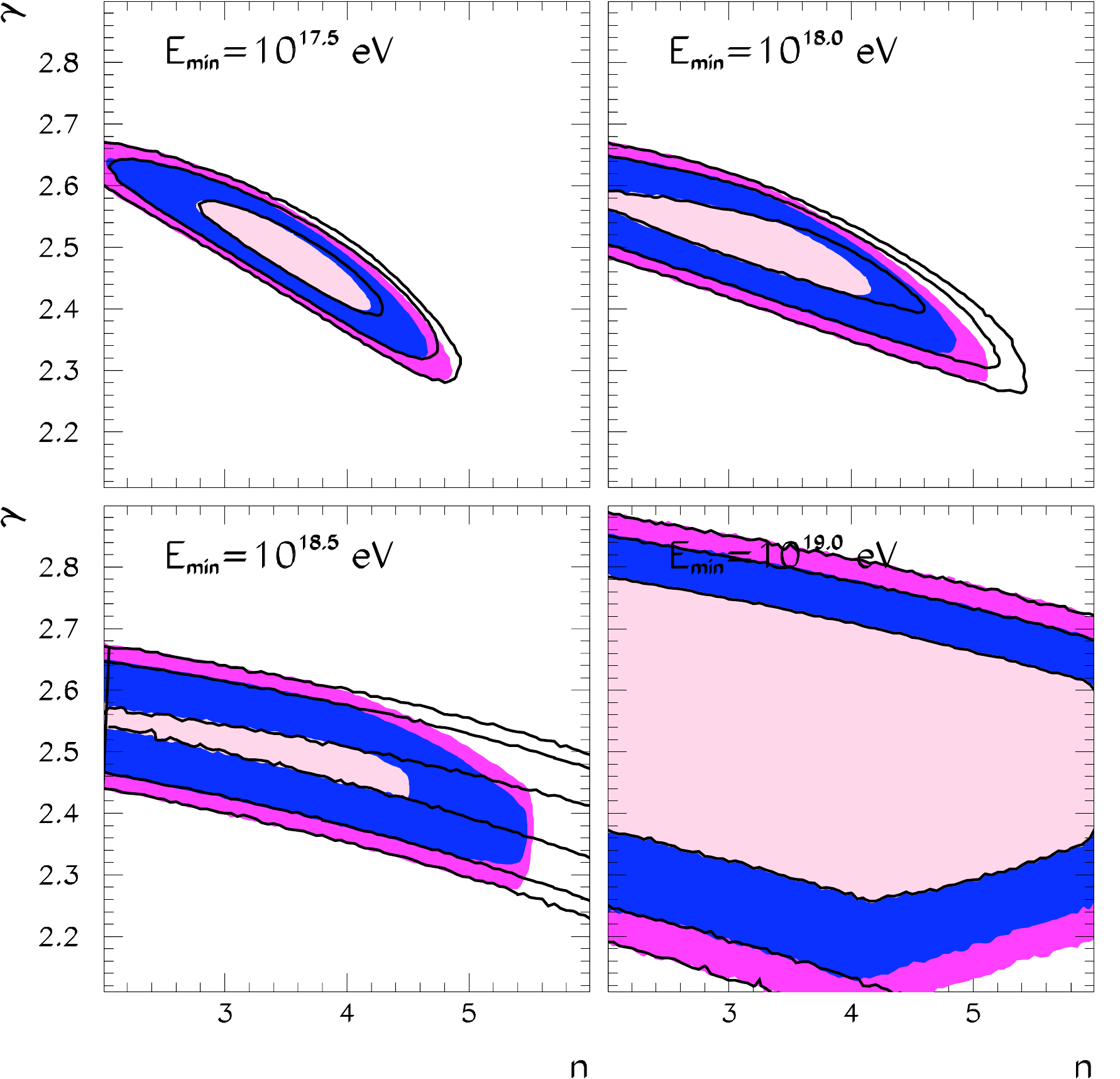}\hfill
\includegraphics[width=0.49\linewidth]{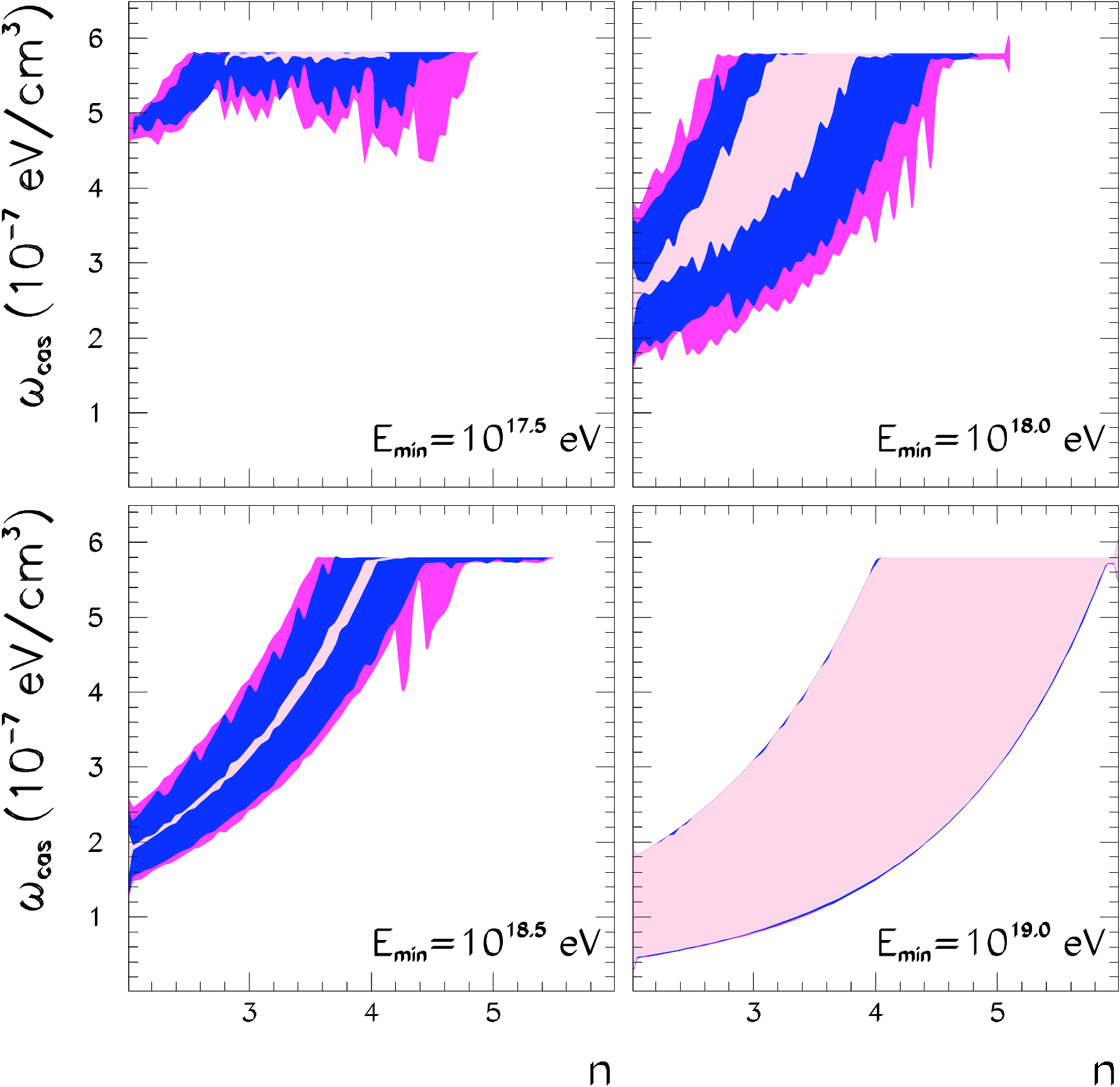}
\end{center}
\vspace{-0.3cm}
\caption[]{{\bf Left Panel:} Goodness-of-fit test of the HiRes data
  \cite{Abbasi:2007sv}. We show the 68\% (pink), 95\% (blue) and 99\%
  (magenta) confidence levels of the injection index $\gamma$ and the
  cosmic evolution index $n$. The black lines indicate the allowed
  regions before the cascade ($\omega_{\rm cas}$) bound is imposed.
  {\bf Right Panel:} The corresponding energy density in the EM
  cascade. From Ref.~\cite{Ahlers:2010fw}.}
\label{fig:gof}
\end{figure}
\begin{figure}[t]
\begin{center}
\includegraphics[width=0.7\linewidth]{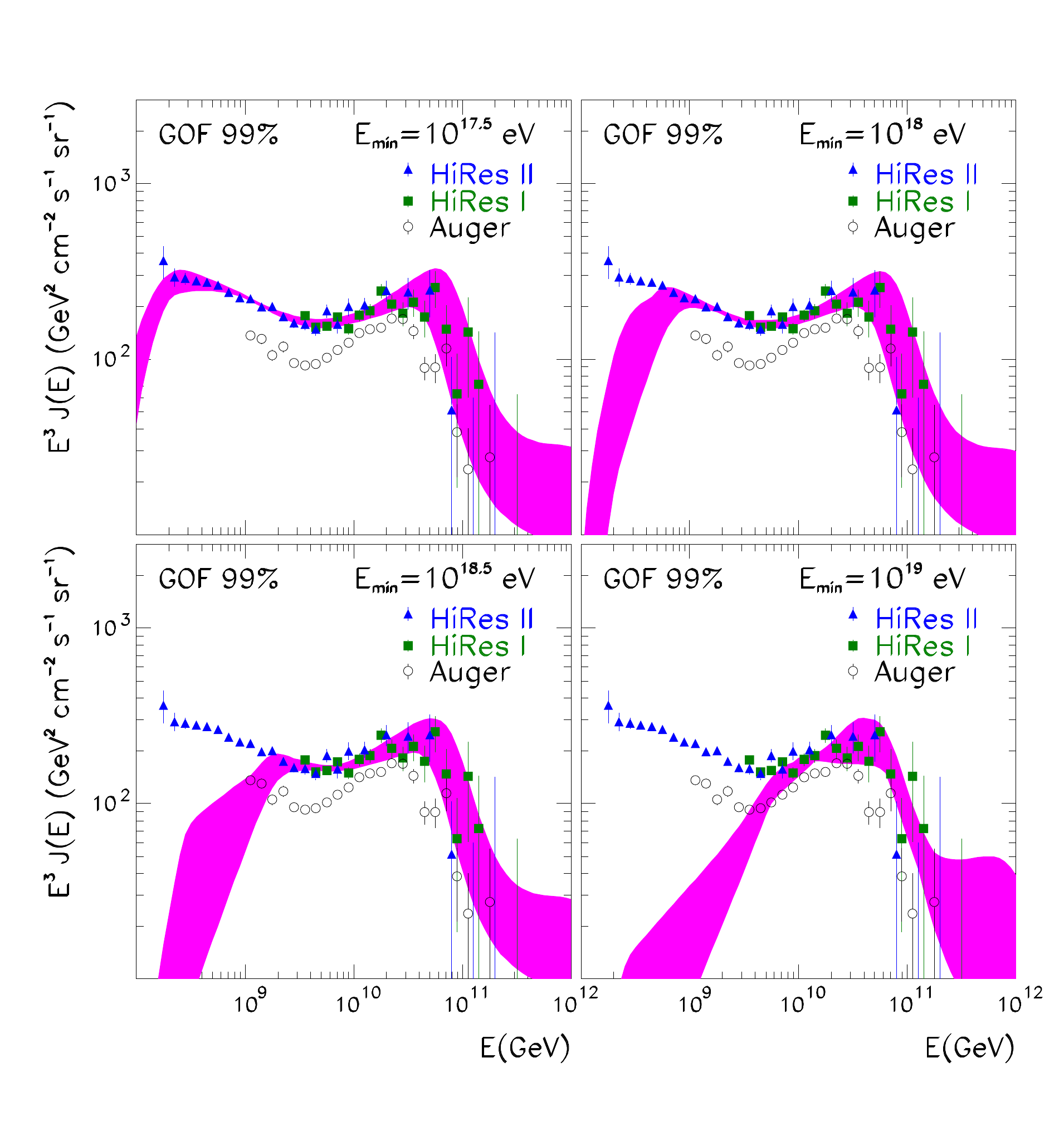}
\end{center}
\vspace{-0.3cm}
\caption[]{The allowed proton flux (at the 99\% confidence level) for
  increasing crossover energy $E_{\rm min}$. Each fit of the proton
  spectrum is marginalized with respect to the experimental energy
  uncertainty and we show the shifted predictions in comparison to the
  HiRes central values~\cite{Abbasi:2007sv}.  For comparison we also
  show the Auger data~\cite{Abraham:2008ru,Abraham:2010mj} which has
  {\em not} been included in the fit. From Ref.~\cite{Ahlers:2010fw}.}
\label{fig:omega}
\end{figure}
\begin{figure}[t]
\begin{center}
\includegraphics[width=0.5\linewidth]{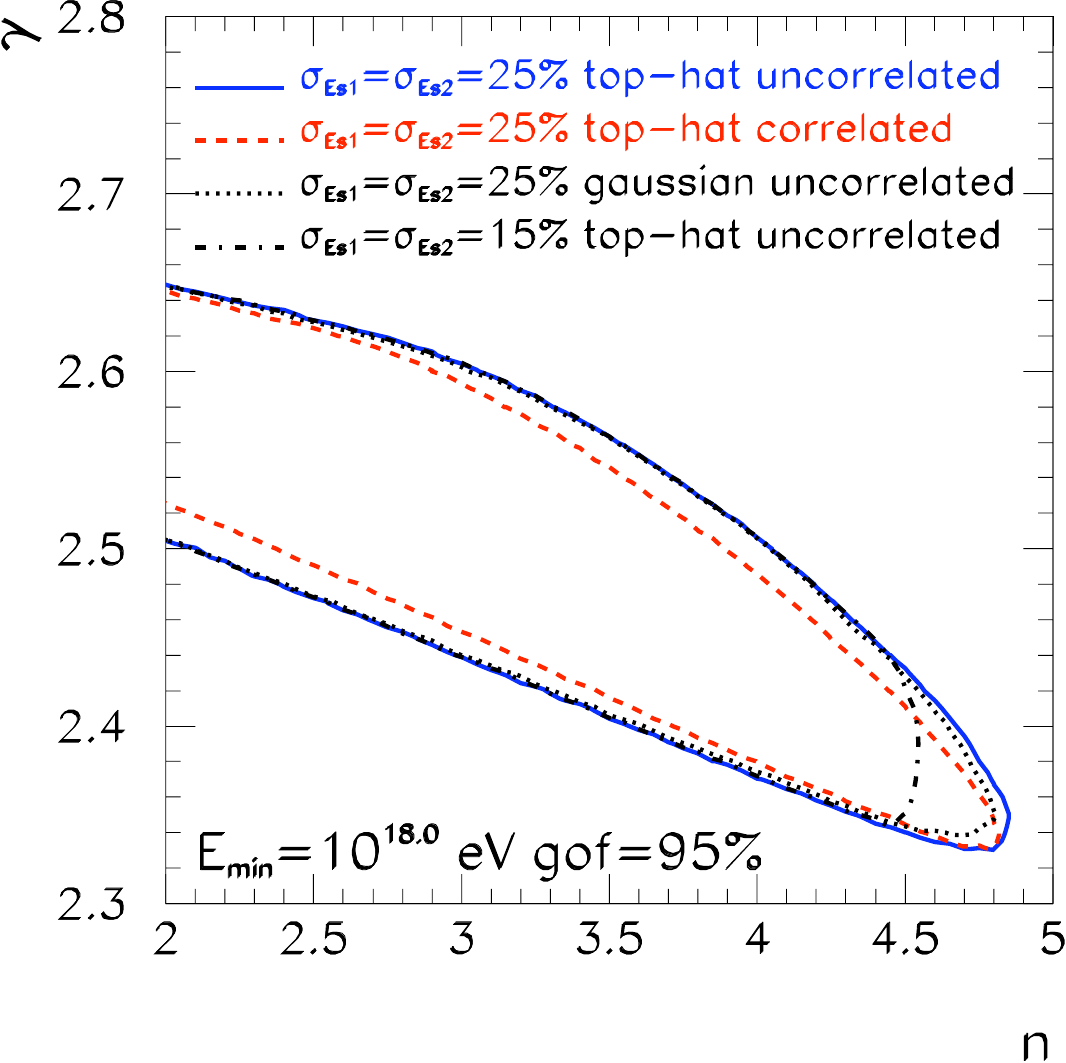}
\end{center}
\vspace{-0.3cm}
\caption[]{Systematic effect of the experimental energy resolution on
  the fitted spectral index $\gamma$ and cosmological evolution
  parameter $n$. For illustration we show the dependence of the 95\%
  C.L. bound for a crossover energy of $10^{18}$~eV. The blue contour
  corresponds to the region shown in Fig.~\ref{fig:gof} assuming an
  uncorrelated energy shift of 25\% in both data sets (HiRes I and
  II)~\cite{Abbasi:2007sv}, for a flat prior (``top-hat''
  distribution). The red dashed curve assumes correlated errors of the
  energy resolution in both data sets. The black dotted curve shows
  the result for uncorrelated errors with a Gaussian prior, and the
  dashed-dotted line shows uncorrelated errors with a flat prior, but
  with a lower uncertainty of 15\%. From Ref.~\cite{Ahlers:2010fw}.}
\label{fig:compa}
\end{figure}

Next, we study the constraint set by the diffuse $\gamma$-ray
background on all-proton models of extragalactic CRs.  We parametrize
our ignorance of the crossover energy -- which marks the transition
between the galactic and extragalactic components -- as a variable
low energy cutoff in the proton injection rate. By fitting only to CR
data above the crossover energy, taken to be between $10^{8.5}$~GeV
and $10^{10}$~GeV, we determine the statistically preferred values of
the spectral index $\gamma$,  cosmic source density evolution index
$n$, and crossover energy $E_{\rm min}$ by a goodness-of-fit (GOF) test of the HiRes data, taking into account the energy resolution of about 25\%.  For each model we check
that the total energy density of the EM cascade is below a critical
value inferred from the recent measurement of the extragalactic $\gamma$-ray background by the Fermi-LAT Collaboration~\cite{Abdo:2010nz}.  

Given the acceptance $A_i$ (in units of area per unit time per unit
solid angle) of the experiment for the energy bin $i$ centered at
$E_i$ and with bin width $\Delta_i$, and the energy scale uncertainty
of the experiment, $\sigma_{E_s}$ the number of expected events in the
bin is given by
\begin{equation}
N_i(n,\gamma,N_0,\delta)=
A_i \int_{E_i(1+\delta)-\Delta_i/2}^{E_i(1+\delta)+\Delta_i/2} 
J^p_{ N_0,n,\gamma}(E) 
dE\;, 
\label{eq:nev}
\end{equation}
where 
\begin{equation}
J^p_{N_0,n,\gamma}(E)=  \frac{c}{4\pi} \  n_p(0,E) 
\end{equation}
is the
proton flux arriving at the detector corresponding to a proton source
luminosity as in Eq.~\eqref{eq:injection}, with the cosmic evolution
of the source density given by Eqs.~\eqref{eq:sourden1}
and~\eqref{eq:sourden2}.  The parameter $\delta$ in Eq.~\eqref{eq:nev}
above is a fractional energy-scale shift that reflects the
energy-scale uncertainty of the experiment, and $N_0$ is the
normalization of the proton source luminosity.

The probability distribution of events in the $i$-th bin is of the
Poisson form with mean $N_i$.  Correspondingly the $r$-dimensional
($r$ being the number of bins of the experiment with $E_i\geq E_{\rm
  min}$) probability distribution for a set of non-negative integer
numbers ${\vec k}=\{k_1,...k_r\}$, $P_{\vec k}(n,\gamma,
  N_0,\delta)$, is just the product of the individual Poisson
distributions. According to this $r$-dimensional probability distribution, the
experimental result 
$
{\vec N^{\rm exp}}=\{N^{\rm exp}_1,...,N^{\rm
  exp}_r\} $
has a probability $P_{\vec N^{\rm exp}}(n,\gamma,
  N_0,\delta)$ and, correspondingly, the experimental probability after
marginalizing over the energy scale uncertainty and normalization is:
\begin{equation}
P_{\rm exp}(n,\gamma)={\rm Max}_{\delta, N_0} 
P_{\vec N^{\rm exp}}(n,\gamma,N_0,\delta)\, , 
\label{eq:margi}
\end{equation}
where the maximization is made within some prior for $\delta$ and
$N_0$.  For the energy shift $\delta$ we have used two forms for
the prior, either a top-hat spanning the energy-scale uncertainty of
the experiment, $\sigma_{E_s}$, or a gaussian prior of width
$\sigma_{E_s}$.

For $N_0$ we impose the prior arising from requiring consistency
with the Fermi-LAT measurements~\cite{Abdo:2010nz} of the diffuse
extra-galactic $\gamma$-ray background. In order to do so, we obtain the
total energy density of EM radiation from the proton propagation using
Eq.~\eqref{eq:omegacas} and we require following
Ref.~\cite{Berezinsky:2010xa}:
\begin{equation}
\omega_{\rm cas}(N_0 ,n,\gamma)\leq 5.8\times 10^{-7}\; {\rm eV}/{\rm cm}^3\,. 
\label{eq:fermilat}
\end{equation}

The marginalization in Eq.~(\ref{eq:margi}) also determines $
  N_0^{\rm best}$ and $\delta^{\rm best}$ for the model, which are the
values of the energy shift and normalization that yield the best
description of the experimental CR data, subject to the constraint
imposed by the Fermi-LAT measurement.

Altogether the model is compatible with the experimental results at
given GOF  if
\begin{equation}
\sum_{\vec k}P_{\vec k}(n,\gamma,N_0^{\rm best},\delta^{\rm best})
\Theta\left[P_{\vec k}(n,\gamma,N_0^{\rm best},\delta^{\rm best})
-P_{\rm exp}(n,\gamma)\right]\leq {\rm GOF} \, .
\end{equation}
Technically, this is computed by generating a large number $N_{\rm
  rep}$ of replica experiments according to the probability
distribution $P_{\vec k}(n,\gamma,N_0^{\rm best},\delta^{\rm
  best})$ and imposing the fraction $F$ of those which satisfy $P_{\vec
  k}(n,\gamma,N_0,\delta^{\rm best})>P_{\rm exp}(n,\gamma)$ to be 
$F\leq{\rm GOF}$.

\begin{table}[t]
\caption[]{Cosmic ray source parameters which best fit the HiRes 
  data~\cite{Abbasi:2007sv}, along with those which yield minimal and maximal 
  contributions to $\omega_\pi$ (i.e. neutrino fluxes) and 
  $\omega_\mathrm{cas} = \omega_\pi + \omega_\mathrm{BH}$ (i.e. 
  $\gamma$-ray fluxes), all at the 99\% C.L.}
\label{tab:parameters}
\centering
\begin{minipage}[t]{0.7\linewidth}\centering\small
\setlength{\tabcolsep}{3pt}\renewcommand{\arraystretch}{1.3}\small
\begin{tabular}{c||cc|c|cc||cc|c|cc}
\hline\hline 
&\multicolumn{5}{c||}{\normalsize $E_{\rm min}=10^{8.5}$~GeV}
&\multicolumn{5}{c}{\normalsize $E_{\rm min}=10^{9}$~GeV}\\
model&$n$&$\gamma$
&$\omega_{\rm cas}$${}^a$\footnotetext[1]
{in units of $10^{-7}$~eV/cm${}^3$}
&$\delta_{I \rm best}$&$\delta_{II\rm best}$
&$n$&$\gamma$&$\omega_{\rm cas}$${}^a$ 
&$\delta_{I\rm best}$&$\delta_{II\rm best}$
\\
\hline &\multicolumn{10}{c}{fit {\it with} Fermi-LAT bound:}\\ \hline
best fit
&$3.50$&$2.49$&$5.8$&0.005&0.
&$3.20$&$2.52$&$5.2$& 0.050&0.045
\\
min.~$\omega_{\rm cas}$
&$4.50$&$2.31$&$4.4$&-0.235&-0.245
&$2.25$&$2.47$&$1.7$&-0.120&-0.150 
\\
max.~$\omega_{\rm cas}$
&$4.60$&$2.36$&$5.8$ &-0.185&-0.175
&$3.35$&$2.55$&$5.8$ &0.050&0.060
\\
min.~$\omega_\pi$
&$2.00$&$2.67$&$4.9$&0.215&0.235 
&$2.00$&$2.51$&$1.8$& -0.070&-0.095
\\
max.~$\omega_\pi$
&$4.80$&$2.29$&$5.8$&-0.220&-0.215
&$5.10$&$2.29$&$5.8$&-0.250&-0.250
\\
\hline &\multicolumn{10}{c}{fit {\it without} Fermi-LAT bound:}\\ 
\hline
max.~$\omega_{\rm cas}$
&$4.45$&$2.44$&$15$ &0.135&0.155 
&$5.25$&$2.36$&$27$ & 0.205&0.205
\\
max.~$\omega_\pi$
&$4.80$&$2.36$&$14$& 0.050&0.055
&$5.30$&$2.35$&$26$& 0.190&0.190
\\\hline\hline
\multicolumn{11}{c}{}\\
\hline\hline 
&\multicolumn{5}{c||}{\normalsize $E_{\rm min}=10^{9.5}$~GeV}
&\multicolumn{5}{c}{\normalsize $E_{\rm min}=10^{10}$~GeV}\\
model
&$n$&$\gamma$ &$\omega_{\rm cas}$${}^a$
&$\delta_{I \rm best}$&$\delta_{II\rm best}$
&$n$&$\gamma$&$\omega_{\rm cas}$${}^a$ 
&$\delta_{I\rm best}$&$\delta_{II\rm best}$
\\
\hline &\multicolumn{10}{c}{fit {\it with} Fermi-LAT bound:}\\ \hline
best fit
&$4.05$&$2.47$&$5.8$& 0.015&0.005
&$4.60$&$2.50$&$4.4$& -0.030&-0.065
\\
min.~$\omega_{\rm cas}$
&$2.00$&$2.45$&$1.4$ & -0.050&-0.060 
&$2.00$&$2.88$&$0.44$ &-0.220&-0.250
\\
max.~$\omega_{\rm cas}$
&$4.95$&$2.37$&$5.8$ & -0.165&-0.160
&$4.45$&$2.13$&$5.8$ & 0.130&0.090
\\
min.~$\omega_\pi$
&$2.00$&$2.63$&$2.1$& 0.075&0.070
&$2.00$&$2.88$&0.44 & -0.220&-0.250
\\
max.~$\omega_\pi$
&$5.35$&$2.28$&$5.8$&-0.240&-0.250
&$4.40$&$2.10$&$5.8$&0.145&0.100
\\
\hline &\multicolumn{10}{c}{fit {\it without} Fermi-LAT bound:}\\ 
\hline
max.~$\omega_{\rm cas}$
&$6.00$&$2.49$&$30$ & 0.120&0.135
&$6.00$&$2.14$&$23$ &0.250&0.210
\\
max.~$\omega_\pi$
&$6.00$&$2.47$&$29$& 0.120&0.125
&$6.00$&$2.10$&$23$&0.250&0.210 \\
\hline\hline
\end{tabular}
\end{minipage}
\end{table}

With this method we determine the value of $(n, \gamma)$ parameters
that are compatible with the HiRes I and HiRes II
experiments~\cite{Abbasi:2007sv}.  In the left panel of
Fig.~\ref{fig:gof}, we plot the regions with GOF 64\%, 95\% and 99\% for four
values of the minimum ({\em i.e.} crossover) energy. In the right
panel, we show the corresponding ranges of $\omega_{\rm cas,best}$ for the
models as a function of the cosmic evolution index $n$. In order to
display explicitly the impact of the constraint from the Fermi-LAT
measurements of the diffuse extra-galactic $\gamma$-ray background
\eqref{eq:fermilat}, we show the corresponding GOF regions without
imposing that constraint.  In Table.~\ref{tab:parameters} we list the
parameters corresponding to the best-fit models and to the models with
minimal and maximal contributions to $\omega_\pi$ and $\omega_{\rm
  cas}=\omega_\pi+\omega_{\rm BH}$ at the 99\% C.L., together with the
corresponding energy shifts which give best fits to the HiRes I and
HiRes II data.  We also show the parameters for the models with
maximum $\omega_\pi$ and $\omega_{\rm cas}$ {\em without} imposition
of the Fermi-LAT constraint.

It is interesting to compare the allowed values of $n$ with those in the cosmological evolution of UHECR candidate sources.  For GRBs~\cite{Yuksel:2006qb},
\begin{equation}\label{HGRB}
\mathcal{H}_{\rm GRB}(z) = (1+z)^{1.4}\, \mathcal{H}_{\rm SFR}(z)\,,
\end{equation}
for AGNs~\cite{Hasinger:2005sb,Stanev:2008un} 
\begin{equation}\label{HAGN}
\mathcal{H}_{\rm AGN}(z) = \begin{cases}(1+z)^{5.0}&z<1.7\,,\\N_{1.7}&1.7<z<2.7\,,\\N_{1.7}\,N_{2.7}^{(2.7-z)}&z>2.7\,,\end{cases}
\end{equation}
with $N_{1.7}=2.7^5$ and $N_{2.7}=10^{0.43}$, and for QSO~\cite{Engel:2001hd}
\begin{equation}
\mathcal{H}_{\rm QSO}  (z) = 
\begin{cases} (1+z)^3 & z<
1.9 \,, \\  (1 + 1.9)^3 &  1.9 < z < 2.7 \,,\\
 (1 + 1.9)^3 \exp\{(2.7-z)/2.7\} \,, & z >
2.7 \, . \end{cases}
\end{equation}

As an illustration of the agreement with the CR data, in
Fig.~\ref{fig:gof} we show the range of proton fluxes corresponding to models
with GOF 99\% or better for increasing crossover energies $E_{\rm
  min}$. As discussed above, each fit of the proton spectra is
marginalized with respect to the experimental energy scale uncertainty
and we show the shifted predictions with $\delta^{\rm best}$ in
comparison to the HiRes data at central value.  In the
 figure  we also show the results from Auger~\cite{Abraham:2008ru,Abraham:2010mj},
though these have not been included in the analysis.

These results are obtained assuming an energy scale uncertainty
$\sigma_{E_s}=5\%$ with a ``top-hat'' prior for the corresponding
energy shifts which are taken to be uncorrelated for HiRes I and HiRes
II.  In Fig.~\ref{fig:compa} we explore the dependence of the results
on these assumptions by using a different form for the prior, assuming
the energy shifts to be correlated between the two experiments, or
reducing the uncertainty to $\sigma_{E_s}=15\%$. As seen in the
figure, the main effect is associated with the reduction of the energy
scale uncertainty which, as expected, results in a worsening of the
GOF for models with larger $n$. This is directly related to the
normalization constraint from Eq.~(\ref{eq:fermilat}).  If one naively
ignores the energy scale uncertainty, the constraint in
Eq.~(\ref{eq:fermilat}) rules out models with $n\gtrsim 3$ (the
precise value depending on the assumed $E_{\rm min}$). However, once
the energy scale uncertainty is included, the constraint of
Eq.~(\ref{eq:fermilat}) plays a weaker role on the determination of
the GOF of the models. It does, however, imply a maximum value of $
  N_0^{\rm best}$ which, as we will see, impacts the corresponding
ranges of neutrino fluxes.

\begin{figure}[t]
\begin{center}
\includegraphics[width=\linewidth]{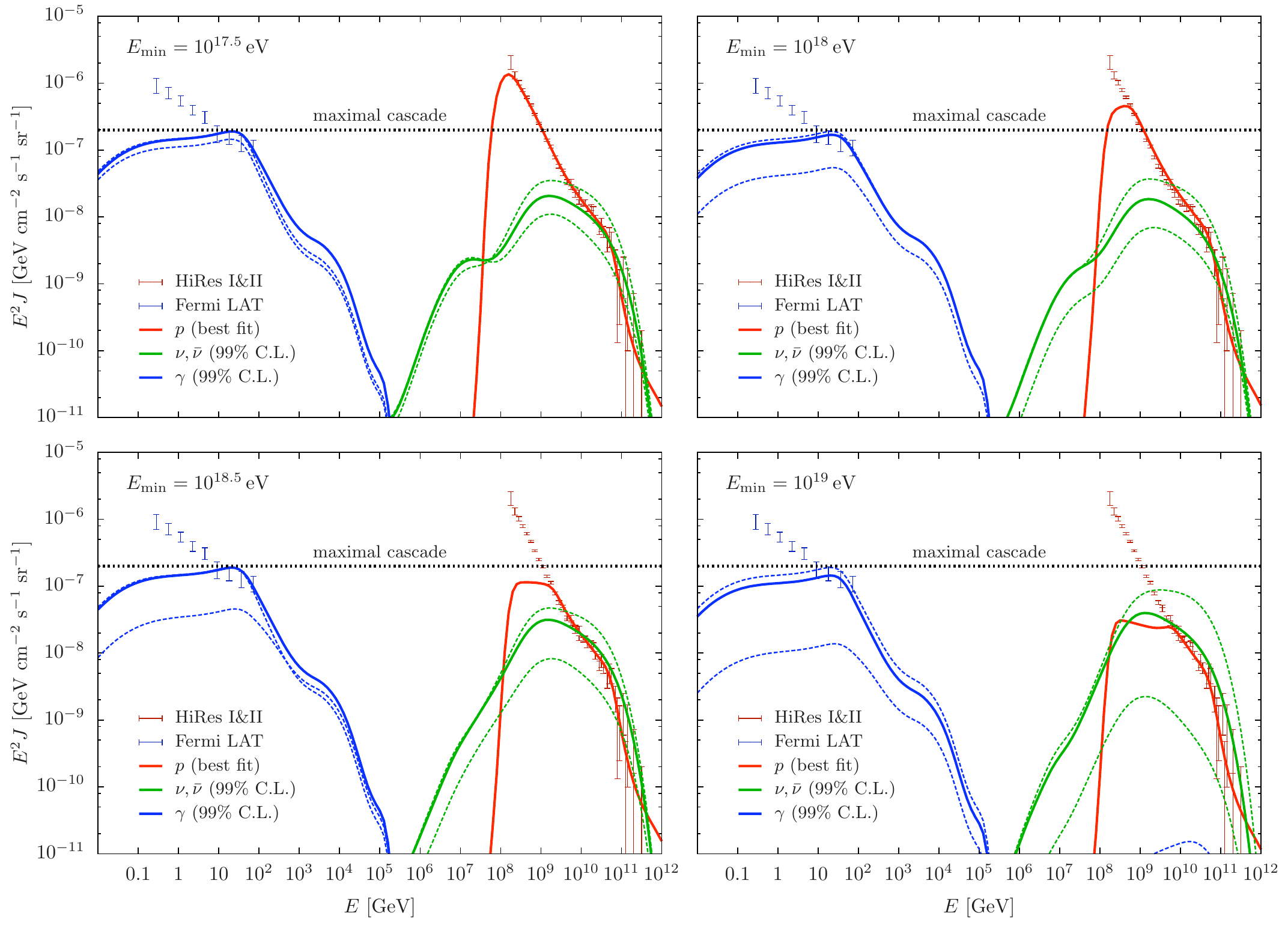}
\end{center}
\vspace{-0.3cm}
\caption[]{Comparison of proton, neutrino and $\gamma$-ray fluxes for
  different crossover energies. We show the best-fit values (solid
  lines) as well as neutrino and $\gamma$-ray fluxes within the 99\%
  C.L. with minimal and maximal energy density (dashed lines). The
  values of the corresponding model parameters can be found in
  Table.~\ref{tab:parameters}. The dotted line labeled ``maximal
  cascade'' indicates the approximate limit $E^2 J_{\rm cas} \lesssim
  c\, \omega_{\rm cas}^{\rm max}/4\pi\log({\rm TeV}/{\rm GeV})$,
  corresponding to a $\gamma$-ray flux in the GeV-TeV range saturating
  the energy density~(\ref{eq:fermilat}). The $\gamma$-ray fluxes are
  marginally consistent at the 99\% C.L. with the highest energy
  measurements by Fermi-LAT. The contribution around 100~GeV is
  somewhat uncertain due to uncertainties in the cosmic infrared
  background. From Ref.~\cite{Ahlers:2010fw}.}
\label{fig:comparison}
\end{figure}

The corresponding range of $\gamma$-ray and cosmogenic neutrino fluxes
(summed over flavor) is shown in Fig.~\ref{fig:comparison} for models
with minimal and maximal energy density at the 99\% C.L.
As expected, the maximum $\gamma$-ray fluxes are consistent with the
Fermi-LAT data within the errors.  For illustration, we also show as a
dotted line the ``naive'' $\gamma$-ray limit $E^2J_{\rm cas} \lesssim
c\,\omega_{\rm cas}^{\rm max}/4\pi\log({\rm TeV}/{\rm GeV})$,
corresponding to a $\gamma$-ray flux in the GeV-TeV range which saturates the
energy density~(\ref{eq:fermilat}).

The cosmogenic neutrino and photon fluxes are also sensitive to the primary composition. For example, if the primaries are iron nuclei one would expect a considerably lower flux on both photons and neutrinos, rendering photon pile-up measurements less helpful in constraining  cosmogenic fluxes. (See Appendix~\ref{AA} for details).

\subsubsection{Upper limits on the cosmic neutrino flux}

High energy neutrino detection is one of the experimental challenges in particle astrophysics for the forthcoming years. It is widely believed that one of the most appropriate techniques for neutrino detection consists of detecting the Cherenkov light from muons or showers produced by the neutrino interactions in underground water or ice. For a  recent review see {\em e.g.}~\cite{Anchordoqui:2009nf}. This allows instrumentation of large enough volumes to compensate for both the low neutrino cross section and the low fluxes expected. There are several projects under way to build sufficiently large detectors to measure the expected signals from a variety of neutrino sources. The IceCube facility, deployed near the Amundsen-Scott station, is the  largest neutrino telescope in the world~\cite{Halzen:2007ip}.  It comprises a cubic-kilometer of ultra-clear ice about a mile below the South Pole surface, instrumented with long strings of sensitive photon detectors which record light produced when neutrinos interact in the ice.

\begin{figure}[htb]
\centering
\includegraphics[width=0.7\textwidth]{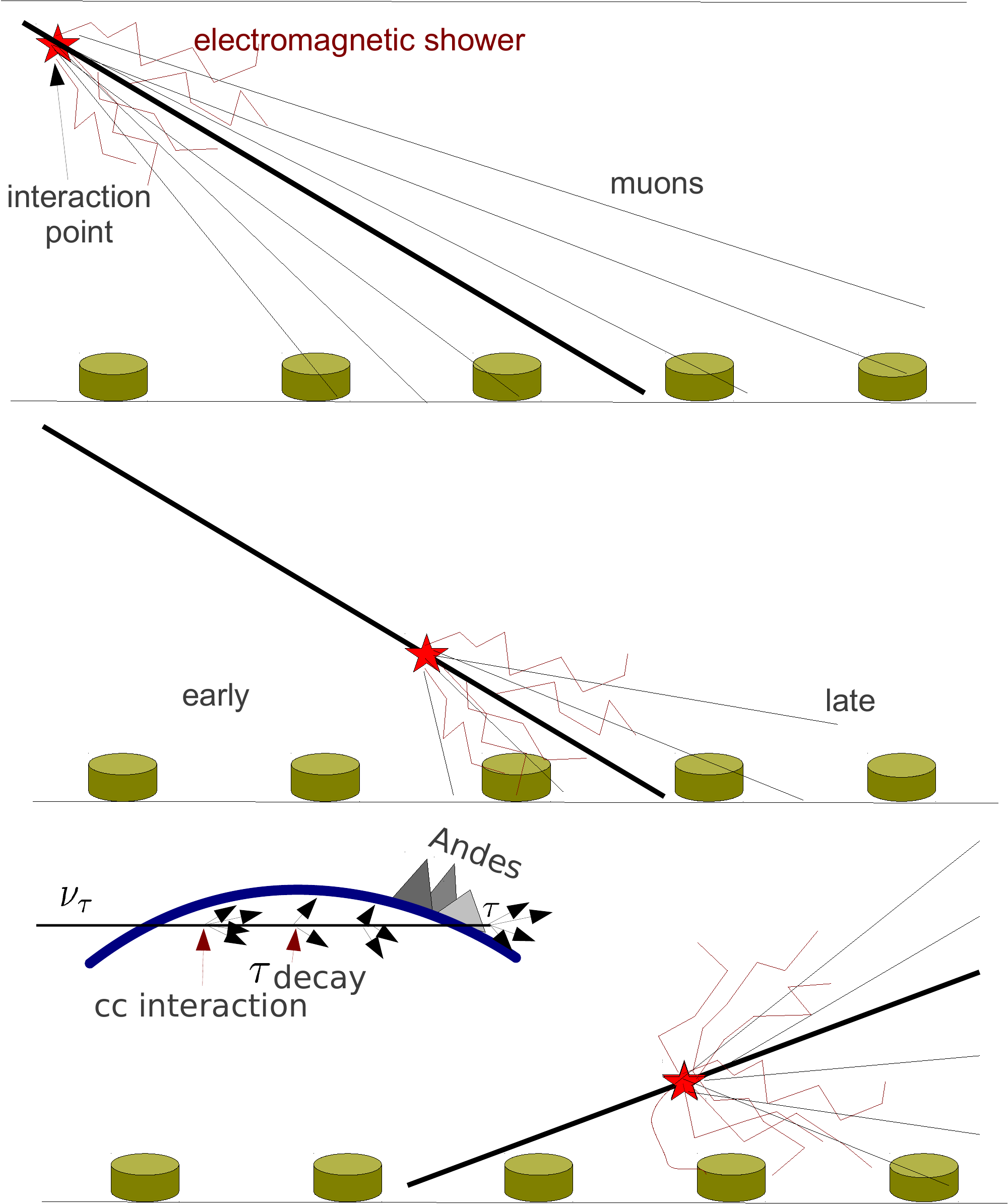}
\caption{Schematic illustration of the properties of a hadron-induced
shower (top), an $\nu$-induced nearly horizontal shower (middle) and a
$\nu_\tau$-induced earth skimming shower (bottom). Note that only up-going showers
resulting from $\tau$ neutrino interactions in the Earth can be 
detected with any efficiency using the surface array.  
In contrast, all three neutrino species can be detected in down-going
showers.  Also note from the inset in the lower panel, that the 
incident $\tau$ can experience several CC interactions and decays and thereby
undergo a regeneration process.  The Andes mountain range lies to the west 
of the observatory, and provides roughly an additional 20\% target volume for
$\nu_\tau$ interactions.}
\label{f:nu-schematic}
\end{figure}

CR experiments, like Auger, provide a complementary technique for  UHEC$\nu$  detection 
by searching for deeply--developing, large zenith angle ($>75^\circ$) 
showers~\cite{Capelle:1998zz}. At these large angles, hadron-induced
showers traverse the equivalent of several atmospheres before
reaching detectors at the ground.  Beyond about 2 atmospheres,
most of the electromagnetic component of a shower is 
extinguished and only very high energy muons survive.
Consequently, a hadron-induced shower front is relatively flat and
the shower particles arrive within a narrow time 
window (Fig.~\ref{f:nu-schematic}, top panel).
In contrast, a neutrino shower exhibits
characteristics similar to those of a vertical shower, 
which has a more curved 
front and a wider distribution in particle arrival times due to the 
large number of lower energy electrons and photons.
Furthermore, the ``early'' part of the shower will tend to be 
dominated by the electromagnetic component, while ``late'' 
portion will be enriched with tightly bunched muons
(Fig.~\ref{f:nu-schematic}, middle panel).
Using these characteristic features, it is possible to
distinguish neutrino induced events from background hadronic
showers. Moreover, because of full flavor mixing, tau neutrinos are
expected to be as abundant as other species in the cosmic
flux. Tau neutrinos can interact in the Earth's crust, producing $\tau$
leptons which may decay above the Auger 
detectors~\cite{Feng:2001ue,Bertou:2001vm,Fargion:2000iz} (Fig.~\ref{f:nu-schematic},  
bottom panel).  Details on how such events can be 
selected at the Auger Observatory are discussed 
in ~\cite{Abraham:2007rj, Abraham:2009uy}.

So far, no neutrino candidates have been observed  resulting in  upper limits on the diffuse flux of neutrinos. 
 For the case of up-going $\tau$ neutrinos  in the energy range 
$2 \times 10^{8}~{\rm GeV} < E_\nu < 2 \times 10^{10}~{\rm GeV}$, assuming a diffuse spectrum of the form $E_\nu^{-2}$, the current 90\%CL bound, is~\cite{Abraham:2009eh}
\begin{equation}
E_\nu^2 \ \Phi^{\nu_\tau} \simeq 4.7^{-2.5}_{+2.2} \times 10^{-8}~{\rm GeV}~{\rm cm}^{-2}~{\rm s}^{-1}~{\rm sr}^{-1} \, .
\end{equation}
Though Auger  was not designed specifically as a neutrino
detector, it is interesting to note that it exhibits good sensitivity in 
an energy regime complementary to those available to other dedicated instruments.
For example, the 90\% CL upper limit from  IceCube, for neutrinos of all flavors in the energy range
$2.0 \times 10^{6}~{\rm GeV} < E_\nu < 6.3 \times 10^{9}~{\rm GeV}$, is~\cite{Abbasi:2011ji}
\begin{equation}
E_\nu^2 \ \Phi^{\nu_{\rm all}} \simeq 3.6 \times 10^{-8}~{\rm GeV}~{\rm cm}^{-2}~{\rm s}^{-1}~{\rm sr}^{-1} \, .
\end{equation}

We now derive model-independent bounds on the total neutrino flux.
The event rate for quasi-horizontal deep showers is
\begin{equation}
N = \sum_{i,X} \int dE_i\, N_A \, \Phi^i(E_i) \, \sigma_{i
N \to X} (E_i) \, {\cal E} (E_i)\ , \label{numevents}
\end{equation}
where the sum is over all neutrino species $i = \nu_e, \bar{\nu}_e,
\nu_{\mu}, \bar{\nu}_{\mu}, \nu_{\tau}, \bar{\nu}_{\tau}$, and all
final states $X$. $N_A = 6.022 \times 10^{23}$ is Avogadro's number,
and $\Phi^i $ is the source flux of neutrino species $i$, $\sigma$ as usual denotes 
the cross section, and ${\cal E}$ is the exposure measured in cm$^3$ w.e. sr time. The Pierre Auger Collaboration has searched for quasi-horizontal showers that
are deeply-penetrating~\cite{Abraham:2009eh}. There are no events  that unambiguously pass all the experimental cuts, with zero events expected from hadronic 
background. This implies an upper bound of 2.4 events at 90\%CL from neutrino fluxes~\cite{Feldman:1997qc}.
Note that if the number of events integrated over
energy is bounded by 2.4, then it is certainly true bin by bin in
energy. Thus, using Eq.~(\ref{numevents}) one obtains
\begin{equation}
\sum_{i,X} \int_{\Delta} dE_i\, N_A \, \Phi^i (E_i) \,
\sigma_{i N \to X} (E_i) \, {\cal E} (E_i)\  < 2.4\ ,
\label{bound}
\end{equation}
at 90\% CL for some interval $\Delta$. Here, the sum over $X$ takes
into account charge and neutral current processes.  In a logarithmic
interval $\Delta$ where a single power law approximation
\begin{equation}
\Phi^i(E_i)\, \sigma_{i N \to X} (E_i) \, {\cal E} (E_i)
\sim E_i^{\alpha}
\end{equation}
is valid, a straightforward calculation shows that
\begin{equation}
\int_{\langle E\rangle e^{-\Delta/2}}^{\langle E\rangle e^{\Delta/2}}
\frac{dE_i}{E_i} \,
E_i\, \Phi^i \, \sigma_{i N \to X}  \, {\cal
E}  =   \langle \sigma_{i N\rightarrow X}\,
{\cal E} \, E_i\, \Phi^i \rangle \, \frac{\sinh \delta}{\delta}\, \Delta \,,
\label{sinsh}
\end{equation}
where $\delta=(\alpha+1)\Delta/2$ and $\langle A \rangle$ denotes the
quantity $A$ evaluated at the center of the logarithmic interval~\cite{Anchordoqui:2002vb}.  The
parameter $\alpha = 0.363 + \beta - \gamma$, where 0.363 is the
power law index of the SM neutrino cross
section~\cite{Gandhi:1998ri} and $\beta$ and $-\gamma$ are the power
law indices (in the interval $\Delta$) of the exposure and flux
$\Phi^i$, respectively.  Since $\sinh \delta/\delta >1$, a
conservative bound may be obtained from Eqs.~(\ref{bound}) and
(\ref{sinsh}):
\begin{equation}
N_A\, \sum_{i,X} \langle \sigma_{i N\rightarrow X} (E_i)
\rangle \, \langle {\cal E} (E_i)\rangle\, \langle E_i
\Phi^i \rangle < 2.4/\Delta\ . \label{avg}
\end{equation}
By taking $\Delta =1$ as a likely interval in which the single power law behavior is
valid (this corresponds  to one $e$-folding of energy), it is straightforward to obtain  upper limits on the neutrino flux. The model-independent upper limits on the total neutrino flux,  derived using an  equivalent of 0.8~yr of full Auger exposure, are collected in Table~\ref{tableMI}~\cite{Abraham:2009eh}.

\begin{table}
\caption{Model-independent upper limits on the neutrino
flux at 90\% CL.}
\begin{tabular}{ccc} \hline \hline
$E_\nu$ (GeV) & \hspace{7.5cm}
&$ \langle E_\nu \ \Phi^{\nu_{\rm all}} \rangle$
(cm$^{-2}$ sr$^{-1}$ s$^{-1})$ \\
\hline $1\times 10^{8}$ & & $4.3 \times 10^{-14}$ \\
$3\times 10^8$ & & $5.3\times 10^{-15}$\\
$1\times 10^{9}$ & & $1.2 \times 10^{-15}$ \\
$3\times 10^9$ & & $4.7\times 10^{-16}$ \\
$1\times 10^{10}$ & &  $2.2 \times 10^{-16}$ \\
$3\times 10^{10}$ & & $1.3 \times 10^{-16}$ \\
$1\times 10^{11}$ & & $7.2 \times 10^{-17}$ \\
$3 \times 10^{11}$ & &$4.3 \times 10^{-17}$ \\
\hline \hline \label{tableMI}
\end{tabular}
\end{table}

\section{Phenomenology of extensive air showers}

\subsection{Systematic uncertainties in air shower measurements from  hadronic interaction models}
\label{hadronic}

Uncertainties in hadronic interactions at ultrahigh energies constitute one of the most problematic sources of systematic error  in the analysis of air showers.  This section will explain the  two principal schools of thought for extrapolating collider data to  ultrahigh energies.

Soft multiparticle production with small transverse momenta with respect to 
the collision axis is a dominant feature of most hadronic events at c.m. energies 
$10~{\rm GeV} < \sqrt{s} < 62~{\rm GeV}$ (see {\it e.g.},~\cite{Capella:yb,Predazzi:1998rp}). 
Despite the fact that strict calculations based on ordinary QCD perturbation 
theory are not feasible, there are some phenomenological models that 
successfully take into account the main properties of the soft diffractive 
processes. These models, inspired by $1/N$ QCD expansion are also 
supplemented with generally accepted theoretical principles like duality, 
unitarity, Regge behavior, and parton structure. The 
interactions are no longer described by single particle exchange, but by 
highly complicated modes known as Reggeons. Up to about 62~GeV, the slow growth of the cross 
section with $\sqrt{s}$ is driven by a dominant contribution of a special 
Reggeon, the Pomeron. 

At higher energies, semihard (SH) interactions arising from 
the hard scattering of partons that carry only a very small fraction of the 
momenta of their parent hadrons can compete successfully with soft 
processes~\cite{Cline:1973kv,Ellis:1973nb,Halzen:1974vh,Pancheri:sr,Gaisser:1984pg,DiasdeDeus:1984ip,
Pancheri:ix,Pancheri:qg}. These semihard interactions lead 
to the ``minijet'' phenomenon, {\em i.e.}  jets with transverse energy 
($E_T = |p_{_T}|$) 
much smaller than the total c.m. energy.  Such low-$p_{_T}$ processes
cannot be identified by jet finding algorithms, but  (unlike soft 
processes) still they can be calculated using perturbative QCD.
The cross section for SH interactions is described  by
\begin{equation}
\sigma_{\rm QCD}(s,p_{{_T}}^{\rm min}) = \sum_{i,j} \int 
\frac{dx_1}{x_1}\, \int \frac{dx_2}{x_2}\,
\int_{Q_{\rm min}^2}^{\hat{s}/2} \, d|\hat t|\,\, 
\frac{d\hat{\sigma}_{ij}}{d|\hat t|}\,\,
x_1 f_i(x_1, |\hat t|)\,\,\, x_2 f_j(x_2, |\hat t|) \,\,\,,
\label{sigmaminijet}
\end{equation}
where $x_1$ and $x_2$ are the fractions of the momenta of the parent hadrons 
carried by the partons which collide,
$d\hat{\sigma}_{ij}/d|\hat t|$ is the cross section for scattering of 
partons of types $i$ and $j$ according to elementary QCD diagrams, 
$f_i$ and $f_j$ are parton distribution functions (PDFs), 
$\hat{s} = x_1\,x_2 s$ 
and $-\hat{t} = \hat{s}\, (1 - \cos \vartheta^*)/2 =  Q^2$ 
are the Mandelstam variables for this parton-parton process,
and the sum is over all parton species. Here,
\begin{equation}
p_{_T} = E_{\rm jet}^{\rm lab} \,\,\sin \vartheta_{\rm jet} = \frac{\sqrt{\hat s}}{2}\,\, 
\sin \vartheta^*\,,
\end{equation}
and
\begin{equation}
p_{_\parallel} = E_{\rm jet}^{\rm lab} \,\,\cos \vartheta_{\rm jet}\,,
\end{equation}
where $E_{\rm jet}^{\rm lab}$ is the energy 
of the jet in the lab frame,   
$\vartheta_{\rm jet}$ the angle of the jet with respect to the beam 
direction in the lab frame, and $\vartheta^*$ is the angle of the jet with respect to the beam direction 
in the c.m. frame of the elastic parton-parton collision. This implies that for 
small $\vartheta^*,$ $p_{_T}^2 \approx Q^2$. The integration limits satisfy 
\begin{equation}
Q_{\rm min}^2 < |\hat t| < \hat{s}/2 \,,
\end{equation}
 where $Q_{\rm min} = 1 - 2~{\rm GeV}$ is the 
minimal momentum transfer. The measured minijet cross sections indicate that the onset of SH interactions has just occurred by CERN SPS  energies ($\sqrt{s} > 200~{\rm GeV}$~\cite{Albajar:1988tt}.

\begin{figure}[tbp]
  \postscript{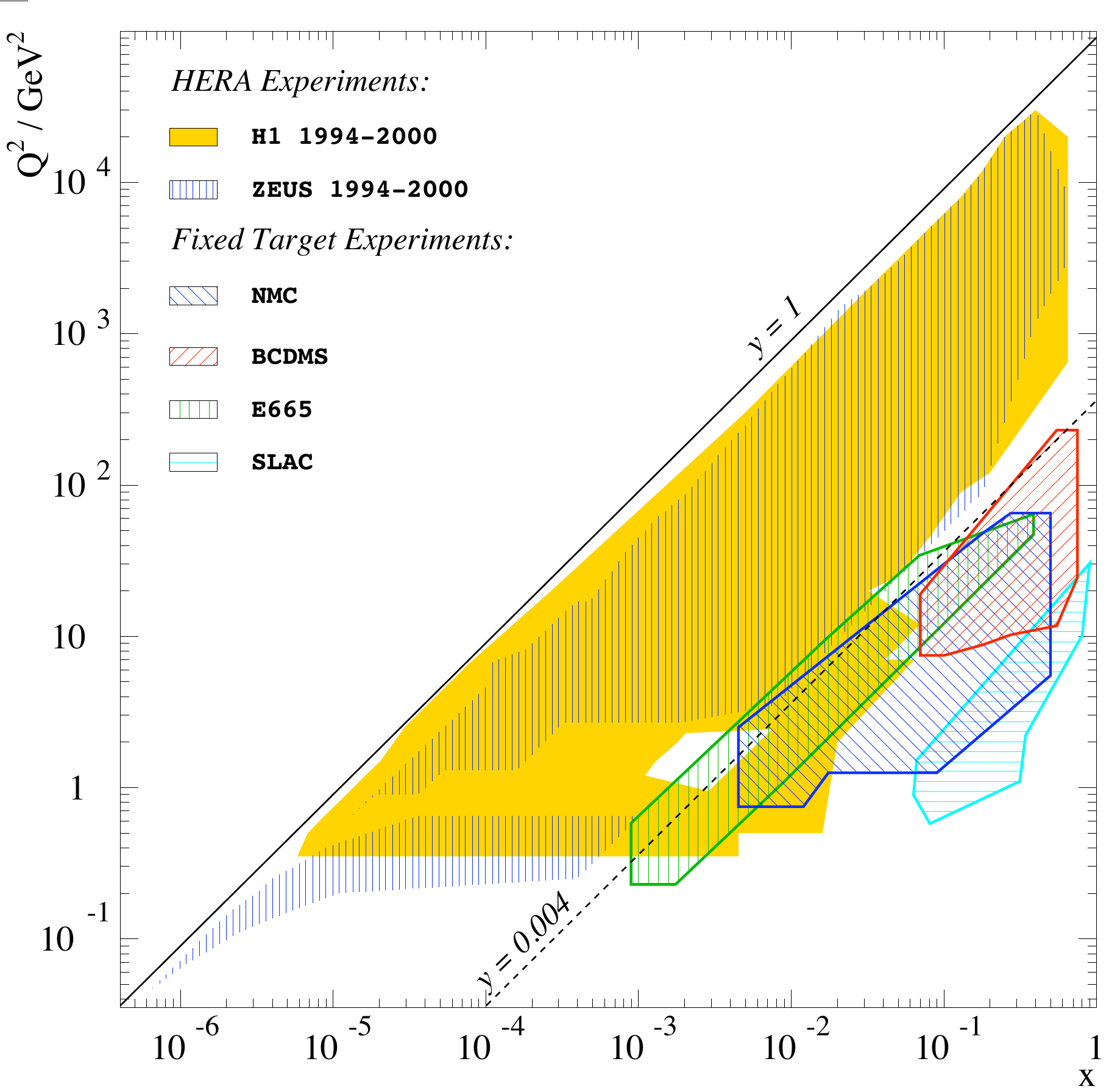}{0.70}
  \caption{Kinematic $x$-$Q^2$ plane accessible to the H1 and ZEUS experiments at HERA 
    and the region accessible to fixed-target experiments. The inelasticity $y = (1 -\cos \vartheta^*)/2$ is also shown. From Ref.~\cite{Anchordoqui:2004xb}.}
\label{hera}
\end{figure}

A first source of  uncertainty in modeling cosmic ray interactions at 
ultrahigh energy is encoded in the extrapolation of the measured 
parton densities  several orders of magnitude down to low $x$. 
Primary protons that impact on the upper atmosphere with energy $\sim 10^{11}$~GeV  
yield partons with $x \equiv 2 p^*_{_\parallel}/\sqrt{s} \sim m_\pi/\sqrt{s} \sim  10^{-7},$ whereas current data on quark and gluon 
densities are only available for $x \gtrsim 10^{-4}$ to within an experimental accuracy of 3\% for 
$Q^2 \approx 20$~GeV$^2$~\cite{Adloff:2000qk}. In Fig.~\ref{hera} we show the region of the 
$x-Q^2$ plane probed by H1, ZEUS,  and fixed target experiments. In addition, extrapolation of HERA data to UHECR interactions assumes universality
of the PDFs. This assumption, based on  the QCD factorization conjecture (the cross section of
Eq.~(\ref{sigmaminijet}) can always be written in a form which factorizes the parton
densities and the hard interaction processes irrespective of the order 
in perturbation theory and the particular hard process)  holds in
the limit $Q^2 \gg \Lambda_{\rm QCD}$, where $\Lambda_{\rm QCD} \sim 200$~MeV is the 
QCD renormalization scale. 

\begin{figure}[tbp]
\postscript{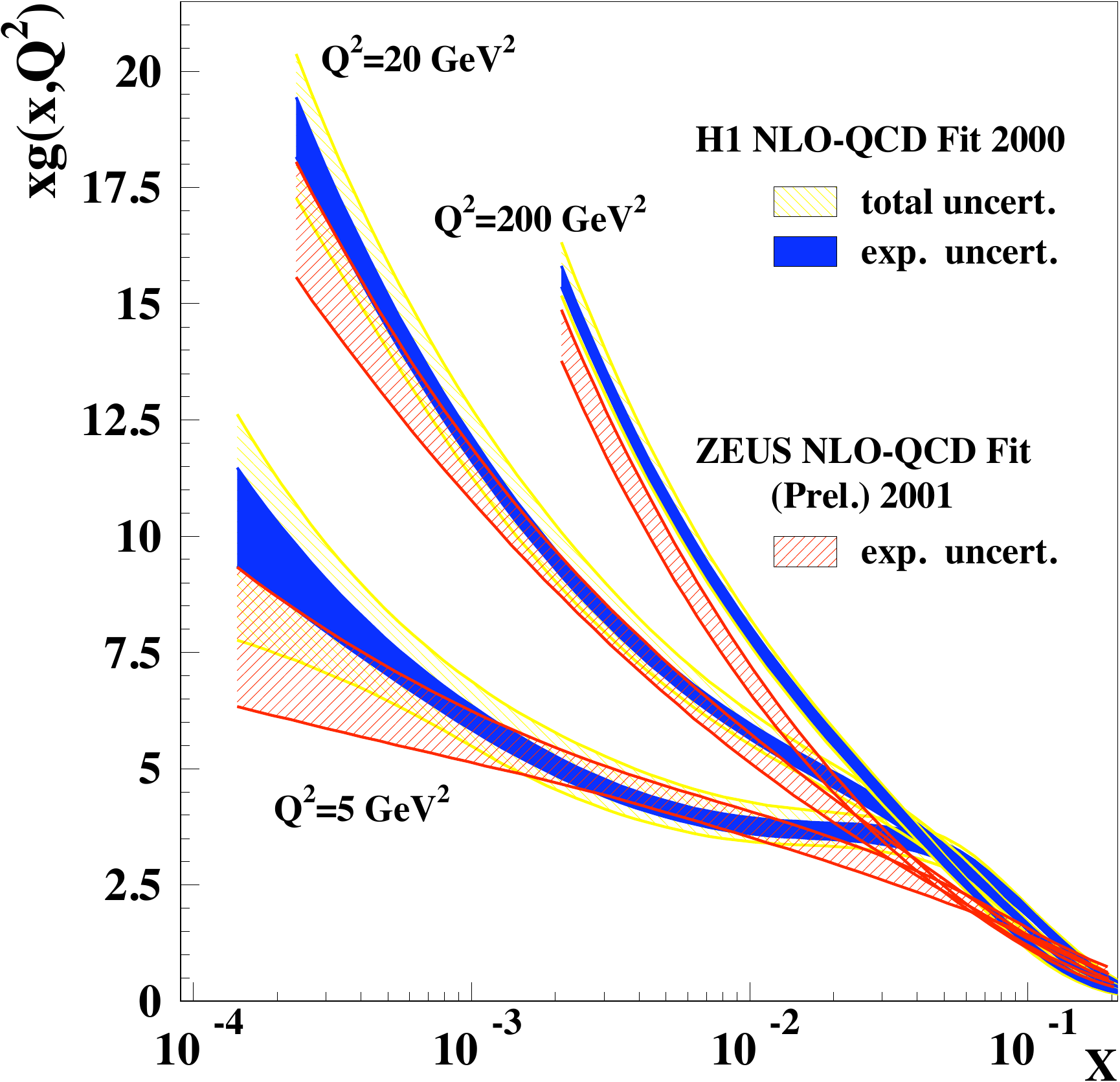}{1.0} 
\caption{Gluon momentum distributions $x g(x,Q^2)$ in the proton
  as measured by the ZEUS and H1 experiments at various $Q^2$. From Ref.~\cite{Dittmar:2009ii}.}
\label{gpdf}
\end{figure}

For large $Q^2$ and not too small $x$, the 
Dokshitzer-Gribov-Lipatov-Altarelli-Parisi (DGLAP) 
equations~\cite{Gribov:rt,Gribov:ri,Dokshitzer:sg,Altarelli:1977zs} 
\begin{equation}
\frac{\partial}{\partial \ln Q^2} {q(x, Q^2) \choose g(x,Q^2)}  =
\frac{\alpha_s(Q^2)}{2 \pi} {P_{qq} \,\,\,\, P_{qg} \choose P_{gq} 
\,\,\,\, P_{gg}}  \otimes  {q(x, Q^2) \choose g(x,Q^2)} 
\end{equation} 
successfully 
predict the $Q^2$ dependence of the quark and gluon densities ($q$ and $g,$ 
respectively). Here, $\alpha_s = g^2/(4\pi),$ with  $g$ the strong 
coupling constant. The splitting functions $P_{ij}$ indicate the probability
of finding a daughter parton $i$ in the parent parton $j$ with a given 
fraction of parton $j$ momentum. This probability will depend on the number 
of splittings allowed in the approximation. In the double-leading-logarithmic approximation, that is $\lim_{x \to
  0} \ln(1/x)$ and $\lim_{Q^2 \to \infty}\ln(Q^2/\Lambda_{\rm QCD})$,
the 
DGLAP equations predict a steeply rising gluon density, $xg \sim x^{-0.4},$
which dominates the quark density at low $x$. This prediction is 
in agreement with the experimental results from HERA shown in
Fig.~\ref{gpdf}~\cite{Chekanov:2002pv,Adloff:2003uh}. HERA data are found to be consistent with a 
power law, $xg(x,Q^2) \sim x^{-\Delta_{\rm H}},$ with an exponent $\Delta_{\rm H}$ 
between 0.3 and 0.4~\cite{Engel:ac}.

The high energy minijet cross section is then 
determined by  the dominant gluon distribution 
\begin{equation}
\sigma_{\rm QCD} (s,p_{{_T}}^{\rm min})   
\approx \int \frac{dx_1}{x_1}\, \int \frac{dx_2}{x_2} \,
\int_{Q_{\rm min}^2}^{\hat{s}/2} \, d|\hat t|\,\,\, 
\frac{d\hat \sigma}{d|\hat t|}\,\,\,
x_1 g(x_1, |\hat t|)\,\,\, x_2 g(x_2, |\hat t|) \,\,,
\label{gluon}
\end{equation}
where the integration limits satisfy
\begin{equation}
x_1 x_2 s > 2 |\hat t| > 2 Q^2_{\rm min} \, .
\end{equation}
Furthermore, because $d\hat \sigma/d|\hat t|$ is peaked at the low end of the $|\hat t|$ integration (see {\it e.g}~\cite{Anchordoqui:2009eg}), the high energy behavior of $\sigma_{\rm QCD}$ is controlled (via the lower limits of the $x_1, \, x_2$ integrations) by the small-$x$ behavior of the gluons~\cite{Kwiecinski:1990tb}
\begin{eqnarray}
\sigma_{\rm QCD} (s)   
\propto  \int_{2\,Q_{\rm min}^2/s}^{1} \frac{dx_1}{x_1}\,\,
x_1^{-\Delta_{\rm H}}\, \int_{2 \,Q_{\rm min}^2/x_1s}^{1} \frac{dx_2}{x_2} 
\,\,x_2^{-\Delta_{\rm H}} \sim 
s^{\Delta_{\rm H}}\, \ln s \, \sim_{_{\!\!\!\!\! \! \!\!\!\! s \to \infty}}  s^{\Delta_{\rm H}} \, .
\label{KH}
\end{eqnarray}
This estimate is, of course, too simplistic. At sufficiently small $x$, gluon shadowing corrections suppress the singular $x^{-\Delta_H}$  behavior of $xg$ and hence suppress the power growth of $\sigma_{\rm QCD}$ with increasing $s$.

Although we have shown that the onset of semihard processes is an unambiguous prediction of QCD, in practice it is difficult to isolate these contributions from the soft interactions. 
Experimental evidence indicates that  SH interactions can essentially be neglected up to and throughout the CERN ISR energy regime, $\sqrt{s} < 62~{\rm GeV}$. Therefore,  measurements made in this energy region can be used to model the soft interactions. A reasonable approach introduced in~\cite{DiasdeDeus:1987yw}  is to base the extrapolation of the soft interactions on the assumption of geometrical scaling~\cite{DiasDeDeus:1987bf}, which is observed to be true throughout the ISR energy range~\cite{Amaldi:1979kd,Castaldi:1985ft}. To this end, we introduce the standard partial-wave amplitude in impact-parameter space $f(s,b)$, which is the Fourier transform of the elastic $pp$ (or $p\bar p$) scattering amplitude. (We neglect any difference between $pp$ and $p \bar p$ for $\sqrt{s} > 200~{\rm GeV}$.) Geometrical scaling (GS) corresponds to the assumption that $f$, which {\em a priori} is a function of two  dimensional variables $b$ and $s$, depends only upon one dimensional  variable $\beta = b/R(s)$, where $R$ is the energy dependent radius of the proton, {\it i.e.}
\begin{equation}
f(s,b) = f_{\rm GS} \left(\beta = b/R(s) \right) \, .
\end{equation}
Physically, this means that the opaqueness of the proton remains constant with rising energy  and that the increase of the total cross section, $\sigma_{\rm tot}$, in the ISR energy range reflects a steady growth of the radius $R(s)$. An immediate obvious consequence of GS is that the partial wave at $b=0$ should be independent of energy
\begin{equation}
f(s,b=0) = f_{\rm GS} (\beta =0) \, .
\end{equation}
Another consequence is that the ratio of elastic scattering to total cross section, $\sigma_{\rm el} (s)/\sigma_{\rm tot} (s)$, should be energy independent. This follows from
\begin{eqnarray}
\sigma_{\rm tot} & = & 8 \pi \int {\rm Im} f(s,b) \, b \, db \nonumber \\
 & = & 8 \pi R^2(s) \, \int {\rm Im}  f_{\rm GS}(\beta) \, \beta \, d\beta, \\
 & ^{\rm GS} & \nonumber
\end{eqnarray}
and
\begin{eqnarray}
\sigma_{\rm el} & = & 8 \pi \int \left|f(s,b) \right|^2 \, b \, db \nonumber \\
 & = & 8 \pi R^2(s) \, \int \left|f_{\rm GS}(\beta) \right|^2 \, \beta \, d\beta \, . \\
 & ^{\rm GS} & \nonumber
\end{eqnarray}
To determine the gross features at high energies  we can assume that the elastic amplitude has a simple form 
\begin{equation}
F(s,t) = i \ \sigma_{\rm tot} (s) \ \ e^{Bt/2} \, ,
\label{giusti}
\end{equation}
with $B$  the  slope parameter that measures the size of the proton~\cite{Block:1984ru}. This is a reasonable assumption: the amplitude is predominantly imaginary, and the exponential behavior observed for $|t| \lesssim 0.5~{\rm GeV}^2$ gives the bulk of the elastic cross section. Now, the Fourier transform $f(s,b)$ of the elastic amplitude $F(s,t)$ given by (\ref{giusti}) has a Gaussian form in impact parameter space
\begin{equation}
f(s,b) = \frac{i \sigma_{\rm tot}(s)}{8 \pi B} e^{-b^2/2B} \, ,
\label{percudani}
\end{equation}
and it follows that
\begin{equation}
{\rm Im}  f(s, b=0) = \frac{\sigma_{\rm tot}}{8 \pi B} = \frac{2 \sigma_{\rm el}}{\sigma_{\rm tot}} \, .
\label{bochini}
\end{equation}
Equation (\ref{bochini}) offers a very clear way to see the breakdown of GS and to identify semihard interactions from the growth of the central partial wave.

In general unitarity requires ${\rm Im}  f(s,b) \leq \frac{1}{2}$, which in turm implies $\sigma_{\rm el}/\sigma_{\rm tot} \leq \frac{1}{2}$~\cite{Block:1984ru}. This seems to indicate that the Gaussian form (\ref{percudani}) may not longer be applicable at ultrahigh energies, but rather it is expected that the proton will approximate a ``black disk'' of radius $b_0$, {\em i.e.}  $f(s,b) = \frac{i}{2}$ for $0<b\lesssim b_0$ and zero for $b \gtrsim b_0$. Then $\sigma_{\rm el} \simeq \frac{1}{2} \sigma_{\rm tot} \simeq \pi b_0^2$. 

In order to satisfy the unitarity constraints, it is convenient to introduce the eikonal $\chi$ defined by
\begin{equation}
f(s,b) = \frac{i}{2} \left\{ 1 - {\rm exp} \left[i \chi (s,b) \right] \right\} \,,
\end{equation}
where ${\rm Im}  \, \chi \geq 0$. If we neglect for the moment the shadowing corrections to the PDFs and take $xg \propto x^{-\Delta_{\rm H}}$ in the small-$x$ limit we obtain, as explained above,  power growth of the cross section for SH interactions, $\sigma_{\rm QCD} \sim s^{\Delta_{\rm H}}$ and 
${\rm Im}  \chi(s, b=0) \gg 1$ as $s \to \infty$. Indeed we expect ${\rm Im}  \, \chi \gg 1$ (and unitarity to be saturated) for a range
of $b$ about $b=0$. Then we have
\begin{eqnarray}
\sigma_{\rm tot} & = & 4 \pi \int_0^\infty b \ db \,  \Theta (b_0 -  b) \nonumber \\
 & \simeq & 4 \pi \int_0^{b_0(s)} b \,  db = 2 \pi b_0^2 \,,
\end{eqnarray}
with $\chi \simeq \chi_{_{\rm SH}}$ and where $b_0(s)$ is such that
\begin{equation}
{\rm Im} \, \chi_{_{\rm SH}} (s, b_0(s)) \simeq 1 \, .
\label{handset}
\end{equation}

Hereafter, we ignore the small real part of the scattering amplitude, which is good approximation at high energies. The unitarized elastic, inelastic, and total cross sections (considering now a real eikonal function) are given by~\cite{Glauber:1970jm,L'Heureux:jk,Durand:prl,Durand:cr}
\begin{equation}
\sigma_{\rm el}=2\pi \int d b\, b\
\left\{1-\exp\left[ -\chi_{_{\rm soft}}(s, b)
-\chi_{_{\rm SH}}(s, b)\right]\right\}^2\ \,,
\label{elastic}
\end{equation}
\begin{equation}
\sigma_{\rm inel}= 2 \pi \int d b\, b \,
\left\{1-\exp\left[ -2\chi_{_{\rm soft}}(s, b)
-2\chi_{_{\rm SH}}(s, b)\right]\right\}\ ,
\label{inelastic}
\end{equation}
\begin{equation}
\sigma_{\rm tot}= 4 \pi \int d b\, b\,
\left\{1-\exp\left[-\chi_{_{\rm soft}}(s, b)
-\chi_{_{\rm SH}}(s, b)\right]\right\}\ ,
\label{total}
\end{equation}
where the scattering is compounded as a sum of QCD ladders
via SH and soft processes through the
eikonals $\chi_{_{\rm SH}}$ and $\chi_{_{\rm soft}}$. 

Now, if the eikonal function, 
$\chi (s, b) \equiv \chi_{_{\rm soft}}(s, b) + 
\chi_{_{\rm SH}}(s, b) =\lambda/2,$
indicates the mean number of partonic interaction pairs at impact parameter 
$b,$ the probability $p_n$ for having $n$ independent partonic 
interactions using Poisson statistics reads, 
$p_n = (\lambda^n/n!) \, e^{-\lambda}$.
Therefore, the factor $1-e^{-2\chi} = \sum_{n=1}^\infty p_n$ in Eq.~(\ref{inelastic}) 
can be interpreted semiclassically as the probability 
that at least 1 of the 2 protons is broken up in a collision at impact 
parameter $b$.
With this in mind, the inelastic cross section is simply the integral
over all collision impact parameters of the probability of having at 
least 1 interaction, yielding a mean minijet multiplicity of 
\mbox{$\langle n_{\rm minijet} \rangle \approx \sigma_{\rm QCD}/
\sigma_{\rm inel}$~\cite{Gaisser:1988ra}.} The leading
contenders to approximate the (unknown) cross sections at
cosmic ray energies, {\sc sibyll}~\cite{Fletcher:1994bd} and 
{\sc qgsjet}~\cite{Kalmykov:te}, share the eikonal
approximation but differ in their {\em ans\"atse} for the
eikonals. In both cases, the core of dominant scattering at
very high energies is the SH cross section given in
Eq.~(\ref{sigmaminijet}),
\begin{equation}
\chi_{_{\rm SH}} = \frac{1}{2} \, 
\sigma_{\rm QCD}(s,p_{{_T}}^{\rm min})\,\, A(s,\vec b) \,,
\label{hard}
\end{equation}
where the normalized profile function, 
$2 \pi \int_0^\infty d b \, b \,A(s, b) = 1,$
 indicates the distribution of partons in the plane transverse to the 
collision axis. 

In the {\sc qgsjet}-like 
models, the  core of the SH eikonal 
is dressed with a soft-pomeron pre-evolution factor. This amounts to
taking a parton distribution which is Gaussian in the transverse
coordinate distance $b,$ 
\begin{equation}
A(s, b) = \frac{e^{- b^2/R^2(s)}}{\pi R^2(s)} \, ,
\label{a}
\end{equation}
with  $R$ being a parameter. For a QCD cross section dependence, $\sigma_{\rm QCD} \sim 
 s^{\Delta_{\rm H}},$ one gets for a Gaussian profile
\begin{equation}
b_0^2(s) \sim R^2 \Delta_{\rm H} \, \ln s
\end{equation}
 and at high energy 
\begin{eqnarray}
\sigma_{\rm inel}  =   2 \pi \int_0^{b_0(s)} d b \, b  \sim  \pi R^2 \Delta_{\rm H} \, \ln s \, .
\label{qsig}
\end{eqnarray}
If the effective radius $R$ (which controls parton shadowing) is energy-independent, the cross section increases only logarithmically with rising energy. However, the parameter $R$ itself depends on the collision energy through a convolution with the parton momentum fractions, $R^2(s) \sim  R_0^2 + 4 \, \alpha^\prime_{\rm eff} \,\ln^2 s$, with 
$\alpha^\prime_{\rm eff} \approx 0.11$~GeV$^{-2}$~\cite{Alvarez-Muniz:2002ne}. Thus, the {\sc qgsjet}  cross section exhibits a faster than $\ln \, s$ rise,
\begin{equation}
 \sigma_{\rm inel } \sim 4\pi \, \alpha^\prime_{\rm eff} \,\,\Delta_{\rm H}\,\,
\ln^2 s \, .
\label{beber}
\end{equation}

In  {\sc sibyll}-like models, the transverse density
distribution is taken as the Fourier transform of the proton electric
form factor, resulting in an energy-independent exponential 
(rather than Gaussian) fall-off of the parton density profile 
 for large $b$,
\begin{eqnarray}
A(b)  =  \frac{\mu^2}{96 \pi} (\mu b)^3 \,   K_3 (\mu b) 
  \sim  e^{-\mu b} \,,
\end{eqnarray} 
where $K_3$ denotes the modified Bessel function of the third kind and $\mu^2 \approx 0.71  {\rm GeV}^{2}$~\cite{Fletcher:1994bd}. Thus, (\ref{hard}) becomes
\begin{equation}
\chi_{_{\rm SH}} \sim e^{-\mu b}  \,  s^{\Delta_{\rm H}} ,
\end{equation}
 and  (\ref{handset}) is satisfied when
\begin{equation}
b_0(s) = \frac{\Delta_{\rm H}}{\mu} \, \ln \, s \, .
\end{equation}
Therefore, for {\sc sibyll}-like models, the  growth of the inelastic
cross section also saturates the $\ln^2s$ Froissart 
bound~\cite{Froissart:ux},
but with a multiplicative constant which is larger 
than the one in {\sc qgsjet}-like models
\begin{equation}
\sigma_{\rm inel} \sim \pi c \,\, \frac{\Delta_{\rm H}^2}{\mu^2} \,  \ln^2 s \,\,,
\label{ssig}
\end{equation}
where the coefficient $c \approx 2.5$ is found numerically~\cite{Alvarez-Muniz:2002ne}.

The main characteristics of the $pp$ cascade spectrum
resulting from these choices are readily predictable: the harder 
form of the {\sc sibyll}
form factor allows a greater retention of energy by the leading
particle, and hence less available for the ensuing 
shower. Consequently, on average {\sc sibyll}-like models predict a smaller  
multiplicity than {\sc qgsjet}-like models~\cite{Anchordoqui:1998nq}. 

There are three event generators, {\sc sibyll}~\cite{Fletcher:1994bd}, 
{\sc qgsjet}~\cite{Kalmykov:te}, 
and {\sc dpmjet}~\cite{Ranft:fd} 
which are tailored specifically for simulation of hadronic interactions up to 
the highest cosmic ray energies. The latest versions of these packages are {\sc sibyll} 
2.1~\cite{Ahn:2009wx}, {\sc qgsjet} II-03~\cite{Ostapchenko:2007qb}, and {\sc dpmjet III}~\cite{Roesler:2000he}; 
respectively.  In {\sc qgsjet}, both the soft and hard 
processes are formulated in terms of Pomeron exchanges. To describe the 
minijets, the soft Pomeron  mutates into a ``semihard Pomeron'', 
an ordinary soft Pomeron with the middle piece replaced by a QCD parton 
ladder, as sketched in the previous paragraph.  This is generally referred to 
as the ``quasi-eikonal'' model.  In
contrast, {\sc sibyll} and {\sc dpmjet} follow a ``two channel'' eikonal model, where the soft and  
the semi-hard regimes are demarcated by a sharp cut in the transverse momentum: 
{\sc sibyll} 2.1 uses a cutoff parametrization inspired in the double leading logarithmic approximation 
of the DGLAP equations,
\begin{equation}
p_{{_T}}^{\rm min} (\sqrt{s}) =  p_{{_T}}^0 + 0.065~{\rm GeV}\, \exp[0.9\,\sqrt{\ln s}]\,,
\end{equation}
whereas {\sc dpmjet} III uses an {\it ad hoc} parametrization for the transverse momentum cutoff
\begin{equation}
p_{{_T}}^{\rm min} (\sqrt{s}) =  p_{{_T}}^0 + 0.12~{\rm GeV}\, [\log_{10} (\sqrt{s}/50{\rm GeV})]^3\,,
\end{equation}
where  $p_{{_T}}^0 = 2.5$~GeV~\cite{Engel:ac}.

The transition process from asymptotically free partons to colour-neutral 
hadrons is described in all codes by string fragmentation 
models~\cite{Sjostrand:1987xj}.  Different choices of fragmentation functions 
can lead to some differences in the hadron multiplicities. 
However, the main difference in the predictions of {\sc qgsjet}-like and 
{\sc sibyll}-like models arises from different assumptions in extrapolation of the parton 
distribution function to low energy.

The proton-air collisions of interest for air shower development cause additional headaches for event generators. 
Both {\sc sibyll} and {\sc qgsjet}  adopt the
Glauber formalism~\cite{Glauber:1970jm}, which is equivalent
to the eikonal approximation in nucleon-nucleon scattering, 
except that the nucleon density functions of the target nucleus are
folded with that of the nucleon.  The inelastic and production
cross sections read:
\begin{equation}
\sigma_{\rm inel}^{p{\rm -air}} \approx 2 \pi \int d b\, b \,
\left\{1-\exp\left[\sigma_{\rm tot} 
\,\, A T_N( b) \right]\right\}\ \,,
\label{inelasticn}
\end{equation}
\begin{equation}
\sigma_{\rm prod}^{p{\rm -air}} \approx 2 \pi \int d b\, b\,
\left\{1-\exp\left[\sigma_{\rm inel} 
\,\, A T_N(b) \right]\right\}\ \,,
\label{prod}
\end{equation}
where  $T_N(b)$ is the transverse distribution function of a nucleon inside a nucleus. Here, 
$\sigma_{\rm inel}$ and $\sigma_{\rm tot}$ are given by Eqs.~(\ref{inelastic}) and 
(\ref{total}), respectively. The $p$-air inelastic cross section is the sum of the
``quasi-elastic'' cross section, which corresponds to 
cases where the target nucleus breaks up without production
of any new particles, and the production cross section,
in which at least one new particle is generated.  Clearly 
the development of  EASs is mainly sensitive to the production
cross section.  Overall, the geometrically large size of 
nitrogen and oxygen nuclei dominates 
the inclusive proton-target cross section, and as a result
the disagreement from model-dependent
extrapolation is not more than about 15\%. More complex nucleus-nucleus interactions are discussed in~\cite{Engel:vf}.\\  

\begin{figure}[tbp]
\begin{minipage}[t]{0.48\textwidth}
\postscript{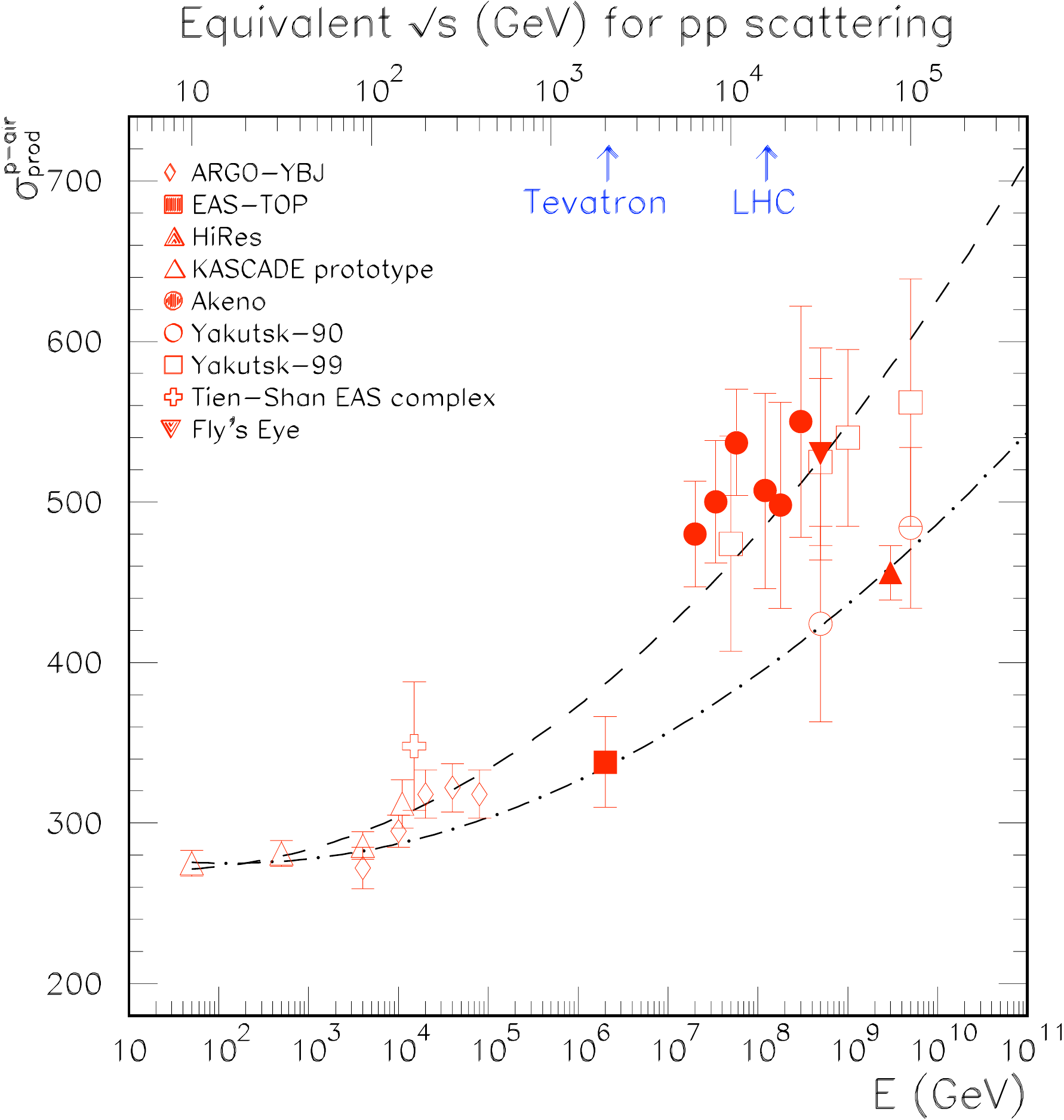}{0.99}
\end{minipage}
\hfill
\begin{minipage}[t]{0.48\textwidth}
\postscript{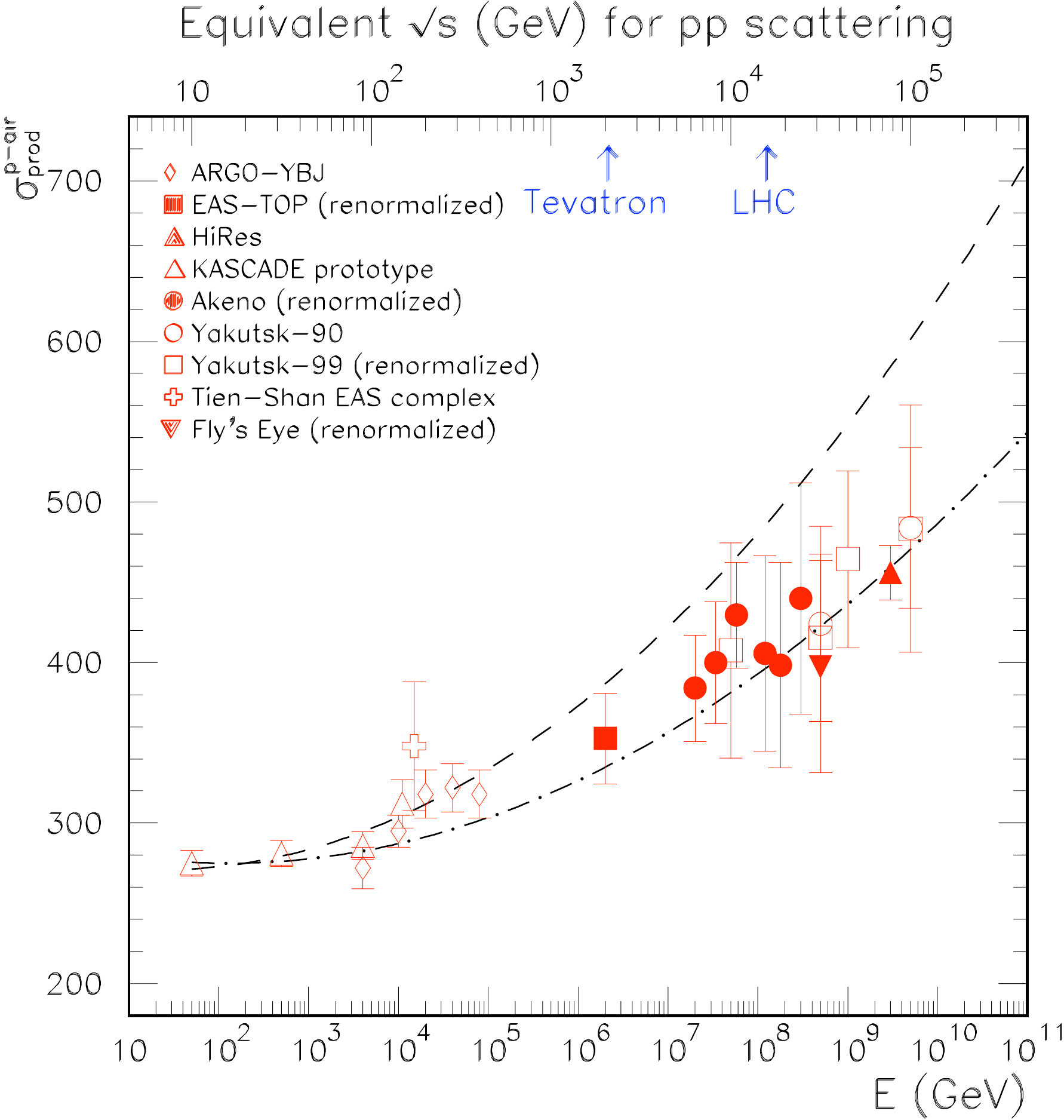}{0.99}
\end{minipage}
\caption{Compilation of proton-air production cross section from cosmic ray measurements (ARGO-YBJ~\cite{Aielli:2009ca}, EAS-TOP~\cite{Aglietta:2009zza}, HiRes~\cite{Belov:2006mb}, KASCADE prototype~\cite{Mielke:1994un}, Akeno~\cite{Honda:1992kv}, Yakutsk-90~\cite{Dyakonov:1990cd}, Yakutsk-99~\cite{Knurenko:1999cr}, Tien-Shan EAS complex~\cite{Nam:1975xk} and Fly's Eye~\cite{Baltrusaitis:1984ka}). The data are compared to the parametrizations discussed in the text; case-$i$ corresponds to the dashed line and case-$ii$ to the dot-dashed line. \label{fig:pair}}
\end{figure}

{\bf EXERCISE 3.1}~~~Consider a typical air nuclei of average $\langle A \rangle = 14.5$  
and calculate the proton-air cross section using the approximated expressions (\ref{beber}) and (\ref{ssig})  together with the $z$-integrated  Woods-Saxon profile~\cite{Woods:1954zz}
\begin{equation}
T_N (b) = \frac{1}{Z} \int_{-\infty}^\infty dz \, \left\{ 1 + \exp \left[(\sqrt{b^2 + z^2} - R_N)/\alpha \right] \right\}^{-1} \,, 
\end{equation}
where 
\begin{equation}
Z = \frac{4 \pi}{3} R_N^3 \left[ 1 + \pi^2 \left(\frac{\alpha}{R_N}\right)^2 \right]
\end{equation}
$\alpha =0.5~{\rm fm}$ and $R_N = 1.1 A^{1/3}~{\rm fm}$~\cite{Portugal:2009xc}. Compare the results with those shown in Fig.~\ref{fig:pair}.\\

Adding a greater challenge to the determination of the proton air cross section at ultrahigh energies is the lack of direct measuremnts in a controlled laboratory environment. The measured shower attenuation length, $\Lambda_m$, is not only
 sensitive to the interaction length of the protons in the atmosphere, $\lambda_{p{\rm -air}}$, with
\begin{equation}
\Lambda_m = k \lambda_{p{\rm -air}} = k { 14.4~m_p \over \sigma_{\rm prod}^{p{\rm -air}}} \,,  \label{eq:Lambda_m}
\end{equation}
but also depends on the rate at which the energy of the primary proton
 is dissipated into EM shower energy observed in the
 experiment. Here, $\Lambda_m$ and $\lambda_{p{\rm -air}}$ are in g\,cm$^{-2}$,  the proton mass $m_p$ is in g, and the inelastic production cross section $\sigma_{\rm prod}^{p{\rm -air}}$ is in mb.
The value of $k$ depends critically on the inclusive  particle production cross section and its energy dependence in nucleon and meson interactions on the light nuclear target of the atmosphere. The measured depth $X_{\rm max}$ at which a shower reaches
 maximum development in the atmosphere has been the basis of 
 cross section measurements from experiments prior to HiRes and Auger. However, $X_{\rm max}$ is a combined measure
 of the depth of the first interaction (which is determined by
 the inelastic cross section) and of the subsequent shower development (which has to be corrected for). 
 The model dependent rate of shower development and its fluctuations
 are the origin of the deviation of $k$ from unity
 in Eq.\,(\ref{eq:Lambda_m}). As can be seen in Table~\ref{ktable}, there is a  large range of $k$ values (from 1.6 for a very old model where the inclusive cross section exhibited Feynman scaling, to 1.15  for modern models with large scaling violations) that make the published  values of $\sigma_{p{-\rm air}}$ unreliable.

\begin{table}
\caption{Different $k$-values used in cosmic ray experiments.\label{ktable}} 
\begin{center}
 \begin{tabular}[b]{l||c}
\hline
\hline
Experiment~~~~~~~~~~&~~~~~~~~~~$k$\\ \hline
Fly's Eye~~~~~~~~~~&~~~~~~~~~~1.6~\\ \hline
Akeno~~~~~~~~~~&~~~~~~~~~~1.5~\\ \hline
Yakutsk-99~~~~~~~~~&~~~~~~~~~~1.4~\\ \hline
EAS-TOP~~~~~~~~~~& ~~~~~~~~~~1.15\\ \hline
\hline
\end{tabular}
\end{center}
\end{table}

Recently, the HiRes Collaboration developed a quasi-model-free method of measuring $\sigma_{\rm prod}^{p{\rm -air}}$ directly~\cite{Belov:2006mb}.  This is accomplished by folding a randomly generated exponential distribution of  first interaction points into the shower development program, and therefore fitting the entire distribution and not just the trailing edge.  
Interestingly, the measured  $k = 1.21^ {+0.14}_{-0.09}$ by the HiRes group is in agreement with the one obtained by tuning the data to the theory~\cite{Block:1999ub,Block:2007rq}.

A compilation of published proton-air cross section measurements is shown in Fig.~\ref{fig:pair}. In the left panel we show the data without any modification. In the right panel, the published values of $\sigma_{\rm prod}^{p{\rm -air}}$ for Fly's Eye~\cite{Baltrusaitis:1984ka}, Akeno~\cite{Honda:1992kv}, Yakutsk-99~\cite{Knurenko:1999cr}, and EAS-TOP~\cite{Aglietta:2009zza} collaborations have been renormalized using the {\em common} value of  $k = 1.264 \pm 0.033 (\rm stat) \pm 0.013 (\rm syst)$~\cite{Block:2007rq}. 
 We have parametrized the rise of the cross section using a functional form that saturates the  
 Froissart bound,
\begin{equation}
\sigma_{\rm prod}^{p{\rm -air}} = {\cal A} - {\cal B} \ln  (E/{\rm GeV}) + {\cal C} \ln^2 (E/{\rm GeV})~{\rm mb} \, .
\end{equation}
The curve with a fast rise, hereafter referred to as case-$i$, corresponds to 
${\cal A} = 280$, ${\cal B} = 5.7$, and ${\cal C} = 0.9$. The slow rise of case-$ii$ has the following parameters: ${\cal A} = 290$, ${\cal B} = 6.2,$ and ${\cal C} = 0.64.$ The behavior of the cross section in case-$i$ roughly matches the one implemented in  {\sc sibyll}-2.1~\cite{Ahn:2009wx}.

\begin{figure}[tbp]
\postscript{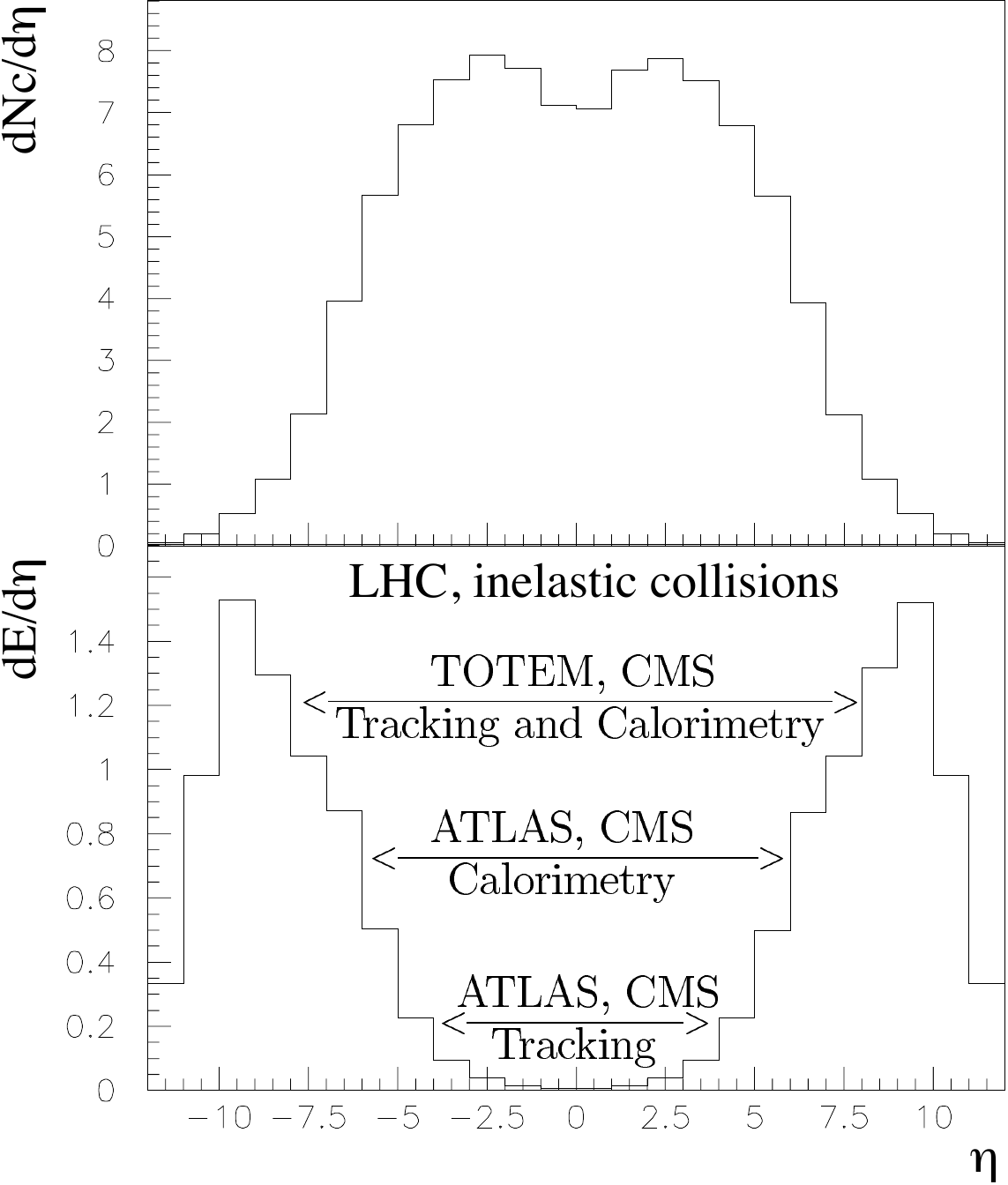}{0.70}
\caption{Pseudorapidity distributions of charged particles (upper panel) and of 
the energy flow (lower panel) for $pp$ collisions at LHC. From Ref.~\cite{Eggert:ca}.} 
\label{totem1}
\end{figure}

Some guidance towards understanding hadronic processes in the forward
direction may come directly from measurements of hadrons in airshowers~\cite{Antoni:2001qj}.
However, the most useful experimental input in the forseeable future will
likely come from the LHC.  Processes with low momentum transfer   tend to populate the region at very small angles $\vartheta$ with respect to the beam direction.  The distribution of pseudorapidity, $\eta = -\ln \tan (\vartheta/2)$, and 
the energy flow distribution are shown in Fig.~\ref{totem1}.  While
the particle multiplicity is greatest in the low $| \eta |$ region, it  
is clearly seen that the energy flow is peaked at small production angles (large $| \eta |$). 
The two general-purpose experiments, ATLAS and CMS,  cover up to $|\eta| < 5$. TOTEM's 
coverage in the very forward rapidity range, $3.1 \leq \eta \leq 6.5$, is committed to measure the
$pp$ total and elastic scattering cross sections, see Appendix~\ref{AC}. 
The fragmentation region that plays a crucial role in the development of EASs tends to populate the region corresponding  to pseudorapidity range  $6\le |\eta|\le 10$.  

\begin{figure}
\centering
\postscript{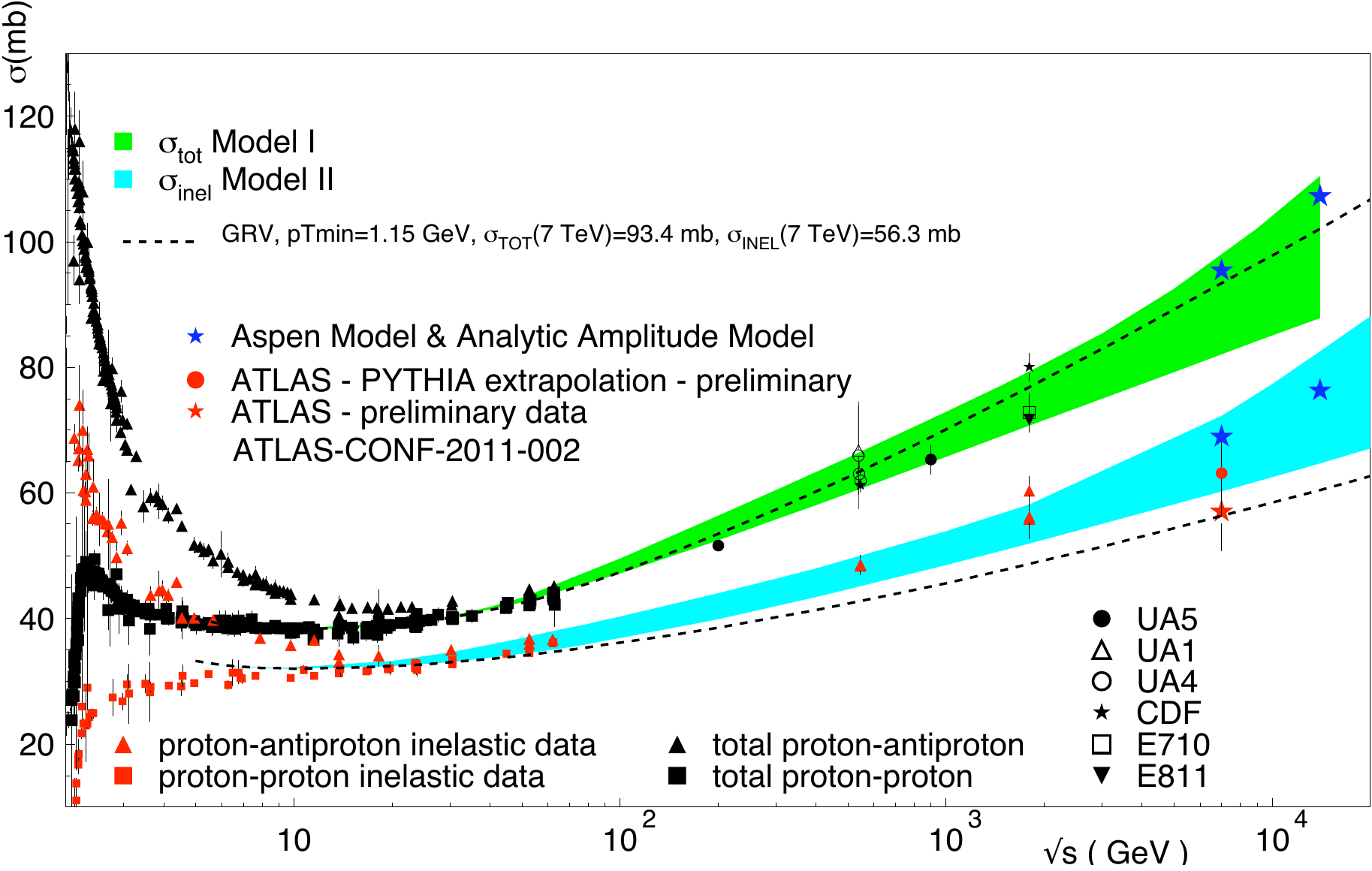}{0.9}
\caption{Compilation of total and inelastic $pp$ and $p {\bar p}$ cross section measurements. 
The data are compared to model predictions. The dashed lines indicate the predictions of Model II~\cite{Achilli:2011sw}. From low to high energy the same eikonal function has been used to compute both the total and the inelastic cross sections. The predictions are calculated using GRV pfd's, with $p_T^{\rm min} = 1.15~{\rm GeV}.$ The green band is obtained with Model I varying $p_T^{\rm min}$ by no more than a few percent and using different partonic densities~\cite{Achilli:2011sw}. The blue band corresponds to Model II using $p_T^{\rm min} = 1.1~{\rm GeV}$ (with MRST)  in the upper edge and  $p_T^{\rm min} = 1.15~{\rm GeV}$ (with GRV94) in the lower edge~\cite{Achilli:2011sw}. Predictions for $\sigma_{\rm tot}$  from the analytic amplitude model and for $\sigma_{\rm inel}$ from the Aspen model are indicated by  blue stars~\cite{Block:2011uy}.
This figure is courtesy of Giulia Pancheri.}
\label{fig:Lia}
\end{figure}

The first batch of data collected by the ATLAS experiment has been used to study the rise of the $pp$ cross section~\cite{ATLAS:sigma2011}. In Fig.~\ref{fig:Lia} we show  a comprehensive investigation into the physics underpinning the rise of $\sigma_{\rm tot}$ and $\sigma_{\rm inel}$.\footnote{The CMS measurements of the $pp$ cross section at $\sqrt{s} = 7~{\rm TeV}$  (not shown in the figure)  are summarized in~\cite{Alcaraz}.} The figure shows several models which employ different assumptions about the evolution of the PDFs at low-$x$ (GRV~\cite{Gluck:1991ng}, GRV94~\cite{Gluck:1994uf}, GRV98~\cite{Gluck:1998xa}, MRST~\cite{Martin:1998sq}, and CTEQ~\cite{Lai:1999wy}). Model I is an eikonal minijet model
incorporating effects induced by the transverse momentum distribution of soft gluons~\cite{Achilli:2007pn}.  In the spirit of the Bloch-Nordsieck study in electrodynamics~\cite{Bloch:1937pw},  the parton distribution in $b$-space is determined as the Fourier transform of the resummed distribution of soft gluons (down to zero gluon momenta) emitted from the initial state during the collision. The very large $b$ limit of the profile function 
\begin{equation}
A (b,s) \sim e^{-(\bar \Lambda \, b)^{2\beta}} \,,
\end{equation}
yields an impact parameter distribution falling at its faster like a Gaussian ($\beta=1$) and at its slowest like an exponential ($\beta=1/2).$ The scale $\bar \Lambda \propto \Lambda_{\rm QCD}$ includes a mild energy dependence as well as a residual dependence upon the parameter $\beta$. This behavior in impact parameter space together with the high energy behavior of the minijet cross section leads to an asymptotic rise of the total cross section~\cite{Grau:2009qx} 
\begin{equation}
\sigma_{\rm tot} \sim \frac{2 \pi}{\bar \Lambda^2} \, \Delta_{\rm H}^{1/\beta} \, \, \ln^{1/\beta} \! s  \,,
\end{equation}
consistent with the Froissart bound. In Model II the effective density overlaping parton distributions in the colliding protons is a {\sc sibyll}-like  exponentially decreasing hyperbolic Bessel function, determined from the proton electric form factor~\cite{Grau:2010ju}. The analyticity-constrained analytic amplitude model of Block and Halzen~\cite{Block:2005pt,Block:2005ka} exploits high quality low energy data to determine an asymptotic growth of the cross section,  $\sigma_{\rm tot} \sim \ln^2 s$, which saturates the Froissart bound. The Aspen model  is a revised version of the eikonal model of Block, Gregores, Halzen, and Pancheri~\cite{Block:1998hu}, which now incorporates analyticity constraints~\cite{Block:2006hy}.

In a complementary study, a comparison of LHC data (on inclusive particle production at energies $\sqrt{s} = 0.9, 2.36,$ and 7~TeV) with predictions of a variety of hadronic interaction models has been elaborated in~\cite{d'Enterria:2011kw}. The results show that while reasonable qualitative agreement has been achieved for some event generators, none of them reproduces the $\sqrt{s}$ evolution of all the observables with particularly compeling precision.

In summary, high energy hadronic interaction models are still being refined and therefore the disparity between them can vary even from version to version. At the end of the day, however, the relevant parameters boil down to two: the mean free path,  
$\lambda_{{\rm CR-air}}  = (\sigma_{\rm prod}^{{\rm CR-air}} \,n_{\rm atm})^{-1},$ and the inelasticity,  
$y_{_{{\rm CR-air}}} = 1 - E_{\rm lead}/E_{\rm proj}$, where 
 $n_{\rm atm}$ is the number density of atmospheric target nucleons, $E_{\rm lead}$ is the 
  energy of the most energetic hadron with a long lifetime, and $E_{\rm proj}$ 
  is the energy of the projectile particle. The first parameter characterizes 
  the frequency of interactions, whereas the second one quantifies the energy 
  lost per collision. Overall,
  {\sc sibyll} has a shorter mean free path and a smaller inelasticity than 
  {\sc qgsjet}.  Since a shorter mean free path tends to 
  compensate for a smaller inelasticity, the two codes generate similar predictions 
  for an air shower which has lived through several generations. 
  Both models predict the same multiplicity below about $10^{7}$~GeV, but 
  the predictions diverge above that energy. Such a divergence readily increases 
  with rising energy. While {\sc qgsjet} predicts a 
  power law-like increase of the number of secondaries up to the highest 
  energy, {\sc sibyll} multiplicity exhibits a logarithmic growth. As it is 
  extremely difficult to observe the first interactions experimentally,
  it is not straightforward to determine which model is closer to reality.

\subsection{Electromagnetic processes}  
\label{EMP}

The evolution of an extensive air shower is dominated by electromagnetic
processes. The interaction of a 
baryonic cosmic ray  with an air nucleus high in the atmosphere leads to a 
cascade of secondary mesons and nucleons. The first few 
generations of charged pions interact again, producing a hadronic core, 
which continues to feed the electromagnetic and muonic components of the 
showers. Up to about $50$~km above sea level,
the density of atmospheric target nucleons is $n_{\rm atm} \sim 
10^{20}$~cm$^{-3},$ and so even for relatively
low energies, say  $E_{\pi^{\pm}}\approx 1$~TeV, the probability 
of decay before interaction falls below 10\%.  Ultimately, the electromagnetic cascade dissipates around 90\%
of the primary particle's energy, and hence the total number of 
electromagnetic particles is very nearly proportional to the shower energy~\cite{Barbosa:2003dc}.

Roughly speaking, at $10^{11}$~GeV, baryons and charged pions have interaction lengths of the order of $40~{\rm g}/{\rm cm}^2,$ increasing to about $60~{\rm g}/{\rm cm}^2$ at $10^7$~GeV.  Additionally, below $10^{10}$~GeV, photons, electrons, and positrons have mean interaction lengths of $37~{\rm g}/{\rm cm}^2$.  Altogether, the atmosphere acts as a natural colorimeter with variable density, providing a vertical thickness of 26 radiation lengths and about 15 interaction lengths. Amusingly, this is not too different from the number of radiation and interaction lengths at the LHC detectors. For example, the CMS electromagnetic calorimeter is $\gtrsim 25$ radiation lengths deep, and the hadron calorimeter constitutes 11 interaction lengths.

By the time a vertically incident $10^{11}$~GeV proton shower
reaches the ground, there are about $10^{11}$ secondaries with energy above
90~keV in the the annular region extending 8~m to 8~km from the shower core.
Of these, 99\%  are photons, electrons, and positrons, with a typical ratio 
of $\gamma$ to $e^+  e^-$ of 9 to 1. Their mean energy  is 
around 10~MeV and they transport 85\% of the total energy at ground level. Of course, 
photon-induced showers are even more dominated by the electromagnetic channel, 
as the only significant muon generation mechanism in this case is the decay of
charged pions and kaons produced in $\gamma$-air  interactions~\cite{Mccomb:tp}. 

It is worth mentioning that these figures dramatically change for the case of 
very inclined showers. For a primary zenith angle, $\theta > 70^{\circ},$ the 
electromagnetic component becomes attenuated exponentially with atmospheric 
depth, being almost completely absorbed at ground level. Note 
that the vertical atmosphere is $\approx 1,000~{\rm g/cm}^{2}$, and is about 36 
times deeper for completely horizontal showers. 
As a result, most of the energy at ground level from an inclined shower is
carried by muons.

In contrast to  hadronic collisions, the electromagnetic 
interactions of shower particles can be calculated very accurately from 
quantum electrodynamics. Electromagnetic interactions are thus not a 
major source of systematic errors in shower simulations. The first 
comprehensive treatment of electromagnetic showers was elaborated by 
Rossi and Greisen~\cite{Rossi}.  This treatment was recently cast in a more 
pedagogical form by Gaisser~\cite{Gaisser:vg}, which we summarize in the 
subsequent paragraphs.

The generation of the electromagnetic component is 
driven by electron bremsstrahlung and pair production~\cite{Bethe:1934za}.
Eventually the average energy per particle drops below a critical energy,
$\epsilon_0$, at which point ionization takes over from bremsstrahlung and pair
production as the dominant energy loss mechanism. The $e^\pm$ energy loss rate due to
bremsstrahlung radiation is nearly proportional to their energy, whereas the 
ionization loss rate varies only logarithmically with the $e^\pm$ energy.
Though several different definitions of the critical energy appear in the 
literature~\cite{Hagiwara:fs}, throughout these lectures  we take the critical energy 
to be that at which the ionization loss per radiation lenght
is equal to the electron energy, yielding $\epsilon_{0} = 710~{\rm MeV}/(Z_{\rm eff} +0.92) \sim 
$~86~MeV~\cite{Rossi:book}.\footnote{For altitudes up to 90~km above sea level, the air is a mixture 
of 78.09\% of N$_2$, 20.95\% of O$_2$, and 0.96\% of other gases~\cite{Weast}. 
Such a mixture is generally modeled as an homogeneous substance with atomic charge
and mass numbers $Z_{\rm eff} = 7.3$ and $A_{\rm eff} = 14.6,$ respectively.}  The changeover 
from radiation losses to ionization losses depopulates the shower.
One can thus categorize the shower development in three phases: the growth phase, in which all the particles 
have energy $> \epsilon_0$; the shower maximum, $X_{\rm max}$; and the shower 
tail, where the particles only lose energy, get absorbed or decay. 

The relevant quantities participating in the development
of the electromagnetic cascade are the probability for an electron of
energy $E$ to radiate a photon of energy $k=y_{_{\rm brem}}E$ and
the probability for a photon to produce a pair $e^+e^-$
in which one of the particles (hereafter $e^-$) has energy $E=y_{_{\rm pair}}k$.
These probabilities are determined by the properties of the air and 
the cross sections of the two processes. 

In the energy range of interest, the impact 
parameter of the electron or photon is larger than an atomic radius, so 
the nuclear field is screened by its electron cloud.  
In the case of complete screening, where the momentum transfer is small, 
the cross section for bremsstrahlung can be approximated by~\cite{Tsai:1973py}
\begin{equation}
\frac{d\sigma_{e \rightarrow \gamma}}{dk} \approx \frac{A_{\rm eff}}{X_{_{\rm EM}} N_A k}
\left(\frac{4}{3}-\frac{4}{3}y_{_{\rm brem}}+y_{_{\rm brem}}^2\right)\,\,,\label{brem}
\end{equation}
where $A_{\rm eff}$ is the effective mass number of the air, $X_{_{\rm EM}}$ is a constant,
and $N_A$ is Avogadro's number. In the infrared limit ({\it i.e.} $y_{_{\rm brem}} \ll 1$) this approximation 
is inaccurate at the level of about 2.5\%, which is small compared to typical experimental 
errors associated with cosmic air shower detectors.  Of course, the approximation fails 
as $y_{_{\rm brem}}  \rightarrow 1$, when nuclear screening becomes incomplete, and as $y_{_{\rm brem}}  \rightarrow 0$, at which 
point the LPM and dielectric suppression effects become important, as we discuss below.

Using similar approximations, the cross section
for pair production can be written as~\cite{Tsai:1973py}
\begin{equation}
\frac{d\sigma_{\gamma \rightarrow e^+e^-}}{dE} \approx \frac{A_{\rm eff}}{X_{_{\rm EM}} N_A}
\left(1-\frac{4}{3}y_{_{\rm pair}}+\frac{4}{3}y_{_{\rm pair}}^2\right) \,. \label{pair}
\end{equation}
The similarities between this expression and Eq.~(\ref{brem}) are to be expected, 
as the Feynman diagrams for pair production and bremsstrahlung are variants of one another.

The probability for an electron to radiate a photon 
with energy in the range $(k,k+dk)$ in traversing $dt=dX/X_{_{\rm EM}}$ of atmosphere is 
\begin{equation} \label{p_b}
\frac{d\sigma_{e \rightarrow \gamma}}{dk}\,\frac{X_{_{\rm EM}} N_A}{A_{\rm eff}}\,dk\,dt
\approx \left(y_{_{\rm brem}} +\frac{4}{3}\,\,\frac{1-y_{_{\rm brem}} }{y_{_{\rm brem}} }\right)\,dy_{_{\rm brem}} \,dt \,\,,
\end{equation}
whereas the corresponding probability density for a photon producing 
a pair, with electron energy in the 
range $(E,E+dE)$, is
\begin{equation} \label{p_pp}
\frac{d\sigma_{\gamma \rightarrow e^+e^-}}{dE}\,\frac{X_{_{\rm EM}} N_A}{A_{\rm eff}}\,dE\,dt
\approx \left(1-\frac{4}{3}y_{_{\rm pair}}+\frac{4}{3}y_{_{\rm pair}}^2\right)\,dy_{_{\rm pair}}\,dt \,\,.
\end{equation}
The total probability for pair production per unit of $X_{_{\rm EM}}$ follows from 
integration of Eq.~(\ref{p_pp}), 
\begin{equation}
\int \frac{d\sigma_{\gamma \rightarrow e^+e^-}}{dE}\,\,\,
\frac{X_{_{\rm EM}} N_A}{A_{\rm eff}}\, dE\, \approx \int_0^1 
\left(1-\frac{4}{3}y_{_{\rm pair}} + \frac{4}{3}y_{_{\rm pair}}^2\right)
\,d y_{_{\rm pair}} = \frac{7}{9}\,.
\end{equation}

As can be seen from Eq.~(\ref{p_b}), the total probability for 
bremsstrahlung radiation is logarithmically divergent.  However, this 
infrared divergence is eliminated by the interference of 
bremsstrahlung amplitudes from multiple scattering centers.  
This collective effect of the electric potential of several
atoms is known as the LPM
effect~\cite{Landau:um,Migdal:1956tc}. Of course,
the LPM suppression of the cross section results in an 
effective increase of the mean free path of electrons and photons. This 
effectively retards the development of the electromagnetic component of the shower. 
It is natural to introduce 
an energy scale, $E_{\rm LPM}$, at which the inelasticity is low enough that the LPM effect becomes 
significant~\cite{Stanev:au}.
Below $E_{\rm LPM}$, the energy loss rate due to bremsstrahlung is roughly
\begin{equation}
\frac{dE}{dX} \approx -\frac{1}{X_{_{\rm EM}}} \, \int_0^1 y_{_{\rm brem}} \ E\,  
\left(y_{_{\rm brem}} +\frac{4}{3}\,\,\frac{1-y_{_{\rm brem}} }{y_{_{\rm brem}} }\right) \,dy_{_{\rm brem}}  \, = -\frac{E}{X_{_{\rm EM}}} \,.
\end{equation}
With this in mind, we now identify the constant $X_{_{\rm EM}} \approx 36.7$~g cm$^{-2}$ 
with the radiation length in air  defined as the mean distance over which a high-energy 
electron loses $1/e$ of its energy, 
or equivalently $7/9$ of the mean free path for pair production by a high-energy 
photon~\cite{Hagiwara:fs}.

\begin{figure} [t]
\postscript{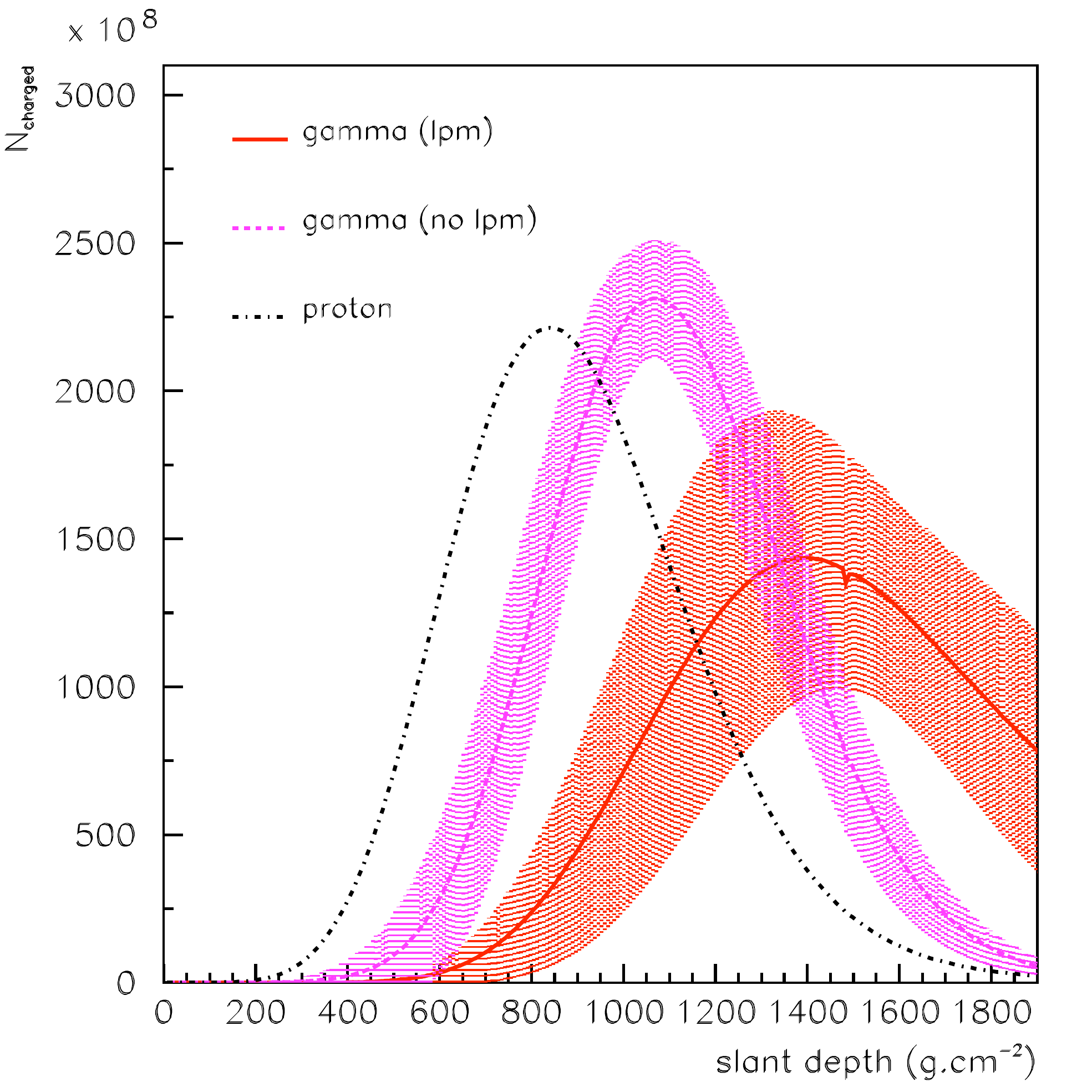}{0.70}
\caption{Average longitudinal shower developments 
of  $10^{11}$~GeV proton (dashed-dotted line) and $\gamma$-rays with 
and without the LPM effect (solid and dotted lines, respectively). The 
primary zenith angle was set to $\theta = 60^{\circ}$. 
The shadow area represents the intrinsic fluctuations of the showers. Larger 
fluctuations can be observed for $\gamma$-ray showers with the LPM effect, 
as expected. From Ref.~\cite{Anchordoqui:2004xb}.} 
\label{lpm} 
\end{figure}

The most evident signatures of the LPM effect on shower development are  
a shift in the position of the shower maximum $X_{\rm max}$ and larger 
fluctuations in the shower development. When considering the LPM effect in the development of air showers produced by  UHECRs, one has to keep in mind that the 
suppression in the cross sections is strongly 
dependent on the atmospheric depth.\footnote{The same occurs for dielectric 
  suppression, although the influence is not as important as 
  for the LPM effect~\cite{Cillis:1998hf}.} Since the upper atmosphere is very thin, the 
LPM effect becomes noticeable only for photons and electrons with energies above
$E_{\rm LPM} \sim 10^{10}$~GeV.  For baryonic primaries the LPM effect does not 
become important until the primary energy exceeds $10^{12}$GeV.  This is 
because the electromagnetic shower does not commence until after a significant
fraction of the primary energy has been dissipated through hadronic interactions.
To give a visual impression of how the LPM effect 
slows down the initial growth of high energy photon-induced showers,
we show the average longitudinal 
shower development of  $10^{10}$~GeV proton and $\gamma$-ray showers (generated using 
{\sc aires} 2.6.0~\cite{Sciutto:1999jh}) with and without the LPM effect in Fig.~\ref{lpm}.

\begin{figure} [t]
\postscript{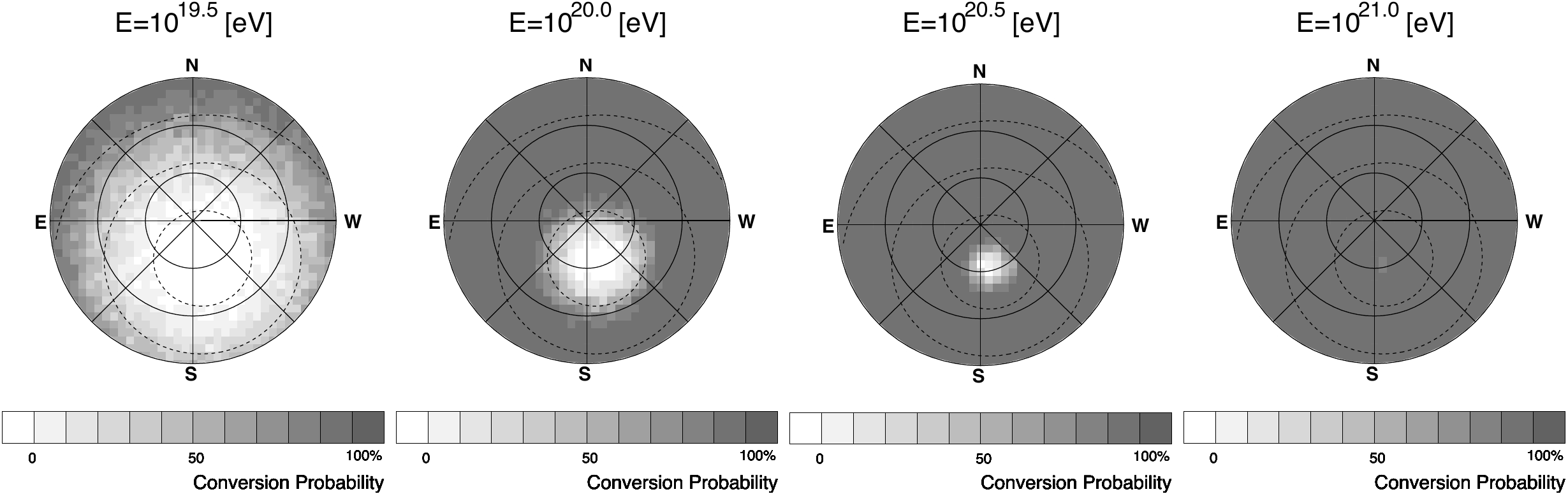}{0.85}
\caption{Maps of gamma ray conversion probability in the geomagnetic field for several primary 
energies. Azimuths are as labeled, ``N'' denotes true north. The inner circles correspond to 
zenith angles $30^\circ$, $60^\circ$ and 
horizon. Dashed curves indicate the opening angles of $30^\circ,$ $60^\circ$ and $90^\circ$ to the 
local magnetic field. From Ref.~\cite{Vankov:2002cb}.}
\label{map} 
\end{figure}

At energies at which the LPM effect is important ({\it viz.} $E > E_{\rm LPM}$), 
$\gamma$-ray showers will have already commenced in the geomagnetic field at almost 
all latitudes: primary photons with $E > 10^{10}$~GeV convert into $e^{+}e^{-}$ 
pairs, which in turn emit synchrotron photons. This reduces 
the energies of the primaries that reach the atmosphere, and thereby  
compensates for  the tendency of the LPM effect to retard the shower development.
The relevant parameter to 
determine both conversion probability and synchrotron emission  is 
$E\times B_{\perp}$, where $E$ is the $\gamma$-ray energy and $B_{\perp}$ the transverse 
magnetic field. This leads to a large 
directional and geographical dependence of shower observables. Thus, each experiment has its own 
preferred direction for identifying primary gamma rays.  For instance,
Fig.~\ref{map} shows a map of the photon 
conversion probability in the geomagnetic field for 
all incident directions evaluated at the location of the 
HiRes experiment ($|\vec{B}| = 0.53$~G, $\iota = 25^{\circ},$ and 
$\delta = 14^{\circ}$)~\cite{Vankov:2002cb}.  The smallest probabilities for conversion are found, 
not surprisingly, around the direction parallel to the local geomagnetic field.  
Note that this conversion-free region shrinks rapidly with increasing primary energy.  
A similar evaluation for the Southern Site of the Pierre Auger 
Observatory  
($|\vec B| = 0.25$~G, $\iota = - 35^{\circ}$, and $\delta = 86^{\circ}$) can
be found in~\cite{Bertou}.

The muonic component of an EAS differs from the electromagnetic component for two main reasons.
First, muons are generated through the decay of cooled 
($E_{\pi^\pm} \lesssim 1$~TeV) charged pions, and thus the muon content is sensitive to the initial
baryonic content of the primary particle.  Furthermore, since there is no ``muonic cascade'', the number
of muons reaching the ground is much smaller than the number of electrons. Specifically, there are
about $5\times 10^{8}$ muons above 10~MeV at ground level for a vertical $10^{11}$~GeV proton 
induced shower.
Second,  the muon has a much smaller cross section for radiation and pair production than the electron, 
and so the muonic component of an EAS develops differently than does the electromagnetic component.
The smaller multiple scattering suffered by muons leads to 
earlier arrival times at the ground for muons than for the electromagnetic component.  

The ratio
of electrons to muons depends strongly on the distance from the core;
for example, the $e^+  e^-$ to  $\mu^+  \mu^-$ ratio for a $10^{11}$~GeV vertical proton 
shower varies from 17 to 1 at 200~m from the core to 1 to 1 at 2000~m.
The ratio between the electromagnetic and muonic shower components behaves somewhat differently
in the case of inclined showers.  For zenith angles greater than $60^{\circ}$, the 
$e^+  e^-$/$\mu^+  \mu^-$  ratio  
remains roughly constant at a given distance from the core.  As the zenith angle grows beyond $60^\circ$,
this ratio decreases, until at $\theta = 75^{\circ}$, it is 400 times smaller than for a vertical shower.
Another difference between inclined and vertical showers is that the average muon energy at ground level changes 
dramatically.  For horizontal showers, the lower energy muons are filtered out by a combination of 
energy loss mechanisms and the finite muon lifetime: for vertical showers, the average muon energy 
is 1~GeV, while for horizontal showers it is about 2 orders of magnitude greater.
 The muon densities obtained
in shower simulations using {\sc sibyll} 2.1 fall more rapidly with lateral 
distance to the shower core
than those obtained using {\sc qgsjet} 01. 
This can be understood as a 
manifestation of the enhanced leading
particle effect in {\sc sibyll}, which can be traced to the relative 
hardness of the electromagnetic form factor profile function. 
The curvature of the distribution
$(d^2\rho_{\mu}/dr^2)$ is measurably different in the two cases, and,
with sufficient statistics, could 
possibly serve as a discriminator between hadronic interaction models, 
provided the primary species can be determined from 
some independent observable(s)~\cite{Anchordoqui:2003gm}.

\begin{figure} [t]
\postscript{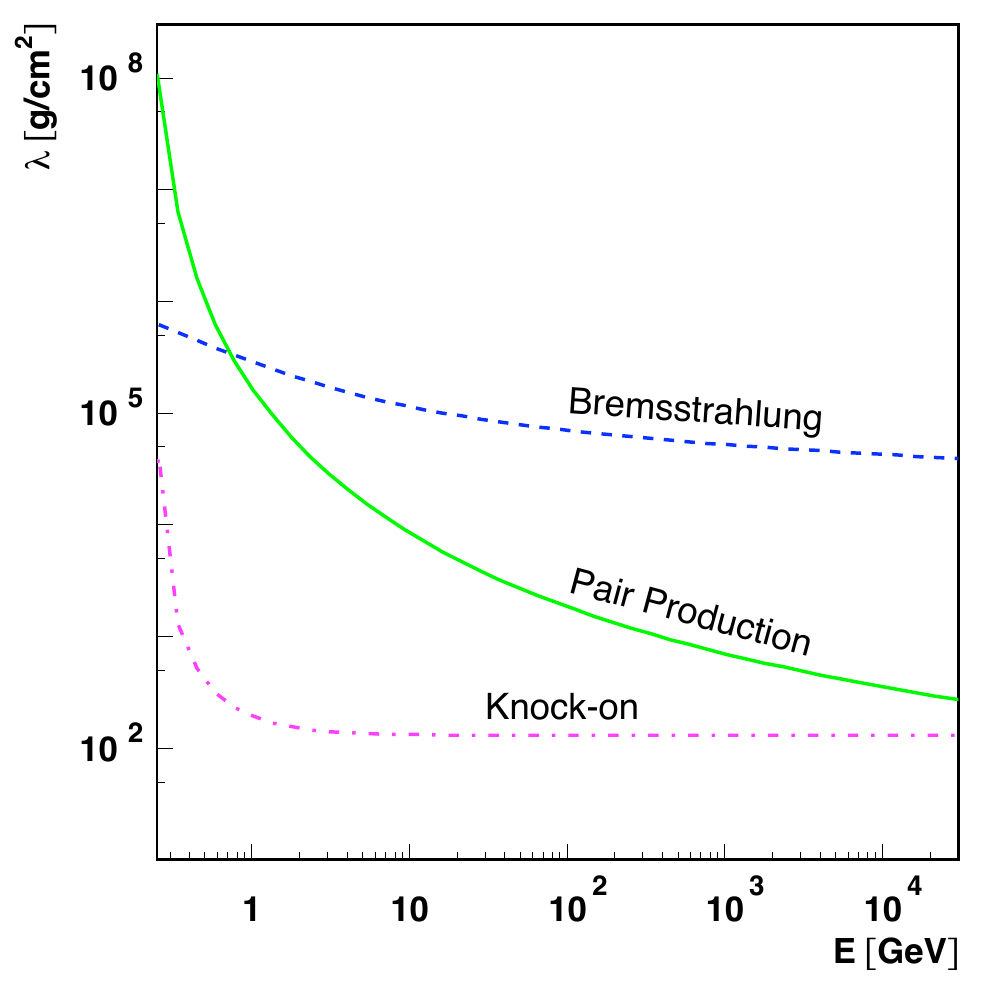}{0.70}
\caption{Mean free path in air for the different muonic interactions as a function of the initial kinetic 
energy. From Ref.~\cite{Anchordoqui:2004xb}.} 
\label{sergio}
\end{figure}

High energy muons lose energy through pair production, 
muon-nucleus 
interaction, bremsstrahlung, and knock-on electron ($\delta$-ray) 
production~\cite{Cillis:2000xc}.  The first three processes 
are discrete in the sense that they are characterized by high inelasticity and a large mean free path.  
On the other hand,  because of its 
short mean free path and its small inelasticity, knock-on electron production can be 
considered a continuous process. The muon bremsstrahlung cross section is suppressed by a factor of 
$(m_e / m_\mu)^2$ with respect to electron bremsstrahlung; see Eq.~(\ref{brem}).  Since the radiation 
length for 
air is about $36.7~\mathrm{g}/\mathrm{cm}^2$, and the vertical atmospheric depth is $1,000~{\rm g/cm}^2$,  muon bremsstrahlung is of negligible importance for vertical air shower development.  Energy loss due to 
muon-nucleus interactions is somewhat smaller than muon bremsstrahlung.    
As can be seen in Fig.~\ref{sergio}, energy  loss by pair production is slightly more important than 
bremsstrahlung at about 1~GeV, and becomes increasingly dominant with energy.  
Finally, knock-on electrons have a 
very small mean free path (see Fig.~\ref{sergio}), but also a very small inelasticity, so that this 
contribution to the energy loss is comparable to that from the hard processes.

\subsection{Paper-and-pencil air shower modeling}
\label{ppm}

\begin{figure}[tbp]
\postscript{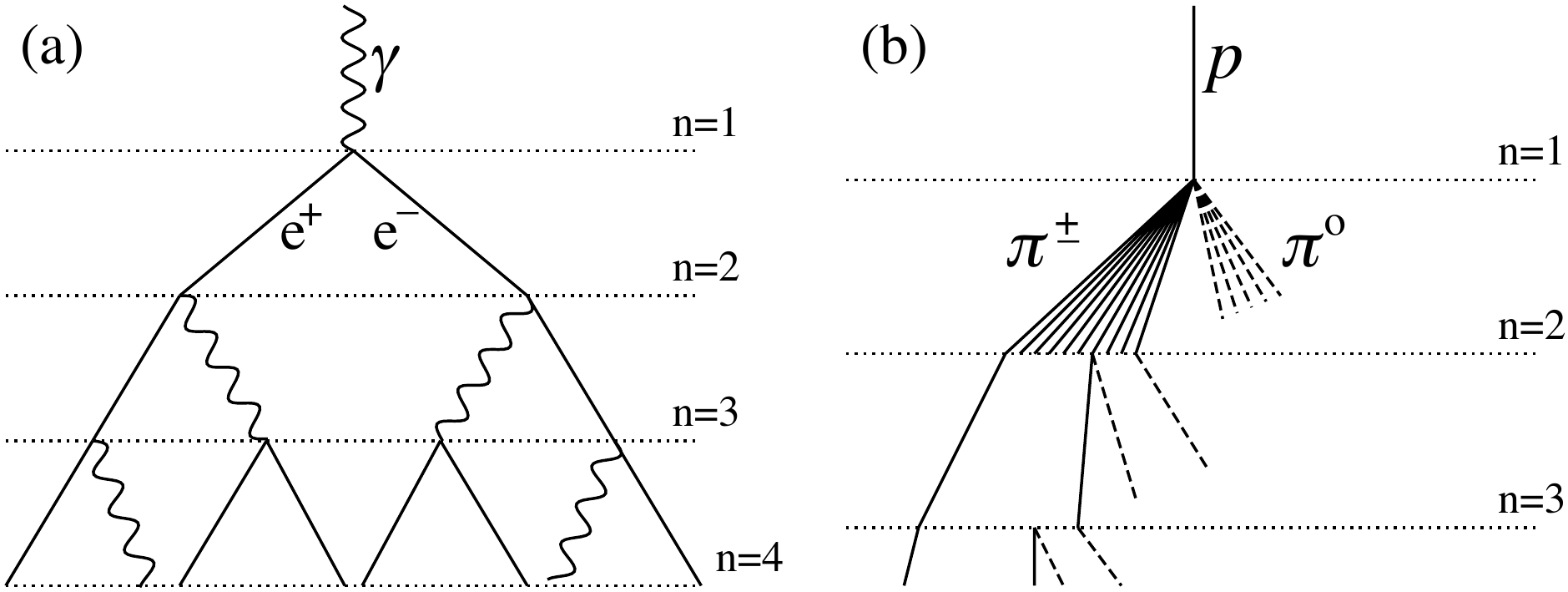}{0.90}
\caption{Schematic views of (a) an electromagnetic cascade and (b) a hadronic shower. In the hadron shower, dashed lines indicate neutral pions which do not re-interact, but quickly decay, yielding electromagnetic subshowers (not shown). Not all pions lines are shown after the $n=2$ level. Neither diagram is to scale. From Ref.~\cite{Matthews:2005sd}.}
\label{cascades2}
\end{figure}

Most of the general features of an electromagnetic cascade can be understood in terms of the toy model 
due to Heitler~\cite{Heitler}. In this model, the shower is imagined to develop exclusively 
via bremsstrahlung and pair production, each of which results in the conversion of one particle into two (see Fig.~\ref{cascades2}).
As was previously discussed, these physical processes are characterized by an 
interaction length $X_{_{\rm EM}} \approx 37.6~{\rm g/cm}^2$. One can thus imagine the shower as a particle tree with branches that bifurcate every $X_{_{\rm EM}}$, until they fall below a critical energy, $\epsilon_0 \approx 86~{\rm MeV}$, at which point energy loss processes dominate. 
Up to $\epsilon_0$, the number of particles grows geometrically, so that after 
$n = X/X_{_{\rm EM}}$ branchings, the total number of particles in the shower is 
$N \approx 2^n$.  At the depth of shower maximum $X_{\rm max}$, all particles are at 
the critical energy, $\epsilon_0$, and the energy of the primary particle, $E_0$, is 
split among all the $N_{\rm max} = E_0 / \epsilon_0$ particles.
Putting this together, we get:
\begin{equation} \label{heitler}
X_{\rm max} \approx X_{_{\rm EM}} \,\, \frac{\ln(E_0/\epsilon_0)}{\ln 2} \,\,.
\end{equation}
Changes in the mean mass composition of the CR flux as a function of energy will 
manifest as changes in the mean values of $X_{\rm max}$. This change of $X_{\rm max}$ 
with energy is commonly known as the elongation rate~\cite{Linsley:P3}:
\begin{equation} 
D_e = \frac{\delta X_{\rm max}}{\delta \ln E} \,\,.
\end{equation}
For purely electromagnetic showers, $X_{\rm max}(E) \approx  X_{_{\rm EM}}\, \ln(E/\epsilon_0)$, and hence $D_e \approx X_{\rm EM}$.
For convenience, the elongation rate is often written in terms of energy decades,
$D_{10} = \partial \langle X_{\rm max} \rangle/\partial \log E$, where $D_{10} = 2.3 D_e.$ 
The elongation rate obtain from the Heitler model,  $D_{10}^\gamma \approx 84~{\rm g/cm}^2$,  is in very good agreement with the results from Monte Carlo simulations. However, the prediction for the particle number at maximum is overestimated by a factor of about 2 to 3. Moreover, Heitler's model predicts a ratio of electron to photons of 2, whereas simulations and direct cascade measurements in the air show a ratio of the order of 1/6. This difference is due to the fact that multiple photons are emitted during bremsstrahlung and that electrons lose energy much faster than photons do.

As we have seen, baryon-induced showers are also dominated by electromagnetic processes, thus  Heitler's toy model is still enlightening for such cases. For proton primaries, the multiplicity rises with energy, and
the resulting elongation rate becomes smaller.  This can be understood by noting that, on average,
the first interaction is determined by the proton mean free path in the atmosphere, $\lambda_{p{\rm -air}} = X_0$.
In this first interaction the incoming proton splits into $\langle n(E) \rangle$ secondary particles, 
each carrying an average energy $E/\langle n(E) \rangle$.  Assuming that $X_{\rm max}(E)$ depends dominantly on the first generation of $\gamma$ subshowers, the depth of maximum is obtained as in Eq.~(\ref{heitler}), 
\begin{equation}
X_{\rm max}(E) \approx X_0 + X_{_{\rm EM}}\, \ln[E/\langle n(E) \rangle]\,\, .
\label{Xmax-Matthews1}
\end{equation}
For a proper evaluation of $X_{\rm max}$, it would be necessary to sum each generation of subshowers carefully from their respective points of origin, accounting for their attenuation near and after the maxima. If we now further assume a multiplicity dependence $\langle n(E) \rangle \approx n_0 E^{\Delta}$, then  
the elongation rate becomes,
\begin{equation} 
\frac{\delta X_{\rm max}}{\delta \ln E}= X_{_{\rm EM}}\,\left[1-\frac{\delta \ln \langle n(E) 
\rangle}{\delta \ln E} \right] + \frac{\delta X_0}{\delta \ln E}
\end{equation}
which corresponds to the form given by Linsley and Watson ~\cite{Linsley:gh},
\begin{equation} 
D_e = X_{_{\rm EM}} \,\left[ 1-\frac{\delta \ln \langle n(E) \rangle}{\delta \ln E} + 
\frac{X_0}{X_{_{\rm EM}}} \frac{\delta \ln(X_0)}{\delta \ln E} \right] =  X_{_{\rm EM}}\,(1-B) \,\,,
\label{ERLW}
\end{equation}
where
\begin{equation}
B \equiv \Delta - \frac{X_0}{X_{_{\rm EM}}} \,\,\frac{\delta \ln X_0}{\delta \ln E} \, .
\label{theB}
\end{equation} 

A precise calculation of a proton shower evolution has been carried out by Matthews~\cite{Matthews:2005sd}, using the simplifying assumption  that hadronic interactions produce exclusively pions,  $2N_\pi$ charged and $N_\pi$ neutral (see Fig.~\ref{cascades2}). 
$\pi^0$'s decay immediately and feed the electromagnetic component of the shower, whereas $\pi^\pm$'s soldier on. The hadronic shower continues to grow,
 feeding the electromagnetic component at each interaction, until charged pions
 reach a characterictic energy at which decay is more likely than a new interaction.
 The interaction length and the pion multiplicity ($3N_\pi$) are
 energy independent in the Heitler-Matthews model. The energy is
 equally shared by the secondary pions. For pion energy between
 1~GeV and 10~TeV, a charged multiplicity of 10 ($N_\pi$ = 5)
 is an appropriate number.

The first interaction diverts 1/3 of the available energy ($E_0/3$) into the EM component while the remaining 2/3 continue
 as hadrons. The number of hadrons increases through subsequent generation of particles and in each generation about 30\% of the energy is transferred to the EM cascade. Therefore the longer it takes for pions to reach the characteristic energy $\xi^\pi_{\rm c} \sim 20~{\rm GeV}$ (below which they
 will decay into muons), the larger will be the EM 
 component. Consequently, in long developing showers  the
 energy of the muons from decaying pions will be smaller. 
 In addition, because of the density profile of the atmosphere,
 $\xi^\pi_{\rm c}$ is larger high above ground than at sea level
 and deep showers will produce fewer muons. 

 This positive correlation  introduces a link between the primary
 cosmic ray interaction cross section on air and the muon
 content at ground level. According to those principles,
 primaries with higher cross
 sections will have a larger muon to electron ratio at ground level.  

 To obtain the number of muons in the shower, one  simply assumes
 that all pions decay into muons when they reach the critical energy:
 $N_\mu = (2N_\pi)^{n_c}$, where $n_c = \ln(E_0/\xi_{\rm c}^\pi)/\ln(3N_\pi)$
 is the number of steps needed for the pions to reach
 $\xi^\pi_{\rm c}$. Introducing $\beta=\ln(2N_\pi)/\ln(3N_\pi)$
 we have
 \begin{equation}\label{hs} 
N_\mu = (E_0/\xi^\pi_{\rm c})^\beta \, . 
\end{equation} 
 For $N_\pi=5$, $\beta = 0.85$. Unlike the electron number, the muon multiplicity does not grow
 linearly with the primary energy, but at a slower rate.
 The precise value of $\beta$ depends on the average pion
 multiplicity used.  It also depends on the inelasticity of the hadronic interactions.
 Assuming that only half of the available energy goes into the pions
 at each step (rather than all of it, as done above) would lead
 to $\beta=0.93$. Detailed simulations give values of $\beta$ in
 the range 0.9 to 0.95~\cite{Alvarez-Muniz:2002ne}.

The first interaction yields $N_\gamma = 2 N_{\pi^0} = N_{\pi^\pm}$. Each photon initiates an EM shower of energy $E_0/(3 N_{\pi^\pm}) = E_0/(6 N_\pi).$ Using $pp$ data ~\cite{Amsler:2008zzb},
we parametrized the charged particle production in the first interaction as
$N_{\pi^{\pm}} = 41.2(E_0/1~{\rm PeV})^{1/5}$.  Now, from  the approximation in (\ref{Xmax-Matthews1}), based on the sole evolution of the EM cascade initiated by the first interaction, we obtain
\begin{eqnarray}
X_{\rm max}^p  & = & X_0 + X_{_{\rm EM}}  \, \ln[E_0/( 6 N_\pi \epsilon_0)]  \nonumber \\
                          & = & (470 + 58 \, {\rm log_{10}} [E_0/1~{\rm PeV}])~{\rm g/cm}^2 \, .
\label{Xmax-Matthews2}
\end{eqnarray}
This falls short of the full simulation value by about 100~${\rm g/cm}^2$~\cite{Matthews:2005sd}. \\

{\bf EXERCISE 3.2}~~~The depth of shower maximum obtained in Eq.~(\ref{Xmax-Matthews2})  is only approximate since it considers just the first interaction as hadronic in nature. Extend the approximation to include hadronic interactions in the second generation of particles. Try the generalization to include all generations of hadronic collisions until the charged pions cool down below the critical energy.\\

A good approximation of the elongation rate can be obtained
 when introducing the cross section and multiplicity energy dependence.
 Using a $p$-air cross section of 550~mb at 10$^{9}$~GeV and a rate
 of change of about 50~mb per decade of energy leads to~\cite{Ulrich:2009zq}
\begin{equation}\label{lambda-i}
 X_0 \simeq 90 - 9 \log{(E_0/{\rm EeV})}~{\rm g/cm}^2 \, .
\end{equation}
 Now, assuming (as in~\cite{Matthews:2005sd}) that the first interaction
 initiates $2N_\pi$ EM cascades of energy $E_0/6N_\pi$, with
 $N_\pi\propto (E_0/{\rm PeV})^{1/5}$ for the evolution of the first interaction
 multiplicity with energy, we can calculate the elongation rate 
\begin{equation}
 D_{10}^ p=\frac{dX_{\rm max}}{d\log{E_0}}= \frac{d(X_0 \, \ln{2}+X_{_{\rm EM}} \, \ln[E_0/(6N_\pi \epsilon_0)]}{d\log{E_0}} \, ,
\end{equation}
or
\begin{equation}
 D_{10}^p=\frac{4}{5}D_{10}^\gamma - 9\ln{2} \simeq 62~{\rm g/cm}^2 \, \ .
\end{equation} 
 This result is quite robust as it only depends on the cross section
 and multiplicity evolution with energy. It is in good agreement with
 Monte Carlo simulation~\cite{Alvarez-Muniz:2002ne}.

To extend this discussion to heavy nuclei, we can apply the superposition principle as 
a reasonable first approximation. In this approximation, we pretend that the nucleus
comprises unbound nucleons, such that the point of first interaction of one nucleon is
independent of all the others.  Specifically, a shower produced by a nucleus with energy 
$E_{_A}$ and mass $A$ is modeled by a collection of $A$ proton showers, each with $A^{-1}$ of the
nucleus energy. Modifying Eq.~(\ref{heitler}) accordingly one easily obtains
$X_{\rm max} \propto \ln (E_0/A)$. Assuming that $B$ is not changing with energy, one obtains for mixed primary 
composition~\cite{Linsley:gh}
\begin{equation}  
D_e =\, X_0\,(1-B)\,
\left[1 - \frac{\partial \langle \ln A \rangle }{\partial \ln E} \right]\, .
\label{er}
\end{equation}
Thus, the elongation rate provides a measurement of the change of the 
mean logarithmic mass with energy.  One caveat of the procedure discussed above is 
that Eq.~(\ref{ERLW}) accounts for the energy dependence of the cross section and violation of Feynman 
scaling only for the first interaction. Note that subsequent interactions are assumed to be 
characterized by Feynman scaling and constant interaction cross sections; see Eq.~(\ref{theB}).  
Above $10^{7}$~GeV, these secondary interactions play a more important role, and thus the predictions of 
Eq.~(\ref{er}), depending on the hadronic interaction model assumed, may
vary by up to 20\%~\cite{Alvarez-Muniz:2002ne}.
In Fig.~\ref{fig:X} we show the variation of $\langle X_{\rm max} \rangle$ with primary energy as measured by HiRes and Auger  together with predictions from a variety of hadronic interaction models. 

\begin{figure}[tbp]
\postscript{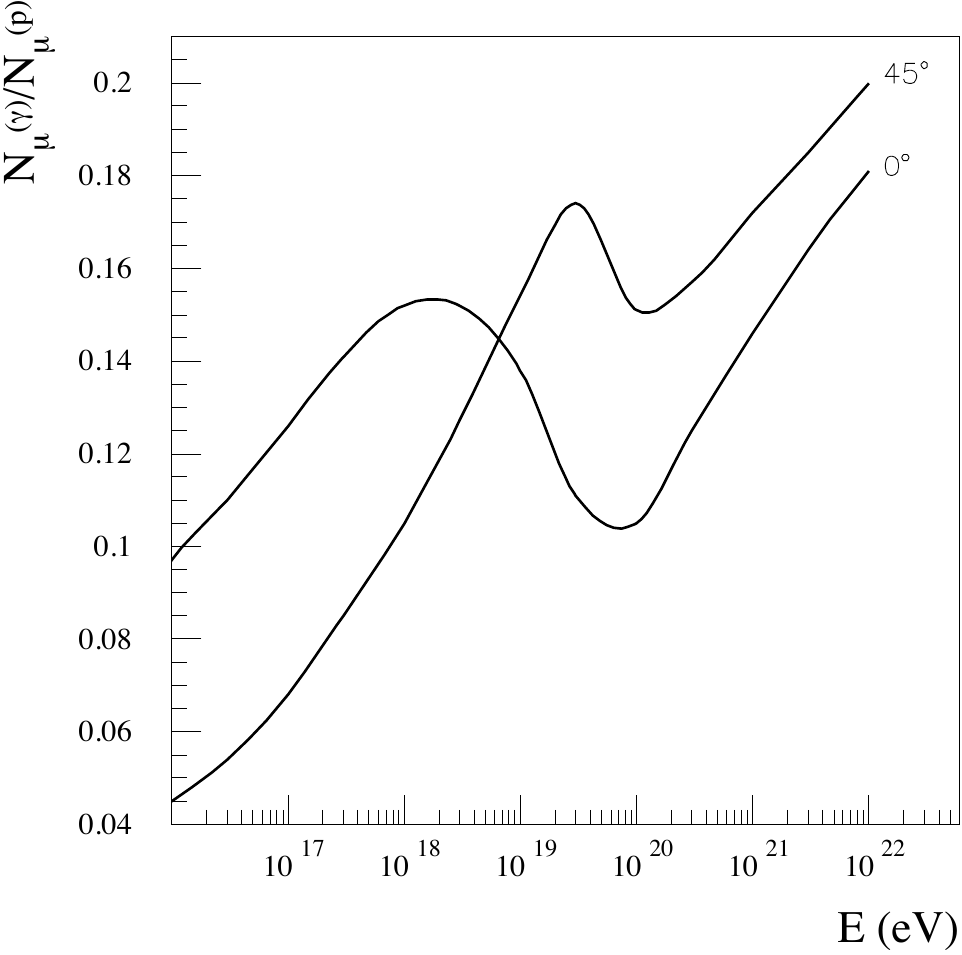}{0.8}
\caption{ Ratio of the muon content for EASs produced by primary gammas and protons. 
The geomagnetic 
field is set to the PAO Southern site. From Ref.~\cite{Plyasheshnikov:2001xw}.}
\label{felix}
\end{figure}

The muon content of an EAS at ground level $N_\mu,$ as well as the ratio $N_{\mu}/N_{e}$, 
are sensitive to primary composition (here, $N_e$ is the electron content at ground level).
To estimate the ratio of the  muon content of nucleus-induced to proton-induced showers, we can resort again to the principle of superposition. 
Using $\beta = 0.93$ we find that the total 
number of muons produced by the superposition of $A$ individual proton showers 
is, $N_{\mu}^A \propto A (E_{_A}/A)^{0.93}$. Consequently, in a vertical shower, 
one expects a cosmic ray nucleus to produce about $A^{0.07}$  more 
muons than a proton.  This implies that a shower initiated by 
an iron nucleus produces about 30\% more muons than a proton shower.  Note, however,
that a change in the hadronic interaction model could produce a much larger effect than a 
change in the primary species.   For example, replacing {\sc qgsjet} 01 with {\sc sibyll} 1.6 
as the hadronic interaction model leads to a prediction of 60\% more muons for an 
iron shower than for a proton shower~\cite{Ave:2000dd}.

The situation for gamma-induced showers is a bit different.  In this case
the muon component of the shower does not simply follow
Eq.~(\ref{hs}) because of the LPM and geomagnetic field effects~\cite{Plyasheshnikov:2001xw}. 
Competition between the two processes leads to a complex behavior in $N_{\mu}^{\gamma}/N_{\mu}^p$, as shown in Fig.~\ref{felix}.  

While these toys models are very useful for imparting a first intuition regarding global shower properties, the details of shower evolution are far too complex to be fully described
by a simple analytical model. Full Monte Carlo simulation of interaction and transport of each individual
particle is required for precise modeling of the shower development. At present, two Monte Carlo  
packages are available to simulate EASs: {\sc corsika} (COsmic Ray SImulation for 
KAscade)~\cite{Heck:1998vt} and {\sc aires} (AIR shower Extended Simulation)~\cite{Sciutto:1999jh}.  
Both programs provide fully 4-dimensional simulations of the air showers initiated by protons, photons, 
and nuclei. To simulate hadronic physics, the programs make use of the event generators described in 
Sec.~\ref{hadronic}. Propagation of particles takes into account the Earth's curvature and geomagnetic field.
For further details on these codes, the reader is referred to~\cite{Knapp:2002vs}.

\section{Searches for new physics beyond the electroweak scale at $\bm{\sqrt{s}} \sim 250~{\rm TeV}$}
\label{beyondSM}

\subsection{General idea}

If new physics interations occur at LHC energies, then CR collisions with c.m. energies ranging up to $250$~TeV would obviously involve new physics as well. The question is,
can new physics be detected by CR experiments?  At ultrahigh energies, the cosmic ray luminosity $ \sim 7 \times 10^{-10}~(E/{\rm PeV})^{-2}$~cm$^{-2}$ s$^{-1}$ (taking a
single nucleon in the atmosphere as a target and integrating over $2
\pi$~sr) is about 50 orders of magnitude smaller than the LHC
luminosity.  This renders the hunt for physics beyond the electroweak scale futile in hadronic cosmic
ray interactions occuring in the atmosphere.  However, there is still a possibility
of uncovering new physics at sub-fermi distances in cosmic neutrino interactions.

Neutrinos are unique probes of new physics, as their interactions are
uncluttered by the strong and electromagnetic forces and, upon arrival
at the Earth, they may experience collisions with c.m. 
energies up to $\sqrt{s} \lesssim 250$~TeV.  However, rates for new
physics processes are difficult to test since the flux of cosmic
neutrinos is virtually unknown.  Interestingly, it is possible in
principle to disentangle the unknown flux and new physics processes by
using multiple
observables~\cite{Kusenko:2001gj,Anchordoqui:2001cg}.

For example, possible deviations of the neutrino--nucleon cross section due to new
non-perturbative interactions\footnote{Throughout these lectures we use
  this term to describe neutrino interactions in which the final state
  energy is dominated by the hadronic component. We are {\em not}
  considering here new ``perturbative'' physics {\em e.g.} (softly broken)
  supersymmetry at the TeV scale which would have quite different
  signatures in cosmic ray showers.} can be uncovered at  Auger
 by combining information from Earth-skimming and
quasi-horizontal showers.  In particular, if an anomalously large rate
is found for deeply developing quasi-horizontal showers, it may be
ascribed either to an enhancement of the incoming neutrino flux, or an
enhancement in the neutrino-nucleon cross section (assuming
non-neutrino final states dominate).  However, these possibilities can
be distinguished by comparing the rates of Earth-skimming and
quasi-horizontal events.  For instance, an enhanced flux will increase
both quasi-horizontal and Earth-skimming event rates, whereas an
enhanced interaction cross-section will also increase the former but
{\em suppress} the latter, because the hadronic decay products cannot
escape the Earth's crust.  Essentially this approach constitutes a
straightforward counting experiment, as the detailed shower properties
are not employed to search for the hypothesized new physics.

The question we would like to answer is then how
many Earth-skimming and quasi-horizontal events would
we need to observe at  Auger  to
make a convincing case for the existence of
non-perturbative physics in which the final state
is dominated by hadrons. The analysis techniques described herein constitute an entirely
general approach to searching for non-perturbative interactions without any dependence
on what hypothetical mechanism might actually cause the
``hadrophilia.'' In Appendix~\ref{AD} we illustrate one possible new physics
process which may be accessible using these techniques at Auger,
and which is {\em beyond} the reach of the LHC.

\subsection{$\bm{\nu}$ acceptance and systematic uncertainties}

Detailed Monte Carlo simulations are used to compute the acceptance
for Earth-skimming (ES) and
quasi-horizontal  (QH) events.   Neutrinos are propagated through the atmosphere, 
the Earth's crust, and the Andes mountains using an extended 
version~\cite{Gora:2007nh} of the code ANIS~\cite{Gazizov:2004va}.
In the simulations, the $\nu N$ cross-sections from
reference~\cite{CooperSarkar:2007cv} are employed.  Particles
resulting from $\nu N$ interactions are fed to 
PYTHIA~\cite{Sjostrand:2006za} and $\tau$ decays are simulated using
TAUOLA ~\cite{Jadach:1993hs}.

\begin{figure}[htb]
\centering
\includegraphics[width=0.70\textwidth]{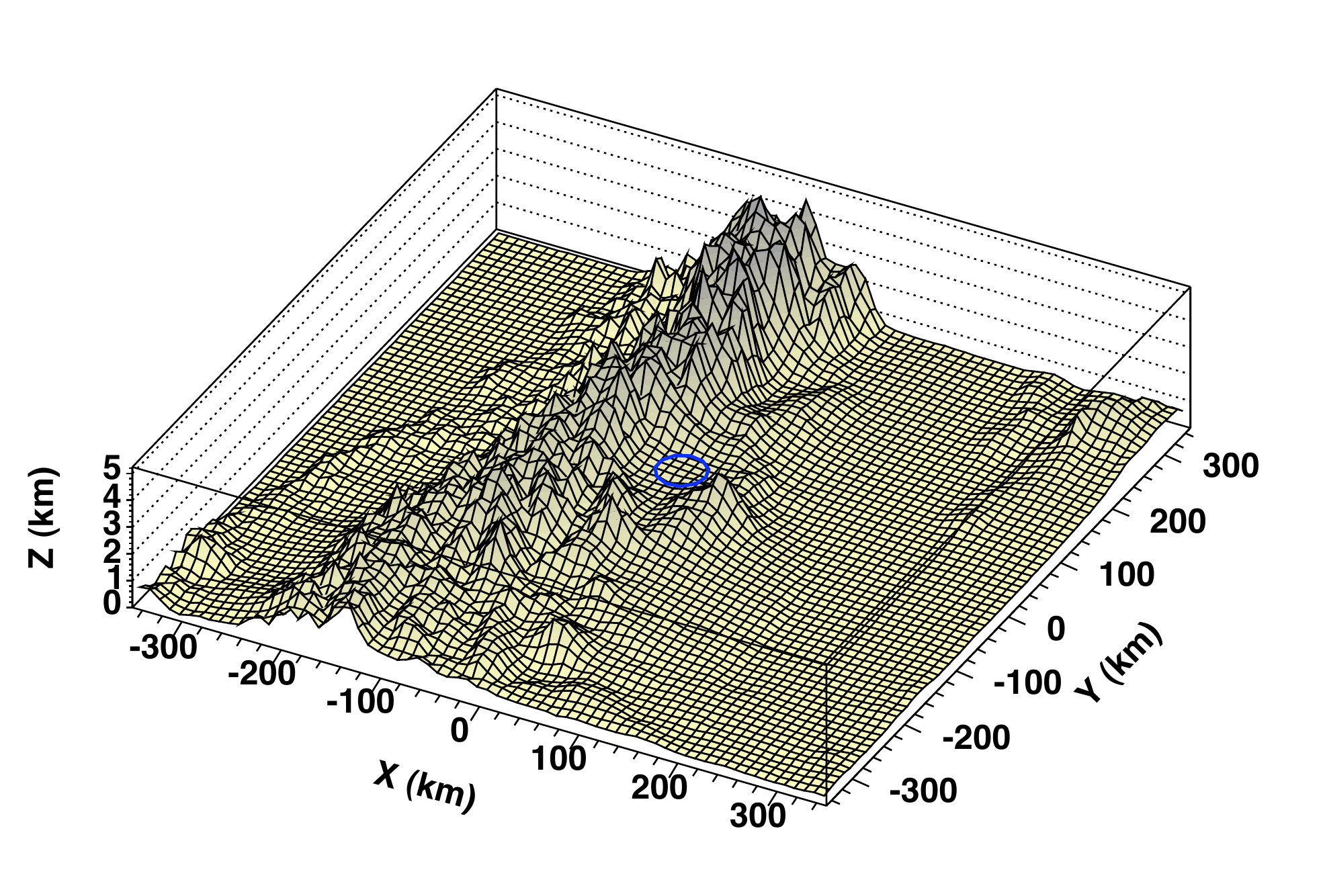}
\caption{Topography in the vicinity of the Auger site.
The surface array is centered at ${\rm X} = {\rm Y} = 0$.  From Ref.~\cite{Anchordoqui:2010hq}.}
\label{f:augermap}
\end{figure}

The flux, energy and decay vertices of outgoing leptons are 
calculated inside an ``active detector'' volume of 
$3,000 \times 10~\rm{km}^3$, including the real
shape of the surface array.
A relief map of the Andes 
mountains was constructed using digital elevation data from the
Consortium for Spatial Information (CGIAR-CSI)~\cite{CGIAR-CSI}.
The map of the area around the Auger site is depicted in 
Fig.~\ref{f:augermap}.

To study the response of the detector, the outputs of
PYTHIA and/or TAUOLA are used as input for the 
AIRES~\cite{Sciutto:2001dn} air shower simulation package.
The response of the surface detector array is simulated
in detail using the Auger \Offline simulation 
package~\cite{Argiro:2007qg}.  Atmospheric background 
muons are also simulated in order to study the impact on 
neutrino identification, as such accidental muons can
be wrongly classified as shower particles.  
The background from hadronic showers above $10^{8}$~GeV
is estimated to be ${\cal O}(1)$ in 20 
years, so for the energy bin 
considered in this analysis, $9.5 < \log_{10} (E_\nu/{\rm GeV})$,
the background is negligible.

To establish benchmark neutrino rates,  we use the 
Waxman-Bahcall bound~\cite{Waxman:1998yy} for 
the flux, 
$\Phi_{\rm WB}^{\nu_{\alpha}} =  2.3 \times 10^{-8}\, E_\nu^{-2}~{\rm GeV}^{-1}\, {\rm s}^
{-1}\, {\rm cm}^{-2}\, {\rm sr}^{-1}$, 
and employ the acceptance computed by the 
simulations described above.  In order to estimate the
systematic uncertainty  associated with our lack 
of knowledge of the dependence of the flux on energy,
we consider several scenarios which plausibly 
bracket the range of possibilities:
\begin{enumerate}
\item $  \Phi_{\rm WB}^{\nu_\alpha}(E_\nu) = ( {\cal C}/E_0) \, E_\nu^{-1}$,
\item $ \Phi_{\rm WB}^{\nu_\alpha}(E_\nu) = {\cal C} \, E_\nu^{-2}$,
\item $\Phi_{\rm WB}^{\nu_\alpha}(E_\nu) =  {\cal C} \, E_0 \, E_\nu^{-3}$,
\item  $\Phi_{\rm WB}^{\nu_\alpha}(E_\nu) = {\cal C} E_\nu^{-2} \, 
{\rm exp} [-\log_{10} (E_\nu/E_0)^2/(2 \sigma^2)]$,
\end{enumerate}
where 
${\cal C} = 2.3 \times 10^{-8}$~${\rm GeV}^{-1}\, {\rm s}^
{-1}\, {\rm cm}^{-2}\, {\rm sr}^{-1} $,  
$E_0 = 10^{10}$~GeV, $\sigma = 0.5$~GeV.  
The expected rates for the entire range over which
Auger is sensitive are given in Table~\ref{table1:rates} and the rates for
the high energy bin  are given in
Table~\ref{t:rates}.

\begin{table}[htb]
\vspace*{-0.1in}
\caption{Expected events per year ($N_i$) at Auger in the energy range $8 < \log_{10} (E_\nu/{\rm GeV})$, for various incident zenith angle ($\theta$) ranges, assuming the Waxman-Bahcall flux.}
\label{table1:rates}
\begin{center}
\begin{tabular}{c|@{}cc|@{}ccccc|c}
\hline
\hline
flux & \multicolumn{2}{@{}c|}{up-going}  & \multicolumn{5}{@{}c|}{down-going} & ratio \\
\hline
  & $\theta$  & $N_{\nu_\tau}$ & $\theta$  & $N_{\nu_e}$ & $N_{\nu_\tau}$  & $N_{\nu_\mu}$ & $N_{\nu_{\rm all} } $ & $N_\tau/N_{\nu_{\rm all}}$ \\
\cline{2-3} \cline{4-9}
$(2)$ &~90-95 & 0.68~&~60-90   & 0.134   & 0.109   & 0.019  & 0.262~& 2.58 \\
$(2)$ &~90-95 & 0.68~&~75-90 & 0.075 & 0.071 & 0.011 & 0.157~& 4.27 \\
\hline
\hline
\end{tabular}
\end{center}
\end{table}

\begin{table}[htb]
\vspace*{-0.1in}
\caption{Expected events per year ($N_i$) at Auger in the energy range $9.5 < \log_{10} (E_\nu/{\rm GeV}) < 10.5$, for various incident zenith angle ($\theta$) ranges and the 4 flux models considered.}
\label{t:rates}
\begin{center}
\begin{tabular}{c|@{}cc|@{}ccccc|c}
\hline
\hline
flux & \multicolumn{2}{@{}c|}{up-going}  & \multicolumn{5}{@{}c|}{down-going} & ratio \\
\hline
  & $\theta$  & $N_{\nu_\tau}$ & $\theta$  & $N_{\nu_e}$ & $N_{\nu_\tau}$  & $N_{\nu_\mu}$ & $N_{\nu_{\rm all}} $ & $N_\tau/N_{\nu_{\rm all}}$ \\
\cline{2-3} \cline{4-9}
$(1)$ &~90-95 & 0.14~&~60-90 & 0.059 & 0.049 &  0.011  & 0.12~& 1.14 \\
$(2)$ &~90-95 & 0.15~&~60-90 & 0.059 & 0.049 & 0.096 & 0.11~& 1.33 \\
$(3)$ &~90-95 & 0.23 &~60-90 & 0.079 & 0.062 & 0.0123 & 0.15~& 1.53 \\
$(4)$ &~90-95 & 0.12 &~60-90 & 0.046 & 0.037 & 0.0080 & 0.091~& 1.33\\
$(1)$ &~90-95 & 0.14 &~75-90 & 0.027 & 0.031 & 0.0056 & 0.064~& 2.14 \\
$(2)$ &~90-95 & 0.15 &~75-90 & 0.026 & 0.029 & 0.0048 & 0.060~& 2.47 \\
$(3)$ &~90-95 & 0.23 &~75-90 & 0.036 & 0.041 & 0.0062 & 0.083~& 2.75\\
$(4)$ &~90-95 & 0.12 &~75-90 & 0.021 & 0.024 & 0.0040& 0.049~& 2.45\\
\hline
\hline
\end{tabular}
\end{center}
\end{table}

Table~\ref{t:systematics} contains a summary of systematic uncertainties
on the ratio of the number of ES to QH events.  The uncertainty in 
spectrum shape is taken from Table~\ref{t:rates}.
The uncertainty on the  parton structure of the nucleon is estimated by considering different PDFs (GRV92NLO~\cite{Gluck:1998js}
and CTEQ66c~\cite{Nadolsky:2008zw}).  Finally, the uncertainty
on the energy loss, $\beta_\tau$, of $\tau$ leptons as they propagate through
the Earth's crust is derived from~\cite{Armesto:2007tg,Bugaev:2002gy,Abramowicz:1997ms,Capella:1994cr}.

\begin{table}[htb]
\caption{Contributions to the systematic uncertainty on the Earth-skimming to 
quasi-horizontal event ratio. We have considered the energy 
range $9.5 < \log_{10} (E_\nu/{\rm GeV}) < 10.5$ and the zenith 
angle range $75^\circ < \theta < 90^\circ$.}
\label{t:systematics}
\vspace*{0.1in}
\begin{center}
\begin{tabular}{c|ccc|c}
\hline
\hline
~~~~ratio~~~~&~~~~flux~~~~&~~~~PDF~~~~&~~~~$\beta_\tau$~~~~~~&~~~~sum~~~~\\
\hline
  & $+ 11\% $ & $0\%$ & $+24\%$ & $\phantom{2.4666666} + 26\%$ \\
2.47 &  &  &  & 2.47 \\
   &$-13\%$ & $-21\%$ & $-25\%$ & $\phantom{2.4666666} - 35\%$\\
\hline
\hline
\end{tabular}
\end{center}
\end{table}

\subsection{Auger discovery reach}

Consider first a flux of Earth-skimming $\tau$ neutrinos
with energy in the range $10^{9.5}~{\rm GeV}<E_\nu< 10^{10.5}~{\rm GeV}$.  
The neutrinos can convert to $\tau$ leptons in the Earth 
via the charged current interaction 
$\nu_{\tau^{\pm}} N \rightarrow \tau^{\pm} X$.
In the (perturbative) SM, the interaction path length for the neutrino is
\begin{equation}
L_{\rm CC}^{\nu} = \left[ N_A \rho_s \sigma^{\nu}_{\rm CC}
\right] ^{-1} \ ,
\end{equation}
where $\sigma^{\nu}_{\rm CC}$ is the charged current cross section for a neutrino energy $E_{\nu} = E_0$. The density 
of the material through which the neutrinos pass, $\rho_s$,
is about $2.65~{\rm g}/{\rm cm}^3$ for the Earth's crust.  Here we have neglected neutral current interactions, which at
these energies only reduce the neutrino energy by
approximately 20\%, which is within the systematic uncertainty.  For
$E_0 \sim 10^{10}~{\rm GeV}$, $L_{\rm CC}^{\nu} \sim {\cal
  O}(100)$~km.  Let us assume some hypothetical non-perturbative physics
process enhances the $\nu N$ cross section.  Then the interaction path length 
becomes
\begin{equation}
L_{\rm tot}^{\nu} = \left[ N_A \rho_s (\sigma^{\nu}_{\rm CC} +
\sigma^{\nu}_{\rm NP}) \right] ^{-1} \ ,
\end{equation}
where $\sigma^{\nu}_{\rm NP}$ is the non-perturbative contribution to the
cross section for $E_{\nu} = E_0$.

Once a $\tau$ is produced by a CC interaction, it can 
be absorbed in the Earth or escape and possibly decay,
generating a detectable air shower.  At these high energies,
the $\tau$ propagation length in the Earth is 
dominated by energy loss rather than the finite $\tau$ 
lifetime.  The energy loss can be expressed as
\begin{equation}
\label{energyloss}
\frac{dE_{\tau}}{dz} = -(\alpha_{\tau} + \beta_{\tau} E_{\tau})
\rho_s \ ,
\end{equation}
where $\alpha_\tau$ characterizes energy loss due to ionization and
$\beta_\tau$ characterizes losses through bremsstrahlung, pair 
production and hadronic interactions.  At these energies, 
energy losses due to ionization turn out to be negligible, 
while   
$\beta_\tau \simeq 0.8 \times 10^{-6}$~${\rm cm}^2 / {\rm g}$~\cite{Dutta:2000hh}.
{}From Eq.~\eqref{energyloss}, we observe that the maximum path 
length for a detectable $\tau$ can be written
\begin{equation}
L^{\tau} = \frac{1}{\beta_{\tau} \rho_s} \ln \left( E_{\rm max} /
E_{\rm min} \right) \ ,
\label{ltau}
\end{equation}
where $E_{\rm max} \approx E_0$ is the energy at which the $\tau$ is
created, and $E_{\rm min}$ is the minimal energy at which a $\tau$ can
produce a shower big enough to be detected.
For $E_{\rm max} / E_{\rm min} = 10$, $L^{\tau} = 11$~km.

\begin{figure}[tb]
\centering
\includegraphics[width=0.90\textwidth]{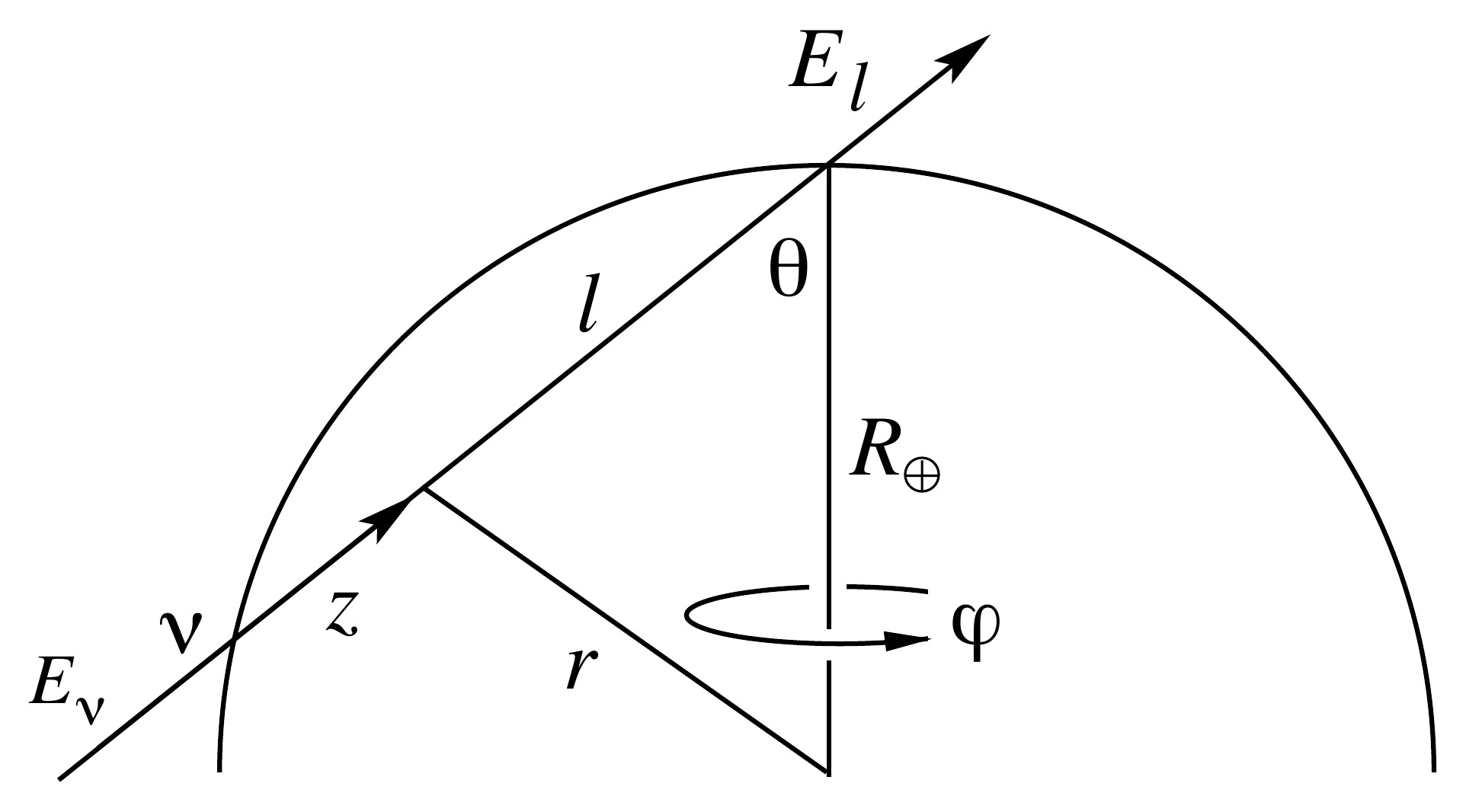}
\caption{The chord length of the intersection of a neutrino
with the Earth is $\ell = 2 R_\oplus \cos\theta$.  In the
figure, the neutrino produces a lepton $l$ after traveling
some distance $z$ inside the Earth's crust.
If $z + L^\tau > \ell$, the lepton will escape the Earth and
can generate an air shower. From Ref.~\cite{Feng:2001ue}.}
\label{f:event}
\end{figure}

The probability for a neutrino with incident nadir angle $\theta$
to emerge as a detectable $\tau$ is
\begin{equation}
P(\theta) = \int_0^{\ell} \frac{dz}{L_{\rm CC}^{\nu}}
e^{-z/L_{\rm tot}^{\nu}} \
\Theta \left[ z - (\ell - L^{\tau} ) \right],
\label{P}
\end{equation}
where $\ell = 2 R_{\oplus} \cos\theta$ is
the chord length of the intersection of the neutrino's trajectory with
the Earth, with $R_{\oplus} \approx 6371$~km the Earth's radius.
Note that we have neglected the possibility that non-perturbative
processes could lead to a detectable signal, since 
the hadrons which dominate the final state will be 
absorbed in the Earth. The step function in Eq.~\eqref{P} reflects the fact that
a $\tau$ will only escape the Earth if $z + L^\tau > \ell$,
as illustrated in Fig~\ref{f:event}.

Assuming an isotropic tau neutrino flux, the number of
taus that emerge from the Earth with sufficient energy to be detected
is proportional to an ``effective solid angle''
\begin{equation}
\Omega_{\rm eff} \equiv \int P(\theta)\, \cos\theta\, d\cos\theta\, d\phi.
\end{equation}
Evaluation of the integrals~\cite{Kusenko:2001gj} yields the unfortunate expression
\begin{eqnarray}
\Omega_{\rm eff}  =  2 \pi
\frac{L_{\rm tot}^{\nu}}{L_{\rm CC}^{\nu}}
\left[ e^{L^{\tau} / L_{\rm tot}^{\nu}} - 1 \right] 
\left[ \left( \frac{L_{\rm tot}^{\nu}}{2 R_{\oplus}} \right)^2 
-  \left( \frac{L_{\rm tot}^{\nu}}{2 R_{\oplus}} +
\left( \frac{L_{\rm tot}^{\nu}}{2 R_{\oplus}} \right)^2 \right)
e^{-2R_{\oplus} / L_{\rm tot}^{\nu}} \right] \ .
\label{Omegaeff}
\end{eqnarray}
At the relevant energies, however, the neutrino interaction length satisfies
$L_{\rm tot}^{\nu} \ll R_{\oplus}$.  In addition, if the hypothesized
non-perturbative cross section enhancement is less than typical
hadronic cross sections, we have $L_{\rm tot}^{\nu} \gg L^{\tau}$.
With these approximations, Eq.~\eqref{Omegaeff} simplifies to~\cite{Anchordoqui:2001cg}
\begin{equation}
\Omega_{\rm eff} \approx
2\pi \frac{L_{\rm tot}^{\nu\, 2} L^{\tau}}{4 R_{\oplus}^2
L_{\rm CC}^{\nu}} \ .
\label{omegaeffapprox}
\end{equation}

Equation~\eqref{omegaeffapprox} describes the functional dependence of the
Earth-skimming event rate on the non-perturbative cross section.  This rate is, of
course, also proportional to the neutrino flux $\Phi^{\nu_{\rm all}}$ at
$E_0$. Thus, the number of Earth-skimming neutrinos is given by
\begin{equation}
N_{\rm ES} \approx
C_{\rm ES}\, \frac{\Phi^{\nu_{\rm all}}}{\Phi^{\nu_{\rm all}}_0}
\frac{\sigma^{\nu\, 2}_{\rm CC}}
{\left( \sigma^{\nu}_{\rm CC} + \sigma^{\nu}_{\rm NP} \right)^2} \ ,
\label{ES}
\end{equation}
where $C_{\rm ES}$ is the number of Earth-skimming events
expected for some benchmark flux $\Phi^{\nu_{\rm all}}_0$ in the absence
of new physics.

In contrast to Eq.~\eqref{ES}, the rate for quasi-horizontal showers
has the form
\begin{equation}
N_{\rm QH} = C_{\rm QH} \frac{\Phi^{\nu_{\rm all}}}{\Phi^{\nu_{\rm all}}_0}
\frac{\sigma^{\nu}_{\rm CC} + \sigma^{\nu}_{\rm NP}}
{\sigma^{\nu}_{\rm CC}} \ ,
\label{QH}
\end{equation}
where  $C_{\rm QH}$ is the number of quasi-horizontal events
expected for flux $\Phi^{\nu_{\rm all}}_0$.

\begin{figure}[tb]
\centering
\includegraphics[width=0.99\textwidth]{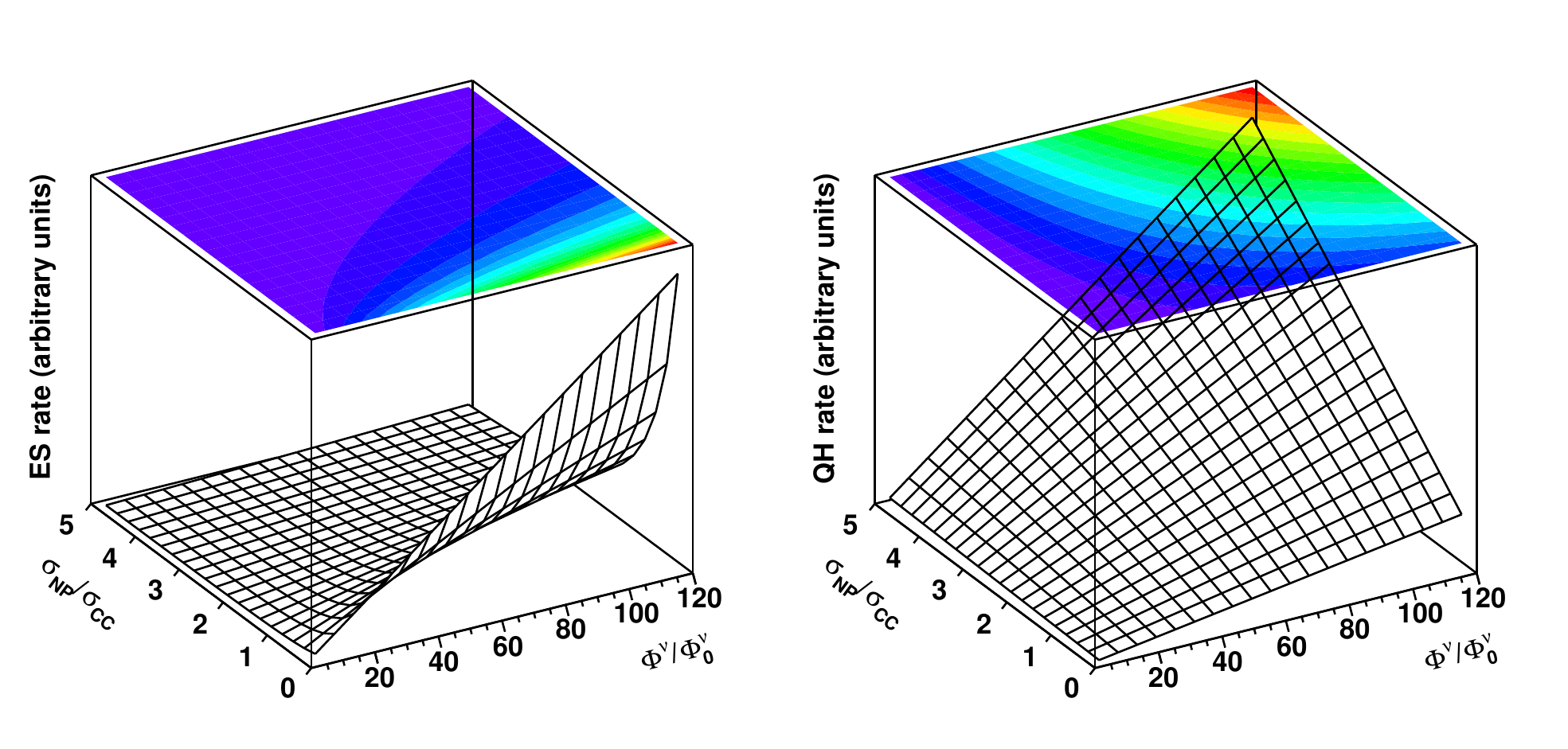}
\caption{Event rates for Earth-skimming (left) and 
quasi-horizontal (right) events in the 
$\Phi^{\nu_{\rm all}}/\Phi_0^{\nu_{\rm all}}-\sigma_{\rm NP}/\sigma_{\rm CC}$ 
plane.  Note that the contours are roughly orthogonal, and so the two
types of event provide complementary information about flux and cross section.}
\label{f:esqh}
\end{figure}

Now, without loss of generality, we normalize the neutrino flux to the Waxmann-Bahcall bound,  
{\em i.e.} $\Phi^{\nu_{\rm all}}_0 \equiv \Phi^{\nu_{\rm all}}_{\rm WB}$.
We use the expected rates for the 
the benchmark flux to determine the values of $C_{\rm ES}$ 
and $C_{\rm QH}$ in Eqs.~\eqref{ES} and~\eqref{QH}, ($C_{\rm ES} = 0.15$ and
$C_{\rm QH} = 0.06$, as shown in Table~\ref{t:rates}).
Given a flux $\Phi^{\nu_{\rm all}}$ and new non-perturbative physics cross section
$\sigma^{\nu}_{\rm NP}$, both $N_{\rm ES}$ and $N_{\rm QH}$ are
determined.  On the other hand, given just a quasi-horizontal event
rate $N_{\rm QH}$, it is impossible to differentiate between an
enhancement of the cross section due to non-perturbative physics and an increase of
the flux.  However, in the region where significant event rates are
expected, the contours of $N_{\rm QH}$ and $N_{\rm ES}$, given by
Eqs.~\eqref{ES} and \eqref{QH}, are more or less orthogonal and
provide complementary information. This is illustrated in Fig.~\ref{f:esqh}.
With measurements of $N_{\rm
  QH}^{\rm obs}$ and $N_{\rm ES}^{\rm obs}$, both $\sigma^{\nu}_{\rm
  NP}$ and $\Phi^{\nu_{\rm all}}$ may be determined independently, and
neutrino interactions beyond the (perturbative) SM may be unambiguously identified.

\begin{figure*}[tbp]
\begin{minipage}[t]{0.32\textwidth}
\postscript{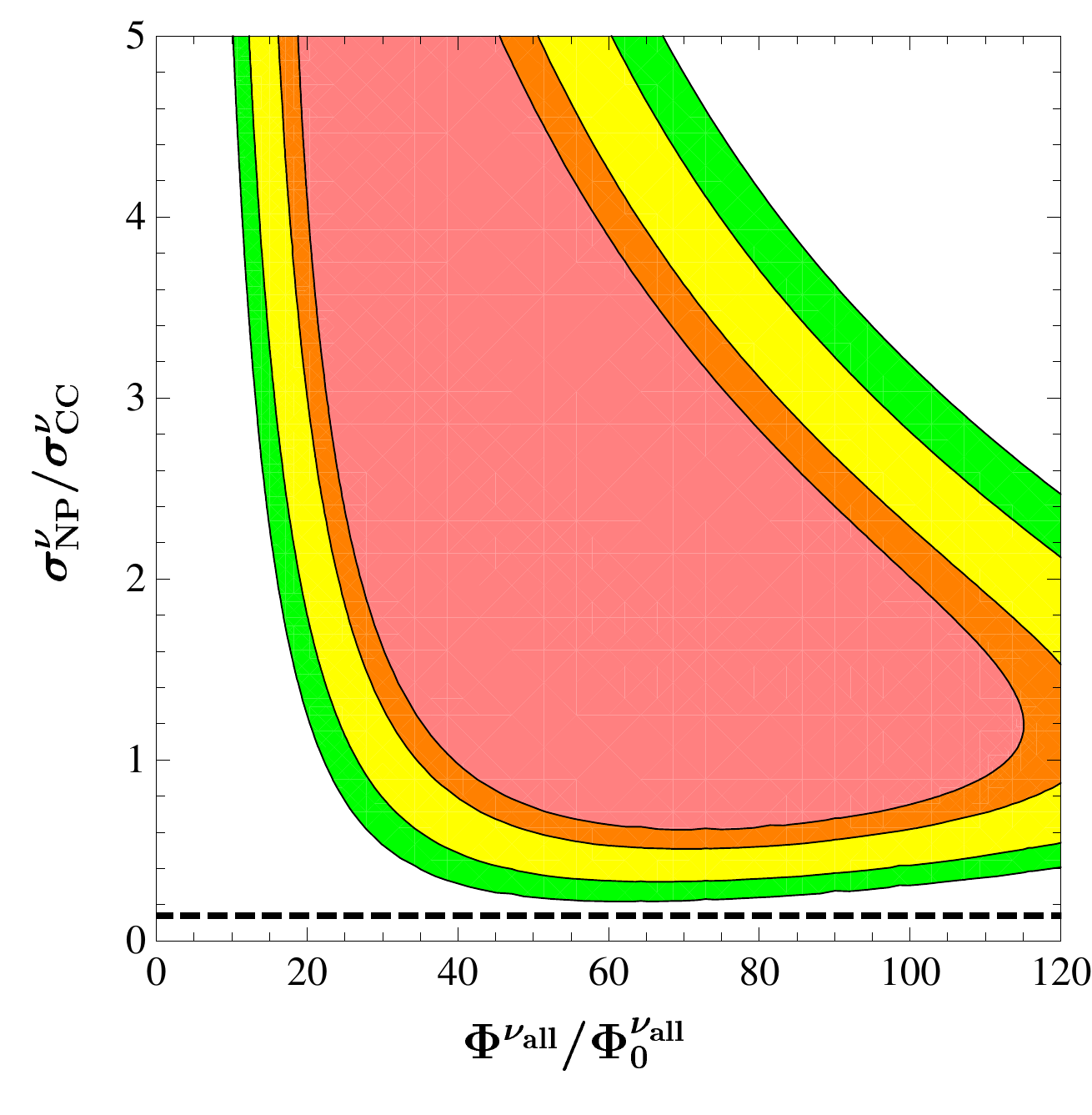}{0.99}
\end{minipage}
\hfill
\begin{minipage}[t]{0.32\textwidth}
\postscript{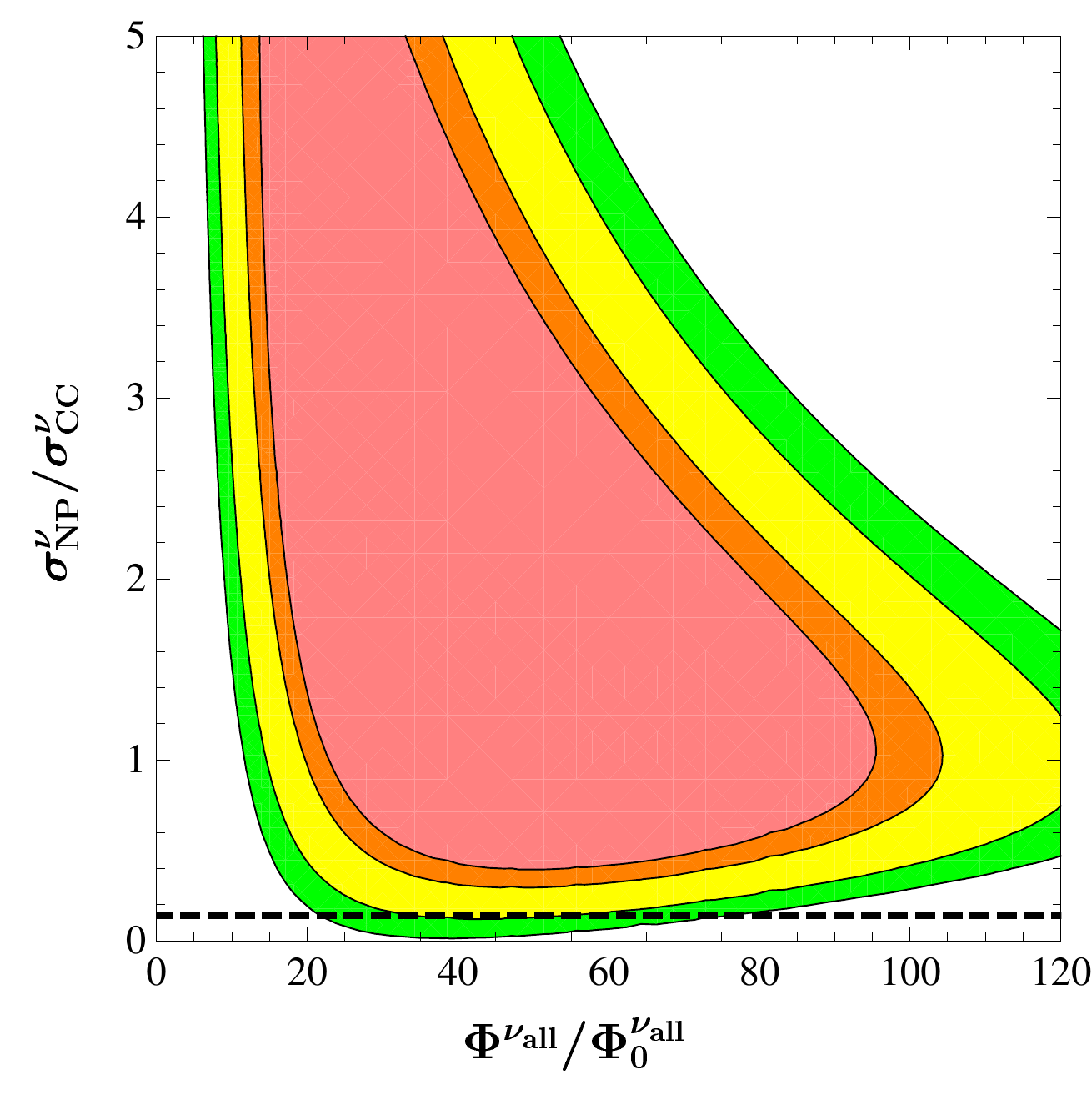}{0.99}
\end{minipage}
\hfill
\begin{minipage}[t]{0.32\textwidth}
\postscript{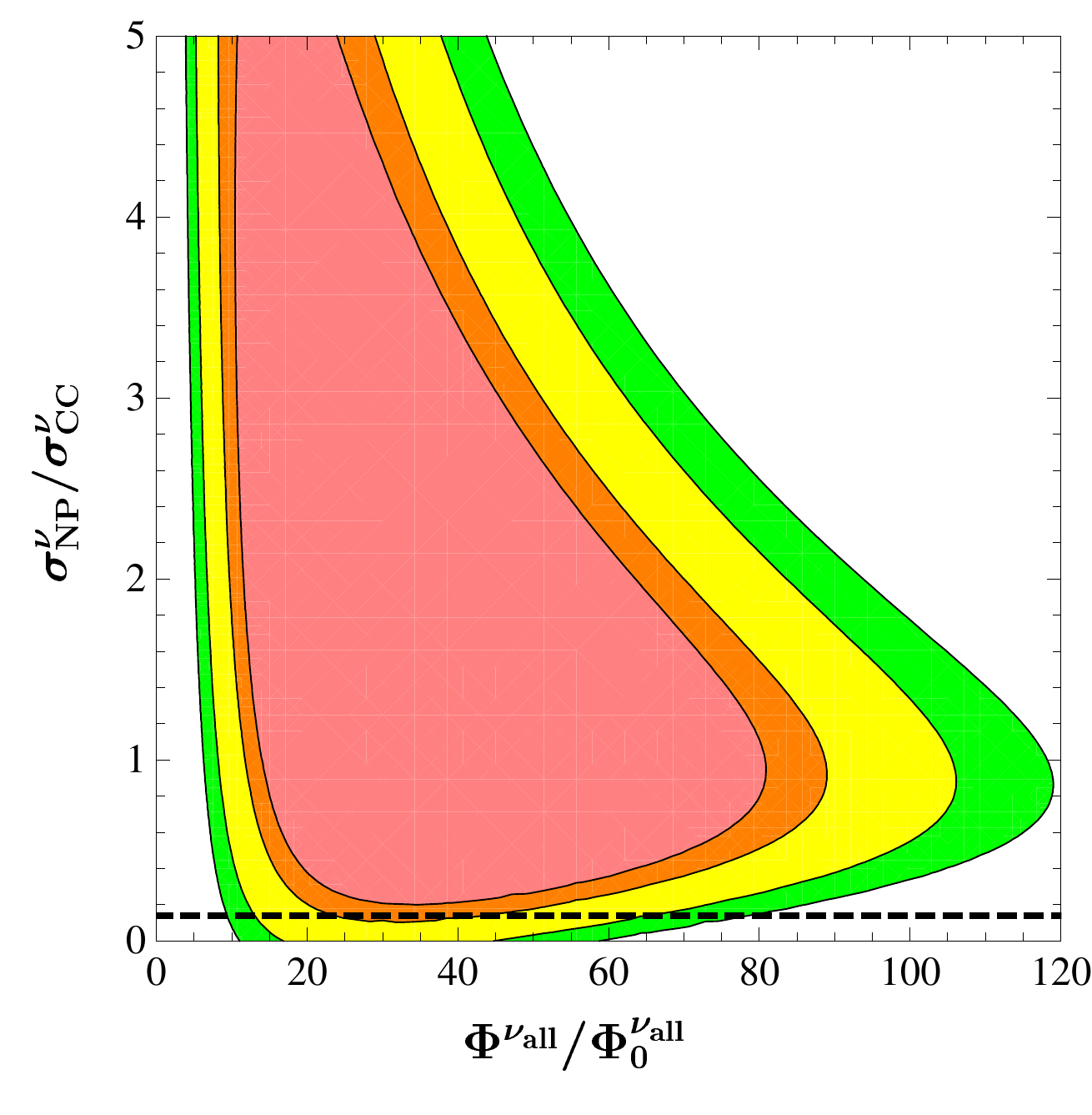}{0.99}
\end{minipage}
\caption{Projected determination of neutrino fluxes and cross sections
  at $\sqrt{s} \approx 250$~TeV from future Auger data. The different
  shaded regions indicate the 90\%, 95\%, 99\% and 3$\sigma$
  confidence level contours in the $\Phi^{\nu_{\rm
      all}}/\Phi_0^{\nu_{\rm all}}-\sigma_{\rm NP}/\sigma_{\rm CC}$
  plane, for $N^{\rm obs}_{\rm ES} = 1$, $N^{\rm obs}_{\rm QH} = 10$
  (left), $N^{\rm obs}_{\rm ES} = 1$, $N^{\rm obs}_{\rm QH} = 7$
  (middle), and $N^{\rm obs}_{\rm ES} = 1$, $N^{\rm obs}_{\rm QH} = 5$
  (right). The dashed line indicates the result of including the
  systematic uncertainty on the NLO QCD CC neutrino-nucleon
  cross section. From Ref.~\cite{Anchordoqui:2010hq}.}
\label{statistics}
\end{figure*}

We now turn to determining the projected sensitivity of Auger to
neutrino fluxes and cross sections. The quantities $N_{\rm ES}$ and
$N_{\rm QH}$ as defined in Eqs.~(\ref{ES}) and (\ref{QH}) can be
regarded as the theoretical values of these events, corresponding to
different points in the $\Phi^{\nu_{\rm all}}/\Phi_0^{\nu_{\rm
    all}}-\sigma_{\rm NP}/\sigma_{\rm CC}$ parameter space. For a
given set of observed rates $N^{\rm obs}_{\rm ES}$ and $N^{\rm
  obs}_{\rm QH}$, two curves are obtained in the two-dimensional
parameter space by setting $N^{\rm obs}_{\rm ES} = N_{\rm ES}$ and
$N^{\rm obs}_{\rm QH} = N_{\rm QH}$.  These curves intersect at a
point, yielding the most probable values of flux and cross section for
the given observations.  Fluctuations about this point define contours
of constant $\chi^2$ in an approximation to a multi-Poisson likelihood
analysis~\cite{Cowan}. The contours are defined by
\begin{equation}
\chi^2  =  \sum_i 2 \,\left[ N_i  -  N_i^{\rm obs}\right] +
2\,  N_i^{\rm obs}\,
\ln \left[ N_i^{\rm obs}/ N_i \right] \,\,,
\label{baker}
\end{equation} 
where $i=$~ES, QH~\cite{Baker:1983tu}. In Fig.~\ref{statistics}, we
show results for three representative cases.  Assuming ($N^{\rm
  obs}_{\rm ES} = 1$, $N^{\rm obs}_{\rm QH} = 10$), ($N^{\rm obs}_{\rm
  ES} = 1$, $N^{\rm obs}_{\rm QH} = 7$), and ($N^{\rm obs}_{\rm ES} =
1$, $N^{\rm obs}_{\rm QH} = 5$) we show the 90\%, 95\%, 99\% and
3$\sigma$ CL contours for 2 d.o.f. ($\chi^2 = 4.61,\ 5.99,\ 9.21,$ and
11.83, respectively).  For $N^{\rm obs}_{\rm ES} = 1$ and $N^{\rm
  obs}_{\rm QH} = 10$, the possibility of a SM interpretation along
the $\sigma^{\nu}_{\rm NP} = 0$ axis (taking into account systematic
uncertainties) would be excluded at greater than 99\% CL for {\em any}
assumed flux. 

In summary, we have found  that observation of 1
Earth-skimming and 10 quasi-horizontal events would exclude the
(perturbative) SM at the 99\% CL.  Thus the expected low
neutrino ``luminosity'' is not at all a show-stopper, and the
observatory has the potential to uncover new physics at scales
exceeding those accessible to the LHC. If new non-perturbative
physics exists, a decade or so would be required to uncover it at Auger in the
best case scenario (cosmic neutrino flux at the Waxman-Bahcall level
and $\nu N$ cross section about an order of magnitude above the
SM prediction). Of course future CR experiments should benefit from a much larger aperture,  making such discovery  conceivable in a short time scale. An optimist may even imagine the possibility of probing QCD dynamics~\cite{Anchordoqui:2006ta}.

\section*{Acknowledgments} I would like to thank Felix Aharonian, Markus Ahlers, Segev BenZvi, Hans Bl\"umer, Mandy Cooper Sarkar, Olivier Deligny, Tere Dova,  Jonathan Feng,  Haim Goldberg, Concha Gonzalez Garcia, Yann Guardincerri, Francis Halzen, Nicole Krohm, Fred Kuehn, Antoine Letessier Selvon, Jim Matthews, Teresa Montaruli, Giulia Pancheri, Tom Paul, Subir Sarkar, Viviana Scherini, Al Shapere, Paul Sommers, John Swain, Diego Torres, Alan Watson, and Tom Weiler for valuable discussions and permission to reproduce some of the figures. I am grateful to Stefanie Pinnow for her proofreading skills and to my sister Gaby for her assistance with slide management. L.A.A.\ is supported by the U.S. National Science Foundation (NSF Grant No PHY-0757598) and the UWM Research Growth Initiative.  Any opinions, findings, and conclusions or recommendations expressed in this material are those of the author and do not necessarily reflect the views of the National Science Foundation.

\appendix

\section{Cosmogenic $\bm{\beta}$-DK and $\bm{A^*}$ processes}
\label{AA}

If UHECRs are heavy nuclei, the relic photons can excite the GDR at nucleus energies $E \gsim 10^{11}$~GeV (corresponding to $\egdr \sim 10~{\rm MeV} - 30~{\rm MeV}$ in the nuclear rest frame), and thus there should be accompanying photo-dissociated free neutrons, themselves a source of $\beta$-decay antineutrinos.  The decay mean free path of a neutron is $c \Gamma_n \overline \tau_n = (E_n/10^{11}$~GeV)~Mpc, the lifetime being boosted from its rest frame value $\overline \tau_n = 886$~s to its lab value via $\Gamma_n = E_n/m_n$.  Compared to cosmic distances $\gsim 100$~Mpc, the decay of even the boosted neutron may be taken as nearly instantaneous, and thus all free neutrons are themselves a source of $\beta$-decay cosmogenic antineutrinos. The neutron emissivity ${\cal Q}_n (E_n),$ defined as the mean number of neutrons emitted per comoving volume per unit time per unit energy as measured at the source can be estimated as follows.  Neutrons with energies above $10^{9.3}$~GeV have parent iron nuclei with $\Gamma > \Gamma_0 \approx 2 \times 10^{9}$ which are almost completely disintegrated in distances of less than 100~Mpc (see Sec.~\ref{nuclei}).  Thus, it is reasonable to define a characteristic time $\tau_{_{\Gamma}}$ given by the moment at which the number of nucleons is reduced to $1/e$ of its initial value $A$, and presume the nucleus, emitted at distance $d$ from the Earth, is a traveling source that at $D \approx (d - c \tau_{_{\Gamma}})$ disintegrates into $A$ nucleons all at once~\cite{Anchordoqui:1997rn}. Then, the number of neutrons with energy $E_n = E_A/A$ can be approximated by the product of 1/2 the number of nucleons generated per nucleus and the number of nuclei emitted, {\em i.e.} ${\cal Q}_n (E_n) = $N$ {\cal Q}_A$, where $N = A - Z$ is the mean neutron number of the source nucleus. Now, to obtain an estimate of the diffuse antineutrino flux one needs to integrate over the population of nucleus-emitting-sources out to the horizon~\cite{Weiler:2004jy,Anchordoqui:2004ma}
\begin{eqnarray}
 \Phi_{\overline \nu_e} & = & \frac{1}{4\pi\,H_0}\, 
   \int dE_n\,{\cal Q}_n (E_n) \,\,
   \left[1-\exp\left(-\frac{D\,m_n}{E_n\,\overline \tau_n}\right) \right] \,\, 
   \int_0^Q d\epsilon_{\overline \nu}\,\frac{dP}{d\epsilon_{\overline \nu}}
   (\epsilon_{\overline \nu})  \nonumber \\
   &  & \times \int_{-1}^1 \frac{d\cos \overline \theta_{\overline\nu}}{2} \; 
   \delta\left[E_{\overline \nu}-E_n\,\epsilon_{\overline \nu}\,(1+\cos 
  \overline \theta_{\overline \nu}) /m_n\right] \,,
\label{nuflux}
\end{eqnarray}
where the
$r^2$ in the volume integral 
is compensated by the usual $1/r^2$ fall-off of flux per source.
Here, $H_0$ is the Hubble constant, $E_{\overline \nu}$ and $E_n$ are the antineutrino and
neutron energies in the lab, $\overline \theta_{\overline \nu}$ is
the antineutrino angle with respect to the direction of the
neutron  momentum in the neutron rest-frame, and $\epsilon_{\overline \nu}$ is
the antineutrino energy in the neutron rest-frame.  The last three variables are not observed
by a laboratory neutrino-detector, and so are integrated over.
The observable $\Enu$ is held fixed.
The delta-function relates the neutrino energy in the lab to the
three integration variables.
The parameters appearing in Eq.~(\ref{nuflux}) are the
neutron mass and rest-frame lifetime ($m_n$ and $\tbar$). Finally, $dP/d\Enubar$ is the
normalized probability that the
decaying neutron produces a $\overline \nu$ with
energy $\Enubar$  in the neutron rest-frame. Note that the maximum $\overline \nu$ energy 
in the neutron 
rest frame is very nearly  $Q \equiv m_n - m_p - m_e = 0.71$~MeV.
Integration of Eq.~(\ref{nuflux})  can be 
easily accomplished, especially when two good approximations are 
applied~\cite{Anchordoqui:2003vc} .
The first approximation is to think of the $\beta$--decay as a $1 \to 2 $ process of 
$\delta m_N \to e^- + \overline \nu,$ in which the neutrino is produced 
monoenergetically in the rest frame, with $\epsilon_{\overline \nu} = \epsilon_0 
\simeq \delta m_N (1  - m_e^2/ \delta^2 m_N)/2 \simeq 0.55$~MeV, where 
$\delta m_N \simeq 1.30$~MeV 
is the neutron-proton mass difference. In the lab,
the ratio of the maximum $\overline \nu$ energy to the neutron energy  
is $2 \epsilon_0/m_n \sim 10^{-3},$ 
and so the boosted $\overline \nu$ has a spectrum with 
$E_{\overline\nu} \in (0, 10^{-3} \, E_n).$ 
 The second approximation is to replace the neutron decay probability 
$1 - e^{-Dm_n/E_n \overline \tau_n}$
with a step function $\Theta (E_n^{\rm max} - E_n)$ at some energy 
$E_n^{\rm max} \sim {\cal O}(D \, m_n/\overline{\tau}_n) = 
(D/10~{\rm Mpc}) \times 10^{12}$~GeV. 
Combining these two approximations we obtain
\begin{equation}
 \Phi_{\overline \nu_e} = \frac{m_n}{8\,\pi\, \epsilon_0 \,H_0} 
\int_{E_A^{\rm min}}^{E_A^{\rm max}} \frac{dE_A}{E_A/A}\,\, {\cal Q}_A (E_A)\,\,,
\label{S1}
\end{equation}
where $E_A^{\rm min} \equiv {\rm max} \{ E_{A,\Gamma_0}, \frac{A\, 
m_n E_{\overline \nu}}{2 \epsilon_0} \},$ and $E_A^{\rm max}$ is the energy 
cutoff at the nucleus-emitting-source $\ll A (D/10~{\rm Mpc}) \times 10^{12}$~GeV.
For ${\cal Q}_A \propto E_A^{-\alpha},$ integration of 
Eq.~(\ref{S1}) leads to 
\begin{equation}
    \Phi_{\overline \nu_e} (\Enu) \approx 10^6 \left(\frac{E_{A,\Gamma_0}}
    {E_A^{\rm max}}\right)^\alpha 
    \left[\left(\frac{E^{\rm max}_{\overline \nu}}{E_{\overline \nu}}
    \right)^\alpha -1 \right] \, \left. 
    {\cal Q}_A \right|_{\Gamma_0}\,\,,
\label{znz}
\end{equation}
where $E_{\overline\nu} \gsim  10^{6.3} \,(56/A)$~GeV and
\begin{equation}
E_{\overline \nu}^{\rm max} = \frac{2 \epsilon_0}{m_n}\,\, 
\frac{E_A^{\rm max}}{A} \sim  10^{7.3}\,\, \left(\frac{56}{A}\right)\,\,
\left(\frac{E_A^{\rm max}}{10^{12}~{\rm GeV}}\right) \,\, {\rm GeV}\,.
\end{equation}
The sub-PeV antineutrino spectrum is flat, as all the free neutrons have sufficient energy
$E_n \gsim E_{\Gamma_0}/A,$ to contribute equally to all the $\overline \nu$ energy 
bins below $10^{6}$~GeV. Taking $\alpha = 2$ as a reasonable example, and inputting the 
observational value for iron nuclei, $\left. E^2_{A,\Gamma_0} \,\ J_{\rm CR}  \right|_{\Gamma_0} 
\approx 10^5$~eV m$^{-2}$ s$^{-1}$ sr$^{-1}$, Eq.~(\ref{znz}) 
becomes
\begin{equation}
    E_{\overline \nu}^2 \   \Phi_{\overline \nu_e}\approx 4 
    \times 10^{1}\,\, \left(\frac{56}{A}\right)\,\,{\rm eV}\, {\rm m}^{-2}\,\, 
    {\rm s}^{-1}\,\, {\rm sr}^{-1} \,. 
\label{buengusto}
\end{equation}
Note that the $\beta$-decay process gives initial antineutrino flavor ratios $1 : 0 : 0$ 
and Earthly ratios nearly $3:1:1.$ Compared to full-blown Monte Carlo 
simulations~\cite{Hooper:2004jc,Ave:2004uj}, 
this back of the envelope calculation underestimates the flux by about 30\%. 
This is because the preceeding calculation does not account for possible neutrons produced in $p\gamma_{\rm CMB}$ collisions.
Of course the situation described above represents the most extreme case, 
in which all cosmic rays at the end of the spectrum are heavy nuclei. 
A more realistic guess would be that the composition at the 
end of the spectrum is mixed. 

Photodisintegration of high energy nuclei is followed by immediate photo-emission from the excited daughter nuclei. For brevity, we label the photonuclear process $A+\gamma\to A'^*+X$, followed by $A'^*\to A' +\gamma$-ray as ``$A^*$''~\cite{Anchordoqui:2006pd}.  The GDR decays dominantly by the statistical emission of a single nucleon, leaving an excited daughter nucleus $(A-1)^*$.  The excited daughter nuclei typically de-excite by emitting one or more photons of energies $\edeex\sim 1-5$~MeV in the nuclear rest frame.  The lab-frame energy of the $\gamma$-ray is then $\elab=\Gamma_A\,\edeex$, where $\Gamma_A=E_A/m_A$ is the boost factor of the nucleus in the lab.

Of interest  here is the $\gamma$-ray flux produced when the
photo-dissociated nuclear fragments produced on the CMB and CIB
de-excite. These $\gamma$-rays create chains of electromagnetic cascades on
the CMB and CIB, resulting in a transfer of the initial energy into
the so-called Fermi-LAT region, which is bounded by
observation to not exceed $\omega_{\rm cas}\sim 5.8\times 10^{-7} {\rm
  eV/cm^3}$~\cite{Berezinsky:2010xa}. Fortunately, we can finesse the details of the calculation by arguing in analogy to work already done.  The photodisintegration chain produces one $\beta$-decay neutrino with energy of order 0.5~MeV in the nuclear rest frame, for each neutron produced. Multiplying this result by 2 to include photodisintegration to protons in addition to neutrons correctly weights the number of steps in the chain.  Each step produces on average one photon with energy $\sim 3$~MeV in the nuclear rest frame.  In comparison, about 12 times more energy is deposited into photons. Including the factor of 12 relating $\omega_\gamma$ to $\omega_{\bar\nu_e}$, we find from (\ref{buengusto}) that cosmogenic photodisintegration/de-excitation energy,
$\omega_{\rm cas}\sim 1.4 \times 10^{-11} \, (56/A)$~eV/cm$^3$,
 is more than three orders of magnitude
below the Fermi-LAT bound.\footnote{These rough calculations are consistent with the more rigurous analysis presented in~\cite{Wang:2011qc}.} This result appears to be
nearly invariant with respect to varying the maximum energy of the Fe
injection spectrum (with a larger $E_{\rm max}$, the additional energy
goes into cosmogenic pion production).  Thus, there is no constraint on
a heavy nuclei cosmic-ray flux from the $A^*$ mechanism.

Neutron emission from a cosmological distribution of CR sources would also lead to a flux of $\overline \nu_e$.  In analogy to our previous estimate, we sum over the neutron-emitting sources out to the edge of the universe; Eq.~(\ref{S1}) becomes~\cite{Anchordoqui:2004eb}
\begin{equation}
\Phi_{\overline \nu_e} = \frac{m_n }{8\,\pi \,\epsilon_0\, H_0}
\int_{\frac{m_n E_{\overline \nu}}{2 \epsilon_0}}^{E_n^{\rm max}}
\frac{dE_n}{E_n} \,\, {\cal Q}_n(E_n) \,\, ,
\label{nDK}
\end{equation}
where $ {\cal Q}_n(E_n)$ is the neutron volume emissivity. An upper limit
can be placed on ${\cal Q}_n$ by assuming an extreme situation in
which all the CRs escaping the source are neutrons, {\em i.e.},
$\dot \epsilon_{_{\rm CR}}^{[10^{10}, 10^{12}]}
 = \int dE_n\,\, E_n\,\, {\cal Q}_n (E_n) .$
With the production rate of ultrahigh energy protons (\ref{professionally}), and an assumed injection spectrum
${\cal Q}_n \propto E_n^{-2},$ Eq.~(\ref{nDK}) gives
\begin{equation}
E_{\nu}^2 \ \Phi_{\overline \nu_e}\approx 3\times 10^{-11} \,\,
{\rm GeV} \,  {\rm cm}^{-2} \, {\rm s}^{-1} \,
{\rm sr}^{-1}\,\,,
\end{equation}
which is about three orders of magnitude below the Waxman-Bahcall
bound.

\section{Energy density of the EM cascade} 
\label{AB}

EM  interactions of photons and leptons with the
extragalactic radiation backgrounds and magnetic field can happen on
time-scales much shorter than their production rates. The relevant
processes with background photons contributing to the differential
interaction rates $\gamma_{ee}$, $\gamma_{\gamma e}$, and $\gamma_{e
  \gamma}$ are inverse Compton scattering (ICS), $e^\pm+\gamma_{\rm
  bgr}\to e^\pm+\gamma$, pair production (PP), $\gamma+\gamma_{\rm
  bgr}\to e^++e^-$, double pair production (DPP) $\gamma+\gamma_{\rm
  bgr}\to e^++e^-+e^++e^-$, and triple pair production (TPP),
$e^\pm+\gamma_{\rm bgr}\to
e^\pm+e^++e^-$~\cite{Blumenthal:1970nn,Blumenthal:1970gc}.

High energetic electrons and positrons may also lose energy via
synchrotron radiation in the intergalactic magnetic field $B$ with a
random orientation $\sin\theta$ with respect to the velocity
vector. We will assume in the following that the field strength $B = 10^{-12}$~G.\footnote{For $B \gtrsim 10^{-12}~{\rm G}$, TPP by electrons can be neglected~\cite{Lee:1996fp}.  For  $E< 10^{12}~{\rm GeV}$,  DPP of photons can also be safely neglect in the calculation~\cite{Demidov:2008az}.} This leads to an efficient transfer of energy into the EM
cascade. The synchrotron power spectrum (W eV${}^{-1}$) has the form
\begin{equation}
\mathcal{P}(E_e,E_\gamma) = \frac{\sqrt{3}\alpha}{2\pi}
\frac{eB\sin\theta}{m_e}F(E_\gamma/E_{\rm c})\,;\qquad F(t) \equiv
t\int_t^\infty  dz\, K_{5/3}(z)\,,
\end{equation}
where $K_{5/3}$ is the modified Bessel function and
$E_{\rm c} = (3eB\sin\theta/2m_e)(E_e/m_e)^2$. 
This can be treated as a continuous energy loss of the electrons and
positrons with a parameter\footnote{Note the identity $\int 
    dE\left[E\,\partial_E(bn_e) + \int dE'
    \mathcal{P}(E',E)n_e\right] = 0$, implying overall energy
  conservation.}
\begin{equation}
b_{\rm syn}(E_e) = \frac{1}{2}\int  d\cos\theta\int
dE_\gamma\mathcal{P}(E_e,E_\gamma) =
\frac{4\alpha}{9}\left(\frac{eB}{m_e}\right)^2\left(\frac{E_e}{m_e}\right)^2\,.
\end{equation}
We will assume in the calculation that the intergalactic magnetic
field is primordial with a (flux-conserving) redshift dependence $B(z) =
(1+z)^2B(0)$. Note that the synchrotron energy loss has then a
redshift dependence similar to BH pair production in the CMB, {\it
  i.e.}~$b_{\rm syn}(z,E) = (1+z)^2b_{\rm syn}(0,(1+z)E)$.  It is also
convenient to define $\gamma_{e\gamma}^{\rm syn}(E_e,E_\gamma) \equiv
\mathcal{P}(E_e,E_\gamma)/E_\gamma$, which has an analogous redshift
dependence, {\it i.e.}~$\gamma_{e\gamma}^{\rm syn}(z,E_e,E_\gamma) =
(1+z)^4\gamma_{e\gamma}^{\rm syn}(0,(1+z)E_e,(1+z)E_\gamma)$.

The fast evolution of the cascade is governed by the set of
differential equations,
\begin{align}
\partial_{\hat{t}}Y_e(E)
=&-\Gamma_e(E)Y_e(E)+\partial_E(b(E)Y_e(E)) + \int d
E'\left[\gamma_{\gamma
e}(E',E)Y_\gamma(E')+\gamma_{ee}(E',E)Y_e(E')\right] \,, \nonumber \\
\partial_{\hat{t}}Y_\gamma(E) =&-\Gamma_\gamma(E)Y_\gamma(E) +
\int dE'\frac{\mathcal{P}(E',E)}{E}Y_e(E')+\int  d
E'\gamma_{e\gamma}(E',E)Y_e(E') \,,\label{eq:CAS1}
\end{align}
which determines the evolution on {\it short} time-scales
$\Delta\hat{t}\,\Gamma_{p\gamma}\ll1$ (the redshift $z$ is
kept {\it fixed} meanwhile). The initial
condition $Y_{\gamma/e}(E)|_{\hat t=0}$ is given by the sum of
previously developed cascades and the newly generated contributions
from proton interactions.

The solution of the system~(\ref{eq:CAS1})  for an
infinitesimally small step $\Delta \hat t$ can be written for a
discrete energy spectrum, $N_i \simeq \Delta E_iY_i$, as
\begin{equation}\label{eq:discrete}
\begin{pmatrix}N_\gamma\\N_e\end{pmatrix}_i(\hat{t}+\Delta\hat{t})
  \simeq \sum_j\begin{pmatrix}T_{\gamma\gamma}(\Delta\hat
  t)&T_{e\gamma}(\Delta\hat t)\\T_{\gamma e}(\Delta\hat
  t)&T_{ee}(\Delta\hat
  t)\end{pmatrix}_{ji}\begin{pmatrix}N_\gamma\\N_e\end{pmatrix}_j(\hat{t})\,.
    \end{equation}
    With the transition matrix $\mathcal{T}(\Delta \hat t)$, defined
    by Eq.~(\ref{eq:discrete}), we can efficiently follow the
    development of the EM cascade over a distance $\Delta
    t = 2^N\Delta \hat t$ via matrix doubling~\cite{Protheroe:1992dx}:
\begin{equation}
\mathcal{T}(2^N\Delta\hat t) \simeq \left[\mathcal{T}(\Delta\hat
t)\right]^{2N}\,.
\end{equation}

The total energy of the cascade can be obtained by integrating
Eq.~(\ref{newdiff}):
\begin{equation}
\label{eq:DE}
\frac{d}{d t}\left[\int d E \ E {\mathscr Z}_{\rm cas}(z,E)\right]
= - \int d E E\partial_{\mathscr{E}}\left[b_{\rm
cas}(z,\mathscr{E}) \mathscr{Z}_{\rm p}(z,E)\right]\,.
\end{equation} 
Integrating the r.h.s.~by parts yields
\begin{equation}
{\rm r.h.s.} = - \int d E\partial_{E}\left[ E\frac{1}{1+z}b_{\rm
cas}(z,\mathscr{E}) {\mathscr Z}_p(z,E)\right] + \int d E
\frac{1}{1+z}b_{\rm cas}(z,\mathscr{E}) \mathscr{Z}_p (z,E)\,.
\end{equation}
The first term vanishes since $b_{\rm cas}=0$ for sufficiently low
energies and $\mathscr{Z}_p=0$ beyond the maximal energy. The time
integration of the l.h.s.~between the present epoch ($t=0$) and the
first sources ($t_{\rm max}$) gives
\begin{equation}
\int_0^{t_{\rm max}} dt [{\rm l.h.s.}] = \int d E E n_{\rm
cas}(E) = \omega_{\rm cas}\, ,
\end{equation}
hence we obtain Eq.~(\ref{eq:omegacas}).

\begin{figure}[t]
  \begin{center}
    \includegraphics[width=0.6\linewidth]{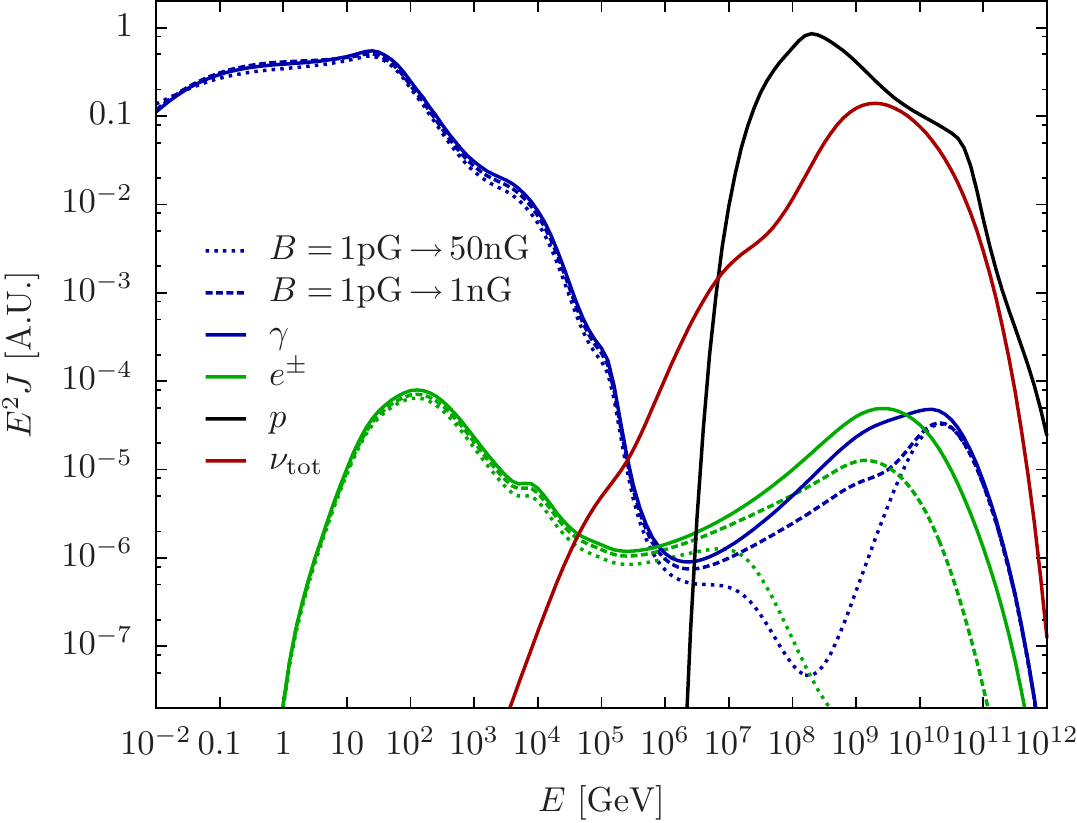}
  \end{center}
  \vspace{-0.5cm}
  \caption[]{Cosmogenic photon (blue lines) and neutrino (red line) fluxes from a "vanilla" proton      test-spectrum (shown as black line) of UHECRs, assuming two exponential cutoffs ($E_{\rm min}=10^{8}~{\rm GeV}$ and  $E_{\rm max} = 10^{11.5}~{\rm GeV}$), power index $\gamma=2$ and evolution with $n=3$ and $0<z<2$. The fluxes are computed for 3 extragalactic magnetic field strengths: 1~pG, 1~nG, and 50~nG. Proton diffusion in extragalactic magnetic fields  $B>>{\rm nG}$  has not yet been included in the calculation. We have neglected radiative losses of muons and pions before decaying into neutrinos; the energy loss in muons is down by a factor $10^6$ compared to electrons. This figure is courtesy of Markus Ahlers.}
\label{fig:Bsamples}
\end{figure}

Synchrotron radiation in strong magnetic fields, $1~{\rm nG} \lesssim B \lesssim 50~{\rm nG}$,  can also by-pass the EM cascade  and transfer energy to sub-GeV photons that are unconstrained by the Fermi-LAT spectrum~\cite{Wdowczyk:1972}. 
This would be relevant  for electrons around $E_e \sim 10^{9}$~GeV, where synchrotron loss starts to dominate over ICS loss in the CMB;  see Fig.~\ref{fig:Bsamples}.
 From Eqs.~(\ref{synch})  and (\ref{Ginzburg})  we can verify that $\langle E_\gamma^{\rm syn} \rangle \lesssim 1~{\rm GeV}$.  Of course, reaching a 50~nG field requires some astrophysical generator to augment the primordial field.\footnote{For $\lambda_B \sim 1$~Mpc, the current upper limit on primordial seed fields is about 1~nG~\cite{Neronov:1900zz,Taylor:2011bn}.} For simplicity, we have taken the redshift evolution of the generator to be the same as the evolution of the primordial field.
If were to assume an evolution which follows SFR, this will not significantly alter our  conclusions.

\section{TOTal Elastic and diffractive cross section Measurement}

\label{AC}

The TOTEM experiment is dedicated to the measurement of the total $pp$ cross section  
with the luminosity independent  method based  on the Optical Theorem, 
\begin{equation}
\sigma_{\rm tot} = {8 \pi \over \sqrt{s} } \; {\rm Im}  f(0)\,\,,
\label{0}
\end{equation}
where $f(\vartheta)$ satisfies
\begin{equation}
{d\sigma_{\rm el}  \over dt} = 
{4 \pi \over s } \; 
{d\sigma_{\rm el}  \over d\Omega} =
{4 \pi \over s } \; |f(\vartheta)|^2 \,\,,
\end{equation}
with $\vartheta$ the angle of the scattered proton with respect to the beam direction.
Squaring Eq.~(\ref{0}) we obtain
\begin{equation}
\sigma^2_{\rm tot} = {16 \pi \,\, {\rm Im}^2 f(0) \over   {\rm Re}^2f(0) + 
  {\rm Im}^2f(0) } \; \left.{d\sigma_{\rm el}  \over dt}\right|_{t=0} = 
{16 \pi \over  1 + \rho^2}  \;\; \frac{[dN_{\rm el}/ dt]_{t=0}}{{\cal L}}\,,
\end{equation}
where ${\cal L}$ is the integrated luminosity. 
Now, following~\cite{Matthiae:gb,Lipari:2003es}, we can obtain 
the total cross section independently from ${\cal L}$, by using $\sigma_{\rm tot} =
(\sigma_{\rm el} + \sigma_{\rm inel})  = (N_{\rm el} + N_{\rm inel})/{\cal L},$
\begin{equation}
\sigma_{\rm tot} = {16 \pi \over 1 + \rho^2} \, { [ d N_{\rm el} / dt  ]_{t = 0} \over (N_{\rm el} + N_{\rm inel}) } \,. 
\end{equation}
Here, $N_{\rm el}$ and $N_{\rm inel}$ are the numbers of elastic and inelastic events, and
$\rho = 0.10 \pm 0.01$ is the ratio between the real and imaginary parts of the forward scattering 
amplitude~\cite{Augier:1993ta}.\footnote{Note that the quoted value of $\rho$ is an extrapolation to 
$\sqrt{s} = 14$~TeV, and may be measured by the LHC experiments. Otherwise, it will contribute to 
the uncertainty in $\sigma_{\rm tot}$.} The difficult aspect of this measurement is obtaining
a good extrapolation of the cross section for low momentum transfer. 
Recall that  $-t = s \,(1-\cos \vartheta) / 2 \simeq
s \, \vartheta^2 / 4$, and so $t \to 0$ implies a measurement in the
extreme forward direction.
The TOTEM experiment aims to measure down to values of $|t| \approx  \times 10^{-3}~\rm{GeV}^{2},$ which
corresponds to $\vartheta \approx 4.5~\mu$rad~\cite{TDR}.

\section{Raiders of  the lost holy grail}

\label{AD}

In 1976 't Hooft observed that the Standard Model 
does not strictly conserve baryon and lepton number~\cite{'tHooft:1976up,'tHooft:1976fv}.
Rather, non-trivial fluctuations in $SU(2)$ gauge
fields generate an energy barrier interpolating between topologically
distinct vacua.  An index theorem describing the
fermion level crossings in the presence of these fluctuations reveals
that neither baryon nor lepton number is conserved during the
transition, but only the combination $B-L.$ Inclusion of the Higgs
field in the calculation modifies the original instanton
configuration~\cite{Klinkhamer:1984di}.  An important aspect of this
modification (called the ``sphaleron'') is that it provides an
explicit energy scale of about 10 TeV for the height of the barrier.
This barrier can be overcome through thermal transitions at high
temperatures~\cite{Kuzmin:1985mm,Fukugita:1986hr,Arnold:1987mh}, providing an important input to any
calculation of cosmological baryogenesis. More speculatively, it has
been suggested~\cite{Aoyama:1986ej,Ringwald:1989ee,Espinosa:1989qn} that the topological transition
could take place in two particle collisions at very high energy. The
anomalous electroweak contribution to the partonic process can be
written as \begin{equation} \hat\sigma_i(\hat s)= 5.3\times 10^3 {\rm
    mb} \cdot e^{-(4\pi/\alpha_W)\ F_W(\epsilon)}\ \
  , \label{sigmahat} \end{equation} where $\alpha_W\simeq 1/30$, the
tunneling suppression exponent $F_W(\epsilon)$ is sometimes called the
``holy-grail function'', and $\epsilon \equiv \sqrt{\hat s} /(4\pi
m_W/\alpha_W)\simeq \sqrt{\hat s}/30$~TeV. Thus, it is even possible
that at or above the sphaleron energy the cross-section could be of
${\cal O}(\rm mb)$~\cite{Ringwald:2003ns}.
Of particular interest here would be an enhancement
of the neutrino cross section over the perturbative SM estimates, say
by an order of magnitude in the energy range $9.5 < \log_{10}
(E_\nu/{\rm GeV}) < 10.5$. 

The argument for strong damping of the anomalous cross section for $\sqrt{\hat s}\gtrsim $~30~TeV was convincingly demonstrated in~\cite {Bezrukov:2003er,Bezrukov:2003qm}, in the case that the classical field providing the saddle point interpolation between initial and final scattering states is dominated by spherically symmetric configurations. This $O(3)$ symmetry allows the non-vacuum boundary conditions to  be fully included in extremizing the effective action. In~\cite{Gould:1993hb} it was shown that a {\em sufficient} condition for the $O(3)$ dominance is that the interpolating field takes the form of a chain of ``lumps'' which are well-separated, so that the each lump lies well into the exponentially damped region of its nearest neighbors. However,
we are not aware of any reason that such lumped interpolating fields should dominate the effective action. It is thus of interest to explore the other extreme, in which non-spherically symmetric contributions dominate the effective action (and let experiment be the arbiter).

It was shown~\cite{Ringwald:2003ns} that for the simple sphaleron
configuration $s$-wave unitarity is violated for $\sqrt{\hat s}> 4\pi
M_W /\alpha_W\sim 36$~TeV. If for $\sqrt{\hat s} > 36$ TeV
we saturate unitarity in each partial wave, then this yields a
geometric parton cross section $\pi R^2$, where $R$ is some average
size of the classical configuration. As a fiducial value we take the
core size of the Manton-Klinkhamer sphaleron, $R\simeq 4/M_W\simeq
10^{-15} $~cm. In this simplistic model, the $\nu N$ cross section is
\begin{equation}
\sigma^{\nu N}_{\rm black \, disk} (E_\nu) 
 = \pi R^2\ \int_{x_{\rm min}}^1 \sum_{\rm partons} f(x)\ dx \,,
\end{equation}
where $x_{\rm min} = \hat s_{\rm min}/ s = (36)^2/ 2ME_\nu \simeq
0.065$.  In the region $0.065 < x < 3(0.065)$ the PDF for the up and
down quarks is well approximated by $f\simeq 0.5/x$, so the expression
for the cross section becomes
\begin{eqnarray}
\sigma^{\nu N}_{\rm black \, disk} (E_\nu) 
&\simeq &\pi R^2\,\, (0.5)\,\, (\ln 3)\,\,  (2/2) \nonumber \\
& \simeq & 1.5\times 10^{-30} {\rm cm}^2  \,,
\end{eqnarray}
where the last factor of 2/2 takes into account the (mostly) 2
contributing quarks $(u,d)$ in this range of $x,$ and the condition
that only the left-handed ones contribute to the scattering. This is
about 80 times the SM
cross section. Of course this
calculation is very approximate and the cross section can easily be
smaller by a factor of 10 ({\em e.g.} if $R$ is 1/3 of the fiducial value
used).

\end{document}